\documentclass[aps,prl,reprint,superscriptaddress,longbibliography,nobibnotes]{revtex4-2}
\usepackage[T1]{fontenc}
\usepackage{lmodern}
\usepackage[utf8]{inputenc}
\usepackage{amsmath,amssymb,bm}
\usepackage{graphicx}
\usepackage[colorlinks=true,allcolors=blue,hypertexnames=false]{hyperref}
\usepackage{orcidlink}
\hypersetup{
  pdftitle={Lieb-Robinson Light Cones in Long-Range Bosonic Systems},
  pdfauthor={Carla Rubiliani, Cheng Shang, Tomotaka Kuwahara}
}
\newcommand{\me}{\mathrm{e}}
\newcommand{\mi}{\mathrm{i}}
\newcommand{\sectionprl}[1]{\par\textit{#1.---}}
\newcommand{\norm}[1]{\left\lVert#1\right\rVert}
\newcommand{\abs}[1]{\left\lvert#1\right\rvert}
\newcommand{\del}[1]{\left(#1\right)}
\newcommand{\sbr}[1]{\left[#1\right]}
\newcommand{\Set}[1]{\left\{#1\right\}}
\newcommand{\cA}{\mathcal A}
\newcommand{\cC}{\mathcal C}
\newcommand{\cD}{\mathcal D}
\newcommand{\cF}{\mathcal F}
\newcommand{\Z}{\mathbb Z}

\newcommand{\md}{\mathrm d}
\newcommand{\Tr}{\mathrm{Tr}}

\newtheorem{theorem}{Theorem}
\usepackage{color}

\usepackage{mathtools}

\makeatletter
\DeclareRobustCommand{\equalcontribution}{%
    \frontmatter@footnote{C.R. and C.S. contributed equally to this work.}%
}
\makeatother

\begin{document}
\title{Lieb-Robinson Light Cones in Long-Range Bosonic Systems}
\author{Carla Rubiliani\,\orcidlink{0009-0000-6169-8658}\equalcontribution\textsuperscript{,}}

\email{carla.rubiliani@uni-tuebingen.de}
\affiliation{Department of Mathematics, University of T\"ubingen, 72076 T\"ubingen, Germany}
\affiliation{Analytical Quantum Complexity RIKEN Hakubi Research Team, RIKEN Center for Quantum Computing (RQC), Wako, Saitama 351-0198, Japan}
\author{Cheng Shang\,\orcidlink{0000-0001-8393-2329}\equalcontribution\textsuperscript{,}}
\email{cheng.shang@riken.jp}
\affiliation{Analytical Quantum Complexity RIKEN Hakubi Research Team, RIKEN Center for Quantum Computing (RQC), Wako, Saitama 351-0198, Japan}
\author{Tomotaka Kuwahara\,\orcidlink{0000-0002-1612-3940}}
\email{tomotaka.kuwahara@riken.jp}
\affiliation{Analytical Quantum Complexity RIKEN Hakubi Research Team, RIKEN Center for Quantum Computing (RQC), Wako, Saitama 351-0198, Japan}
\affiliation{RIKEN Pioneering Research Institute (PRI), Wako, Saitama 351-0198, Japan}

\begin{abstract}

Lieb-Robinson bounds formulate emergent causality in quantum many-body dynamics. In this Letter, we study interacting bosonic systems with long-range interactions, in which both the standard assumptions of short-range interactions and uniformly bounded local energy scales are absent.
We derive upper and lower bounds whose light-cone exponents depend jointly on the spatial decay $\alpha$ and bosonic interaction order $p$.
The key physical mechanism balances enhancement by local boson concentration with coherent amplification of long-range interactions.
Our state-transfer protocol realizes this balance by concentrating bosons within mesoscopic blocks while preserving a spatial array of occupied sites.
For integer $p\ge4$ and $2d<\alpha\le2d+1$, the leading exponents coincide, determining the optimal light cone under controlled-density assumptions, up to arbitrarily small upper-bound exponent losses.
Our results identify a new mechanism for accelerating information propagation through the combined effects of bosonic particle accumulation and long-range interactions.

\end{abstract}

\maketitle

\sectionprl{Introduction}
In nonrelativistic quantum many-body systems, causality is not imposed by spacetime kinematics but emerges dynamically from the Hamiltonian~\cite{ref:LR-bound72,nachtergaele2010lieb,Chen_2023}. Determining the resulting Lieb-Robinson (LR) light cone is therefore a fundamental problem. Locality bounds underlie results on correlation decay~\cite{ref:Hastings2006-ExpDec,ref:Nachtergaele2006-LR}, quantum phases of matter~\cite{PhysRevB.69.104431,PhysRevB.72.045141,bravyi2010topological}, and entanglement growth~\cite{PhysRevLett.97.050401,PhysRevLett.97.150404,PhysRevLett.111.170501}. They also inform classical and quantum simulation~\cite{PhysRevLett.97.157202,HHKL2021,PhysRevX.9.031006,PhysRevX.11.011020,McDonough2025Volume}, thermalization and prethermalization~\cite{Muller2015,KUWAHARA201696,Abanin2017}, and dynamical complexity~\cite{PhysRevLett.121.030501,maskara2019complexity}. The standard setting combines short-range interactions with uniformly bounded local energy scales~\cite{Nachtergaele2006,doi:10.1063/1.5095769}.

\begin{figure}[!t]
  \centering
  \includegraphics[width=0.96\columnwidth]{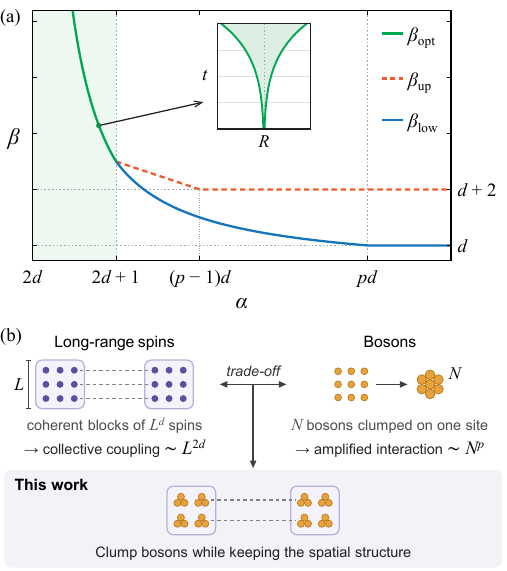}
\caption{Light-cone bounds and amplification mechanisms.
(a) Upper (orange dashed) and lower (blue solid) leading exponents for $R\sim t^\beta$ at $p=4$, $d=2$. Green shading marks their matching regime $2d<\alpha\le2d+1$; the inset shows the optimal cone at $\alpha=4.8$.
(b) In long-range spin protocols, a GHZ-like state of $L^d$ spins in each of two separated blocks allows the $O(L^{2d})$ pairwise couplings to add coherently to one effective interaction. In bosonic systems, concentrating $N$ bosons at each of two sites enhances $(n_i n_j)^{p/2}$ by $N^p$. Our protocol combines both enhancements by concentrating bosons within mesoscopic blocks while maintaining groups of spatially distributed occupied sites.}
  \label{fig:Overview}
\end{figure}

Relaxing spatial locality produces a hierarchy of light cones. For power-law interactions, successive improvements sharpened algebraic tails and established polynomial propagation bounds~\cite{PhysRevLett.113.030602,PhysRevLett.114.157201,Matsuta2017,PhysRevX.9.031006,PhysRevA.101.022333,chen2019finite}, followed by sharp leading-power constraints for $\alpha>2d$ and strictly linear cones for $\alpha>2d+1$~\cite{PhysRevX.10.031009,PhysRevX.10.031010,PhysRevLett.127.160401}. Fast state-transfer and entanglement-generation protocols provide the complementary lower bounds~\cite{PhysRevLett.119.170503,PhysRevX.11.031016,hong2020fast}. Quasi-locality results also extend to long-range lattice fermions~\cite{Teufel2025-am}. The propagation diagnosed by Frobenius norms and infinite-temperature out-of-time-order correlators need not coincide with worst-case operator-norm propagation~\cite{PhysRevX.10.031009,kuwahara2020absence,PhysRevA.104.062420}. Experiments with tunable long-range interactions probe correlation spreading and scrambling~\cite{richerme2014non,jurcevic2014quasiparticle,PhysRevLett.124.240505}.

Removing bounded local energy scales poses a different obstacle: bosons can accumulate at a site, thereby enhancing interaction strengths. Locality estimates for harmonic, anharmonic, and spin-boson models exploit their particular structure~\cite{cramer2008locality,Nachtergaele2009,PhysRevLett.111.230404}, whereas unbounded local degrees of freedom can in general permit accelerated communication~\cite{PhysRevLett.102.240501}. For interacting lattice bosons, state-dependent bounds recover locality under appropriate density, stationarity, or particle-free-region assumptions~\cite{PhysRevLett.127.070403,PhysRevX.12.021039,Faupin2022}. For general controlled-density initial states and short-range Bose-Hubbard-type dynamics, a polynomial cone $R\sim t^d$, up to subleading corrections, can be both bounded and attained~\cite{kuwahara2022optimal}; a recent review gives an alternative proof of a weaker polynomial bound~\cite{LR26}. Translation symmetry and repulsive interactions provide further constraints on propagation and on-site energy accumulation~\cite{KuwaharaLemm2026Symmetry,KuwaharaLemmRubiliani2025Energy}. These developments complement observations of correlation spreading in ultracold bosonic gases~\cite{cheneau2012light,langen2013local,Takasueaba9255} and rigorous approaches to bosonic Hilbert-space truncation and simulation~\cite{Tong2022,kuwahara2022optimal}.

What happens when both constraints are removed? Long-range interacting bosonic systems naturally realize this situation: algebraically decaying couplings coexist with unbounded local degrees of freedom. There are two distinct goals. Establishing a robust polynomial light cone already gives a nontrivial notion of locality and supports many consequences of LR theory. Determining the sharp light cone---the fastest information propagation compatible with the Hamiltonian---is substantially harder, making even partial optimality informative.

Extended Bose-Hubbard models and long-range tunneling have experimental realizations~\cite{Baier201,Meinert1259}. Here, it is essential to distinguish particle transport from general information propagation. Bounds on microscopic and macroscopic transport~\cite{PhysRevA.84.032309,PhysRevLett.128.150602,Vu2024Transport,LemmRubilianiZhang2026Microscopic}, including stronger suppression of macroscopic transport~\cite{Faupin2025Suppression} and particle-number propagation after perturbing stationary states~\cite{Eisert2026Sound}, do not by themselves determine the spreading of arbitrary local observables~\cite{kuwahara2022optimal}. Long-range bosonic information-propagation bounds have been obtained under particle-free-region assumptions~\cite{Lemm2023Information,Lemm2025-ov}; the latter work also controls local particle-number moments, a key input to our proof. Related recent developments address infinite-volume bosonic dynamics with moment-controlled states~\cite{Deuchert2025Dynamics} and macroscopic transport in dissipative long-range systems~\cite{Li2026Dissipative}.

An LR bound must constrain every admissible protocol, including highly optimized dynamics that need not resemble typical experiments. Its optimality is therefore decided by the fastest allowed protocol~\cite{PhysRevX.11.031016,PhysRevLett.127.160401,kuwahara2022optimal}.
For long-range bosons, this worst-case problem depends on both the decay exponent $\alpha$ and the bosonic interaction order $p$, as in density terms $n_i^p$ or $(n_i n_j)^{p/2}$. In long-range spin protocols, GHZ-like encoding of $L^d$ spins in spatially separated blocks makes their $L^{2d}$ pairwise couplings contribute coherently to one effective interaction. By contrast, bosonic interactions can be enhanced by concentrating particles on selected sites. These mechanisms compete: local concentration strengthens occupation-dependent couplings, but reduces the number of spatially distributed occupied sites available for coherent amplification [Fig.~\ref{fig:Overview}(b)]. Their interplay produces a richer causality diagram than in either long-range bounded systems or short-range bosonic systems alone. Establishing optimality requires an upper bound valid throughout the specified initial-state class and a matching propagation protocol.

Our upper bound and propagation protocol lead to explicit exponent formulas, $\beta_{\rm up}(\alpha,p,d)$ and $\beta_{\rm low}(\alpha,p,d)$, for effective cones $R=t^\beta$ on a $d$-dimensional lattice with $\alpha>2d$.
We state leading powers, suppressing subleading corrections and arbitrarily small losses in upper-bound exponents. The formulas match, $\beta_{\rm up}=\beta_{\rm low}=\beta_{\rm opt}$, throughout $2d<\alpha\le{2d+1}$ for integer $p\ge4$, with
\begin{equation}
\beta_{\rm opt}=\frac{pd-2d+1}{\alpha-2d}.
\label{eq:optimal_exponent}
\end{equation}
Since $1/(\alpha-2d)$ is the optimal light-cone exponent for long-range spin systems in this regime, the factor $pd-2d+1$ quantifies the enhancement of the exponent due to bosonic occupation-dependent interactions.
The upper and lower light-cone exponents and their matching regime are illustrated in Fig.~\ref{fig:Overview}(a). The fastest propagation is thus controlled jointly by long-range coupling and bosonic nonlinear amplification.
For $2d<\alpha<2d+1$, the upper bound uses the fixed-$k$ extension of Tran \textit{et al.}~\cite{PhysRevLett.127.160401} supplied in the Supplementary Material (SM).
The point $\alpha=2d+1$ follows with arbitrarily small exponent loss, while $\alpha>2d+1$ uses the existing many-body theorem of Kuwahara and Saito~\cite{PhysRevX.10.031010}.

The key idea is an information-mediating sublattice. For the lower bound, bosons are concentrated within mesoscopic blocks. The selected, highly occupied sites form a {$d$-dimensional} sublattice on which long-range interactions are amplified, effectively realizing a spin system with enhanced couplings. Combining this amplification with fast long-range spin protocols~\cite{PhysRevLett.119.170503,PhysRevX.11.031016} yields $\beta_{\rm low}$. In contrast, the short-range bosonic construction concentrates bosons along an essentially one-dimensional channel~\cite{kuwahara2022optimal}. The same sublattice viewpoint guides the upper bound: controlling the effective Hamiltonians associated with potentially amplified propagation channels rules out faster spreading. Matching exponents in the optimal regime identifies these sublattices as the organizing principle for long-range bosonic causality.

\sectionprl{Setup}
Consider a finite lattice $\Lambda\subset\mathbb Z^d$ and its bosonic Fock space $\cF$. Distances $|i-j|$ use the supremum norm, so $B_r(x)$ is a cube of radius $r$; $d(X,Y)$ denotes the distance between regions. The operators $b_i,b_i^\dagger$ annihilate and create a boson at site $i$, with $n_i=b_i^\dagger b_i$ and $n_X=\sum_{i\in X}n_i$.

Following the SM, write the hopping and density terms as $T_{ij}(t)=J_{ij}(t)b_i^\dagger b_j$ and $V_{ij}(t)=U_{ij}(t)V(n_i,n_j)$, with $V$ a real density potential. Finite sums of such terms {with independently tunable coefficients} allow general polynomial density interactions of total degree at most $p$, including lower-degree terms. In the main text, we focus on the long-range nonlinear term $V(n_i,n_j)=n_i^{p/2}n_j^{p/2}$, which fixes the occupation enhancement in the exponent formulas below. The Hamiltonian restricted to $X$ is
\begin{equation}
H_X(t)=\sum_{i,j\in X}\bigl[T_{ij}(t)+V_{ij}(t)\bigr],
\ \text{and} \
H(t)=H_\Lambda(t).
\label{eq:main_Hamiltonian}
\end{equation}
The coefficients are continuous in time, chosen so that $H(t)$ is Hermitian, and obey $|J_{ij}(t)|\le C_J(1+|i-j|)^{-\alpha}$ and $|U_{ij}(t)|\le {C_U}(1+|i-j|)^{-\alpha}$ uniformly in time, system size, and occupation. The same uniform coefficient bounds apply to every polynomial component and to all density-control terms used in the protocol. Here, $p\ge1$ controls the occupation dependence. The explicit lower-bound sequence uses an integer $p\ge2$, with a nonempty matching regime for $p\ge4$.

The infinite local Hilbert space requires a state-dependent formulation~\cite{PhysRevLett.127.070403,PhysRevX.12.021039,kuwahara2022optimal,LR26}. A low-boson-density assumption is essential: without uniform occupation control, arbitrarily large initial occupations can amplify interactions without bound, precluding an occupation-independent information-propagation speed limit for this class of models. We denote by $\cD_\xi^\lambda$ the initial density matrices with finite energy that satisfy
\begin{equation}
\Tr\!\left[n_{B_r(x)}^\zeta\rho\right]
 \le (\lambda\zeta r^d)^\zeta,
 \quad 1\le\zeta\le\xi,
\label{eq:controlled_density_main}
\end{equation}
for every $x\in\Lambda$ and $r\ge1$, with fixed $\lambda>0$ and sufficiently large $\xi$. This controlled-density condition bounds local occupation fluctuations without fixing the total number of particles as the lattice grows~\cite{LemmRubilianiZhang2026Microscopic,Lemm2025-ov,Deuchert2025Dynamics}.

Write $\cA_X$ for bounded operators supported on $X$, and $A\in\cA_X^{q_0}$ when $A$ changes the particle number in $X$ by at most a fixed $q_0$. This condition preserves the moment control needed after inserting $A$. For evolution from $s$ to $t$, let $U(s,t)$ be the propagator and $\tau_{s,t}(A)=U(s,t)^\dagger A U(s,t)$. The local approximation takes the form
\begin{equation}
\norm{\rho\bigl[\tau_{s,t}(A)-\tilde A\bigr]}_1
 \le\varepsilon(|t-s|,R)\norm{A},
\end{equation}
for $\rho\in\cD_\xi^\lambda$ and some $\tilde A\in\cA_{X[R]}$ with $\|\tilde A\|\le\|A\|$. The approximant can be chosen uniformly throughout the specified state class. Here $X[R]$ is the radius-$R$ neighborhood of $X$ and $\|\cdot\|_1$ is the trace norm; $\|\cdot\|$ denotes the operator norm. By trace-norm duality, this controls the error in correlations with any bounded probe. The error is small outside $R\gg|t-s|^{{\beta_{\rm up}}}$. More precisely, for every $\epsilon>0$, sufficiently high finite moments $\xi\ge\xi_0(\epsilon)$ give an error bounded by $C F_\xi(X)(1+|t-s|)^{-\nu}$ at $R={C_R}(1+|t-s|)^{\beta_{\rm up}+\epsilon}$, for some $\nu>0$ and $F_\xi(X)=|X|^{(1+\xi)/2}+|X|^2$. Constants are independent of lattice size and occupation cutoff. A single fixed finite $\xi$ does not permit an arbitrarily small loss in this estimate. The End Matter gives the parameter choices. In the scaling discussion below, we set $s=0$ and use $t$ for elapsed time.

\sectionprl{Lower-bound protocol}
Double the physical lattice to $\widetilde\Lambda=\Lambda\times\{0,1\}$, with layers $\Lambda_a=\Lambda\times\{a\}$, $a=0,1$. Equivalently, each physical site carries two bosonic modes. Initially, $\Lambda_0$ is in a Mott state with one boson per site, and $\Lambda_1$ is empty. This initial state satisfies the density condition with constants independent of the eventual block size. The protocol has three stages: preparing the concentrated packets, transferring the encoded signal, and decoding it locally (see also Fig.~S2 in the SM). Each stage is assigned a fixed fraction of the available time.

For integer $p\ge3$, partition the lattice into cubes $P_x$ of radius {$\ell\asymp t$}, with centers $x\in\cC$. Concentration uses short-range hopping together with occupation-selective nonlinear on-site control, as in Ref.~\cite{kuwahara2022optimal}. Organize the transport into parallel sweeps along the $d$ coordinate directions, merging neighboring packets in ${\cal{O}}(\ell)$ steps per direction. For a merge containing $M$ bosons, the successive hopping matrix elements are proportional to $\sqrt{(m+1)({M}-m)}$, where $m$ is the occupation of the receiving mode. Their inverse sum over $0\le m<{M}$ is ${\cal{O}}(1)$, explaining the constant ideal transfer time per merge once the nonlinear control isolates the desired transitions. Thus, the ideal parallel concentration time is ${\cal{O}}(\ell)={\cal{O}}(t)$, rather than time proportional to the block volume. The bounded-control implementation in the SM employs hierarchical merges and accounts for transition selection, encoding, and decoding within ${\cal{O}}(\ell)$, with finite coefficients that obey the Setup. The accumulated error of the full protocol, including leakage, changes each ideal destination state by at most a fixed $e_*<1$ in trace norm. A unit-norm readout {with expectations $\pm1$ on the ideal outputs} then retains a signal {of} at least $2-2e_*>0$.
The inverse-matrix-element sum alone does not establish this accuracy estimate; an infinite-amplitude transition-selection limit is not an admissible implementation.

The occupied centers form a coarse-grained sublattice with $N\sim t^d$ bosons per center. Its distance is $d_{\mathrm{cg}}(P_x,P_y)=|x-y|/(2\ell+1)$, so physical separations are of order $t\,d_{\mathrm{cg}}(P_x,P_y)$. At each center $x$, encode $|0\rangle_x=|N,0\rangle_x$ and $|1\rangle_x=|0,N\rangle_x$. To encode a signal initially localized at one original site, first spread its layer qubit coherently through the source block using nearest-neighbor controlled gates, then concentrate each layer. Reversing these operations decodes the destination packet onto one original site. {These operations take ${\cal{O}}(\ell)$ time}, so input and readout do not change the leading time exponent.

Packet rotations require both nonlinear phases and hopping. For even integer $p$, the on-site phase $S_p=\me^{-\mi\pi n_0^p/2}$ equals $\me^{-\mi\pi n_0^2/2}$: it acts as $1$ on even occupations and $-\mi$ on odd ones. A 50:50 layer beam splitter, $S_p$, and the inverse beam splitter therefore produce a logical $\pi/2$ rotation, up to an overall phase. Together with linear number phases, these give arbitrary encoded one-qubit rotations in ${\cal{O}}(1)$ time, independently of $N$; in particular, this applies directly to $p=4$. For odd integer $p\ge3$, choose $N=2\pmod 8$ and use $\me^{-\mi\pi(n_0^p+n_1^p)/4}$ instead. At most seven unused bosons per block suffice to adjust $N$, without changing $N\sim t^d$.

Choose equal-density interactions in the two layers, with no interlayer terms. On the packet code, the induced interaction is proportional to the sum of projectors onto $|0_x0_y\rangle$ and $|1_x1_y\rangle$, equal to $(I+Z_xZ_y)/2$. Up to an irrelevant identity term, this is an Ising interaction whose strength is enhanced by the occupations to
\begin{equation}
J_{\rm eff}(P_x,P_y)\sim
 \frac{t^{pd-\alpha}}{d_{\mathrm{cg}}(P_x,P_y)^\alpha}.
\end{equation}
Both packets contribute a factor $N^{p/2}$, while their physical separation contributes $t^{-\alpha}$. The packet encoding is essential: adjacent-occupation codes such as $|N,N\rangle$ and $|N-1,N+1\rangle$ produce a $ZZ$ coefficient {with an occupation factor only of order $N^{p-2}$}. Preparation, rotations, and readout must therefore use the same packet code. This explains the scale $t^{pd-\alpha}$ without changing the decay exponent on the coarse lattice. The state-transfer construction of Ref.~\cite{PhysRevX.11.031016} then assembles encoded GHZ states recursively, couples them by controlled phases, and decodes the signal. The required one-qubit operations are precisely the packet rotations just described.

For long-range spins with $\alpha>2d$, the optimal leading-power exponent is $\beta_{\rm LR}=\max\{1,{(\alpha-2d)^{-1}}\}$~\cite{PhysRevX.10.031009,PhysRevX.10.031010,PhysRevLett.127.160401}. The formal rescaled spin time is
$g_{\mathrm{lr}}t$, with $g_{\mathrm{lr}}=t^{pd-\alpha}$, but only the density couplings have been amplified by $g_{\mathrm{lr}}$. The local-control time must therefore be counted separately: the spin construction allows arbitrarily fast on-site gates~\cite{PhysRevX.11.031016}.

For $2d<\alpha\le2d+1$, this counting can be made explicit. Let $u(L)$ be the physical time to encode already prepared packet qubits in a coarse cube of side $L$, and write $\chi=\alpha-2d$. Merging $m^d$ subcubes uses preparation, decoding, and reencoding, giving $u(L)\le3u(L/m)+C_mL^\chi/g_{\mathrm{lr}}+{\cal{O}}(1)$. The phase term follows from the $L^{2d}$ pair enhancement competing with $L^{-\alpha}$ decay; the scale $L=m$ costs ${\cal{O}}(1+g_{\mathrm{lr}}^{-1})$. For fixed sufficiently large $m$, iteration gives $u(L)={\cal{O}}(L^\chi/g_{\mathrm{lr}}+L^\sigma)$, with $\sigma=\log_m3<\chi$. Choosing $\sigma<\chi/(pd-\alpha+1)$ makes the rotation overhead subleading at the target $L\sim(g_{\mathrm{lr}}t)^{1/\chi}$, whenever $g_{\mathrm{lr}}t$ grows. Together with the preparation and readout estimate above, this counts the full time of the recursive branch, including the endpoint.

For $\alpha>2d+1$, interaction enhancement alone does not establish the spin-linear branch $L_{\rm cg}\sim g_{\mathrm{lr}}t$. A nearest-neighbor sequence using the ${\cal{O}}(1)$ packet rotations above {incurs local-control time of  ${\cal{O}}(L_{\rm cg})$}, which is not absorbed into ${\cal{O}}(t)$ at $L_{\rm cg}\sim g_{\mathrm{lr}}t$ when $g_{\mathrm{lr}}$ grows. The corresponding transfer estimate is $t_{\rm tr}(L_{\rm cg})\le C L_{\rm cg}(1+g_{\mathrm{lr}}^{-1})$, including all interaction and control time. For $pd\le\alpha<pd+1$, we have $g_{\mathrm{lr}}\le1$, so choosing $L_{\rm cg}\sim g_{\mathrm{lr}}t$ and restoring the spacing gives $R\sim t(g_{\mathrm{lr}}t)$ and $\beta_{\rm low}^{(\mathrm{conc})}=2+pd-\alpha$. For $2d+1<\alpha<pd$, we instead use the collective-relay construction in the SM, which gives $R\gtrsim t^{2+(pd-\alpha)/(\alpha-2d)}/(\log t)^{\eta_{\mathrm{rel}}}$, with a fixed $\eta_{\mathrm{rel}}>0$ and all implementation times included. Local decoding then turns an $\cal{O}$(1) input change into an $\cal{O}$(1) change of a distant observable.

Concentration helps only when occupation enhancement compensates for the separation introduced by coarse-graining. For $\alpha \ge pd+1$, the effective spin time no longer grows with $t$, so this construction does not improve propagation. The competing short-range bosonic protocol attains the leading-power exponent $d$~\cite{kuwahara2022optimal}; its implementation here obeys the same coefficient, preparation, and readout requirements. Taking the largest leading power among these alternative constructions gives, for integer $p\ge3$,
\begin{equation}
\beta_{\rm low}
=
{
\begin{cases}
\displaystyle
1+\frac{pd-\alpha+1}{\alpha-2d},
& 2d<\alpha\le2d+1,\\[7pt]
\displaystyle
\max\!\left\{d,\,2+\frac{pd-\alpha}{\alpha-2d}\right\},
& 2d+1<\alpha\le pd,\\[7pt]
\displaystyle
\max\{d,\,2+pd-\alpha\},
& \alpha>pd.
\end{cases}
}
\label{eq:lower_exponent}
\end{equation}
For $p=2$, the separate microscopic protocols give $\beta_{\rm low}=\beta_{\rm LR}$.
Unlike the essentially one-dimensional concentration channel of short-range bosons, the long-range protocol uses an array of mesoscopic blocks. Its $d$-dimensional mediating sublattice and occupation-enhanced couplings produce the joint dependence on $\alpha$, $p$, and $d$.

\sectionprl{Upper-bound mechanism}
The upper bound uses the same block geometry to control the largest possible interaction strengths. Its main subtlety is that short-range density interactions generate large local energies but commute with one another. They produce rapid local phases without independently mediating long-distance operator growth~\cite{LemmWessel2025Commuting,LR26}.

Fix $A\in\cA_X^{q_0}$ and take $t\ge1$. Take blocks $P_x$ of radius $\ell=\lceil vt\rceil$, with centers $x\in\cC$. Here \(v>0\) is fixed, and the constants in the moment bounds may depend on \(v\). Density interactions commute with {local} particle numbers, so moment propagation controls accumulation independently of the information cone being proved~\cite{Lemm2025-ov}. For radius-${\cal{O}}(t)$ blocks, the {evolved particle-number moments are bounded by $C_\zeta t^{d\zeta}$}. Markov's inequality bounds the weight above a cutoff $\bar q$ by $C_\zeta(t^d/\bar q)^\zeta$.

As in the SM, $\Pi_{P_x,\le q_x}$ projects onto occupations $n_{P_x}\le q_x$. We take $q_x=\lfloor\bar q[1+d(P_x,X)]^\delta\rfloor$, with small $\delta>0$ and $\bar q$ slightly above the leading scale $t^d$. The cutoff increases slowly with distance from $X$; sufficiently high moments ensure that the weighted errors are summable independently of the system size. Duhamel's formula controls the error from truncating the dynamics~\cite{kuwahara2022optimal,LR26}. Bars denote particle-truncated operators, and $\bar\tau_{s,t}$ denotes their dynamics. We further restrict the truncated dynamics to $X_{2R^\eta}$, with a sufficiently large fixed $\eta>1$, controlling the resulting error as described in the End Matter and SM.
Here $X_{2R^\eta}$ denotes the $2R^\eta$-enlargement of $X$, rounded out to whole blocks.
The cutoffs in this region satisfy $q_x\le Q\lesssim\bar q R^{\eta\delta}$. The small exponent losses from these cutoffs are suppressed in the scaling estimates below.

For $p\ge2$, $\sum_{i\in P_x}n_i^{p/2}\le n_{P_x}^{p/2}$ bounds a density coupling between blocks by ${\cal{O}}({Q^p})$, without an extra volume factor. More explicitly, with block occupations bounded by $Q$, the far-block density norm is at most $C Q^p/d(P_x,P_y)^\alpha$. The hopping norm is at most $C Q\ell^d/d(P_x,P_y)^\alpha\le C Q^p/d(P_x,P_y)^\alpha$ for $Q\gtrsim\ell^d$. This block-total estimate removes the extra factor $t^d$ that a sitewise norm sum would introduce. Here $\bar H_{P_x,P_y}$ denotes the interactions connecting the blocks $P_x$ and $P_y$, sandwiched between the particle-number projectors. Thus neighboring-block terms can reach $\|\bar H_{P_x,P_y}\|\lesssim t^{pd}$, whereas well-separated blocks obey $\|\bar H_{P_x,P_y}\|\lesssim t^{pd-\alpha}/d_{\mathrm{cg}}(P_x,P_y)^\alpha$. A direct spin bound~\cite{PhysRevX.10.031010,PhysRevLett.127.160401} would use the largest scale and yield $\beta_{\rm naive}=1+(pd+1)\beta_{\rm LR}$, overestimating the ability of diagonal density terms to propagate information.

As in the SM, we decompose the block Hamiltonian into on-site, neighboring-block, and long-range parts. Write $V_{\mathrm{short}}=V_{\mathrm{near}}+V_{\mathrm{on\text{-}site}}$ for the first two density contributions, $H^0_{\mathrm{short}}$ for the corresponding hopping terms, and $H_{\mathrm{long}}$ for all remaining couplings. The terms of $\bar V_{\mathrm{short}}$ commute even at different times. With $U^V(0,u)$ its propagator, the interaction picture~\cite{kuwahara2022optimal,LR26} gives
\begin{equation}
\begin{split}
\!\!H^{\mathrm{int}}(u)\!=\!U^V(0,u)^\dagger
 \!\times\!\bigl[\bar H^0_{\mathrm{short}}(u)\!+\!\bar H_{\mathrm{long}}(u)\bigr]U^V(0,u).\!\!
\end{split}
\end{equation}
Only density terms touching the original support dress an operator; distant commuting factors cancel. Arbitrarily rapid phase evolution, therefore, enlarges the support by only one coarse-lattice layer~\cite{LemmWessel2025Commuting}. Conjugation preserves term norms, and finite overlap of the enlarged supports preserves the algebraic decay. A dressed two-block hopping term may nevertheless act on several blocks. Thus $H^{\mathrm{int}}$ is fixed $k$-local: $k$ counts blocks, independently of $t$ and system size, and differs from the bosonic order $p$.

At the level of leading powers, the remaining hopping and long-range density scales are $g_{\mathrm{hop}}\asymp t^d$ and $g_{\mathrm{lr}}\asymp t^{pd-\alpha}$. The first follows because the one-particle hopping matrix has bounded row sum for $\alpha>d$, hence bounded norm; its second quantization on one or two retained blocks with at most ${\cal{O}}(Q)$ particles has norm ${\cal{O}}({Q})$. For $p\ge2$, the far-block hopping tail is no larger than the density-interaction estimate. The effective interaction scale is therefore
\begin{equation}
g=\max\{t^d,t^{pd-\alpha}\}.
\end{equation}
For $2d<\alpha<2d+1$, we use the supplied fixed-$k$ extension of Tran \textit{et al.}~\cite{PhysRevLett.127.160401}. Finite support enlargement preserves the pair-summed decay bound
\begin{equation}
\sup_u\sum_{Z\ni x,y}\|\Phi_Z(u)\|
\le
{
\frac{C\max\{Q,Q^p(2\ell+1)^{-\alpha}\}}
     {[1+d_{\mathrm{cg}}(P_x,P_y)]^\alpha}
},
\end{equation}
for $x\ne y$ and $p\ge2$, where $\Phi_Z(u)$ are the interaction terms of $H^{\mathrm{int}}(u)$. The extension applies to time-dependent interactions with constants independent of the truncated local dimensions. At $\alpha=2d+1$, slightly lowering the decay exponent gives the leading linear power with arbitrarily small loss. For $\alpha>2d+1$, the fixed-$k$ theorem in Appendix A of Ref.~\cite{PhysRevX.10.031010} gives a strictly linear cone. {In leading powers, these} inputs yield a coarse-grained radius $\widetilde R_{\rm LC}(t)\lesssim(gt)^{\beta_{\rm LR}}$. Restoring the spacing gives $R_{\rm LC}(t)\lesssim t\,(gt)^{\beta_{\rm LR}}$ and
\begin{equation}
\beta_{\rm up}
 =1+\max\{d+1,pd-\alpha+1\}\beta_{\rm LR}.
\label{eq:upper_exponent}
\end{equation}
The SM treats $1\le p<2$ separately and obtains the same leading upper exponent.

For $\alpha\le(p-1)d$, the amplified long-range scale dominates the hopping scale, and Eq.~\eqref{eq:upper_exponent} reduces to $1+(pd-\alpha+1)\beta_{\rm LR}$. For integer $p\ge4$ and $2d<\alpha\le2d+1$, this exponent exceeds the competing lower exponent $d$ and matches Eq.~\eqref{eq:lower_exponent}. Equation~\eqref{eq:optimal_exponent} gives {the leading power of the optimal light cone in this regime}.

\sectionprl{Outlook}
What is the optimal light-cone shape throughout the full parameter space, including regimes where it need not be polynomial? For integer $p\ge4$, our upper and lower bounds match for $2d<\alpha\le2d+1$, but do not generally coincide for $\alpha>2d+1$. Establishing optimality more generally may require a finer treatment of information spreading to obtain sharper upper bounds, faster propagation protocols, or both. To improve the upper bound, a more flexible coarse-graining scheme, including optimized block partitions, is one possible direction. In the strict short-range limit, the present block estimate gives $\beta_{\rm up}=d+2$, whereas Kuwahara, Vu, and Saito established the leading exponent $d$~\cite{kuwahara2022optimal}. Recovering that limit requires incorporating their more refined localization and concatenation of short-time evolutions, rather than applying a worst-case spin bound to a fixed coarse-graining scale. As a separate extension beyond the Bose--Hubbard-type Hamiltonian class considered here, on-site packet mixing between $|N,0\rangle$ and $|0,N\rangle$, with transition matrix elements growing with $N$, could accelerate local packet rotations and enable faster propagation protocols.

Another direction is to impose additional structure on the initial state. For time-independent $H$, a stationary background satisfying $[H,\rho]=0$ and suitable local occupation-tail conditions suppresses the accumulation mechanism exploited by the protocol. In established short-range settings, bosonic information propagation is linear or almost linear---the latter allowing logarithmic corrections---under appropriate state assumptions~\cite{PhysRevLett.127.070403,PhysRevX.12.021039}, alongside the linear cone of bounded spin systems. It remains to determine how stationarity and related moment constraints reshape long-range bosonic cones, and whether they recover the spin-system exponents. Stationarity alone, without occupation control, is not the condition used in these results. 

Finally, the prospects and challenges of realizing the proposed propagation mechanism using optical-lattice bosons~\cite{Will2010,PhysRevA.90.021601,4l1s-kkw7} and superconducting microwave cavities~\cite{Heeres2015SNAP,Gao2019Exchange,Wang2021PND} are briefly discussed in the End Matter.

\sectionprl{Acknowledgments}
C. S. thanks Hongchao Li for valuable suggestions.~This work was supported by the RIKEN Hakubi Fellows Program.~The research of C. R. is supported by the DFG through the grant TRR 352 -- Project-ID 470903074.
T. K. is supported by the Japan Science and Technology Agency through its Exploratory Research for Advanced Technology program (JST ERATO Grant Number JPMJER2302) and by JSPS KAKENHI Grant Numbers JP23K25796, JP25K24674, JP26K00019, JP26K00020, and JP26H02015.
This research was partially conducted during the internship of C. R. at RIKEN, under the supervision of T. K. 

The research project, central questions, and main mathematical arguments were developed by the authors without AI assistance between 2025 and early 2026.
LLMs were subsequently used to assist in refining arguments, proofreading the manuscript, and improving its clarity and presentation.
The authors take full responsibility for the content of the manuscript.

\makeatletter
\immediate\write\@auxout{\string\citation{PRLReferencesControl}}
\makeatother
\bibliography{Reference}

\setcounter{equation}{0}
\renewcommand{\theequation}{A\arabic{equation}}
\renewcommand{\theHequation}{EndMatter.\arabic{equation}}

\onecolumngrid
\begin{center}
{\large\bfseries {End Matter}}
\end{center}
\twocolumngrid

\textit{Appendix A: Upper-bound proof strategy.---} The following theorem gives the upper bound.
We state it for \(p\ge2\), where the formulas are simpler.
The result also holds for \(p\ge1\), and it's proven in the Supplementary Material.

\begin{theorem}[Upper bound]\label{thm up}
Let $p\ge 2$. For every $\epsilon>0$, there exists $\xi_0>0$ such that for every
 $\lambda>0$, $\xi\ge \xi_0$, and integer
$q_0\ge 0$ 
there exist constants $C,C_R,\nu>0$ such that the following holds.
For every  nonempty
$X\subseteq\Lambda$, every $A\in\mathcal A_X^{q_0}$,
all $s,t\in\mathbb R$, and every $R\ge C_R(1+|t-s|)^{\beta_{\mathrm{up}}+\epsilon}$,
there exists an operator $\widetilde A\in\mathcal A_{X[R]}$ satisfying 
\(
\|\widetilde A\|\le\|A\|
\)
and
\begin{align*}
   &\sup_{\rho\in\mathcal D_\xi^\lambda}
\left\|
\rho\bigl[\tau_{s,t}(A)-\widetilde A\bigr]
\right\|_1\\
&\qquad
\le
C\|A\|
\left(|X|^{(1+\xi)/2}+|X|^2\right)
(1+|t-s|)^{-\nu} .
\end{align*}
Here $X[R]:=\Set{i\in\Lambda\,:\, d(i,X)\le R}$.

\end{theorem}

We sketch the proof of this theorem.
For ease of notation, we assume the Hamiltonian in \eqref{eq:main_Hamiltonian} is time-independent, and we evolve from time 0 to time $t\ge1$.
We write $\tau_{0,t}$ as $\tau_t$ for the evolution induced by the full Hamiltonian. 
We treat the more general case of time-dependent Hamiltonians and small times in the SM.

Consider a bounded observable $A\in\cA_X^{q_0}$, with $q_0\ge 0$, supported on a finite set $X\subset\Lambda$. Set $\ell=\lceil vt\rceil$, with \(v>0\) fixed. 
We define the set of block centers as $\cC:=\Set{x\in(2\ell+1)\,\Z^d\,:\, d(x,\Lambda)\le \ell}$.
For each center $x\in \cC$, define the block $P_x:= \Set{y\in \Lambda\,:\, \abs{x-y}\le \ell}$.
Choose the block cutoffs
\begin{align}\label{qi def}
    q_x:=\lfloor\bar{q}\, (d(P_x,X)+1)^{\delta}\rfloor,
\end{align}
with $0<\delta<1$ small and $\bar q$ to be determined later.
Define the spectral projectors
\begin{align}\label{proj}
     \Pi=\prod_{x\in \cC}\Pi_{P_x,\le q_x},
\end{align}
where $\Pi_{P_x,\le q_x}$ is the projector onto the sectors with $n_{P_x}\le q_x$.
The projector in \eqref{proj} truncates the particle number in each block $P_x$. The cutoff increases with the distance from the support $X$ of $A$, so the truncation becomes milder farther from $X$.
Let \(R>1\) be the localization radius, and let \(\eta>1\) be a parameter to be chosen later. 
This choice of truncation ensures that the bound is independent of system size.
We denote by $\bar \tau_t$ and $\bar \tau_t^{X_{2R^\eta}}$ the dynamics generated, respectively, by the particle-truncated Hamiltonian $\bar H:=\Pi H\Pi$ and its restriction $\bar H_{X_{2R^\eta}}$ to $X_{2R^\eta}$.
Notice that here the enlargement of sets is done block-wise, namely $X_r:=\bigcup_{\substack{x\in \cC\\d(P_x, X)\le  r}}P_x,r>0$. Thus, in general $X[R]$ and $X_R$ are different sets.
The main steps of the proof are as follows:
\begin{align}
    \tau_t (A)\,\xrightarrow{} \,\tau_t (\bar A)\,\xrightarrow{} \,\bar\tau_t (\bar A)\,\xrightarrow{} \,\bar\tau^{X_{2R^\eta}}_t (\bar A)\,\xrightarrow{} \,\tilde A,
\end{align}
where $\bar A:=\Pi A \Pi$ and $\tilde A$ is some operator supported on $X_R$.

First, we approximate $\tau_t(A)$ by $\tau_t(\bar A)$.
This step ensures the commutation relations needed later.
The main challenge is to control particle fluctuations. We use long-range particle propagation bounds:
For \(\alpha>d+1\), \(\xi\ge1\), and \(\lambda>0\), there is a constant \(C>0\) such that $\Tr\del{\tau_t(n_{B_{r_1}(x)}^\xi)\,\rho}\le C(r_2)^{d\xi}$, for all initial states satisfying \eqref{eq:controlled_density_main} with parameters $\lambda$ and $\xi$, all $x\in\Lambda$, and all radii \(r_2>r_1\ge1\) with \(vt\le r_2-r_1\).
The constant \(C\) is independent of system size, the initial state, and the length scales.
Such bounds are proved in Ref.~\cite{Lemm2025-ov} in a slightly less general setting. In the Supplementary Material, we extend them to time-dependent Hamiltonians and to initial states satisfying \eqref{eq:controlled_density_main}.
Set $\bar q =R^\delta t^d$.
Together with Markov's inequality, these bounds provide particle fluctuations above the cut-off and yield
\begin{align*}
		\abs{\Tr\sbr{\rho\;	 \tau_t \del{A-\bar A}B}}\le  C\norm{A}\norm{B}\abs{X}^{(1+\xi)/2}R^{-\delta\xi/2},
	\end{align*}
    for every bounded operator $B$ and every initial state $\rho$ satisfying \eqref{eq:controlled_density_main}.
    
Next, we approximate the dynamics $\tau_t$ by $\bar \tau_t$.
The main ingredients are again the particle propagation bounds.
They yield
\begin{align}
    &\abs{\Tr\sbr{\rho\del{\tau_t(\bar A)-\bar \tau_t(\bar A)}B}}\nonumber\\
    &\quad\le C\norm{A}\norm{B} |X|^{(1+\xi)/2}\frac{t^{d+1+\delta}}{R^{\delta(\xi-3)/2}}.
\end{align}
For \(\delta<\alpha-2d\), a sufficient condition on the moment parameter for both inequalities to hold is \(\xi>3+2(d+2)/\delta\).

The next step is to restrict $\bar H$ to the large block neighborhood $X_{2R^\eta}$, with $\eta>1$ large but independent of system size.
Duhamel's formula bounds the difference between the two dynamics as follows
\begin{align*}
		\!&\norm{\bar\tau_t(\bar A)-\bar \tau_t^{X_{2R^\eta}} (\bar A)}\le 2\int_0^t\!\!\md u \;\bigg \Vert\bigg[\;\!\sum_{\substack{i\in X_{2R^\eta}\\ j\in X^c_{2R^\eta}}}\bar\tau^{X_{2R^\eta}}_{u-t}\big(\bar h_{ij}\big),\bar{A}\bigg]\bigg \Vert\,.
	\end{align*}
Here $\bar H=\sum_{i,j\in\Lambda}\bar h_{ij}$, with $\bar h_{ij}=\Pi h_{ij}\Pi$ and $h_{ij}=h_{ji}=h_{ij}^\dagger$.
To complete this step, we bound the commutators in this inequality.
We distinguish two cases: $i$ close to $X$ and $i$ farther away.
In the first case, we trivially bound the norm of the commutator by twice the product of the norms of the two operators and use \eqref{qi def} to control the sum over $i$ and $j$.
In the second case, we apply Lieb--Robinson bounds for long-range bounded interactions~\cite{PhysRevX.10.031010,PhysRevLett.127.160401}, with the extension to fixed $k$-body interactions given in the Supplementary Material, to the dynamics generated by $\bar H_{X_{2R^\eta}}$ on the truncated Hilbert space $\operatorname{Ran}\Pi$, which factorizes over the disjoint blocks $P_x$.
Writing $h=(\eta+1)\delta$, we have
\begin{equation}
  Q:=\max_{P_x\subseteq X_{2R^\eta}}q_x
\lesssim \bar qR^{\eta\delta}
=t^dR^h, \label{Q def}
\end{equation}
so $\bar H_{X_{2R^\eta}}$ has interaction strength of order $g_{\mathrm{big}}\sim t^{pd}R^{p(\eta+1)\delta}$.
Therefore, we can apply the cited Lieb-Robinson bounds and for every $0<\omega<\min\{1,\alpha-2d\}$ there exist exponents $\gamma_1>1$ and $\gamma_2=\omega\gamma_1>d$ such that
\begin{align}
&\norm{\sbr{
\bar\tau^{X_{2R^\eta}}_{u-t}(\bar h_{ij}),\bar A
}}
\le
C\norm{A}\norm{\bar h_{ij}}
\frac{(g_{\mathrm{big}}t)^{\gamma_1}
      \ell^{\gamma_2}}
     {R^{\eta\gamma_2}},
\end{align}
for all $i\in X_{2R^\eta}\setminus X_{R^\eta}$, $j\in X^c_{2R^\eta}$, $0\le u\le t$ with $R^\eta\ge \ell\, C(g_\mathrm{big}t)^{1/\omega}$.
The constants are independent of the dimension of the truncated local Hilbert space.
Summing the two contributions and integrating over $u$, we find
that, for $R\gtrsim t^{d+2}$, choosing $\eta$ sufficiently large
and then $\delta>0$ sufficiently small gives
\begin{equation}
\norm{\bar\tau_t(\bar A)
-\bar\tau_t^{X_{2R^\eta}}(\bar A)}
\le C\norm{A}|X|\,t^{-\nu},
\end{equation}
for some $\nu>0$.
The detailed estimates and parameter choices are given in the
Supplementary Material.

It remains to approximate $\bar \tau_t^{X_{2R^\eta}}(\bar A)$ by a local observable supported in $X_R$.
We use the interaction picture to improve the light-cone exponent and its dependence on the order of the potential $p$.
Within $X_{2R^\eta}$, let $\bar V_{\mathrm{short}}$ and $\bar H^0_{\mathrm{short}}$ denote, respectively, the density and hopping terms acting within one block or between neighboring blocks. Let $\bar H_{\mathrm{long}}$ contain the remaining couplings between blocks.
Passing to the interaction picture with respect to
$\bar V_{\mathrm{short}}$, we obtain
\begin{equation}
H^{\mathrm{int}}(u)=\me^{\mi u\bar V_{\mathrm{short}}}( \bar H^0_{\mathrm{short}}+\bar H_{\mathrm{long}})\,\me^{-\mi u\bar V_{\mathrm{short}}}.
\end{equation}
We denote by $\tau_t^{\mathrm{int}}$ the time evolution it generates.
By \eqref{Q def}, its interaction strength is of order
\begin{equation}\label{g int def}
    g_{\mathrm{int}}\sim\max\{t^dR^h,t^{pd-\alpha}R^{hp}\}.
\end{equation}

On the particle-truncated Hilbert space $\operatorname{Ran}\Pi$, write
\(
\bar A_V
=\me^{\mi t\bar V_{\mathrm{short}}}
\bar A\,
\me^{-\mi t\bar V_{\mathrm{short}}},
\)
so that
$\bar\tau_t^{X_{2R^\eta}}(\bar A)
=\tau_t^{\mathrm{int}}(\bar A_V)$.
On this space, the commuting density evolution enlarges the support of $\bar A$
by only a fixed number of blocks and preserves its norm.

We apply Lieb--Robinson bounds to this evolution.
Taking the conditional expectation gives an operator supported in $X_R$ on the
truncated Hilbert space.
The Lieb--Robinson bound controls the error of this approximation.
We extend its factor inside $X_R$
by zero on the discarded occupation sectors and tensor it with
the full identity outside $X_R$.
This defines $\widetilde A\in\cA_{X_R}$ on the full Fock
space, with $\|\widetilde A\|\le\|A\|$.

We obtain
\begin{align}
&\norm{\bar\tau_t^{X_{2R^\eta}}(\bar A)
-\Pi\widetilde A\Pi}\le
C\norm{A}|X|^2
\frac{(g_{\mathrm{int}}t)^{\gamma_1}\ell^{\gamma_2}}{R^{\gamma_2}}\nonumber\\
& \ \le 
C\|A\||X|^2\left[
\frac{t^{(d+1)\gamma_1+\gamma_2}}
     {R^{\gamma_2-h\gamma_1}}
+
\frac{t^{(pd-\alpha+1)\gamma_1+\gamma_2}}
     {R^{\gamma_2-hp\gamma_1}}
\right],
\end{align}
provided
\(
R\ge C \ell(g_{\mathrm{int}}t)^{1/\omega}.
\)
Substituting \eqref{g int def} into this condition and choosing $\delta$ sufficiently small that $hp<\omega$, we find that the localization bound holds for $R\ge Ct^\beta$, where
\begin{align}
\beta=\max\left\{
 \frac{\omega+d+1}{\omega-h},
 \frac{\omega+pd-\alpha+1}{\omega-hp}
\right\}.
\end{align}
Fix $\epsilon>0$. 
We can choose $\omega$ sufficiently close to
$\min\{1,\alpha-2d\}$ and then $\delta$ sufficiently small so that the denominators are positive and $\beta<\beta_{\mathrm{up}}+\epsilon$.
Thus, at $R=Ct^{\beta_{\mathrm{up}}+\epsilon}$, after making $\nu$ smaller if necessary, we have
\begin{equation}
\norm{\bar\tau_t^{X_{2R^\eta}}(\bar A)
-\Pi\widetilde A\Pi}
\le C\norm{A}|X|^2t^{-\nu}.
\end{equation}

Finally, we compare $\Pi\widetilde A\Pi$ with $\widetilde A$.
By construction, $\Pi_{X_R}\widetilde A\Pi_{X_R}=\widetilde A$ and
\(
    \Pi\widetilde A\Pi-\widetilde A
=-\widetilde A(I-\Pi_{X_R^c}).
\)
For a union \(Y\) of blocks, \(\Pi_Y\) denotes the product of their occupation projectors.
Using the initial controlled-density assumption and Markov's inequality, we obtain
\begin{align}
&\norm{\rho\del{
\Pi\widetilde A\Pi-\widetilde A}}_1\le C\norm{A}|X|^{1/2}R^{-\delta\xi/2}.
\end{align}

Since each block has diameter at most $2\ell$,
$X_R\subseteq X[R+2\ell]$. 
As $\ell=\mathcal O(t)$,
the additional radius can be absorbed into the fixed
prefactor $C_R$.
Combining all the estimates and making $\nu$ smaller if necessary concludes the proof.


\bigbreak

\textit{Appendix B: Possible experimental prospects.---} Experimental tests {of the proposed propagation mechanism} require both nonlinear occupation enhancement and controlled spatial couplings. Here, $p=4$ means density-quartic terms, including $n_i^4$ and $n_i^2n_j^2$; the usual intermode Kerr term $n_in_j$ has $p=2$. Optical-lattice bosons offer one candidate: coherent on-site multibody interactions have been observed~\cite{Will2010}, and independent control of effective four-body interactions has been proposed~\cite{PhysRevA.90.021601}. Their four-body term is proportional to $n(n-1)(n-2)(n-3)$ and hence has leading growth $n^4$. Driven cavity-assisted lattices additionally offer proposals for programmable nonlocal multibody operations~\cite{4l1s-kkw7}, although an on-site four-body interaction alone does not supply the required intersite $n_i^2n_j^2$ coupling.

Superconducting microwave cavities provide a complementary route. Photon-number-selective phases and entangling operations between bosonic modes have been demonstrated~\cite{Heeres2015SNAP,Gao2019Exchange}, and photon-number-dependent Hamiltonian engineering has been proposed~\cite{Wang2021PND}. Together, these controls suggest synthesizing the joint Fock-state phase $\me^{-\mi\theta n_i^2n_j^2}$ in a finite occupation window. Realizing its power-law spatial kernel, maintaining coherence as occupations grow, and accounting for control overhead remain experimental tasks. A first test could measure the distance-dependent response to a localized perturbation in a finite array; observing the asymptotic optimal cone would additionally require a sufficiently broad window before loss, occupation cutoffs, and finite size dominate.




\end{document}


\renewcommand\thefootnote{*\arabic{footnote}}

\addtocounter{section}{0}

\setcounter{equation}{0}
\setcounter{section}{0}

\renewcommand{\theequation}{S.\arabic{equation}}
\renewcommand{\thesection}{S.\Roman{section}}
\setcounter{secnumdepth}{3}

\title{Supplementary Material for \\``Lieb-Robinson Light Cones in Long-Range Bosonic Systems''}

\author{Carla Rubiliani$^{1,2,}$\equalcontribution}
\email{carla.rubiliani@uni-tuebingen.de}

\author{Cheng Shang$^{2,}$\equalcontribution}
\email{cheng.shang@riken.jp}

\author{Tomotaka Kuwahara$^{2,3}$}
\email{tomotaka.kuwahara@riken.jp}

\affiliation{$^{1}$
Department of Mathematics, University of T\"ubingen, 72076 T\"ubingen, Germany
}

\affiliation{$^{2}$
Analytical Quantum Complexity RIKEN Hakubi Research Team, RIKEN Center for Quantum Computing (RQC), Wako, Saitama 351-0198, Japan
}

\affiliation{$^{3}$
RIKEN Pioneering Research Institute (PRI), Wako, Saitama 351-0198, Japan
}

\maketitle
This Supplementary Material provides detailed proofs and technical analyses that support the results presented in the main text and End Matter. In particular, we detail the bosonic propagation protocols underlying the lower bounds, including their implementation times and error estimates. We then establish an extension of the Lieb--Robinson bound of Tran \emph{et al.}~\cite{PhysRevLett.127.160401} to fixed-\(k\)-local Hamiltonians. Finally, we provide the technical derivations required for the upper bound, including particle-number moment estimates, occupation truncation, and the interaction-picture construction. Throughout, we follow the notation and assumptions of the main text.
\tableofcontents

\section{Setup and main results}
\label{S-I}

For convenience, the notation used most frequently throughout the main text and the Supplementary Material is summarized in Table~\ref{tab:SM_notation}.

\begin{table}[b]
\centering
\small
\setlength{\tabcolsep}{5pt}
\renewcommand{\arraystretch}{1}
\begin{tabular}{c|l}
Symbol & Meaning \\
\hline
$\alpha,\;p,\;d$
& Spatial decay exponent, bosonic interaction order, and spatial dimension. \\

$\beta_{\mathrm{up}},\;\beta_{\mathrm{low}},\;\beta_{\mathrm{opt}}$
& Leading upper-bound, lower-bound, and optimal light-cone exponents. \\

$\Lambda,\;n_X$
& Physical lattice and boson-number operator in a region $X\subseteq\Lambda$. \\

$\mathcal D_\xi^\lambda$
& Controlled-density states with density scale $\lambda$ and moments up to order $\xi$. \\

$\mathcal A_X^{q_0}$
& Bounded operators on $X$ changing its particle number by at most $q_0$. \\

$\tau_{s,t},\;T$
& Heisenberg dynamics from $s$ to $t$; $T=1+|t-s|$ in the upper-bound estimates. \\

$R,\;X_R,\;\widetilde A$
& Localization radius, blockwise enlargement of $X$, and a local approximant to $\tau_{s,t}(A)$. \\

$v,\;\ell,\;P_x,\;d_{\mathrm{cg}}$
& velocity in particle propagation bound, block radius, blocks, and the corresponding coarse-grained distance. \\

$N$
& Boson number in an encoded packet, with $N\sim t^d$. \\

$\beta_{\mathrm{LR}}$
& Reference bounded-interaction exponent, $\max\{1,(\alpha-2d)^{-1}\}$. \\

$\omega,\;\gamma_1,\;\gamma_2$
& Exponents in the bounded-interaction localization bound, with $\gamma_2=\omega\gamma_1>d$. \\

$g_{\mathrm{lr}},\,g$ &
{Long-range amplification scale $g_{\mathrm{lr}}=t^{pd-\alpha}$ and leading-power effective scale $g=\max\{t^d,t^{pd-\alpha}\}$.} \\

$g_*$
& Effective interaction scale including occupation-cutoff powers. \\

$\bar q,\;q_x,\;\Pi$
& Base and block-dependent occupation cutoffs, and their joint projector. \\

$\delta,\;\eta,\;h$
& Cutoff-growth and auxiliary spatial-enlargement exponents, with $h=(\eta+1)\delta$ in the upper-bound proof. \\

$\bar A,\;\bar H,\;\bar\tau_{s,t}$
& Particle-truncated observable and Hamiltonian, and the corresponding Heisenberg dynamics. \\

$U^V,\;H^{\mathrm{int}}$
& Propagator generated by $\bar V_{\mathrm{short}}$ and the corresponding interaction-picture Hamiltonian. \\

$k$ &
Maximum number of coarse-grained blocks in an interaction term. \\

\end{tabular}
\caption{Summary of frequently used notation.}
\label{tab:SM_notation}
\end{table}

\subsection{Hamiltonian and dynamics}
\label{S-I-A}

Consider a finite lattice $\Lambda\subset\mathbb{Z}^d$, where $d\ge1$ is the spatial dimension, and let $\mathcal F$ denote the corresponding bosonic Fock space.
For $i,j\in\Lambda$, $|i-j|$ denotes their distance in the supremum norm.
For $i\in\Lambda$ and $r\ge0$, we write $B_r(i)\coloneqq\{j\in\Lambda:|i-j|\le r\}$ for the cube of radius $r$ centered at $i$.
For nonempty regions $X,Y\subset\Lambda$, let
\begin{equation}
d(X,Y)\coloneqq
\min_{i\in X,\,j\in Y}|i-j|.
\end{equation}

Given $i\in\Lambda$, let $b_i$ and $b_i^\dagger$ denote the bosonic annihilation and creation operators, respectively.
They satisfy the canonical commutation relations $[b_i,b_j^\dagger]=\delta_{ij}$ and $[b_i,b_j]=[b_i^\dagger,b_j^\dagger]=0$.
We define the local number operator at $i\in\Lambda$ and the total number operator in a region $X\subset\Lambda$ by
\begin{equation}
n_i\coloneqq b_i^\dagger b_i,
\quad
n_X\coloneqq\sum_{i\in X}n_i.
\end{equation}

For $t\in \R$ and $i,j\in\Lambda$ define the hopping and density-interaction terms as
\begin{equation}
T_{ij}(t)\coloneqq J_{ij}(t)b_i^\dagger b_j,
\quad
V_{ij}(t)\coloneqq U_{ij}(t)V(n_i,n_j),
\end{equation}
where $J_{ij}(t)$ is the hopping amplitude, $U_{ij}(t)$ is the density-interaction coefficient, and $V$ is a real function of the commuting occupation operators $n_i$ and $n_j$, with bosonic interaction order $p\ge1$.
A representative density interaction is
\begin{equation}
V(n_i,n_j)=n_i^{p/2}n_j^{p/2}.
\label{eq:SM_density_potential}
\end{equation}
The allowed density interactions are fixed finite sums of the displayed term and polynomial terms of total degree at most $p$, with independently tunable real coefficients satisfying the bounds below, as used in the lower-bound protocols of Sec.~\ref{S-II}.
The upper bound allows $p\ge1$, while the explicit lower-bound protocols use integer $p\ge2$.
The coefficients are continuous in time and satisfy $J_{ji}(t)=\overline{J_{ij}(t)}$ and $U_{ij}(t)\in\mathbb R$ for all $i,j\in\Lambda$ and $t\in\mathbb R$.
Furthermore, we assume that, for a fixed decay exponent
$\alpha>2d$, there exist positive constants $C_{J}$ and $C_{U}$ independent of time, system size, and total particle number, such that
\begin{equation}
|J_{ij}(t)|
\le C_J(1+|i-j|)^{-\alpha},
\quad
|U_{ij}(t)|
\le C_U(1+|i-j|)^{-\alpha},
\label{eq:SM_power_law_bounds}
\end{equation}
for all $i,j\in\Lambda$ and $t\in \R$.
All hopping and density-interaction coefficients used in the propagation protocol satisfy the bounds above with the same fixed constants $C_J$ and $C_U$.

\begin{figure}[t]
\includegraphics[width=0.92\linewidth]{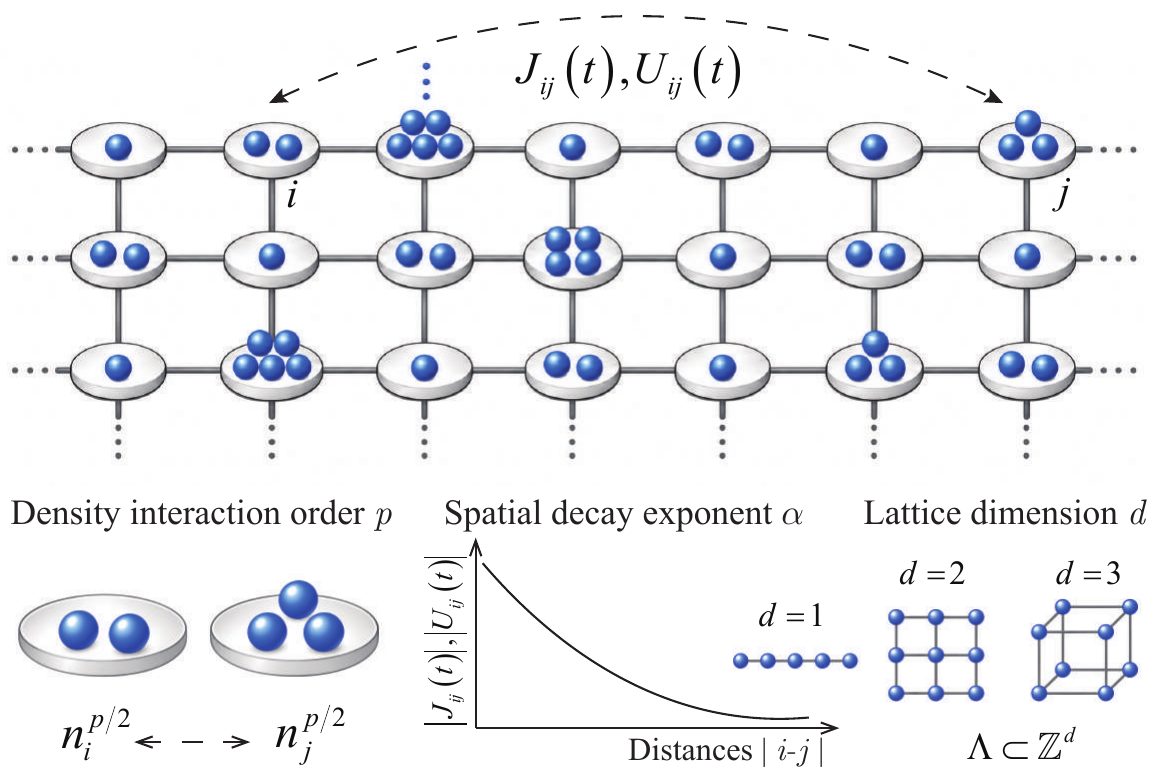}
\caption{Schematic of the long-range bosonic model and its key
physical parameters.
(Top) Blue spheres represent bosons on a finite lattice $\Lambda\subset\mathbb Z^d$. Gray links indicate the lattice geometry, while the dashed arrow illustrates long-range hopping
and density-density interactions between sites $i$ and $j$, with
respective coupling strengths $|J_{ij}(t)|$ and $|U_{ij}(t)|$. The bosonic interaction order $p$ controls the occupation dependence of the representative density term
$n_i^{p/2}n_j^{p/2}$. The power-law exponent $\alpha$ specifies the uniform decay bounds $|J_{ij}(t)|,|U_{ij}(t)|\lesssim C(1+|i-j|)^{-\alpha}$. The $d$-dimensional lattice is illustrated by the examples $d=1,2,3$.}
\label{fig:SM_bosonic_setup}
\end{figure}

For $X\subset\Lambda$, the Hamiltonian restricted to $X$ is
\begin{equation}
H_X(t)\coloneqq
\sum_{i,j\in X}
\bigl[T_{ij}(t)+V_{ij}(t)\bigr],
\quad
H(t)\coloneqq H_\Lambda(t).
\label{eq:SM_restricted_Hamiltonian}
\end{equation}
\begingroup
For the two-mode realization of $H(t)$ introduced in Sec.~\ref{S-II-A}, each physical site $i\in\Lambda$ carries two bosonic modes $a=0,1$, with annihilation operators $b_{i,a}$ and number operators $n_{i,a}=b_{i,a}^\dagger b_{i,a}$.
We denote the corresponding local Fock states by $\ket{m,n}_i$. In this realization, the total occupation of a physical region $X\subseteq\Lambda$ is $n_X=\sum_{i\in X}(n_{i,0}+n_{i,1})$.
The Bose--Hubbard-type model and its key physical parameters $p$, $\alpha$, and $d$ are illustrated in Fig.~\ref{fig:SM_bosonic_setup}.
\endgroup

Notice that the Hamiltonian conserves the total particle number, and the density terms satisfy
\begin{equation}
[V_{ij}(t),n_k]=0,
\quad
[V_{ij}(t),V_{kl}(s)]=0
\end{equation}
for all $i,j,k,l\in\Lambda$ and $s,t\in\mathbb{R}$.

Since $\Lambda$ is finite, each subspace of $\mathcal F$ with a fixed total particle number is finite-dimensional.
Particle-number conservation ensures that $H(t)$ preserves these subspaces. Within each one, $H(t)$ is a Hermitian matrix that depends continuously on time and therefore defines a unitary evolution.
Together, these evolutions define a unitary propagator $U(s,t)$ on the full Fock space $\mathcal F$, for $s,t\in\mathbb R$.
On each fixed-particle-number subspace, it is the unique solution of
\begin{equation}
\mi\partial_t U(s,t)=H(t)U(s,t),
\quad U(s,s)=I.
\end{equation}
We define the corresponding Heisenberg dynamics $\tau_{s,t}$ for every $s,t\in\R$ and every operator $A$ acting on Fock space as
\begin{equation}
\tau_{s,t}(A)\coloneqq U(s,t)^\dagger A U(s,t).
\label{eq:SM_dynamics}
\end{equation}
If $A$ is unbounded, the formula applies to vectors $\psi$
for which $A\,U(s,t)\psi$ is well defined.

\subsection{Initial states and local observables}
\label{S-I-B}

The infinite local Hilbert space requires a state-dependent formulation~\cite{PhysRevLett.127.070403,PhysRevX.12.021039,
kuwahara2022optimal,LR26}.
We consider initial density matrices with controlled local particle-number moments. 
For fixed $\lambda>0$ and $\xi\ge1$, let $\mathcal{D}_\xi^\lambda$ denote the class of density matrices $\rho$ on $\mathcal{F}$ with finite energy for the Hamiltonian under consideration that satisfy
\begin{equation}
\operatorname{Tr}\!\big[
n_{B_r(i)}^\zeta \rho
\big]
\le
\bigl(\lambda \zeta r^d\bigr)^\zeta,
\quad \text{ for all }
1\le \zeta\le\xi,
\quad
i\in\Lambda,
\quad
r\ge1.
\label{eq:SM_controlled_density}
\end{equation}
Here $\lambda$ sets the density scale, while $\xi$ specifies the highest particle-number moment controlled in the class of states.
Both are fixed independently of the lattice size.
Under this controlled-density assumption, local particle-number
fluctuations can be bounded independently of the total number
of {particles}~\cite{LemmRubilianiZhang2026Microscopic,Lemm2025-ov,
Deuchert2025Dynamics}.

Define $\mathcal{B}(\mathcal F)$ {as} the algebra of bounded operators on $\mathcal F$. 
For a finite region $X\subset\Lambda$, let $\mathcal{A}_X\subset\mathcal{B}(\mathcal{F})$ denote the subalgebra of bounded operators supported on $X$. 
We write $A\in\mathcal{A}_X^{q_0}$ when $A$ changes the particle number in $X$ by at most a fixed nonnegative integer $q_0$. 
Specifically, if
$\Pi_{X,m}$ denotes the spectral projector of $n_X$ with eigenvalue $m$, then
\begin{equation}
\Pi_{X,m}A\Pi_{X,n}=0
\quad
\text{for all }m,n\in\{0,1,2,\ldots\}
\text{ with }|m-n|>q_0.
\label{eq:SM_q0_observable}
\end{equation}
The restriction to $\mathcal{A}_X^{q_0}$ ensures that insertion of the local observable preserves the particle-number moment control required later on.
We denote by $\norm{\cdot}$ and $\norm{\cdot}_1$ the operator and trace norm, respectively.

\subsection{Block geometry and local approximation}
\label{S-I-C}

For $i\in\mathbb{Z}^d$ and nonempty $Y\subset\Lambda$, we use $d(i,Y)\coloneqq\min_{j\in Y}|i-j|$. For a block radius $\ell\in\{0,1,2,\ldots\}$, define the set of block centers and the corresponding blocks by
\begin{equation}
\mathcal{C}\coloneqq
\left\{
x\in(2\ell+1)\mathbb{Z}^d:
d(x,\Lambda)\le\ell
\right\},
\quad
P_x\coloneqq
\left\{
i\in\Lambda:
|i-x|\le\ell
\right\},
\quad x\in\mathcal{C}.
\label{eq:SM_block_geometry}
\end{equation}
The collection $\{P_x\}_{x\in\mathcal{C}}$ forms a disjoint partition of $\Lambda$, with $|P_x|\le(2\ell+1)^d.$
We equip the resulting coarse-grained lattice with the distance
\begin{equation}
d_{\mathrm{cg}}(P_x,P_y)
\coloneqq
\frac{|x-y|}{2\ell+1}, \quad \text{for every }x,y\in\cC.
\end{equation}
Thus $d_{\mathrm{cg}}$ measures separations in units of the block spacing $2\ell+1$.

For a nonempty finite region \(X\subset\Lambda\) and \(R\ge0\), its blockwise enlargement is
\begin{equation}
X_R
\coloneqq
\bigcup_{\substack{x\in\mathcal C\\ d(P_x,X)\le R}}
P_x.
\label{eq:SM_block_enlargement}
\end{equation}
Thus, $X_R$ is the radius-$R$ neighborhood of $X$ rounded out to whole blocks.
For the sitewise enlargement, we use the notation
\begin{align}
    X[R]:=\Set{i\in\Lambda \,:\, d(i,X)\le R}.
\end{align}
The lower- and upper-bound arguments use the same block geometry,
but choose the block radius $\ell$ differently.
In both cases,
$\ell=\bigO{ t}$.

\subsection{Main results}
\label{S-I-D}

\begingroup

We establish bounds on information propagation for the Bose--Hubbard-type Hamiltonian $H(t)$ in Eq.~\eqref{eq:SM_restricted_Hamiltonian}.
Throughout this subsection, $d\ge1$ is an integer and $\alpha>2d$. We use the state and observable classes of Sec.~\ref{S-I-B} and the block geometry of Sec.~\ref{S-I-C}.
The reference bounded-interaction exponent is
\begin{equation}
\beta_{\mathrm{LR}}
\coloneqq
\max\left\{1,\frac{1}{\alpha-2d}\right\}.\notag
\end{equation}

\begin{restatable}[Upper bound on information propagation]{theorem}{SMUpperBound}
\label{thm:SM_main_upper_bound}
Consider $p\ge1$ and define 
\begin{equation}
\beta_{\mathrm{up}}
\coloneqq
1+\max\{d+1,pd-\alpha+1\}\beta_{\mathrm{LR}}.\notag
\end{equation}
For every $\epsilon>0$ 
there exist positive $\nu$ and $\xi_0$ such that for every finite
 $\lambda>0$, $\xi\ge \xi_0$, and integer
$q_0\ge 0$ 
there exist constants $C,C_R>0$ such that the following holds.
For every  nonempty
$X\subseteq\Lambda$, every $A\in\mathcal A_X^{q_0}$,
all $s,t\in\mathbb R$, and every
\[
R\ge C_R(1+|t-s|)^{\beta_{\mathrm{up}}+\epsilon},
\]
there exists an operator $\widetilde A\in\mathcal A_{X[R]}$  satisfying 
\(
\|\widetilde A\|\le\|A\|
\)
and
\[
\left\|
\rho\bigl[\tau_{s,t}(A)-\widetilde A\bigr]
\right\|_1
\le
C\|A\|
\left(|X|^{(1+\xi)/2}+|X|^2\right)
(1+|t-s|)^{-\nu}
\]
for every $\rho\in\mathcal D_\xi^\lambda$.

\end{restatable}
In particular, one may take $\nu=\epsilon$.
The constants may depend on the fixed model parameters and on $\epsilon,\nu,\lambda,q_0,\xi$, but are independent of the lattice size, $s,t,X,A$, and $\rho$.

This result essentially shows that information can at most propagate to a distance $R\sim t^{\beta_{\mathrm{up}}}$ under the assumptions provided. 
The proof first imposes a cutoff on the total particle number
in each block, allowing the cutoff to increase sublinearly
with the block's distance from $X$.
Particle-propagation bounds~\cite{Lemm2025-ov} control the
occupation tails and the resulting truncation error.
We then restrict the truncated dynamics to a sufficiently large neighborhood of $X$ using the Lieb--Robinson bound for bounded interactions~\cite{PhysRevX.10.031010,
PhysRevLett.127.160401}.
An interaction picture on the level of blocks reduces the effective
interaction strength entering such bounds and yields a better light cone shape.
The proof is given in Sec.~\ref{S-IV}.

\begin{theorem}[Constructive lower bound on information propagation]
\label{thm:SM_main_lower_bound}
Fix an integer $p\ge2$, a finite moment order $\xi\ge1$,
and $0<e_*<1$.
For $p=2$, let $\beta_{\mathrm{low}}=\beta_{\mathrm{LR}}$.
For integer $p\ge3$, let
\begin{equation}
\beta_{\mathrm{low}}
=
\begin{cases}
\displaystyle
1+\frac{pd-\alpha+1}{\alpha-2d},
& 2d<\alpha\le2d+1,\\[7pt]
\displaystyle
\max\left\{d,\,2+\frac{pd-\alpha}{\alpha-2d}\right\},
& 2d+1<\alpha\le pd,\\[7pt]
\displaystyle
\max\{d,\,2+pd-\alpha\},
& \alpha>pd.\notag
\end{cases}
\end{equation}
There exist constants $\lambda,c,c_R>0$ and a function
$h(t)\ge1$, with
\[
\lim_{t\to\infty}\frac{\log h(t)}{\log t}=0,
\]
such that, for every sufficiently large $t$, a finite lattice with two bosonic modes per physical site supports an admissible Hamiltonian and two initial states
$\rho_{\mathrm{in}}^{(b)}\in\mathcal D_\xi^\lambda$,
$b=0,1$, related by a number-preserving unitary supported
at a single input site $i_{\mathrm{in}}$.

The same Hamiltonian acts on both inputs.
Let $\rho_{\mathrm{out}}^{(b)}$ be the reduced state at a readout site $i_{\mathrm{out}}$ at time $t$.
There is a Hermitian observable $A_{\mathrm{out}}\in\mathcal A_{\{i_{\mathrm{out}}\}}^0$, with $\|A_{\mathrm{out}}\|=1$, such that
\begin{equation}
|i_{\mathrm{in}}-i_{\mathrm{out}}|
\ge c\,\frac{t^{\beta_{\mathrm{low}}}}{h(t)},
\quad
\left|
\operatorname{Tr}\!\left[
A_{\mathrm{out}}
\bigl(\rho_{\mathrm{out}}^{(0)}
-\rho_{\mathrm{out}}^{(1)}\bigr)
\right]
\right|
\ge2-2e_*.\notag
\end{equation}

For $X=\{i_{\mathrm{out}}\}$, one can choose a block partition as specified in Sec.~\ref{S-II-E} and a radius $R(t)\ge c_Rt^{\beta_{\mathrm{low}}}/h(t)$, with
$i_{\mathrm{in}}\notin X_{R(t)}$, such that every common $\widetilde A\in\mathcal A_{X_{R(t)}}$ satisfies
\begin{equation}
\max_{b=0,1}
\left\|
\rho_{\mathrm{in}}^{(b)}
\bigl[\tau_{0,t}(A_{\mathrm{out}})-\widetilde A\bigr]
\right\|_1
\ge1-e_*.\notag
\end{equation}

All constants are independent of $t$ and the ambient lattice size. The function $h$ accounts for subpolynomial corrections of the selected construction and is independent of the lattice size. The total time $t$ includes preparation, signal transfer, and microscopic decoding.
\end{theorem}

The constructions and their implementation estimates are given in Sec.~\ref{S-II}; the combined statement is proved in Proposition~\ref{prop:SM_lower_bound_protocol}.

\textit{Optimality of the light-cone exponent.---}
For integer $p\ge4$ and $2d<\alpha\le2d+1$, Theorems~\ref{thm:SM_main_upper_bound}
and \ref{thm:SM_main_lower_bound} give
\begin{equation}
\beta_{\mathrm{opt}}
=\beta_{\mathrm{up}}
=\beta_{\mathrm{low}}
=
1+\frac{pd-\alpha+1}{\alpha-2d}
=
\frac{(p-2)d+1}{\alpha-2d}.\notag
\end{equation}
In this regime, the recursive packet-transfer construction has $h(t)=1$.
Its Mott inputs satisfy the controlled-density condition for every finite moment order with a fixed density parameter.
Since $\beta_{\mathrm{opt}}>1$, the input--output separation grows faster than $t$, so the localization obstruction also applies to the upper-bound block scale $\ell=\lceil vt\rceil$.

Thus, exponents below $\beta_{\mathrm{opt}}$ are excluded by the lower bound, while every exponent above $\beta_{\mathrm{opt}}$ is supported by the upper bound under
the corresponding finite-moment condition. This determines the optimal leading exponent but does not establish an upper bound with zero exponent loss.
Outside the displayed matching regime, the two theorems provide bounds without asserting optimality.
\endgroup

\section{Lower bound on information propagation}
\label{S-II}

\begin{figure}[!tbp]
\centering
\includegraphics[width=0.879\linewidth]{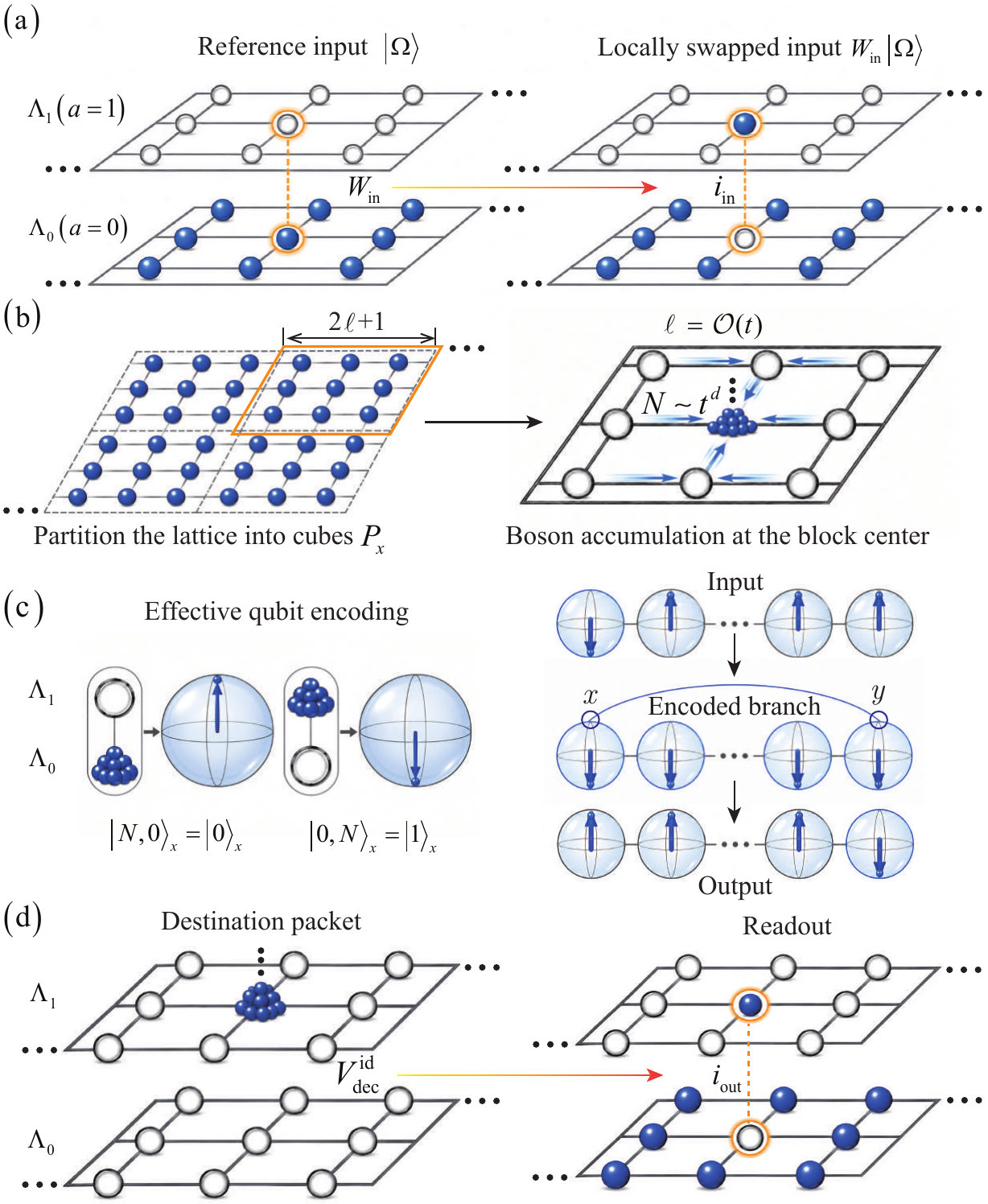}
\caption{Schematic of the state-transfer protocol.
(a) Each physical site carries two bosonic modes. For the reference input, $\Lambda_0$ contains one boson per site and $\Lambda_1$ is empty. The second input is obtained by exchanging the layer occupation at a single site $i_{\mathrm{in}}$.
(b) The physical lattice is partitioned into blocks $P_x$ of radius $\ell\asymp t$. In the source block, nearest-neighbor-controlled gates first coherently spread the layer qubit. Successive nearest-neighbor sweeps with nonlinear on-site control then concentrate particles at the block centers, producing packets with $N\asymp\ell^d\asymp t^d$ bosons.
(c) At each center, the packet states $\ket{0}_x=\ket{N,0}_x$ and $\ket{1}_x=\ket{0,N}_x$ encode an effective qubit. Same-layer long-range density interactions induce effective Ising couplings with strength $J_{\mathrm{eff}}(P_x,P_y)\asymp t^{pd-\alpha}/d_{\mathrm{cg}}(P_x,P_y)^\alpha$, where $d_{\mathrm{cg}}(P_x,P_y)=|x-y|/(2\ell+1)$. Together with local packet rotations, these couplings implement recursive encoding and decoding to transfer the signal across the coarse-grained lattice. The arc indicates a representative effective $ZZ$ coupling.
(d) The destination packet is decoded onto a single original site $i_{\mathrm{out}}$ by reversing the preparation sequence, {restoring all other active sites in the destination block to the reference local state $\ket{1,0}$}.
Panels (c,d) only illustrate the logical $\ket{1}$ branch of the ideal protocol.}
\label{fig:SM_lower_protocol}
\end{figure}

{
We construct explicit bosonic propagation protocols that yield lower bounds on information propagation, accounting for their implementation times and accumulated errors.} The {packet-based protocol} comprises three stages:
preparation of bosonic packets at the centers of mesoscopic blocks,
transfer of the encoded signal between these packets, and final decoding onto a single original site.
Each stage is assigned a fixed fraction of the available protocol time.
{
We also consider alternative propagation protocols in the parameter regimes specified below.
}
Throughout this section, we set the initial time to zero and use $t>0$ for the elapsed protocol time.
All Hamiltonians and controls used below obey the assumptions of Sec.~\ref{S-I}.
{
The packet-based protocol is summarized in Fig.~\ref{fig:SM_lower_protocol}.}

\textit{Overview of the lower bound.---}
Let $p\ge2$ and $d\ge1$ be integers denoting the density-interaction order and spatial dimension, respectively, and let $\alpha>2d$ be the spatial decay exponent.
The reference spin exponent is
\[
\beta_{\mathrm{LR}}
=\max\left\{1,\frac{1}{\alpha-2d}\right\}.
\]
For $p=2$, the microscopic protocols give
\[
\beta_{\mathrm{low}}=\beta_{\mathrm{LR}}.
\]
For $p\ge3$, the constructive lower bounds give
\[
\beta_{\mathrm{low}}=
\begin{cases}
\displaystyle
\frac{(p-2)d+1}{\alpha-2d},
&2d<\alpha\le2d+1,\\[7pt]
\displaystyle
\max\left\{
d,\,
2+\frac{pd-\alpha}{\alpha-2d}
\right\},
&2d+1<\alpha<pd,\\[9pt]
\displaystyle
\max\{d,\,2+pd-\alpha\},
&\alpha>2d+1,\quad\alpha\ge pd.
\end{cases}
\]
The first branch follows from recursive packet transfer.
The second compares collective packet transfer with short-range propagation, while the third compares sequential packet transfer with short-range propagation.
In each case, the maximum selects the larger leading power attained within the same total time.
The sequential construction is used only when $pd-\alpha+1>0$; otherwise, short-range propagation supplies the bound.
The crossover between these alternatives is not an optimality boundary.

Fix a finite moment order $\xi\ge1$ and a total trace-norm error allowance $0<e_*<1$ at readout. There exist constants $\lambda,c,c_R>0$ and a function $h(t)\ge1$, with
\[
\lim_{t\to\infty}
\frac{\log h(t)}{\log t}=0,
\]
such that, for every sufficiently large $t$, a finite lattice $\Lambda$ with two bosonic modes per physical site supports an admissible Hamiltonian and two initial states $\rho_{\mathrm{in}}^{(b)}\in\mathcal D_\xi^\lambda$, $b=0,1$.
Here $\lambda$ is the density parameter and $\xi$ is the highest controlled particle-number moment in the state class of Sec.~\ref{S-I-B}.
The function $h$ accounts for the subpolynomial corrections in the respective constructions.

The two inputs are related by a number-preserving unitary supported on a single site $i_{\mathrm{in}}$ and evolve under the same Hamiltonian.
Let $i_{\mathrm{out}}$ be the readout site and $\rho_{\mathrm{out}}^{(b)}$ its reduced state at time $t$.
There is a bounded Hermitian observable $A_{\mathrm{out}}\in\mathcal A_{\{i_{\mathrm{out}}\}}^0$, supported on that site and commuting with its total occupation, with $\|A_{\mathrm{out}}\|=1$, such that
\[
|i_{\mathrm{in}}-i_{\mathrm{out}}|
\ge c\,\frac{t^{\beta_{\mathrm{low}}}}{h(t)},
\quad
\left|\operatorname{Tr}\!\left[
A_{\mathrm{out}}
\bigl(\rho_{\mathrm{out}}^{(0)}
-\rho_{\mathrm{out}}^{(1)}\bigr)
\right]\right|
\ge2-2e_*.
\]
The time $t$ includes preparation, signal propagation, and microscopic readout. The constants are independent of $t$ and the lattice size. The function $h$ is independent of the lattice size.

A localization radius $R(t)$, measured in the original lattice and chosen as a sufficiently small fixed fraction of the input--output separation, satisfies $R(t)\ge c_Rt^{\beta_{\mathrm{low}}}/h(t)$.
For every common bounded approximant supported in the blockwise radius-$R(t)$ enlargement of $\{i_{\mathrm{out}}\}$, the maximum trace-norm approximation error over the two inputs is at least $1-e_*$.
This is the local-approximation obstruction stated in Proposition~\ref{prop:SM_lower_bound_protocol}.
The blockwise enlargement and the error criterion are defined in Sec.~\ref{S-I-C}.

For the Bose--Hubbard-type model and integer $p\ge4$, the lower and upper exponents match throughout
$2d<\alpha\le2d+1$, giving
\[
\beta_{\mathrm{opt}}
=\beta_{\mathrm{up}}
=\beta_{\mathrm{low}}
=\frac{(p-2)d+1}{\alpha-2d}.
\]
This matching holds in the leading-power sense and requires
no additional implementation assumptions.
The upper-bound estimate is taken as a separate input
to this comparison.
No optimality for this model is inferred outside this interval.

\subsection{Initial state and boson concentration}
\label{S-II-A}

For the lower-bound construction, we augment each physical site by a two-valued mode label,
\begin{equation}
\widetilde{\Lambda}
\coloneqq
\Lambda\times\{0,1\},
\quad
\Lambda_a
\coloneqq
\Lambda\times\{a\},
\quad a=0,1.
\end{equation}
Each physical site $i\in\Lambda$ thus carries two bosonic modes.
The layer index is an internal mode label.
{
Spatial distances and blocks are defined by the physical coordinate $i\in\Lambda$.}
{
For every pair of modes $(i,a)$ and $(j,b)$, with $a,b\in\{0,1\}$, the hopping and density coefficients satisfy Eq.~\eqref{eq:SM_power_law_bounds} with the physical distance $|i-j|$ and the same constants $C_J,C_U$.
The coefficients may depend on the mode labels.}
We write $\ket{n_0,n_1}_i$ for the local Fock state and $n_{i,a}\coloneqq b_{i,a}^\dagger b_{i,a}$ for the occupation of mode $i,a$.
The operators now act on the bosonic Fock space over $\widetilde{\Lambda}$, while spatial regions remain subsets
of $\Lambda$.
For $Y\subset\Lambda$, the total occupation includes both modes:
\begin{equation}
n_Y
\coloneqq
\sum_{i\in Y}\sum_{a=0,1}n_{i,a}.
\end{equation}

\textit{Controlled-density input.---}
We start from the product state
\begin{equation}
\ket{\Omega}
\coloneqq
\bigotimes_{i\in\Lambda}\ket{1,0}_i,
\quad
\rho_{\mathrm{in}}^{(0)}
\coloneqq
\ket{\Omega}\!\bra{\Omega},
\end{equation}
with one boson per site in $\Lambda_0$ and vacuum in $\Lambda_1$, as illustrated in Fig.~\ref{fig:SM_lower_protocol}(a).
For all $i\in\Lambda$, $r\ge1$, and $\xi\ge1$,
\begin{equation}
\operatorname{Tr}\!\left[
n_{B_r(i)}^\xi\rho_{\mathrm{in}}^{(0)}
\right]
=
|B_r(i)|^\xi
\le
(2r+1)^{d\xi}
\le
(3^d r^d)^\xi.
\label{eq:SM_lower_density_check}
\end{equation}
Thus Eq.~\eqref{eq:SM_controlled_density} holds for every finite $\xi$ with any fixed $\lambda\ge3^d$, independently of the block size and system size.
The energy expectation and total particle-number moments are finite because the state lies in a fixed finite-particle sector on each finite lattice.

To define a localized input, choose a physical site $i_{\mathrm{in}}$ in the source block.
Let $W_{\mathrm{in}}$ exchange $\ket{1,0}_{i_{\mathrm{in}}}$ and $\ket{0,1}_{i_{\mathrm{in}}}$ and act as the identity on all other local particle-number sectors.
This defines a number-preserving unitary supported on
$i_{\mathrm{in}}$.
The second input state is
\begin{equation}
\rho_{\mathrm{in}}^{(1)}
\coloneqq
W_{\mathrm{in}}\rho_{\mathrm{in}}^{(0)}
W_{\mathrm{in}}^\dagger.
\end{equation}
The two states differ only at $i_{\mathrm{in}}$ and obey the same bound in Eq.~\eqref{eq:SM_lower_density_check}.

Let $\Pi_{\mathrm{in}}$ denote the orthogonal projector onto $\operatorname{span}\{\ket{\Omega}, W_{\mathrm{in}}\ket{\Omega}\}$.
Every state in this subspace has exactly one boson per physical site, since
$(n_{i,0}+n_{i,1})\Pi_{\mathrm{in}}=\Pi_{\mathrm{in}}$, $i\in\Lambda$.
Consequently, Eq.~\eqref{eq:SM_lower_density_check} also holds for any density matrix in this subspace.
The same bound holds for the reduced lattice state of any
joint state supported on $\operatorname{Ran}(\Pi_{\mathrm{in}}) \otimes\mathcal H_{\mathrm{ref}}$, since tracing out the reference preserves its support on $\operatorname{Ran}(\Pi_{\mathrm{in}})$.

\textit{Ideal block concentration.---}
The packet-concentration construction below is used for integer $p\ge3$.
For $p=2$, microscopic recursive transfer and nearest-neighbor propagation already attain $\beta_{\mathrm{low}}=\beta_{\mathrm{LR}}$ without boson concentration.
Moreover, the concentration construction below relies on cubic density controls, which are not available for $p=2$.
We use the block decomposition of Sec.~\ref{S-I-C}, with $\ell=\lfloor c_\ell t\rfloor$ for a fixed $c_\ell>0$ and sufficiently large elapsed time $t$, as illustrated in Fig.~\ref{fig:SM_lower_protocol}(b).
For each sufficiently large $t$, choose the finite lattice $\Lambda$ large enough to contain all complete blocks required by the protocol.
We call these selected blocks active, while all other blocks, including incomplete boundary blocks, are inactive and remain decoupled throughout the protocol.
Each active block contains its center $x$ and satisfies $|P_x|=(2\ell+1)^d$.

We first describe an ideal (error-free) concentration sequence assuming perfect isolation of the required occupation transitions, following the short-range strategy of Ref.~\cite{kuwahara2022optimal}.
The bounded-control implementation, including the time and errors associated with transition isolation, is established in Sec.~\ref{S-II-B-preparation}.

Particles are collected at the block centers by processing the $d$ coordinate directions successively.
Along each coordinate row, first collect particles from one end toward the central site, and then repeat from the other end.
This requires at most $2\ell$ successive merges per row.
Rows belonging to the same sweep and different blocks are processed in parallel.

Consider a merge involving two neighboring sites $i,j$ in the same layer $a$, and write $b_1=b_{i,a}$ and $b_2=b_{j,a}$ for this calculation.
Let $M\ge1$ be their total particle number.
Choose a fixed amplitude $0<J_0\le2^{-\alpha}C_J$, independent of $M$, $\ell$, and system size.
The hopping term
\begin{equation}
H_{\mathrm{hop}}
=
J_0\left(
b_1^\dagger b_2+b_2^\dagger b_1
\right)
\label{eq:SM_merge_hopping}
\end{equation}
then satisfies the coefficient bound of Eq.~\eqref{eq:SM_power_law_bounds}.
Its matrix elements in the fixed-$M$ sector are
\begin{equation}
\bra{m+1,M-m-1}H_{\mathrm{hop}}\ket{m,M-m}
=
J_0\sqrt{(m+1)(M-m)},
\quad 0\le m<M.
\end{equation}
Conditioned on the exact isolation of a selected transition, a constant resonant pulse transfers the population between these two states in time
$\pi/[2J_0\sqrt{(m+1)(M-m)}]$.
The corresponding ideal hopping time required to move all particles to one mode is bounded by
\begin{equation}
t_{\mathrm{merge}}^{\mathrm{id}}
\le
\frac{\pi}{2J_0}
\sum_{m=0}^{M-1}
\frac{1}{\sqrt{(m+1)(M-m)}}
\le
\frac{2\pi}{J_0}.
\end{equation}
To obtain the last inequality, pair terms on opposite sides of the midpoint:
\begin{align}
\sum_{m=0}^{M-1}
\frac{1}{\sqrt{(m+1)(M-m)}}
\le
2\sqrt{\frac{2}{M+1}}
\sum_{k=1}^{K}\frac{1}{\sqrt{k}}
\le
4\sqrt{\frac{2K}{M+1}}
\le4,
\quad
K\coloneqq
\left\lfloor\frac{M+1}{2}\right\rfloor.
\label{eq:SM_inverse_matrix_element_sum}
\end{align}
Here we used $\sum_{k=1}^{K}k^{-1/2}\le2\sqrt K$. A merge starting from a prescribed intermediate occupation traverses only part of this sequence and satisfies the same bound.

Continuity of the hopping envelope does not change this ideal time scale.
For an isolated transition, choose
\begin{equation}
J_m(u)=J_0\sin^2\!\left(\frac{\pi u}{\tau_m}\right),
\quad
\tau_m=
\frac{\pi}{J_0\sqrt{(m+1)(M-m)}},
\quad 0\le u\le\tau_m,
\end{equation}
and set $J_m(u)=0$ outside this interval.
Its pulse area satisfies
$\sqrt{(m+1)(M-m)}
\int_0^{\tau_m}J_m(u)\,\md u=\pi/2$.
This accounts only for the hopping time.
The hopping Hamiltonian also couples other neighboring occupation states, so the selected transition is not isolated by the pulse envelope alone.
Section~\ref{S-II-B-preparation} constructs the required selective dynamics using bounded hopping and density controls, including their implementation times and errors.

Since there are at most $2\ell$ successive rounds per coordinate direction, the ideal concentration time satisfies
\begin{equation}
t_{\mathrm{conc}}^{\mathrm{id}}
=
{\cal O}(\ell)
=
{\cal O}(t)
\end{equation}
for fixed $d$.
Each complete active block initially contains $|P_x|=(2\ell+1)^d$ bosons, one per physical site. For even $p$, all these bosons participate in the concentration, so $N=|P_x|$.
For odd $p$, the packet rotation of Sec.~\ref{S-II-B-rotations} requires $N\equiv2\pmod8$, which can be achieved by leaving at most seven bosons per block as spectators. Consequently, in both cases, the central packet contains
\begin{equation}
N\asymp\ell^d\asymp t^d
\end{equation}
bosons. For odd $p$, the spectator sites are chosen at the outer end of a row in the first coordinate sweep.
They remain in their reference states and are decoupled from the protocol.
This choice only shortens that row and does not affect the concentration at the block center. In the source block, the localized layer qubit is first coherently spread across the active sites, after which the particles in the two layers are separately concentrated.
The corresponding encoding and its inverse decoding are described in Sec.~\ref{S-II-B}.

\textit{Inverse concentration and uniform implementation.---}
Let $H_{\mathrm c}(u)$ denote the time-dependent control Hamiltonian implementing a concentration sequence over $0\le u\le\tau$, and let $V$ denote the resulting unitary.
The reversed control Hamiltonian
\begin{equation}
H_{\mathrm c}^{\mathrm{inv}}(u)
\coloneqq-H_{\mathrm c}(\tau-u),
\quad 0\le u\le\tau,
\end{equation}
implements $V^\dagger$ in the same duration.
Time reversal and sign reversal preserve continuity, spatial support, and the coefficient bounds.
Let $V^{\mathrm{id}}$ be a unitary extension of the ideal concentration map on its prescribed input subspace, and let $\Pi$ project onto that subspace.
Then
\begin{equation}
\left\|
\left[V^\dagger-(V^{\mathrm{id}})^\dagger\right]
V^{\mathrm{id}}\Pi
\right\|
=
\left\|(V-V^{\mathrm{id}})\Pi\right\|.
\label{eq:SM_inverse_concentration_error}
\end{equation}
Thus, reversing the implemented sequence preserves the approximation error on the corresponding ideal output subspace $\operatorname{Ran}(V^{\mathrm{id}}\Pi)$.
We now state the preparation and decoding estimates proved in Proposition~\ref{prop:SM_A1_bounded_preparation} for integer $p\ge3$.

Let $V_{\mathrm{prep}}^{\mathrm{id}}$ denote the ideal preparation operation of Sec.~\ref{S-II-B}, combining coherent encoding and concentration in all active blocks.
Let $V_{\mathrm{dec}}^{\mathrm{id}}$ denote the inverse of the ideal local encoder at the destination block.
The projector $\Pi_{\mathrm{dec}}^{\mathrm{id}}$ specifies the ideal input subspace of this decoder: an arbitrary superposition of $\ket{N,0}$ and $\ket{0,N}$ at the destination, with all other packets in $\ket{N,0}$ and the vacuum and spectator factors
fixed as in the ideal preparation.
This is the output subspace of the ideal packet-transfer protocol.

For fixed $b_{\mathrm{arr}}\ge0$ and $C_B>0$, let $\Gamma_{\mathrm{act}}\subset\mathcal C$ denote the centers of the complete blocks participating in the protocol, with $B_{\mathrm{act}}\coloneqq|\Gamma_{\mathrm{act}}|
\le C_B\ell^{b_{\mathrm{arr}}}$.
This polynomial bound ensures that the accumulated transition errors can be controlled within ${\cal O}(\ell)$ preparation time.
Fix $0<e_*<1$ and positive error allowances $e_{\mathrm{prep}}$ and $e_{\mathrm{dec}}$
satisfying $e_{\mathrm{prep}}+e_{\mathrm{dec}}<e_*$. Any transfer error introduced below is chosen within the remaining error budget.

For all sufficiently large $\ell$, Proposition~\ref{prop:SM_A1_bounded_preparation}
constructs preparation and decoding unitaries $V_{\mathrm{prep}}$ and $V_{\mathrm{dec}}$ approximating the corresponding ideal operations.
They are generated by time-dependent hopping and density-interaction coefficients that are continuous, satisfy Eq.~\eqref{eq:SM_power_law_bounds}, and vanish at the endpoints of each stage. Their implementation times and errors satisfy
\begin{equation}
t_{\mathrm{prep}}+t_{\mathrm{dec}}
\le C_{\mathrm{pd}}\ell
\label{eq:SM_preparation_assumption_time}
\end{equation}
and
\begin{equation}
\begin{aligned}
\left\|
(V_{\mathrm{prep}}-V_{\mathrm{prep}}^{\mathrm{id}})
\Pi_{\mathrm{in}}
\right\|
\le \frac{e_{\mathrm{prep}}}{2}, \quad
\left\|
(V_{\mathrm{dec}}-V_{\mathrm{dec}}^{\mathrm{id}})
\Pi_{\mathrm{dec}}^{\mathrm{id}}
\right\|
\le \frac{e_{\mathrm{dec}}}{2}.
\end{aligned}
\label{eq:SM_preparation_assumption_accuracy}
\end{equation}
Preparation acts within the selected active blocks, whereas decoding acts only on the destination block containing the readout site $i_{\mathrm{out}}$.
The superscript $\mathrm{id}$ denotes the ideal, error-free operations on the prescribed input subspaces.
The unitaries $V_{\mathrm{prep}}$ and $V_{\mathrm{dec}}$ are their bounded-control implementations, with errors controlled by Eq.~\eqref{eq:SM_preparation_assumption_accuracy}.
The controls are independent of the input amplitudes $c_0,c_1$ in $c_0\ket{1,0}_{i_{\mathrm{in}}}
+c_1\ket{0,1}_{i_{\mathrm{in}}}$.
The constant $C_{\mathrm{pd}}$ may depend on the fixed microscopic parameters, $b_{\mathrm{arr}}$, $C_B$, and error allowances, but not on $t$, $\ell$, $N$, the number of active blocks, or the ambient lattice size.
The ideal maps include the active-site choices and spectator particles for odd $p$.
The bounded-control implementation uses a hierarchy of merges instead of the ideal row sweeps and approximates the same prescribed input--output maps on the relevant subspaces.

The preparation bound in Eq.~\eqref{eq:SM_preparation_assumption_accuracy} holds on
$\operatorname{Ran}(\Pi_{\mathrm{in}})
=\operatorname{span}\{\ket{\Omega},
W_{\mathrm{in}}\ket{\Omega}\}$.
The decoding bound holds on $\operatorname{Ran}(\Pi_{\mathrm{dec}}^{\mathrm{id}})$, where the logical qubit is encoded in the destination packet and all other packet, vacuum, and spectator factors are fixed as specified above.
Both bounds apply to arbitrary coherent superpositions in their respective input subspaces, rather than to the entire Fock space.
They remain valid after adjoining an arbitrary reference system and imply trace-norm errors at most $e_{\mathrm{prep}}$ and $e_{\mathrm{dec}}$, respectively.
Their contribution to the full error budget is evaluated in Sec.~\ref{S-II-D-error}.
The proof of Proposition~\ref{prop:SM_A1_bounded_preparation} accounts for the time required for isolating selected transitions and the accumulated error across the active array.
These implementation bounds do not follow from Eq.~\eqref{eq:SM_inverse_matrix_element_sum} alone.

\subsection{Packet encoding and local operations}
\label{S-II-B}

At the center $x$ of each active block, we use the packet code
\begin{equation}
\ket{0}_x\coloneqq\ket{N,0}_x,
\quad
\ket{1}_x\coloneqq\ket{0,N}_x,
\label{eq:SM_packet_code}
\end{equation}
as illustrated in Fig.~\ref{fig:SM_lower_protocol}(c).
Let $\mathcal Q_x$ denote the span of these two states.
The logical Pauli operators on $\mathcal Q_x$ are
\begin{equation}
Z_x\coloneqq
\ket{0}_x\!\bra{0}-\ket{1}_x\!\bra{1},
\quad
X_x\coloneqq
\ket{0}_x\!\bra{1}+\ket{1}_x\!\bra{0}.
\label{eq:SM_packet_Pauli}
\end{equation}
Both code states contain the same total number $N$ of bosons.
They differ only in whether these bosons occupy mode $a=0$ or mode $a=1$, so the logical information is encoded in the mode occupation rather than the total particle number.

\subsubsection{Coherent encoding and decoding}
\label{S-II-B-encoding}

Choose $N$ sites in each selected complete block $P_x$ and denote their set by $P_x^{\mathrm{act}}$.
For even $p$, take $N=|P_x|$.
For odd $p$, choose
\begin{equation}
N\equiv2\pmod8,
\quad |P_x|-7\le N\le|P_x|,
\label{eq:SM_odd_packet_number}
\end{equation}
which is possible for sufficiently large blocks.
This condition is required for the odd-$p$ packet rotation in Eq.~\eqref{eq:SM_odd_phase_congruence}.
Here a complete block has $|P_x|=(2\ell+1)^d$ sites, whereas $P_x^{\mathrm{act}}$ contains only the sites participating in the protocol.
All selected complete blocks have the same size, so the same $N$ can be used throughout the active array.
The unused sites remain in their initial states and act as spectators.

For odd $p$, place at most seven spectator sites consecutively at the outer end of a boundary row along the first coordinate direction, avoiding the block center and the designated input or readout site. The remaining sites form a connected nearest-neighbor
graph, and each can reach the center by a coordinate path of at most $d\ell$ edges without crossing the spectator sites. Choose a root $i_x\in P_x^{\mathrm{act}}$: $i_{\mathrm{in}}$ in the source block, the readout site $i_{\mathrm{out}}$ in the destination block, and $x$ {in} a block different from the first two.
The graph of active sites therefore admits a nearest-neighbor spanning tree rooted at $i_x$, with depth at most $2d\ell$ and degree at most $2d$.

Before concentration, each active physical site contains one boson in its two modes.
On the subspace $\operatorname{span}\{\ket{1,0}_i,\ket{0,1}_i\}$, the intermode hopping
$b_{i,0}^\dagger b_{i,1}+b_{i,1}^\dagger b_{i,0}$ acts as Pauli $X$, while $n_{i,0}-n_{i,1}$ acts as Pauli $Z$.
Their pulses generate arbitrary single-qubit rotations in ${\cal O}(1)$ time with bounded coefficients.
On the same subspace, $n_{i,0}+n_{i,1}=I$, so a pulse proportional to this operator can correct a common phase when needed.
For neighboring physical sites $i,j$, choose a fixed $0<U_0\le2^{-\alpha}C_U$ and a continuous real pulse envelope $h(u)$ supported on $[0,\tau_{\mathrm{CZ}}]$,
with $h(0)=h(\tau_{\mathrm{CZ}})=0$ and $|h(u)|\le U_0$.
During this interval, the interaction Hamiltonian is $h(u)n_{i,1}n_{j,1}$, with pulse area
\begin{equation}
\int_0^{\tau_{\mathrm{CZ}}}h(u)\,\md u=\pi.
\end{equation}
The resulting unitary is
\begin{equation}
\exp\!\left(-\mi\pi n_{i,1}n_{j,1}\right),
\end{equation}
which acts as a controlled-$Z$ gate when each site contains exactly one boson.
The ordered sum in Eq.~\eqref{eq:SM_restricted_Hamiltonian}
includes both $(i,j)$ and $(j,i)$.
Assigning half of $h(u)$ to each ordered contribution
produces the required interaction.
The term $n_{i,1}n_{j,1}$ belongs to the allowed
polynomial density interactions because its total
occupation degree is $2\le p$.
For example, choosing $\tau_{\mathrm{CZ}}=2\pi/U_0$
and
$h(u)=U_0\sin^2(U_0u/2)$ for $0\le u\le\tau_{\mathrm{CZ}}$,
with $h(u)=0$ otherwise, implements the gate in time $2\pi/U_0$. Conjugating this gate by a Hadamard on the target gives a controlled-NOT gate. These operations preserve one particle per physical site.

Apply the controlled-NOT gates along the nearest-neighbor spanning tree rooted at $i_x$, where $i_x=i_{\mathrm{in}}$ in the source block, $i_x=i_{\mathrm{out}}$ in the
destination block, and $i_x=x$ in the other blocks.
At each tree level, group the parent--child edges according to their $2d$ possible lattice directions.
Within each group, the edges are disjoint, so the corresponding gates can be applied in parallel.
Since the tree has depth at most $2d\ell$, the resulting unitary $\mathsf F_x$ takes ${\cal O}(\ell)$ time.
Starting with all non-root sites in $\ket{1,0}$, the sequence maps either root basis state to the corresponding product state over all active sites.
For an arbitrary superposition at the root, it produces a coherent superposition of these two product states. We denote the two resulting product states, with all active bosons occupying mode $a=0$ or mode $a=1$, respectively, by
\begin{equation}
\ket{\Phi_{0,x}}\coloneqq
\bigotimes_{i\in P_x^{\mathrm{act}}}\ket{1,0}_i,
\quad
\ket{\Phi_{1,x}}\coloneqq
\bigotimes_{i\in P_x^{\mathrm{act}}}\ket{0,1}_i.
\label{eq:SM_block_layer_states}
\end{equation}
Then
\begin{equation}
\mathsf F_x
\left[
\left(c_0\ket{1,0}+c_1\ket{0,1}\right)_{i_x}
\otimes
\bigotimes_{i\in P_x^{\mathrm{act}}\setminus\{i_x\}}
\ket{1,0}_i
\right]
=
c_0\ket{\Phi_{0,x}}+c_1\ket{\Phi_{1,x}},
\label{eq:SM_microscopic_fanout}
\end{equation}
where $c_0,c_1\in\mathbb C$ and $|c_0|^2+|c_1|^2=1$. The spectator sites in $P_x\setminus P_x^{\mathrm{act}}$ remain in $\ket{1,0}$ and are decoupled from the protocol. Their tensor factors are omitted from
Eq.~\eqref{eq:SM_microscopic_fanout}.

We define the vacuum state of the remaining active sites by
\begin{equation}
\ket{\mathrm{vac}}_x\coloneqq
\bigotimes_{i\in P_x^{\mathrm{act}}\setminus\{x\}}\ket{0,0}_i.
\end{equation}
The vacuum factor includes only the active sites other than the block center. Let $\mathsf C_x^{\mathrm{id}}$ denote the ideal concentration operation of Sec.~\ref{S-II-A}.
For the two input states $\ket{\Phi_{0,x}}$ and $\ket{\Phi_{1,x}}$ defined above, this operation concentrates all $N$ bosons into mode $a=0$ or $a=1$ at the block center, respectively, leaving the other active sites empty.
Identical controls act independently on the two modes, with intermode and spectator couplings switched off.
Exchanging the two modes on the active sites interchanges these two input states and commutes with the concentration sequence.
Consequently, both branches acquire the same phase. Fixing this common phase gives
\begin{equation}
\mathsf C_x^{\mathrm{id}}
\left(c_0\ket{\Phi_{0,x}}+c_1\ket{\Phi_{1,x}}\right)
=
\left(c_0\ket{0}_x+c_1\ket{1}_x\right)
\otimes\ket{\mathrm{vac}}_x.
\label{eq:SM_coherent_packet_encoding}
\end{equation}
The empty modes and spectator particles carry no information about
$c_0,c_1$.

The ideal encoder is $\mathsf C_x^{\mathrm{id}}\mathsf F_x$.
Its inverse, $\mathsf F_x^\dagger(\mathsf C_x^{\mathrm{id}})^\dagger$,
returns the signal to $i_x$ and restores the other active sites to $\ket{1,0}$.
At the destination, this gives the local decoding shown in Fig.~\ref{fig:SM_lower_protocol}(d).
These identities also hold for an input entangled with an external system, since the same linear map acts on both input basis states.

Applying the ideal encoders in parallel to all active blocks defines $V_{\mathrm{prep}}^{\mathrm{id}}$ of Sec.~\ref{S-II-A}; all roots except the source initially have state $\ket{1,0}$.
The inverse encoder on the destination block defines $V_{\mathrm{dec}}^{\mathrm{id}}$.
The fanout gates above are exact and require no transition selection.
Uniform implementation of concentration and its inverse, including the
active-site choices for odd $p$, is established for integer $p\ge3$
in Proposition~\ref{prop:SM_A1_bounded_preparation} below.
The implementation times and errors are specified in Eqs.~\eqref{eq:SM_preparation_assumption_time} and
\eqref{eq:SM_preparation_assumption_accuracy}.
The packet rotations constructed below are exact and have
bounded implementation time.

\subsubsection{Exact packet rotations}
\label{S-II-B-rotations}

We next construct rotations on an already prepared packet.
Fix a center $x$ and suppress its position index by writing $b_a={b_{x,a}}$ and $n_a=n_{x,a}$.
Define the number-conserving 50:50 beam splitter
\begin{equation}
\mathsf B\coloneqq
\exp\!\left[
\frac{\pi}{4}\left(b_1^\dagger b_0-b_0^\dagger b_1\right)
\right].
\end{equation}
In the ordered one-particle basis $(\ket{1,0},\ket{0,1})$, its matrix is $2^{-1/2}\left(\begin{smallmatrix}1&-1\\1&1\end{smallmatrix}\right)$.
The parity operator $\me^{\mi\pi n_0}$ has matrix $\operatorname{diag}(-1,1)$, so its conjugation by $\mathsf B$ is the mode-swap matrix.
{The conjugated parity operator $\mathsf B^\dagger \me^{\mi\pi n_0}\mathsf B$ exchanges $b_0^\dagger$ and $b_1^\dagger$ under conjugation and leaves the vacuum invariant.}
Thus, the same identity holds in every particle-number sector:
\begin{equation}
\mathsf B^\dagger \me^{\mi\pi n_0}\mathsf B\ket{m,N-m}_x
=\ket{N-m,m}_x,
\quad 0\le m\le N.
\label{eq:SM_parity_mode_swap}
\end{equation}
On the packet code $\mathcal Q_x$ defined after Eq.~\eqref{eq:SM_packet_code}, this operator acts as the logical Pauli $X_x$ of Eq.~\eqref{eq:SM_packet_Pauli}.

For even $p\ge2$, the fact $m^p\equiv m^2\pmod4$ gives
\begin{equation}
S_p\coloneqq \me^{-\mi\pi n_0^p/2}
=\me^{-\mi\pi n_0^2/2}
=\frac{1-\mi}{2}I+\frac{1+\mi}{2}\me^{\mi\pi n_0}.
\label{eq:SM_even_packet_phase}
\end{equation}
Thus $S_p$ acts as $1$ on even occupations and as $-\mi$ on odd occupations. Combining Eqs.~\eqref{eq:SM_parity_mode_swap} and \eqref{eq:SM_even_packet_phase} yields
\begin{equation}
\left.\mathsf B^\dagger S_p\mathsf B\right|_{\mathcal Q_x}
=\frac{1-\mi}{2}I+\frac{1+\mi}{2}X_x
=\me^{-\mi\pi/4}\me^{\mi\pi X_x/4}.
\end{equation}
This is a logical rotation through an angle of magnitude $\pi/2$, up to an overall phase. Although the beam splitter populates intermediate occupations, the full sequence returns exactly to the packet code.

For odd $p\ge3$, use the occupation choice in Eq.~\eqref{eq:SM_odd_packet_number} and replace $S_p$ by
\begin{equation}
S_p^{\mathrm{odd}}\coloneqq
\exp\!\left[-\frac{\mi\pi}{4}\left(n_0^p+n_1^p\right)\right].
\label{eq:SM_odd_packet_phase}
\end{equation}
On the fixed-$N$ sector, $n_1=N-n_0$.
An even integer raised to a power $p\ge3$ vanishes modulo eight, whereas $m^p\equiv m\pmod8$ for odd $m$ and odd $p$.
Since $N\equiv2\pmod8$, it follows that
\begin{equation}
m^p+(N-m)^p\equiv
\begin{cases}
0\pmod8, & m\text{ even},\\
2\pmod8, & m\text{ odd}.
\end{cases}
\label{eq:SM_odd_phase_congruence}
\end{equation}
The phase in Eq.~\eqref{eq:SM_odd_packet_phase} therefore has the same parity dependence as Eq.~\eqref{eq:SM_even_packet_phase} on the entire fixed-$N$ sector. Since $\mathsf B$ preserves $N$,
\begin{equation}
\left.\mathsf B^\dagger S_p^{\mathrm{odd}}\mathsf B\right|_{\mathcal Q_x}
=\me^{-\mi\pi/4}\me^{\mi\pi X_x/4}.
\end{equation}
The omission of at most seven particles per block leaves $N\asymp\ell^d$ unchanged.

\subsubsection{Number phases and operation times}
\label{S-II-B-times}

Arbitrary logical $Z$ rotations follow from linear number phases:
\begin{equation}
\left.\me^{-\mi\phi n_0/N}\right|_{\mathcal Q_x}
=\me^{-\mi\phi/2}\me^{-\mi\phi Z_x/2}.
\label{eq:SM_packet_Z_rotation}
\end{equation}
The angle $\phi$ may be reduced modulo $2\pi$ on the code.

To make the number of pulses explicit, define $Y_x\coloneqq-\mi\ket0_x\!\bra1+\mi\ket1_x\!\bra0$.
Conjugation by $\me^{\mi\pi X_x/4}$ takes $Z_x$ to $Y_x$.
The Euler decomposition of an arbitrary logical unitary, up to a common phase, can therefore be written as
\begin{equation}
\mathsf U_x
=\me^{-\mi\theta_1Z_x/2}
 \me^{\mi\pi X_x/4}
 \me^{-\mi\theta_2Z_x/2}
 \me^{-\mi\pi X_x/4}
 \me^{-\mi\theta_3Z_x/2},
\label{eq:SM_packet_Euler_decomposition}
\end{equation}
where $\theta_1,\theta_2,\theta_3$ are real angles.
Thus, three $Z$ rotations and two fixed $X$ rotations suffice.

In particular, the logical Hadamard gate is
\begin{equation}
\frac{X_x+Z_x}{\sqrt2}
=-\mi\,\me^{\mi\pi Z_x/4}\me^{\mi\pi X_x/4}\me^{\mi\pi Z_x/4}.
\end{equation}

All these gates use controls allowed by Sec.~\ref{S-I}.
The Hermitian generator $G_{\mathrm B}\coloneqq \mi(b_1^\dagger b_0-b_0^\dagger b_1)$ produces $\mathsf B$ with pulse area $\pi/4$.
Its two hopping coefficients are complex conjugates with equal modulus.
The nonlinear phases have generators $n_0^p$ or $n_0^p+n_1^p$ and pulse areas $\pi/2$ or $\pi/4$, respectively.
These areas are independent of $N$.
The linear phase in Eq.~\eqref{eq:SM_packet_Z_rotation} requires area $\phi/N$, which does not grow with occupation.

For a fixed Hermitian generator $G$, a pulse $f(u)G$ produces
\begin{equation}
U_f=\exp\!\left[-\mi G\int f(u)\,\md u\right].
\end{equation}
A prescribed real pulse area $\vartheta$ is realized exactly by
\begin{equation}
f_{\vartheta}(u)
=\frac{2\vartheta}{\tau_0}
 \sin^2\!\left(\frac{\pi u}{\tau_0}\right),
\quad 0\le u\le\tau_0,
\label{eq:SM_packet_continuous_pulse}
\end{equation}
with $f_{\vartheta}=0$ outside this interval.
The envelope is continuous, vanishes at both endpoints, and satisfies $|f_{\vartheta}(u)|\le2|\vartheta|/\tau_0$.
Reduce the number-phase angles modulo $2\pi$ and choose a fixed
\begin{equation}
\tau_0\ge
\max\left\{
\frac{\pi}{2C_{J}},\frac{4\pi}{C_{U}}
\right\}.
\label{eq:SM_packet_pulse_duration}
\end{equation}
Then every hopping and density coefficient in these pulses satisfies Eq.~\eqref{eq:SM_power_law_bounds}, since the two modes are at the same physical site.
The two terms in $n_0^p+n_1^p$ are driven simultaneously, with the bound applied to each coefficient.
Each fixed $X$ rotation uses three pulses, including the two beam splitters. Equation~\eqref{eq:SM_packet_Euler_decomposition} thus uses
at most nine pulses.
If needed, one further pulse proportional to $n_0+n_1$ fixes the overall phase; its required area is also of order $1/N$.
Consequently,
\begin{equation}
t_{\mathrm{rot}}\le C_{\mathrm{rot}},
\quad C_{\mathrm{rot}}\coloneqq10\tau_0,
\end{equation}
where $C_{\mathrm{rot}}$ is independent of $N$, $\ell$, and system size.
Reversing the signs and order of the pulses implements the inverse gates.

During each packet rotation, all couplings to other physical sites are kept zero. Rotations at different block centers can therefore be performed in parallel.
The construction uses no increase in the control coefficient with occupation and no approximation that selects individual transitions.
These gate identities are exact in the packet code; any errors inherited from preparation remain part of the full error budget.
The finite times of the local rotations must be counted separately from the evolution under the enhanced density interactions.

\subsubsection{{Uniform preparation and microscopic decoding}}
\label{S-II-B-preparation}

For integer $p\ge3$, uniform preparation and microscopic decoding are realized with bounded nearest-neighbor hopping and density-polynomial controls of degree at most three. We retain the active sets, packet code, and ideal input--output maps of Secs.~\ref{S-II-A} and \ref{S-II-B}.
The construction below replaces the ideal row sweeps by a hierarchy of merges; this changes their scheduling, but not the encoding or its local inverse on the specified subspaces.

\begin{proposition}[Uniform preparation on polynomial-size arrays]
\label{prop:SM_A1_bounded_preparation}
Fix an integer $p\ge3$, $d\ge1$, $\alpha>2d$, and constants $b_{\mathrm{arr}}\ge0$ and $C_B>0$.
Let $\Gamma_{\mathrm{act}}\subset\mathcal C$ be the active block centers, with
\begin{equation}
B_{\mathrm{act}}\coloneqq|\Gamma_{\mathrm{act}}|
\le C_B\ell^{b_{\mathrm{arr}}}.
\end{equation}
Fix $0<e_*<1$ and positive error allowances
$e_{\mathrm{prep}}$ and $e_{\mathrm{dec}}$ satisfying
$e_{\mathrm{prep}}+e_{\mathrm{dec}}<e_*$.
For all sufficiently large $\ell$, there exist preparation and microscopic
decoding unitaries satisfying
Eqs.~\eqref{eq:SM_preparation_assumption_time} and
\eqref{eq:SM_preparation_assumption_accuracy}.
Preparation acts separately within the active blocks, and decoding acts only within the destination block.
All controls are independent of the input amplitudes, obey Eq.~\eqref{eq:SM_power_law_bounds}, and vanish at the endpoints of each stage. The constant $C_{\mathrm{pd}}$ may depend on the fixed microscopic parameters, $b_{\mathrm{arr}}$, $C_B$, and the error allowances, but not on $t$, $\ell$, $N$, $B_{\mathrm{act}}$, or the size of the ambient lattice.
The estimates hold for inputs entangled with an arbitrary reference system. Spectator particles remain unchanged.
\end{proposition}

\begin{proof}
\textit{Admissible pulses.---} For each merge, $b_1,b_2$ denote the two neighboring modes in one layer and $n_1,n_2$ their occupations, as in Sec.~\ref{S-II-A}.
Use the fixed hopping amplitude $J_0$ of Eq.~\eqref{eq:SM_merge_hopping} and choose
$0<U_*\le 2^{-\alpha}{C_U}/4$.
For a density generator $G$ whose displayed monomials have coefficients of modulus one, the pulse
\begin{equation}
f_{\vartheta,\tau}(u)G,
\quad
f_{\vartheta,\tau}(u)
=\frac{2\vartheta}{\tau}\sin^2\!\left(\frac{\pi u}{\tau}\right),
\quad
0\le u\le\tau,
\quad
\tau\ge\frac{2|\vartheta|}{U_*},
\label{eq:SM_A1_continuous_pulse}
\end{equation}
implements $\me^{-\mi\vartheta G}$ exactly; the coefficient is zero outside this interval. A zero-area pulse is omitted. The analogous envelope with dimensionless amplitude at most one implements $\me^{-\mi\upsilon K}$, where $K=J_0(b_1^\dagger b_2+b_2^\dagger b_1)$, in physical time $2|\upsilon|$.
Negative areas are implemented by reversing the corresponding coefficients. During concentration, all other couplings are zero.
These operations are permitted by Sec.~\ref{S-I}; no bound on the time derivatives of the coefficients is imposed there. Each finite protocol uses finitely many continuous pulses, although the number of pulses and the required precision of their areas may grow with the scale.

\textit{Fast density phases.---} For a neighboring pair in one layer, let
\begin{equation}
\mathcal H_M\coloneqq
\operatorname{span}\{\ket{m,M-m}:0\le m\le M\},
\quad
\mathcal H_0=\operatorname{span}\{\ket{0,0}\},
\label{eq:SM_A1_number_sectors}
\end{equation}
where $M\ge1$ is known from the prescribed ideal occupation pattern.
On $\mathcal H_M\oplus\mathcal H_0$, define
\begin{equation}
D_M\coloneqq
\left.\exp\!\left[\frac{\mi\pi}{2}n_1(n_1-1)\right]
\right|_{\mathcal H_M\oplus\mathcal H_0}.
\label{eq:SM_A1_sign_operator}
\end{equation}
In the basis $\{\ket{m,M-m}\}_{m=0}^{M}$, $D_M$ is diagonal with eigenvalues
\begin{equation}
(-1)^{m(m-1)/2},
\quad 0\le m\le M,
\end{equation}
which repeat with period four as $+1,+1,-1,-1$.
Choose $\theta_M,\gamma_M\in[-\pi,\pi]$ such that
\begin{equation}
\theta_M\equiv\frac{\pi(M-1)}{2}\pmod{2\pi},
\quad
\gamma_M\equiv\frac{\pi M^2}{6}-\frac{\theta_M M}{2}
\pmod{2\pi}.
\label{eq:SM_A1_phase_angles}
\end{equation}
The physical sequence
\begin{equation}
\mathsf D_M\coloneqq
 \me^{-\mi\gamma_M(n_1+n_2)/M}
 \me^{\mi\theta_M(n_1^2-n_2^2)/(2M)}
 \me^{\mi\pi(n_1^3+n_2^3)/(6M)}
\label{eq:SM_A1_phase_realization}
\end{equation}
restricts exactly to $D_M$.
Indeed, substituting $n_2=M-n_1$ gives the exponent
\begin{equation}
\mi\left[
\frac{\pi}{2}n_1^2+
\left(\theta_M-\frac{\pi M}{2}\right)n_1+
\frac{\pi M^2}{6}-\frac{\theta_M M}{2}-\gamma_M
\right].
\end{equation}
Equation~\eqref{eq:SM_A1_phase_angles} and the integer spectrum of $n_1$ reduce this to the phase in Eq.~\eqref{eq:SM_A1_sign_operator}. Every factor is the identity in a vacuum. The area of each polynomial coefficient is bounded by a constant divided by $M$. Hence $\mathsf D_M$ and its inverse each take at most $C_D/M$ time, with $C_D$ independent of $M$.
This identity is required only on $\mathcal H_M\oplus\mathcal H_0$; the same pulse sequence need not realize it on other number sectors.

On $\mathcal H_M$, the hopping has norm $\|K\|=J_0M$.
Conjugation by $D_M$ multiplies the matrix element from $m$ to $m+1$ by $(-1)^m$. Thus
\begin{equation}
K_\pm\coloneqq\frac12(K\pm D_MKD_M^\dagger)
\label{eq:SM_A1_selected_hoppings}
\end{equation}
retain, respectively, the even and odd starting values of $m$. Each is a direct sum of isolated two-state transitions and possibly isolated endpoints. These are auxiliary generators on the specified sectors, not additional instantaneous Hamiltonians.
Every exponential of a conjugated hopping is implemented as $\mathsf D_M \me^{-\mi\upsilon K}\mathsf D_M^\dagger$, with both density sequences included in its duration.

\textit{Product-formula error and control time.---} We use the even-order Suzuki formulas~\cite{Suzuki1991Fractal,PhysRevX.11.011020}.
For Hermitian matrices $A_1,\ldots,A_{s_0}$ on an invariant finite-dimensional sector, write $A_*=\sum_j\|A_j\|$ and define
\begin{align}
\mathsf S_2(\upsilon)
=\left(\prod_{j=1}^{s_0}\me^{-\mi\upsilon A_j/2}\right)
  \left(\prod_{j=s_0}^{1}\me^{-\mi\upsilon A_j/2}\right), \quad
\mathsf S_{2q}(\upsilon)
=\mathsf S_{2q-2}(a_q\upsilon)^2
  \mathsf S_{2q-2}((1-4a_q)\upsilon)
  \mathsf S_{2q-2}(a_q\upsilon)^2,
\label{eq:SM_A1_Suzuki_recursion}
\end{align}
where $a_q=(4-4^{1/(2q-1)})^{-1}$, $q\ge2$. The second product in $\mathsf S_2$ is ordered in decreasing $j$.
The order $2q$ is independent of the interaction degree $p$.
The formula contains at most $L_q=2s_0 5^{q-1}$ factors $\me^{-\mi c_\mu\upsilon A_{j_\mu}}$, with $|c_\mu|\le1$, and agrees with $\me^{-\mi\upsilon\sum_jA_j}$ through order $2q$.
Set $\lambda_q=\sum_\mu|c_\mu|\|A_{j_\mu}\|$, so that $A_*\le\lambda_q\le L_qA_*$.
For every real $\upsilon$ and integer $r_{\mathrm{PF}}\ge1$,
\begin{equation}
\left\|
\mathsf S_{2q}(\upsilon/r_{\mathrm{PF}})^{r_{\mathrm{PF}}}
-\me^{-\mi\upsilon\sum_j A_j}
\right\|
\le
\frac{2(\lambda_q|\upsilon|)^{2q+1}}
{(2q+1)!\,r_{\mathrm{PF}}^{2q}}.
\label{eq:SM_A1_product_formula_error}
\end{equation}
To prove Eq.~\eqref{eq:SM_A1_product_formula_error}, we regard both the product formula and the exact evolution as functions of the real parameter $\upsilon$. Since all exponential factors are unitary for real $\upsilon$, repeated application of the Leibniz rule gives
\begin{equation}
\left\|
\frac{d^k}{d\upsilon^k}\mathsf S_{2q}(\upsilon)
\right\|
\le \lambda_q^k, 
\quad {\text{whereas}} \quad
\left\|
\frac{d^k}{d\upsilon^k}
\me^{-\mi\upsilon\sum_j A_j}
\right\|
\le A_*^k.
\end{equation}
Since the two Taylor expansions agree through order $2q$, Taylor's theorem with integral remainder yields the corresponding one-step error bound. Applying this bound to the $r_{\mathrm{PF}}$ unitary steps and telescoping their difference gives Eq.~\eqref{eq:SM_A1_product_formula_error}.
In particular, the estimate is independent of the sector dimension.
The density generators in $\mathsf D_M$ are exponentiated exactly; their norms do not enter $A_*$, but their physical pulse durations are counted below.

\textit{A prescribed merge.---} For a prescribed input $\ket{m_0,M-m_0}$, move along the number ladder from $m_0$ to $M$. At the step $m\to m+1$, choose the sign in Eq.~\eqref{eq:SM_A1_selected_hoppings} that retains this transition and use the ideal duration
\begin{equation}
\upsilon_m=\frac{\pi}{2J_0\sqrt{(m+1)(M-m)}},
\quad \mathfrak a_m\coloneqq J_0M\upsilon_m.
\end{equation}
The exact two-state evolution gives the desired transfer with phase $-\mi$. Apply Eq.~\eqref{eq:SM_A1_product_formula_error} to $A_1=K/2$ and $A_2=\pm D_MKD_M^\dagger/2$.
For a fixed $q\ge1$, a step error at most $0<\varepsilon\le1$ is attained with an integer number of repetitions satisfying
\begin{equation}
r_m\le
C_q\mathfrak a_m^{1+1/(2q)}
\varepsilon^{-1/(2q)}+1.
\end{equation}
Here and below $C_q$ denotes constants that may depend on $q$ and the fixed microscopic bounds, but not on $M$ or $\varepsilon$.
The absolute hopping time is at most $C_q\upsilon_m$.
There are at most $C_qr_m$ density conjugations, each of duration at most $C_D/M$.
For $\nu_q=1+1/(2q)<2$, symmetry about the midpoint of the ladder gives
\begin{equation}
\sum_{m=0}^{M-1}\mathfrak a_m^{\nu_q}
\le C_q M^{\nu_q/2}
\sum_{k=1}^{\lceil M/2\rceil}k^{-\nu_q/2}
\le C_q M,
\quad
\sum_{m=0}^{M-1}\upsilon_m\le\frac{2\pi}{J_0}.
\end{equation}
The last bound follows from Eq.~\eqref{eq:SM_inverse_matrix_element_sum}.
Summing all hopping and density-pulse durations, therefore, gives
\begin{equation}
t_{\mathrm{merge}}\le C_q(1+\varepsilon^{-1/(2q)}),
\quad
\left\|(V_{\mathrm{merge}}-V_{\mathrm{merge}}^{\mathrm{id}})
\big|_{\mathcal H_M\oplus\mathcal H_0}\right\|
\le (M-m_0)\varepsilon\le M\varepsilon.
\label{eq:SM_A1_merge_estimate}
\end{equation}
The ideal merge sends the prescribed input to $\ket{M,0}$ up to a known phase. Both operations leave the vacuum unchanged.
If $M=1$, hopping alone suffices; an empty pair requires no operation.
The sequence depends on the known $M$ and $m_0$, not on the input amplitudes. It is not a unitary that sends distinct Fock inputs to the same output. Also, $\varepsilon$ is the error per ladder step, not the error of the entire merge; the latter is allocated globally below.

\textit{Spatial scheduling and uniform accuracy.---} Process the $d$ coordinate directions successively.
Within a row, split each interval into two children whose lengths differ by at most one. After each child has been collected, move its packet through the empty sites until the two packets are adjacent, perform the merge above, and move the result to the chosen site of the parent interval. The routing uses an exact swap of the two bosonic modes,
\begin{equation}
\mathsf{SWAP}_{12}
=\me^{\mi\pi(n_1+n_2)/2}
 \me^{-\mi\pi(b_1^\dagger b_2+b_2^\dagger b_1)/2},
\quad
\mathsf{SWAP}_{12}\ket{m_1,m_2}=\ket{m_2,m_1}.
\end{equation}
Its physical duration is bounded independently of the occupations.
At a level whose interval lengths are at most $s$, routing takes ${\cal O}(s)$ time. Disjoint intervals, different rows, and different blocks are processed in parallel. The interval lengths grow geometrically, so all routing costs ${\cal O}(\ell)$, while the number of layers of nonlinear merges is at most $C_d\log(2+\ell)$.
After each row has been collected, its packet is placed at the central coordinate for the next sweep. For odd $p$, the omitted sites of Sec.~\ref{S-II-B-encoding} only shorten one boundary row in the first sweep; the center remains active, and no route crosses a spectator.

At any merge layer, the sum of the known particle numbers is at most a fixed multiple of $B_{\mathrm{act}}N$, also counting both layers.
Choose an integer upper bound $G_{\mathrm{tot}}\ge1$ on the total number of selected ladder steps such that
\begin{equation}
G_{\mathrm{tot}}\le C_dB_{\mathrm{act}}N\log(2+\ell),
\quad N\le C_d\ell^d.
\label{eq:SM_A1_total_transitions}
\end{equation}
Fix $0<{\varepsilon_{\mathrm{conc}}}<1$ and use $\varepsilon={\varepsilon_{\mathrm{conc}}}/G_{\mathrm{tot}}$ in every step. Shorter simultaneous merges are padded with zero controls.
Equation~\eqref{eq:SM_A1_merge_estimate} bounds the longest merge in a layer, independently of its occupation. Thus
\begin{equation}
t_{\mathrm{conc}}
\le C\ell+C_q\log(2+\ell)
\left(\frac{G_{\mathrm{tot}}}{{\varepsilon_{\mathrm{conc}}}}\right)^{1/(2q)}.
\label{eq:SM_A1_concentration_time}
\end{equation}
Choose one fixed order $2q>b_{\mathrm{arr}}+d$. The second term is then
\begin{equation}
{\cal O}\!\left(
\ell^{(b_{\mathrm{arr}}+d)/(2q)}
[\log(2+\ell)]^{1+1/(2q)}
\right)=o(\ell),
\label{eq:SM_A1_subleading_control_time}
\end{equation}
so $t_{\mathrm{conc}}={\cal O}(\ell)$, including the cost of suppressing all accumulated transition errors.

To check coherence, consider the tensor product of the spaces $\operatorname{span}\{\ket{\Phi_{0,x}},\ket{\Phi_{1,x}}\}$ from Eq.~\eqref{eq:SM_block_layer_states}, with all spectator and inactive factors fixed, and denote its projector by $\Pi_{\mathrm{pol}}$.
Apply identical concentration controls in the two layers.
At each ideal step, a layer is in the prescribed occupied sector or in a vacuum. Concentration preserves the particle number in each layer of each block, so different layer-polarized branches remain orthogonal.
The step estimate therefore holds on their coherent direct sum and with an arbitrary reference system, not only on each basis input separately. Telescoping the errors along the ideal subspaces bounds the total operator-norm error by $G_{\mathrm{tot}}\varepsilon={\varepsilon_{\mathrm{conc}}}$.
Earlier leakage is carried by unitaries and need not lie in the ideal subspace of the next step.

The ideal phase acquired in a block is the same in both of its polarized branches. Correct it at the center with a pulse proportional to $n_{x,0}+n_{x,1}$, with its angle reduced modulo $2\pi$ and divided by $N$. These corrections have duration ${\cal O}(1/N)$ and run in parallel. After correction, the ideal concentration has precisely the action of Eq.~\eqref{eq:SM_coherent_packet_encoding}.
Writing $\mathsf C$ for the implemented concentration and $\mathsf C^{\mathrm{id}}=\bigotimes_{x\in\Gamma_{\mathrm{act}}}\mathsf C_x^{\mathrm{id}}$, with identities on all other factors, gives
\begin{equation}
\left\|(\mathsf C-\mathsf C^{\mathrm{id}})\Pi_{\mathrm{pol}}\right\|
\le{\varepsilon_{\mathrm{conc}}}.
\label{eq:SM_A1_coherent_concentration_error}
\end{equation}
The same argument gives a local error at most ${\varepsilon_{\mathrm{conc}}}$ in each block for an arbitrary root qubit, including a block whose root is initialized in $\ket{1,0}$ in the global protocol.

Prepending the exact tree fanout of Sec.~\ref{S-II-B-encoding} gives $V_{\mathrm{prep}}$ and adds only ${\cal O}(\ell)$ time. For decoding, reverse the local encoder at the destination, including its fanout and phase corrections.
Equation~\eqref{eq:SM_inverse_concentration_error} transfers the local encoder error to its ideal output subspace and hence to $\Pi_{\mathrm{dec}}^{\mathrm{id}}$.
Taking ${\varepsilon_{\mathrm{conc}}}\le\min\{e_{\mathrm{prep}},e_{\mathrm{dec}}\}/2$ proves Eq.~\eqref{eq:SM_preparation_assumption_accuracy}; adding preparation and decoding times proves Eq.~\eqref{eq:SM_preparation_assumption_time}.
Only the ideal restrictions to the specified subspaces must agree; the unitary extensions outside those subspaces need not coincide.
All physical pulses have zero endpoints, so their concatenations and the inverse sequence satisfy the required continuity.
\end{proof}

\textit{Application to the protocol scales.---}
For $2d<\alpha\le2d+1$, the recursive packet construction of Sec.~\ref{S-II-D} uses $B_{\mathrm{act}}=L^d$.
For $\alpha>2d+1$ with $\alpha\ge pd$ and $pd-\alpha+1>0$, the sequential packet construction uses $B_{\mathrm{act}}=L_{\mathrm{cg}}+1$.
In their respective parameter ranges, the array sizes satisfy
\begin{equation}
\begin{aligned}
L^d
={\cal O}\!\left(
t^{d(pd-\alpha+1)/(\alpha-2d)}
\right),\quad
L_{\mathrm{cg}}+1
={\cal O}\!\left(t^{pd-\alpha+1}\right).
\end{aligned}
\end{equation}
For $2d+1<\alpha<pd$, the collective relay uses the active array counted in Eq.~\eqref{eq:SM_A2_relay_array_size}.
That estimate, including its logarithmic factor, gives a fixed polynomial envelope for every $b_{\mathrm{arr}}>1+d(pd-\alpha)/(\alpha-2d)$.
Since $\ell\asymp t$, each array has a fixed polynomial envelope in $\ell$.
For each construction, choose $b_{\mathrm{arr}}$ strictly above the corresponding array-size power.
Then $C_B=1$ is valid at sufficiently large times for every fixed $c_\ell>0$. The order $q$ and $C_{\mathrm{pd}}$ can be fixed before choosing $c_\ell$ small enough for the time budget of Sec.~\ref{S-II-D}.
Reducing $c_\ell$ changes only the large-time threshold for the envelope.
Consequently, Eqs.~\eqref{eq:SM_preparation_assumption_time} and \eqref{eq:SM_preparation_assumption_accuracy} apply to all packet arrays used in the stated propagation protocols for integer $p\ge3$.
No uniform global-error estimate for arbitrarily many independent blocks at a fixed $\ell$ is asserted.

{\textit{The quadratic case.---} When $p=2$, the cubic phase in Eq.~\eqref{eq:SM_A1_phase_realization} is not allowed by the interaction-order restriction.
Instead, concentration can be omitted: for $2d<\alpha\le2d+1$ the input is propagated directly using the microscopic one-particle-layer qubits.}
Apply the exact recursion of Sec.~\ref{S-II-D-recursive} directly to the microscopic one-particle layer qubits of Sec.~\ref{S-II-B-encoding}, with density couplings of fixed strength. For a physical cube of side $r=m^k$, its encoder time obeys
\begin{equation}
u_{\mathrm{mic}}(r)
\le3u_{\mathrm{mic}}(r/m)+C_mr^{\alpha-2d}+C_0.
\end{equation}
A fixed $m$ with $\log_m3<\alpha-2d$ gives $u_{\mathrm{mic}}(r)={\cal O}(r^{\alpha-2d})$, including all local gate times. Encoding and inverse decoding at different roots, therefore, transfer the microscopic input exactly over $r\asymp t^{1/(\alpha-2d)}$.
For $2d<\alpha\le2d+1$, this gives the propagation exponent $1/(\alpha-2d)$ without concentration.
For $\alpha\ge2d+1$, the exact nearest-neighbor swap chain along a one-dimensional path in the $d$-dimensional lattice, described in Sec.~\ref{S-II-E-short-range}, provides a distance of order $t$.
Together, these microscopic constructions yield $\beta_{\mathrm{low}}=\beta_{\mathrm{LR}}$ for $p=2$ and $\alpha>2d$, without additional implementation assumptions.

\subsection{Effective long-range interaction}
\label{S-II-C}

We now couple the packets prepared at distinct block centers {in the packet-based constructions for integer $p\ge3$}, as illustrated in Fig.~\ref{fig:SM_lower_protocol}(c).
The block radius $\ell$ and packet occupation $N$ are fixed by the chosen protocol duration $t$.
We retain the packet code of Eq.~\eqref{eq:SM_packet_code} throughout.

\subsubsection{Density interaction on the packet code}
\label{S-II-C-density}

For distinct active centers $x,y$, choose equal density interactions in the two layers:
\begin{equation}
H_{xy}^{\mathrm{dens}}(u)
\coloneqq
\frac{C_{U}f_{xy}(u)}{(1+|x-y|)^\alpha}
\left[
n_{x,0}^{p/2}n_{y,0}^{p/2}
+
n_{x,1}^{p/2}n_{y,1}^{p/2}
\right],
\end{equation}
where $f_{xy}(u)$ is a real continuous envelope satisfying $|f_{xy}(u)|\le1$ and vanishing at the pulse endpoints.
Here $u$ denotes time during the pulse.
Each unordered pair of centers is included once.
In the ordered sum of Eq.~\eqref{eq:SM_restricted_Hamiltonian}, this
choice is implemented by assigning half of the displayed coefficient
to each of $U_{(x,a),(y,a)}(u)$ and $U_{(y,a),(x,a)}(u)$, for $a=0,1$.
The coefficients therefore satisfy Eq.~\eqref{eq:SM_power_law_bounds}.
During a density pulse, all hopping terms are kept zero, and only the
selected interactions between block centers are retained.

On the code space $\mathcal Q_x$, the local density operators satisfy
\begin{equation}
\left.n_{x,0}^{p/2}\right|_{\mathcal Q_x}
=
\frac{N^{p/2}}{2}(I+Z_x),
\quad
\left.n_{x,1}^{p/2}\right|_{\mathcal Q_x}
=
\frac{N^{p/2}}{2}(I-Z_x).
\end{equation}
Define the available Ising coupling strength by
\begin{equation}
J_{\mathrm{eff}}(P_x,P_y)
\coloneqq
\frac{C_{U}N^p}{2(1+|x-y|)^\alpha}.
\label{eq:SM_effective_coupling_definition}
\end{equation}
Writing $\ket{ab}_{xy}\coloneqq\ket a_x\otimes\ket b_y$ for $a,b\in\{0,1\}$, we obtain
\begin{align}
\left.H_{xy}^{\mathrm{dens}}(u)\right|_{\mathcal Q_x\otimes\mathcal Q_y}
&=
2f_{xy}(u)J_{\mathrm{eff}}(P_x,P_y)
\left(
\ket{00}_{xy}\!\bra{00}
+
\ket{11}_{xy}\!\bra{11}
\right)
\notag\\
&=
f_{xy}(u)J_{\mathrm{eff}}(P_x,P_y)(I+Z_xZ_y).
\label{eq:SM_exact_packet_Ising}
\end{align}
The density interaction preserves every mode occupation, so $\mathcal Q_x\otimes\mathcal Q_y$ is invariant.
Equation~\eqref{eq:SM_exact_packet_Ising} is therefore an exact restriction, not a perturbative effective Hamiltonian.
The identity term contributes only a common phase to the code.

The occupied centers have separation $|x-y|=(2\ell+1)d_{\mathrm{cg}}(P_x,P_y)$.
For distinct centers, $|x-y|\ge1$ and hence $|x-y|\le1+|x-y|\le2|x-y|$.
Consequently,
\begin{equation}
J_{\mathrm{eff}}(P_x,P_y)
\asymp
\frac{N^p}
{(2\ell+1)^\alpha d_{\mathrm{cg}}(P_x,P_y)^\alpha}
\asymp
\frac{t^{pd-\alpha}}{d_{\mathrm{cg}}(P_x,P_y)^\alpha},
\label{eq:SM_effective_coupling_scaling}
\end{equation}
where we used $\ell\asymp t$ and $N\asymp\ell^d$.
The comparison constants are independent of $t$, the pair separation, and system size. They may depend on the fixed microscopic bounds and on $p,d,\alpha,c_\ell$.
With the main-text notation $g_{\mathrm{lr}}\coloneqq t^{pd-\alpha}$, Eq.~\eqref{eq:SM_effective_coupling_scaling} reads $J_{\mathrm{eff}}\asymp g_{\mathrm{lr}}/d_{\mathrm{cg}}^\alpha$.
This scale is fixed for a given protocol; it is not varied during a pulse or at successive recursion levels.

Each packet contributes a factor $N^{p/2}$, whereas the physical separation of the centers contributes $(2\ell+1)^{-\alpha}$.
Thus, concentration changes the interaction scale but not its decay exponent on the coarse-grained lattice.
The distance used in this calculation is the separation between the occupied centers, not the minimum distance between the spatial blocks.
The scaling in Eq.~\eqref{eq:SM_effective_coupling_scaling} describes the strength available in the explicit construction above; it is not a lower bound on arbitrary allowed interaction coefficients.

\subsubsection{Controlled-phase gates}
\label{S-II-C-phases}

The Ising interaction supplies the entangling phases used in the transfer protocol.
On $\mathcal Q_x\otimes\mathcal Q_y$, define
\begin{equation}
\mathsf{CP}_{xy}(\theta)
\coloneqq
\exp\!\left[-\mi\theta\ket{11}_{xy}\!\bra{11}\right]
=
\me^{-\mi\theta/4}
\me^{\mi\theta Z_x/4}
\me^{\mi\theta Z_y/4}
\me^{-\mi\theta Z_xZ_y/4},
\label{eq:SM_packet_controlled_phase}
\end{equation}
where $\theta\in\mathbb R$ is the controlled phase.
The decomposition follows from $\ket{11}_{xy}\!\bra{11}=(I-Z_x-Z_y+Z_xZ_y)/4$.
In particular, $\theta=\pi$ gives a controlled-$Z$ gate.
A density pulse satisfying
\begin{equation}
J_{\mathrm{eff}}(P_x,P_y)
\int_{\mathrm{pulse}}f_{xy}(u)\,\md u
=
\frac{\theta}{4}
\label{eq:SM_controlled_phase_area}
\end{equation}
produces the scalar factor and the Ising factor in Eq.~\eqref{eq:SM_packet_controlled_phase}.

The phase $\theta$ can be reduced modulo $2\pi$ to $[-\pi,\pi]$.
For $\theta\ne0$, choose $\tau_{\mathrm{dens}}\ge|\theta|/[2J_{\mathrm{eff}}(P_x,P_y)]$ and set
\begin{equation}
f_{xy}(u)
=
\frac{\theta}
{2J_{\mathrm{eff}}(P_x,P_y)\tau_{\mathrm{dens}}}
\sin^2\!\left(\frac{\pi u}{\tau_{\mathrm{dens}}}\right),
\quad 0\le u\le\tau_{\mathrm{dens}},
\label{eq:SM_density_pulse_envelope}
\end{equation}
with $f_{xy}=0$ outside the pulse interval.
This envelope is continuous, obeys $|f_{xy}|\le1$, and has integral $\theta/[4J_{\mathrm{eff}}(P_x,P_y)]$.
When $\theta=0$, the gate is omitted.

The local corrections use the linear number phases of Sec.~\ref{S-II-B-times}.
An implementation without an additional common phase is
\begin{equation}
\left.
\exp\!\left[
\frac{\mi\eta}{N}(n_{x,0}-n_{x,1})
\right]
\right|_{\mathcal Q_x}
=
\me^{\mi\eta Z_x},
\quad \eta\in\mathbb R.
\label{eq:SM_packet_Z_correction}
\end{equation}
Indeed, $(n_{x,0}-n_{x,1})/N$ restricts to $Z_x$.
Use the envelope in Eq.~\eqref{eq:SM_packet_continuous_pulse} with area $-\eta/N$ and generator $n_{x,0}-n_{x,1}$.
For $|\eta|\le\pi$, each linear coefficient has magnitude at most $2\pi/(N\tau_0)$, which satisfies Eq.~\eqref{eq:SM_power_law_bounds} with the fixed $\tau_0$ of Eq.~\eqref{eq:SM_packet_pulse_duration}.
Taking $\eta=\theta/4$ at both centers gives the two local factors in Eq.~\eqref{eq:SM_packet_controlled_phase} exactly.
They are applied in parallel after the density pulse.
Therefore, choosing $\tau_{\mathrm{dens}}=|\theta|/[2J_{\mathrm{eff}}(P_x,P_y)]$ gives
\begin{equation}
t_{\mathrm{CP}}(\theta;P_x,P_y)
\le
\frac{|\theta|}{2J_{\mathrm{eff}}(P_x,P_y)}+\tau_0
=
{\cal O}\!\left(J_{\mathrm{eff}}(P_x,P_y)^{-1}+1\right),
\end{equation}
with constants independent of $N$, $\ell$, and system size.

For simultaneous gates, let $\mathcal E$ be a finite nonempty set of distinct unordered pairs, with nonzero phase $\theta_{xy}$ assigned to each pair.
All pair Hamiltonians commute at all times, including when pairs share a center.
Choose a common density-pulse duration at least $\max_{\{x,y\}\in\mathcal E} |\theta_{xy}|/[2J_{\mathrm{eff}}(P_x,P_y)]$ and use Eq.~\eqref{eq:SM_density_pulse_envelope} on each pair.
Each microscopic coefficient still obeys its individual bound.

Combine the local corrections before applying them.
At each participating center, choose $\eta_x$ satisfying
\begin{equation}
\eta_x\equiv
\frac14\sum_{y:\{x,y\}\in\mathcal E}\theta_{xy}
\pmod{2\pi},
\quad |\eta_x|\le\pi.
\end{equation}
Since $\me^{\mi2\pi Z_x}=I$ on $\mathcal Q_x$, this reduction leaves the required logical correction unchanged.
Applying Eq.~\eqref{eq:SM_packet_Z_correction} with $\eta=\eta_x$ at all participating centers completes the product of controlled-phase gates in one further interval of duration $\tau_0$.
Thus, neither the local coefficient bound nor the correction time grows with the number of pairs meeting a center.

No transition selection is needed during these density pulses or local corrections, and the packet code is preserved exactly.
However, the enhanced scale $g_{\mathrm{lr}}$ applies only to the density evolution. The local operations in Sec.~\ref{S-II-B} have their separate time cost, which must be included in the complete transfer protocol.

\subsubsection{Comparison with adjacent occupations}
\label{S-II-C-adjacent}

The entangling strength depends on how strongly the density operators distinguish the logical states. Large occupations alone do not guarantee the $N^p$ enhancement.
For comparison, consider the alternative local basis $\ket{N,N}$ and $\ket{N-1,N+1}$ mentioned in the main text.
Both states contain $2N$ bosons in total.
On this basis, assign the logical $Z$ eigenvalues $+1$ and $-1$ to these two states, respectively.
A diagonal factor with eigenvalues $v_0,v_1$ then has Pauli-$Z$ coefficient $(v_0-v_1)/2$.
For the same density interaction at $f_{xy}=1$, the coefficient of the product of these logical $Z$ operators is
\begin{equation}
J_{\mathrm{eff}}^{\mathrm{adj}}(P_x,P_y)
=
\frac{C_{U}}{4(1+|x-y|)^\alpha}
\Bigl\{
\left[N^{p/2}-(N-1)^{p/2}\right]^2
+
\left[(N+1)^{p/2}-N^{p/2}\right]^2
\Bigr\}.
\end{equation}
Each term is the product of the density contrast at the two centers; the factors of $1/2$ arise from the local Pauli decompositions.

For fixed $p\ge2$ and large $N$,
\begin{equation}
N^{p/2}-(N-1)^{p/2}
\sim
(N+1)^{p/2}-N^{p/2}
\sim
\frac p2 N^{p/2-1}.
\end{equation}
These relations follow from the mean value theorem applied to $s\mapsto s^{p/2}$ on $[N-1,N]$ and $[N,N+1]$.
Consequently,
\begin{equation}
J_{\mathrm{eff}}^{\mathrm{adj}}(P_x,P_y)
\asymp
\frac{N^{p-2}}{(1+|x-y|)^\alpha},
\quad
\frac{J_{\mathrm{eff}}^{\mathrm{adj}}(P_x,P_y)}
{J_{\mathrm{eff}}(P_x,P_y)}
=
{\cal O}(N^{-2}).
\end{equation}
The remaining terms in this restricted Hamiltonian are scalar or act on a single logical qubit. They do not contribute to its $ZZ$ coefficient.

By contrast, the states $\ket{N,0}_x$ and $\ket{0,N}_x$ differ by $N$ particles in either mode, giving a density contrast $N^{p/2}$ at each center and hence the $N^p$ enhancement in Eq.~\eqref{eq:SM_effective_coupling_definition}.
Therefore, the packet-based long-range constructions use this code for preparation, local rotations, interpacket interactions, and decoding. The adjacent-occupation encoding is used in the separate short-range construction of Sec.~\ref{S-II-E-short-range}, whose hopping-based mechanism is not constrained by the density-coupling comparison above.

\subsection{State-transfer protocol, implementation time, and error control}
\label{S-II-D}

We construct state-transfer protocols on the packet code using the
exact operations of Secs.~\ref{S-II-B} and \ref{S-II-C}.
Their preparation and microscopic decoding are implemented by Proposition~\ref{prop:SM_A1_bounded_preparation}.
Throughout this subsection, $p\ge3$ is an integer and $\alpha>2d$.
The interaction scale $g_{\mathrm{lr}}=t^{pd-\alpha}$ is fixed by the
chosen total duration $t$ and is held fixed throughout one protocol.
We first analyze recursive transfer and the sequential packet circuit, then construct a collective relay for $2d+1<\alpha<pd$.
Every construction has a finite duration for its local operations.

The explicitly constructed control pulses vanish at their endpoints.
For completeness, an admissible continuous stage that does not have
this property can be reparametrized without changing its unitary.
For a stage generated by a continuous admissible Hamiltonian $H(u)$
over $0\le u\le\tau$, with $\tau>0$, define
\begin{equation}
\varphi(w)
=\frac{w}{2}
-\frac{\tau}{2\pi}\sin\!\left(\frac{\pi w}{\tau}\right),
\quad
\widetilde H(w)
=\varphi'(w)H(\varphi(w)),
\quad 0\le w\le2\tau.
\end{equation}
Then
\begin{equation}
\varphi(0)=0,\quad
\varphi(2\tau)=\tau,\quad
\varphi'(w)=\frac{1-\cos(\pi w/\tau)}{2}\in[0,1],
\quad
\varphi'(0)=\varphi'(2\tau)=0.
\end{equation}
The corresponding propagators satisfy
$\widetilde U(0,w)=U(0,\varphi(w))$ and hence
$\widetilde U(0,2\tau)=U(0,\tau)$.
Thus, the final unitary and its implementation error are unchanged.
Every Hamiltonian coefficient is multiplied by $\varphi'(w)$, so the
coefficient bounds and spatial supports are preserved, and all
coefficients vanish at both endpoints. This ensures that the control coefficients can be concatenated continuously across successive stages.
If the reparametrization is used, its fixed factor-of-two increase in duration is absorbed into the corresponding implementation-time constant.

\subsubsection{Recursive encoding on the coarse lattice}
\label{S-II-D-recursive}

Let $\Gamma_L$ be the centers of a coarse cube containing $L$ blocks
in each direction.
For each chosen $t$ and $L$, take the finite lattice $\Lambda$ large enough to contain these complete active blocks.
All packets contain $N$ bosons.
For a nonempty set $\Gamma$ of centers, write
\begin{equation}
\ket{\mathbf a}_{\Gamma}
\coloneqq\bigotimes_{x\in\Gamma}\ket a_x,
\quad a=0,1.
\label{eq:SM_collective_code_states}
\end{equation}

For a distinguished center $z\in\Gamma_L$, we construct an encoder
$\mathsf E_{\Gamma_L,z}$ with the action
\begin{equation}
\mathsf E_{\Gamma_L,z}
\left[
(c_0\ket0+c_1\ket1)_z\otimes
\ket{\mathbf0}_{\Gamma_L\setminus\{z\}}
\right]
=
c_0\ket{\mathbf0}_{\Gamma_L}
+c_1\ket{\mathbf1}_{\Gamma_L}.
\label{eq:SM_coarse_encoder}
\end{equation}
Here $|c_0|^2+|c_1|^2=1$.
Empty modes and spectator particles are omitted from the notation
and remain unchanged.
The empty tensor product at $L=1$ is understood as a scalar.
Let $u(L)$ bound the physical duration of this encoder, uniformly in $z$.

We adapt the state-transfer construction of Tran \emph{et al.}~\cite{PhysRevX.11.031016}, retaining the finite time of every
packet rotation.
Fix an integer $m\ge2$, take $L=m^n$ with $n\ge1$, and split $\Gamma_L$ into $m^d$ subcubes $\Gamma^{(0)},\ldots,\Gamma^{(m^d-1)}$ of side $L/m$,
with $z\in\Gamma^{(0)}$.
Write
\begin{equation}
\ket{\pm}_{\Gamma}
\coloneqq
\frac{\ket{\mathbf0}_{\Gamma}
\pm\ket{\mathbf1}_{\Gamma}}{\sqrt2}.
\end{equation}

First, encode the input in $\Gamma^{(0)}$ and prepare $\ket+_{\Gamma^{(j)}}$ in every other subcube.
The latter operation uses a Hadamard on one root packet followed
by the smaller-cube encoder.
These operations run in parallel and take at most $u(L/m)+C_{\mathrm{loc}}$ time.
We may take $C_{\mathrm{loc}}=C_{\mathrm{rot}}$, which bounds a packet Hadamard and also the simultaneous number-phase corrections of Sec.~\ref{S-II-C-phases}.

Next, apply a controlled-$Z$ between the collective qubit of $\Gamma^{(0)}$ and that of every other subcube.
Choose a fixed $c_g>0$ such that
\begin{equation}
a_L\coloneqq c_g\frac{g_{\mathrm{lr}}}{L^\alpha}
\le J_{\mathrm{eff}}(P_x,P_y)
\end{equation}
for every participating pair.
This follows from Eq.~\eqref{eq:SM_effective_coupling_scaling}, since all coarse distances inside $\Gamma_L$ are at most $L-1$.
For pairs connecting $\Gamma^{(0)}$ to another subcube, set $f_{xy}(u)=a_Lr_L(u)/J_{\mathrm{eff}}(P_x,P_y)$; set all other pair envelopes to zero. The common envelope $r_L$ is continuous, satisfies $0\le r_L\le1$, and vanishes at its endpoints.
Thus, every microscopic coefficient obeys Eq.~\eqref{eq:SM_power_law_bounds}.

Each pair of subcubes contains $(L/m)^{2d}$ interacting packet pairs.
Their phases add to the collective code.
For $j\ge1$, $a,b\in\{0,1\}$, and $\theta\in\mathbb R$,
\begin{equation}
\begin{aligned}
\left[
\prod_{x\in\Gamma^{(0)}}\prod_{y\in\Gamma^{(j)}}
\mathsf{CP}_{xy}\!\left(\frac{\theta}{(L/m)^{2d}}\right)
\right]
\left(
\ket{\mathbf a}_{\Gamma^{(0)}}\otimes
\ket{\mathbf b}_{\Gamma^{(j)}}
\right)
=
\me^{-\mi\theta ab}
\ket{\mathbf a}_{\Gamma^{(0)}}\otimes
\ket{\mathbf b}_{\Gamma^{(j)}}.
\end{aligned}
\label{eq:SM_collective_controlled_phase_action}
\end{equation}
Thus $\theta=\pi$ implements a controlled-$Z$ between the two collective qubits.
The factor $(L/m)^{2d}$ counts packet pairs; the occupation factor $N^p$ is already included in $a_L$ through $J_{\mathrm{eff}}$.
By Eq.~\eqref{eq:SM_controlled_phase_area}, the required density-pulse area is
\begin{equation}
a_L\int_0^{\tau_{\mathrm{ph}}}r_L(u)\,\md u
=\frac{\pi}{4(L/m)^{2d}}.
\end{equation}

Taking $r_L(u)=\sin^2(\pi u/\tau_{\mathrm{ph}})$ gives
\begin{equation}
\tau_{\mathrm{ph}}
=\frac{\pi}{2a_L(L/m)^{2d}}
=\frac{\pi m^{2d}}{2c_g}
\frac{L^{\alpha-2d}}{g_{\mathrm{lr}}}.
\label{eq:SM_recursive_phase_time}
\end{equation}
The local corrections are also bounded uniformly in $L$.
Every pair has phase $\pi/(L/m)^{2d}$, so the angles multiplying
$Z_x$ in Eq.~\eqref{eq:SM_packet_Z_correction} are
\begin{equation}
\eta_x=
\begin{cases}
\displaystyle\frac{\pi(m^d-1)}{4(L/m)^d},
&x\in\Gamma^{(0)},\\[6pt]
\displaystyle\frac{\pi}{4(L/m)^d},
&x\in\Gamma^{(j)},\quad j\ge1.
\end{cases}
\end{equation}

Reducing these angles modulo $2\pi$ and applying the corrections in parallel incur at most $C_{\mathrm{loc}}$ cost.
Together with the density pulse, they give the exact product of
controlled-phase gates in Eq.~\eqref{eq:SM_collective_controlled_phase_action}.
The state becomes
\begin{equation}
c_0\ket{\mathbf0}_{\Gamma^{(0)}}
\bigotimes_{j=1}^{m^d-1}\ket+_{\Gamma^{(j)}}
+
c_1\ket{\mathbf1}_{\Gamma^{(0)}}
\bigotimes_{j=1}^{m^d-1}\ket-_{\Gamma^{(j)}}.
\end{equation}
In each target subcube, undo its encoder, apply a Hadamard to its root packet, and reapply the encoder.
These operations send $\ket+_{\Gamma^{(j)}}$ to $\ket{\mathbf0}_{\Gamma^{(j)}}$ and
$\ket-_{\Gamma^{(j)}}$ to $\ket{\mathbf1}_{\Gamma^{(j)}}$.
They take $2u(L/m)+C_{\mathrm{loc}}$ time in parallel.
The source subcube is unchanged.
This completes Eq.~\eqref{eq:SM_coarse_encoder}.
Linearity gives the same action for an input entangled with
an external reference.
Set $\chi\coloneqq\alpha-2d>0$.
The three stages above give
\begin{equation}
u(L)\le3u(L/m)+C_m\frac{L^\chi}{g_{\mathrm{lr}}}+C_0,
\quad u(1)=0,
\label{eq:SM_transfer_recurrence}
\end{equation}
where one may take $C_m=\pi m^{2d}/(2c_g)$ and
$C_0=3C_{\mathrm{loc}}$.
The encoder at $L=1$ is the identity; the first nontrivial scale $L=m$ costs
${\cal O}(1+g_{\mathrm{lr}}^{-1})$.
Every pulse has zero endpoints, so concatenation and reversal preserve continuity of the coefficients.
The recursion uses no further particle concentration: its decoding and re-encoding act on already prepared packets.
Unlike the spin protocol of Ref.~\cite{PhysRevX.11.031016}, which permits arbitrarily fast single-site gates, the constant $C_0$ retains their physical time here.

\subsubsection{Solution of the recurrence and total time}
\label{S-II-D-recursive-time}

Choose $m$ so that $\sigma\coloneqq\log_m3<\chi$.
For $L=m^n$, iteration of Eq.~\eqref{eq:SM_transfer_recurrence} gives
\begin{align}
u(L)
\le
\frac{C_mL^\chi}{g_{\mathrm{lr}}}
\sum_{j=0}^{n-1}\left(\frac3{m^\chi}\right)^j
+C_0\sum_{j=0}^{n-1}3^j
\le
C_1\frac{L^\chi}{g_{\mathrm{lr}}}+C_2L^\sigma,
\end{align}
with $C_1=C_m/(1-3/m^\chi)$ and $C_2=C_0/2$ as sufficient choices.
The constants may depend on the fixed microscopic bounds, $p,d,\alpha,c_\ell$, and $m$, but not on $t,L,N$, or the ambient lattice size.
The first term is the density-evolution cost; the second is the accumulated time of the local operations.

Choose source and destination centers
$z_{\mathrm{in}},z_{\mathrm{out}}\in\Gamma_L$ on opposite faces,
with
$d_{\mathrm{cg}}(P_{z_{\mathrm{in}}},P_{z_{\mathrm{out}}})=L-1$.
The microscopic input and readout sites lie in the corresponding
blocks.
By Eq.~\eqref{eq:SM_coarse_encoder},
\begin{equation}
\mathsf T_{\Gamma_L}
\coloneqq
\mathsf E_{\Gamma_L,z_{\mathrm{out}}}^{\dagger}
\mathsf E_{\Gamma_L,z_{\mathrm{in}}}
\label{eq:SM_encoded_transfer_unitary}
\end{equation}
transfers the source qubit to the destination and returns every
other packet to $\ket0$.
Inverting an encoder has the same duration and coefficient bounds.
Consequently,
\begin{equation}
t_{\mathrm{tr}}(L)\le2u(L)
\le2C_1\frac{L^\chi}{g_{\mathrm{lr}}}+2C_2L^\sigma.
\end{equation}
This transfer is exact for an arbitrary source-qubit state,
including one entangled with a reference, with the other packets
initialized in $\ket0$.
The vacuum and spectator factors are unchanged.

For integer $p\ge3$ and $2d<\alpha\le2d+1$,
we have $pd-\alpha+1>0$.
Choose $m$ large enough that
\begin{equation}
0<\sigma<
\min\left\{\chi,\frac{\chi}{pd-\alpha+1}\right\}.
\end{equation}

For a fixed $c_L>0$, let $L$ be the largest power of $m$
not exceeding $c_L(g_{\mathrm{lr}}t)^{1/\chi}$.
For all sufficiently large $t$ this gives
$L\asymp(g_{\mathrm{lr}}t)^{1/\chi}$ and
\begin{equation}
\frac{L^\chi}{g_{\mathrm{lr}}}\le c_L^\chi t,
\quad
L^\sigma=
{\cal O}\!\left(t^{(pd-\alpha+1)\sigma/\chi}\right)
=o(t).
\end{equation}
The second relation follows from
$(pd-\alpha+1)\sigma/\chi<1$.
Rounding to a power of $m$ changes the length by at most
the fixed factor $m$.
The number of active blocks is
$B_{\mathrm{act}}=L^d={\cal O}(t^{d(pd-\alpha+1)/\chi})$.
Choose a fixed $b_{\mathrm{arr}}>d(pd-\alpha+1)/\chi$.
For every fixed $c_\ell,c_L>0$, the array then satisfies
$B_{\mathrm{act}}\le\ell^{b_{\mathrm{arr}}}$ at sufficiently large $t$.
Proposition~\ref{prop:SM_A1_bounded_preparation}, with $C_B=1$,
provides a constant $C_{\mathrm{pd}}$ that can be fixed before
$c_\ell$ and $c_L$ are chosen, and gives
\begin{equation}
t_{\mathrm{prep}}+t_{\mathrm{dec}}\le C_{\mathrm{pd}}\ell.
\end{equation}

Choose $c_\ell$ so that $C_{\mathrm{pd}}c_\ell<1/3$.
With $c_\ell$ and $m$ fixed, fix $c_g$ and the resulting constants
$C_1,C_2$, then choose $c_L$ so that $2C_1c_L^\chi<1/3$.
The local-operation term $2C_2L^\sigma$ is below $t/3$
for all sufficiently large $t$, and hence
\begin{equation}
t_{\mathrm{prep}}+t_{\mathrm{tr}}+t_{\mathrm{dec}}\le t.
\label{eq:SM_full_protocol_time}
\end{equation}
If the protocol finishes before time $t$, we set $H(u)=0$
for the remainder of the interval.
Thus, the signal is evaluated at the same total time $t$
used to choose $\ell$ and $N$.
The physical separation is of order $(2\ell+1)L$.
This argument includes the endpoint $\alpha=2d+1$ and proves the
near-critical packet-transfer distance without any further
implementation assumption.

\subsubsection{{Sequential packet transfer and local-control cost}}
\label{S-II-D-linear}

For $\alpha>2d+1$, choose a straight path of adjacent coarse
sites, with source and destination at coarse distance
$L_{\mathrm{cg}}\ge1$.
Three controlled-NOT gates implement a swap of adjacent
packet qubits.
By Secs.~\ref{S-II-B} and \ref{S-II-C}, each swap takes at most
a fixed multiple of $1+g_{\mathrm{lr}}^{-1}$.
Therefore, a sequence along the path gives the exact transfer bound
\begin{equation}
t_{\mathrm{tr}}(L_{\mathrm{cg}})
\le C L_{\mathrm{cg}}(1+g_{\mathrm{lr}}^{-1}).
\label{eq:SM_sequential_packet_transfer_time}
\end{equation}
Every packet other than the destination is returned to $\ket0$.
When $g_{\mathrm{lr}}\le1$, the right-hand side is at most
$2C L_{\mathrm{cg}}/g_{\mathrm{lr}}$.
For $\alpha\ge pd$ with $pd-\alpha+1>0$, use
$L_{\mathrm{cg}}=\lfloor c_{\mathrm{seq}}g_{\mathrm{lr}}t\rfloor$.
This path contains
$B_{\mathrm{act}}=L_{\mathrm{cg}}+1={\cal O}(t^{pd-\alpha+1})$
active blocks.
Choose a fixed $b_{\mathrm{arr}}>pd-\alpha+1$ and apply
Proposition~\ref{prop:SM_A1_bounded_preparation} with $C_B=1$.
As above, its constant $C_{\mathrm{pd}}$ can be fixed before
$c_\ell$ and $c_{\mathrm{seq}}$ are chosen.
First choose $c_\ell$ so that $C_{\mathrm{pd}}c_\ell<1/3$.
After fixing the constant $C$ in
Eq.~\eqref{eq:SM_sequential_packet_transfer_time}, choose
$c_{\mathrm{seq}}>0$ so that $2Cc_{\mathrm{seq}}<1/3$.
Since $g_{\mathrm{lr}}\le1$, this gives $t_{\mathrm{tr}}<t/3$
and $t_{\mathrm{prep}}+t_{\mathrm{dec}}<t/3$ for sufficiently large $t$.
Therefore, the full protocol satisfies Eq.~\eqref{eq:SM_full_protocol_time}.
For $pd-\alpha+1\le0$, this choice does not give a growing coarse distance, and no growing-distance claim is based on it.

For $2d+1<\alpha<pd$, however, $g_{\mathrm{lr}}$ grows with $t$. The local operation cost of this sequential circuit is of order $L_{\mathrm{cg}}$ and cannot be absorbed into ${\cal O}(t)$ at $L_{\mathrm{cg}}\asymp g_{\mathrm{lr}}t$.
The constructive collective relay below gives a different admissible
transfer scale in this regime.

The finite time of an isolated packet Hadamard can also be checked directly.
In the fixed-$N$ sector of its two modes, let
\begin{equation}
H_{\mathrm{loc}}(u)
=J(u)b_0^\dagger b_1+J(u)^*b_1^\dagger b_0+D(u),
\quad [D(u),n_0]=0,
\end{equation}
where $|J(u)|\le C_J$ and the total number $N$ is conserved.
For $z(u)=\langle n_0\rangle_u/N$, Cauchy--Schwarz gives
$|\langle b_0^\dagger b_1\rangle_u|
\le\sqrt{\langle n_0\rangle_u\langle n_1\rangle_u}$.
The Heisenberg equation then yields
\begin{equation}
|\dot z(u)|
\le2C_J\sqrt{z(u)[1-z(u)]}.
\end{equation}
A Hadamard sends $\ket{N,0}$ to a state with $z=1/2$,
starting from $z=1$.
Integrating the bound on the total variation of
$2\arcsin\sqrt{z(u)}$ gives
\begin{equation}
t_{\mathrm H}
\ge\frac{1}{2C_J}
\int_{1/2}^{1}\frac{\md z}{\sqrt{z(1-z)}}
=\frac{\pi}{4C_J}.
\end{equation}
This estimate is restricted to the isolated two-mode operation.
It explains why increasing diagonal couplings does not justify
rescaling every gate time in the displayed swap circuit by
$g_{\mathrm{lr}}$. It is not a bound on general bosonic transfer protocols involving other sites.

\subsubsection{{Constructive collective relay}}
\label{S-II-D-relay}

For $2d+1<\alpha<pd$, we use the recursive encoder of Sec.~\ref{S-II-D-recursive} with subdivision factors depending on the recursion level. The elementary gates are unchanged. Larger collective codes are then joined by swaps. Every packet rotation retains its finite physical duration; no arbitrarily fast local gate is used.

\begin{proposition}[Constructive transfer with a polylogarithmic loss]
\label{prop:SM_A2_constructive_transfer}
Fix an integer $p\ge3$, $d\ge1$, $2d+1<\alpha<pd$,
a finite $\xi\ge1$, and $0<e_*<1$.
Define
\begin{equation}
\eta_{\mathrm{rel}}
\coloneqq
\left(1-\frac{1}{\alpha-2d}\right)
\frac{\log3}{\log(\alpha/(2d))}>0.
\end{equation}
There is a constant $c>0$ such that, for every sufficiently large $t$, a finite lattice supports an admissible protocol with the same ideal
endpoint action as the transfer in
Eq.~\eqref{eq:SM_encoded_transfer_unitary}, microscopic output error
satisfying Eq.~\eqref{eq:SM_destination_error}, and separation
\begin{equation}
|i_{\mathrm{in}}-i_{\mathrm{out}}|
\ge c\,\frac{t^{\,2+(pd-\alpha)/(\alpha-2d)}}
{(\log t)^{\eta_{\mathrm{rel}}}}.
\end{equation}
Preparation and microscopic decoding are included in $t$.
The inputs are the Mott pair of Sec.~\ref{S-II-A}, and the controls
are independent of the input amplitudes.
The packet transfer is exact on its ideal input subspace.
The preparation and microscopic decoding errors are uniform for
coherent inputs entangled with an arbitrary reference system.
The ideal transfer returns every other packet to $\ket0$ and leaves
the vacuum and spectator factors unchanged.
The constants are independent of $t$ and the system size.
This construction uses only the explicitly specified bounded controls.
\end{proposition}

\begin{proof}
\textit{Choice of the recursion scales.---}
Write $\chi=\alpha-2d>1$ and take $t$ sufficiently large that
$g_{\mathrm{lr}}>\me$.
Let $r_0=1$ and $r_j=m_jr_{j-1}$, with integer $m_j\ge1$.
Denote by $u_j$ the encoder-time bound at side $r_j$,
uniformly in the chosen root.
The gate sequence of Sec.~\ref{S-II-D-recursive} has the action
in Eq.~\eqref{eq:SM_coarse_encoder} at each level.
For $m_j\ge2$, Eqs.~\eqref{eq:SM_recursive_phase_time} and
\eqref{eq:SM_transfer_recurrence} give
\begin{equation}
u_j\le3u_{j-1}
+\frac{\pi}{2c_g}
\frac{m_j^\alpha r_{j-1}^{\chi}}{g_{\mathrm{lr}}}+C_0,
\quad u_0=0.
\label{eq:SM_A2_encoder_recurrence}
\end{equation}
Here $m_j^{2d}r_j^\chi=m_j^\alpha r_{j-1}^\chi$.
The constants $c_g$ and $C_0=3C_{\mathrm{loc}}$
are those of Sec.~\ref{S-II-D-recursive}.
In particular, $C_0$ is independent of $m_j$: combine the number-phase corrections at each packet, reduce their angles modulo $2\pi$, and apply them in parallel.
If $m_j=1$, omit the level and reuse the preceding encoder;
Eq.~\eqref{eq:SM_A2_encoder_recurrence} remains an upper estimate.
Set
\begin{equation}
\vartheta_{\mathrm{rec}}\coloneqq\frac{2d}{\alpha}\in(0,1),
\quad
\kappa_{\mathrm{enc}}\coloneqq
\frac{\log3}{\log(1/\vartheta_{\mathrm{rec}})},
\quad
n_{\mathrm{lev}}\coloneqq
\left\lceil
\frac{\log\log g_{\mathrm{lr}}}
{\log(1/\vartheta_{\mathrm{rec}})}
\right\rceil.
\end{equation}
For $0\le j\le n_{\mathrm{lev}}$, define
$w_j=3^j\vartheta_{\mathrm{rec}}^{\,n_{\mathrm{lev}}-j}$.
Starting from $r_0=1$, choose
\begin{equation}
m_j=\max\!\left\{1,
\left\lfloor
\left(
\frac{g_{\mathrm{lr}}w_j}{r_{j-1}^{\chi}}
\right)^{1/\alpha}
\right\rfloor
\right\},
\quad
r_j=m_jr_{j-1},
\quad 1\le j\le n_{\mathrm{lev}}.
\label{eq:SM_A2_collective_scale}
\end{equation}
At each nontrivial level, $m_j^d$ congruent child cubes tile their parent exactly.
The entire sequence of subdivisions is chosen before the protocol and is independent of the input state.
For sufficiently large $g_{\mathrm{lr}}$,
$g_{\mathrm{lr}}w_0\ge1$.
The invariant
\begin{equation}
r_j^\chi\le g_{\mathrm{lr}}w_j
\label{eq:SM_A2_scale_invariant}
\end{equation}
holds by induction.
Indeed, $w_j/w_{j-1}=3/\vartheta_{\mathrm{rec}}>1$, so the quantity inside the floor in Eq.~\eqref{eq:SM_A2_collective_scale} is at least one, and
\begin{equation}
r_j\le
(g_{\mathrm{lr}}w_j)^{1/\alpha}
r_{j-1}^{\vartheta_{\mathrm{rec}}}
\le(g_{\mathrm{lr}}w_j)^{1/\chi}.
\end{equation}
At each nontrivial level, the choice of $m_j$ gives $m_j^\alpha r_{j-1}^\chi/g_{\mathrm{lr}}\le w_j$, so the density duration is at most $\pi w_j/(2c_g)$.
If $m_j=1$, the omitted level has zero duration.
Iterating Eq.~\eqref{eq:SM_A2_encoder_recurrence} gives
\begin{equation}
u_{n_{\mathrm{lev}}}
\le
C\sum_{j=1}^{n_{\mathrm{lev}}}
3^{n_{\mathrm{lev}}-j}(1+w_j)
\le C_{\mathrm{enc}}3^{n_{\mathrm{lev}}}.
\end{equation}
For the weighted part of this sum,
$3^{n_{\mathrm{lev}}-j}w_j
=3^{n_{\mathrm{lev}}}
\vartheta_{\mathrm{rec}}^{\,n_{\mathrm{lev}}-j}$, so the bound is uniform in the number of levels.

Integer rounding also preserves the required group size.
Using $\lfloor x\rfloor\ge x/2$ for $x\ge1$ gives
\begin{equation}
\log r_j\ge
\vartheta_{\mathrm{rec}}\log r_{j-1}
+\frac{\log g_{\mathrm{lr}}+\log w_j}{\alpha}
-\log2.
\end{equation}
Writing $r=r_{n_{\mathrm{lev}}}$, iteration yields
\begin{align}
\log r\ge{}
\frac{
(\log g_{\mathrm{lr}}+n_{\mathrm{lev}}\log3)
(1-\vartheta_{\mathrm{rec}}^{\,n_{\mathrm{lev}}})
}{\chi}
-\frac{\log3-\log\vartheta_{\mathrm{rec}}}{\alpha}
\sum_{l=0}^{n_{\mathrm{lev}}-1}
l\vartheta_{\mathrm{rec}}^l
-\log2\sum_{l=0}^{n_{\mathrm{lev}}-1}
\vartheta_{\mathrm{rec}}^l.
\end{align}
Both sums are bounded independently of $n_{\mathrm{lev}}$.
Moreover,
$(\log g_{\mathrm{lr}})
\vartheta_{\mathrm{rec}}^{\,n_{\mathrm{lev}}}\le1$
and
$n_{\mathrm{lev}}
\vartheta_{\mathrm{rec}}^{\,n_{\mathrm{lev}}}$
is bounded.
Combining this estimate with
Eq.~\eqref{eq:SM_A2_scale_invariant} gives
\begin{equation}
c_r(g_{\mathrm{lr}}3^{n_{\mathrm{lev}}})^{1/\chi}
\le r\le
(g_{\mathrm{lr}}3^{n_{\mathrm{lev}}})^{1/\chi},
\quad
(\log g_{\mathrm{lr}})^{\kappa_{\mathrm{enc}}}
\le3^{n_{\mathrm{lev}}}
\le3(\log g_{\mathrm{lr}})^{\kappa_{\mathrm{enc}}},
\label{eq:SM_A2_collective_scale_bounds}
\end{equation}
with $c_r>0$ independent of $t$.

\textit{Collective gates and the relay.---}
Take adjacent disjoint coarse cubes $\Gamma,\Gamma'$,
each with $r^d$ packets, and use the collective states
$\ket{\mathbf a}_\Gamma$ of
Eq.~\eqref{eq:SM_collective_code_states}.
Every inter-cube coarse distance is at most $2r$.
Choose equal available Ising strengths
$a_r=c_g g_{\mathrm{lr}}/(2r)^\alpha$,
decreasing the fixed $c_g$ if necessary.
The pulse construction of Sec.~\ref{S-II-C-phases} implements
\begin{equation}
\left[
\prod_{x\in\Gamma}\prod_{y\in\Gamma'}
\mathsf{CP}_{xy}\!\left(\frac{\pi}{r^{2d}}\right)
\right]
\left(
\ket{\mathbf a}_\Gamma\otimes\ket{\mathbf b}_{\Gamma'}
\right)
=
(-1)^{ab}
\ket{\mathbf a}_\Gamma\otimes\ket{\mathbf b}_{\Gamma'},
\quad a,b\in\{0,1\}.
\end{equation}
Its density duration is
$\pi/(2a_r r^{2d})=C r^\chi/g_{\mathrm{lr}}$.
Each packet participates in $r^d$ pairs, giving a local
$Z$ correction of angle $\pi/(4r^d)$.
These corrections run within a single fixed parallel interval.
Thus, the full collective controlled-$Z$ takes at most
$C(1+r^\chi/g_{\mathrm{lr}})$ time and obeys the microscopic
coefficient bounds throughout.

For a root $z\in\Gamma$, write
$\mathsf H_z=(X_z+Z_z)/\sqrt2$.
Let $\mathsf E_{\Gamma,z}$ now denote the encoder with
the subdivision schedule above;
its action is still Eq.~\eqref{eq:SM_coarse_encoder}.
On the collective code, a Hadamard is realized by
\begin{equation}
\mathsf H_\Gamma=
\mathsf E_{\Gamma,z}
(\mathsf H_z\otimes I_{\Gamma\setminus\{z\}})
\mathsf E_{\Gamma,z}^\dagger
\quad\text{on }\operatorname{span}
\{\ket{\mathbf0}_\Gamma,\ket{\mathbf1}_\Gamma\}.
\end{equation}
It costs at most
$2u_{n_{\mathrm{lev}}}+C_{\mathrm{rot}}$.
Conjugating the collective controlled-$Z$ by the target
Hadamard gives a collective CNOT.
Three such CNOTs give an exact collective swap, including
all decoding, rotation, and reencoding intervals, in time
\begin{equation}
t_{\mathrm{swap}}^{\mathrm{coll}}
\le
C\left(
1+u_{n_{\mathrm{lev}}}
+\frac{r^\chi}{g_{\mathrm{lr}}}
\right)
\le C_{\mathrm{coll}}3^{n_{\mathrm{lev}}}.
\end{equation}

Arrange $K+1$ such cubes in a row, with roots in the same
relative position.
Encode the source packet into the first cube;
all other cubes are initially in their collective zero state.
Apply $K$ successive collective swaps and decode the last
cube at its root.
This transfers an arbitrary packet qubit exactly, also when
it is entangled with a reference, and returns all other
packets to $\ket0$.
Vacuum and spectator factors are unchanged.
Including both endpoint encoders in a fixed constant gives
\begin{equation}
L_{\mathrm{cg}}=Kr,
\quad
t_{\mathrm{tr}}\le
C_{\mathrm{step}}(K+1)3^{n_{\mathrm{lev}}}.
\end{equation}
For these separations,
Eq.~\eqref{eq:SM_A2_collective_scale_bounds} implies
a sufficient time estimate
\begin{equation}
t_{\mathrm{tr}}\le
C\left[
(\log g_{\mathrm{lr}})^{\kappa_{\mathrm{enc}}}
+
\frac{L_{\mathrm{cg}}}{g_{\mathrm{lr}}^{1/\chi}}
(\log g_{\mathrm{lr}})^{\eta_{\mathrm{rel}}}
\right],
\quad
\eta_{\mathrm{rel}}
=\kappa_{\mathrm{enc}}(1-1/\chi).
\end{equation}

\textit{Preparation, microscopic readout, and the final distance.---}
Fix positive $e_{\mathrm{prep}},e_{\mathrm{dec}}$ with
$e_{\mathrm{prep}}+e_{\mathrm{dec}}<e_*$.
Choose $b_{\mathrm{arr}}>1+(pd-\alpha)d/\chi$ and set $C_B=1$.
The array-size estimate below verifies this polynomial envelope for
all sufficiently large $t$ after the remaining constants are fixed.
Apply Proposition~\ref{prop:SM_A1_bounded_preparation} to obtain
$C_{\mathrm{pd}}$, and choose $c_\ell>0$ so that
$C_{\mathrm{pd}}c_\ell<1/3$.
With $c_\ell$ fixed, fix $c_g$ and the resulting relay constants,
including $C_{\mathrm{step}}$.
Now choose
\begin{equation}
K=
\left\lfloor
\frac{t}{6C_{\mathrm{step}}3^{n_{\mathrm{lev}}}}
\right\rfloor.
\label{eq:SM_A2_relay_length}
\end{equation}
Since $3^{n_{\mathrm{lev}}}$ grows only as a power of $\log g_{\mathrm{lr}}=(pd-\alpha)\log t$, one has $K\to\infty$, and the packet transfer takes less than $t/3$ for sufficiently large $t$.
The number of active physical blocks satisfies
\begin{equation}
B_{\mathrm{act}}=(K+1)r^d
\le
C t\,g_{\mathrm{lr}}^{d/\chi}
\left(3^{n_{\mathrm{lev}}}\right)^{d/\chi-1}.
\label{eq:SM_A2_relay_array_size}
\end{equation}
The last factor must be retained;
it can grow when $d>\chi$.
Nevertheless, for any fixed $b_{\mathrm{arr}}>1+(pd-\alpha)d/\chi$, the array obeys $B_{\mathrm{act}}\le\ell^{b_{\mathrm{arr}}}$ at sufficiently large $t$ for every fixed $c_\ell>0$.
This verifies the envelope used above without changing the chosen $C_{\mathrm{pd}}$.
As in Sec.~\ref{S-II-B-preparation}, decreasing the fixed $c_\ell$ changes the threshold for this envelope, not its exponent.
The relay constants may depend on $c_\ell$, but not on $t$.
Hence $t_{\mathrm{prep}}+t_{\mathrm{dec}}<t/3$ and $t_{\mathrm{tr}}<t/3$ for sufficiently large $t$, so the full protocol satisfies Eq.~\eqref{eq:SM_full_protocol_time}.

Choose the microscopic input and readout sites at the first and last packet roots.
Their separation is exactly $(2\ell+1)Kr$, and Eqs.~\eqref{eq:SM_A2_collective_scale_bounds} and
\eqref{eq:SM_A2_relay_length} give
\begin{align}
|i_{\mathrm{in}}-i_{\mathrm{out}}|
\ge
c t^2g_{\mathrm{lr}}^{1/\chi}
\left(3^{n_{\mathrm{lev}}}\right)^{-(1-1/\chi)}
\ge
c'\frac{t^{\,2+(pd-\alpha)/\chi}}
{(\log t)^{\eta_{\mathrm{rel}}}}.
\end{align}
The packet transfer is exact, so $e_{\mathrm{tr}}=0$.
With the error allowances fixed above,
Proposition~\ref{prop:SM_A1_bounded_preparation} and Sec.~\ref{S-II-D-error} then give the microscopic output estimate, uniformly over the coherent input subspace and with a reference adjoined.
The ideal packet endpoint action is the same as for recursive and sequential transfers, though the subdivision schedule differs.
All pulses have zero endpoints.
Setting all coefficients to zero for the remaining time
preserves the final state and continuity.

The initial Mott pair satisfies Eq.~\eqref{eq:SM_lower_density_check} for every fixed finite $\xi$ and any fixed $\lambda\ge3^d$.
The separation divided by $\ell$ tends to infinity.
Taking $R$ to be half this separation therefore ensures Eq.~\eqref{eq:SM_input_outside_enlargement} for sufficiently large $t$.
The readout and trace-norm argument of Sec.~\ref{S-II-D-witness} gives a signal at least
$2-2e_*$ and the same uniform local-approximation lower bound $1-e_*$.
\end{proof}

\subsubsection{Preparation, decoding, and accumulated error}
\label{S-II-D-error}

{
We propagate the implementation errors through the full packet protocol.
For each of the recursive, sequential, and collective-relay constructions,
let $V_{\mathrm{tr}}$ denote the implemented packet-transfer unitary.
Let $V_{\mathrm{tr}}^{\mathrm{id}}$ be an ideal unitary with the specified
transfer action on the ideal input subspace.
This subspace has a projector
}
\begin{equation}
\Pi_{\mathrm{tr}}^{\mathrm{id}}
\coloneqq
V_{\mathrm{prep}}^{\mathrm{id}}\Pi_{\mathrm{in}}
(V_{\mathrm{prep}}^{\mathrm{id}})^\dagger.
\label{eq:SM_transfer_input_projector}
\end{equation}
Only the action on this ideal input subspace is needed.
Every constructed packet transfer satisfies
$(V_{\mathrm{tr}}-V_{\mathrm{tr}}^{\mathrm{id}})
\Pi_{\mathrm{tr}}^{\mathrm{id}}=0$.
No agreement outside that subspace is required.
Let $V_j^{\mathrm{id}}$ and $V_j$ be ideal and implemented
unitaries, respectively, for $1\le j\le n_{\mathrm{op}}$.
For $b=0,1$, define
\begin{equation}
\rho_j^{(b),\mathrm{id}}
=
V_j^{\mathrm{id}}\cdots V_1^{\mathrm{id}}
\rho_{\mathrm{in}}^{(b)}
(V_1^{\mathrm{id}})^\dagger\cdots(V_j^{\mathrm{id}})^\dagger,
\quad
\rho_0^{(b),\mathrm{id}}=\rho_{\mathrm{in}}^{(b)},
\end{equation}
and
\begin{equation}
\rho_j^{(b)}
\coloneqq
V_j\cdots V_1\rho_{\mathrm{in}}^{(b)}
V_1^\dagger\cdots V_j^\dagger,
\quad
\rho_0^{(b)}=\rho_{\mathrm{in}}^{(b)}.
\end{equation}
The same implemented sequence is used for both inputs.
Set
\begin{equation}
\epsilon_j\coloneqq
\max_{b=0,1}
\left\|
V_j\rho_{j-1}^{(b),\mathrm{id}}V_j^\dagger
-
V_j^{\mathrm{id}}\rho_{j-1}^{(b),\mathrm{id}}
(V_j^{\mathrm{id}})^\dagger
\right\|_1.
\end{equation}
All states include the other blocks and ancillary modes.
Adding and subtracting
$V_j\rho_{j-1}^{(b),\mathrm{id}}V_j^\dagger$ gives
\begin{equation}
\|\rho_j^{(b)}-\rho_j^{(b),\mathrm{id}}\|_1
\le
\|\rho_{j-1}^{(b)}-\rho_{j-1}^{(b),\mathrm{id}}\|_1
+\epsilon_j.
\end{equation}
Iteration yields
\begin{equation}
\max_{b=0,1}
\left\|
\rho_{n_{\mathrm{op}}}^{(b)}
-\rho_{n_{\mathrm{op}}}^{(b),\mathrm{id}}
\right\|_1
\le
\sum_{j=1}^{n_{\mathrm{op}}}\epsilon_j.
\label{eq:SM_full_error_telescope}
\end{equation}
This argument does not require the implemented state to remain in the ideal subspace.
Unitary invariance preserves the trace norm of an error accumulated earlier, including leakage outside the packet code.
Use $\Pi_{\mathrm{in}}$ from Sec.~\ref{S-II-A}, and let $\Pi_{j-1}^{\mathrm{id}}$ be its image under the preceding ideal operations.
A sufficient condition is
\begin{equation}
\left\|
(V_j-V_j^{\mathrm{id}})\Pi_{j-1}^{\mathrm{id}}
\right\|
\le\delta_j,
\quad \epsilon_j\le2\delta_j.
\end{equation}

Indeed, for any density matrix $\rho$ supported on this
subspace,
\begin{equation}
V_j\rho V_j^\dagger
-V_j^{\mathrm{id}}\rho(V_j^{\mathrm{id}})^\dagger
=
(V_j-V_j^{\mathrm{id}})\rho V_j^\dagger
+
V_j^{\mathrm{id}}\rho
\bigl[V_j^\dagger-(V_j^{\mathrm{id}})^\dagger\bigr],
\end{equation}
and each term has trace norm at most $\delta_j$.
The estimate remains valid after tensoring with an arbitrary reference system.
It requires no approximation on the full Fock space.

Apply Proposition~\ref{prop:SM_A1_bounded_preparation} by taking
preparation, transfer, and decoding as three operations in
Eq.~\eqref{eq:SM_full_error_telescope}.
Their ideal input projectors are $\Pi_{\mathrm{in}}$,
$\Pi_{\mathrm{tr}}^{\mathrm{id}}$, and $\Pi_{\mathrm{dec}}^{\mathrm{id}}$,
respectively.
The definition in Eq.~\eqref{eq:SM_transfer_input_projector} applies to all three packet-transfer constructions.
Each ideal transfer returns every packet other than the destination to $\ket0$, so
\begin{equation}
V_{\mathrm{tr}}^{\mathrm{id}}
\Pi_{\mathrm{tr}}^{\mathrm{id}}
(V_{\mathrm{tr}}^{\mathrm{id}})^\dagger
=
\Pi_{\mathrm{dec}}^{\mathrm{id}}.
\end{equation}
The preparation and decoding bounds in
Eq.~\eqref{eq:SM_preparation_assumption_accuracy}
therefore give errors at most $e_{\mathrm{prep}}$ and
$e_{\mathrm{dec}}$.
For the recursive transfer, take
$V_{\mathrm{tr}}=V_{\mathrm{tr}}^{\mathrm{id}}=\mathsf T_{\Gamma_L}$
on the specified subspace.
The exact sequential packet circuit and the constructive collective relay also agree with their ideal transfers on $\operatorname{Ran}\Pi_{\mathrm{tr}}^{\mathrm{id}}$.
Thus $e_{\mathrm{tr}}=0$ for each packet transfer considered here.
Consequently, the sufficient total budget is
\begin{equation}
e_{\mathrm{prep}}+e_{\mathrm{tr}}+e_{\mathrm{dec}}
\le e_*<1.
\label{eq:SM_protocol_error_budget}
\end{equation}
The constructive preparation estimate and the exact transfer circuits therefore place the final full state within trace norm $e_*$ of its ideal counterpart, uniformly over the specified coherent inputs.

A bound on an entire stage is not a bound on the sum of the errors
of an arbitrary subdivision of that stage.
The proof of Proposition~\ref{prop:SM_A1_bounded_preparation}
controls the accumulated error of all selected transitions by
Eqs.~\eqref{eq:SM_A1_total_transitions}--\eqref{eq:SM_A1_coherent_concentration_error}.
Its $O(\ell)$ preparation and decoding time includes the cost of
transition isolation under the uniform coefficient bounds.
No separate preparation assumption is used in this error estimate.

\subsubsection{Local readout and the localization witness}
\label{S-II-D-witness}

Let $\rho_{\mathrm{out}}^{(b)}$ be the reduced state at
$i_{\mathrm{out}}$ after the full protocol.
The ideal outputs are
\begin{equation}
\rho_{\mathrm{out}}^{(0),\mathrm{id}}
=
\ket{1,0}_{i_{\mathrm{out}}}\!\bra{1,0},
\quad
\rho_{\mathrm{out}}^{(1),\mathrm{id}}
=
\ket{0,1}_{i_{\mathrm{out}}}\!\bra{0,1}.
\end{equation}
By Eq.~\eqref{eq:SM_protocol_error_budget}
and contractivity under partial trace,
\begin{equation}
\left\|
\rho_{\mathrm{out}}^{(b)}
-\rho_{\mathrm{out}}^{(b),\mathrm{id}}
\right\|_1
\le e_*<1,
\quad b=0,1.
\label{eq:SM_destination_error}
\end{equation}
We use the full trace norm, without a factor of $1/2$.
Define the bounded readout
\begin{equation}
A_{\mathrm{out}}\coloneqq
\ket{1,0}_{i_{\mathrm{out}}}\!\bra{1,0}
-
\ket{0,1}_{i_{\mathrm{out}}}\!\bra{0,1},
\label{eq:SM_destination_readout}
\end{equation}
extended by zero on the other local particle-number sectors.
It has norm one and commutes with the total occupation
at $i_{\mathrm{out}}$, so $A_{\mathrm{out}}\in\mathcal A_{\{i_{\mathrm{out}}\}}^0$.
Its two ideal expectations are $+1$ and $-1$.
Trace-norm duality and Eq.~\eqref{eq:SM_destination_error} then give
\begin{equation}
\left|
\operatorname{Tr}\!\left[
A_{\mathrm{out}}
(\rho_{\mathrm{out}}^{(0)}-\rho_{\mathrm{out}}^{(1)})
\right]\right|
\ge2-2e_*.
\label{eq:SM_nonvanishing_destination_signal}
\end{equation}

Set $X=\{i_{\mathrm{out}}\}$.
Every block entering $X_R$ contains a site within distance $R$
of $i_{\mathrm{out}}$ and has diameter at most $2\ell$.
Thus,
\begin{equation}
X_R\subset B_{R+2\ell}(i_{\mathrm{out}}),
\end{equation}
and hence
\begin{equation}
R+2\ell<|i_{\mathrm{in}}-i_{\mathrm{out}}|
\quad\Longrightarrow\quad
i_{\mathrm{in}}\notin X_R.
\label{eq:SM_input_outside_enlargement}
\end{equation}
Choose $R$ to satisfy this condition.
The two inputs differ by a unitary supported on
$i_{\mathrm{in}}$, so they have identical reduced states
on $X_R$.
Every $\widetilde A\in\mathcal A_{X_R}$ therefore satisfies
\begin{equation}
\operatorname{Tr}\!\left[
\widetilde A
(\rho_{\mathrm{in}}^{(0)}-\rho_{\mathrm{in}}^{(1)})
\right]=0.
\label{eq:SM_same_reduced_initial_states}
\end{equation}
Write $\Delta=\tau_{0,t}(A_{\mathrm{out}})-\widetilde A$
for this calculation.
Passing Eq.~\eqref{eq:SM_nonvanishing_destination_signal}
to the Heisenberg picture and using
Eq.~\eqref{eq:SM_same_reduced_initial_states},
\begin{equation}
2-2e_*
\le
\left|
\operatorname{Tr}\!\left[
\Delta(\rho_{\mathrm{in}}^{(0)}-\rho_{\mathrm{in}}^{(1)})
\right]\right|
\le
\sum_{b=0,1}
\left\|\rho_{\mathrm{in}}^{(b)}\Delta\right\|_1.
\end{equation}
Consequently, for every common local approximant,
\begin{equation}
\max_{b=0,1}
\left\|
\rho_{\mathrm{in}}^{(b)}
\left[
\tau_{0,t}(A_{\mathrm{out}})-\widetilde A
\right]
\right\|_1
\ge1-e_*.
\end{equation}
The inequality holds for every bounded $\widetilde A$ on $X_R$, and in particular for those with $\|\widetilde A\|\le1$.
The two inputs lie in the same $\mathcal D_\xi^\lambda$ for every fixed finite $\xi$ and any fixed $\lambda\ge3^d$.
For the constructed packet protocols, the error in Eq.~\eqref{eq:SM_final_uniform_upper_bound} cannot tend to zero at these radii.

For recursive transfer, $|i_{\mathrm{in}}-i_{\mathrm{out}}|\asymp(2\ell+1)L$ as $L\to\infty$.
For sequential packet transfer or collective relay, the corresponding
coarse source--destination distance is $L_{\mathrm{cg}}$, and
$|i_{\mathrm{in}}-i_{\mathrm{out}}|\asymp(2\ell+1)L_{\mathrm{cg}}$.
Each protocol used for a growing-distance lower bound has
$L\to\infty$ or $L_{\mathrm{cg}}\to\infty$.
One may therefore take $R$ to be a sufficiently small fixed fraction
of the physical separation.
The correction $2\ell$ is negligible relative to that separation, so Eq.~\eqref{eq:SM_input_outside_enlargement} holds for large scales.
Rounding to whole blocks does not change the leading power.
This proves the required witness for uniform local approximation, without an operator-norm approximation on the full bosonic Fock space.

\subsection{{Lower-bound exponents and short-range propagation}}
\label{S-II-E}

We now extract the leading light-cone powers from the preceding estimates.
Throughout this subsection, $p\ge2$ is an integer and $\alpha>2d$.
We use the main-text notation
\begin{equation}
\beta_{\mathrm{LR}}
\coloneqq
\max\left\{1,\frac{1}{\alpha-2d}\right\}.
\end{equation}
All lower bounds below concern the Bose--Hubbard-type model and are obtained from explicitly constructed Hamiltonians and controls satisfying Sec.~\ref{S-I}.
For $p\ge3$, the packet-based protocols use the uniform preparation estimate of Proposition~\ref{prop:SM_A1_bounded_preparation}.
We compare their recursive, collective-relay, and sequential transfer distances with independently constructed short-range propagation.
For $p=2$, microscopic protocols avoid macroscopic packet concentration.
The alternatives are not successive stages of one protocol.

\subsubsection{{Physical distance and packet-transfer exponents}}
\label{S-II-E-distance}

The source and destination centers in Sec.~\ref{S-II-D-recursive-time} are separated by $(2\ell+1)(L-1)$.
The microscopic input and readout sites are each within distance $\ell$ of their respective centers.
Therefore, the triangle inequality gives
\begin{equation}
(2\ell+1)(L-1)-2\ell
\le |i_{\mathrm{in}}-i_{\mathrm{out}}|
\le (2\ell+1)(L-1)+2\ell.
\end{equation}

For $L\to\infty$, the radius used in Sec.~\ref{S-II-D-witness} can be chosen to satisfy
\begin{equation}
R\asymp |i_{\mathrm{in}}-i_{\mathrm{out}}|
\asymp (2\ell+1)L
\asymp tL,
\end{equation}
with a sufficiently small fixed ratio $R/|i_{\mathrm{in}}-i_{\mathrm{out}}|$ to ensure Eq.~\eqref{eq:SM_input_outside_enlargement}.
For the sequential and collective-relay protocols, the corresponding coarse separation is $L_{\mathrm{cg}}$, and the same geometric comparison holds with $L-1$ replaced by $L_{\mathrm{cg}}$.

For integer $p\ge3$ and $2d<\alpha\le2d+1$, Sec.~\ref{S-II-D-recursive-time} constructs
$L\asymp(g_{\mathrm{lr}}t)^{1/(\alpha-2d)}$ with $t_{\mathrm{prep}}+t_{\mathrm{tr}}+t_{\mathrm{dec}}\le t$.
Since $g_{\mathrm{lr}}=t^{pd-\alpha}$ and $\ell\asymp t$, this gives
\begin{equation}
R_{\mathrm{rec}}(t)
\asymp t(g_{\mathrm{lr}}t)^{1/(\alpha-2d)}
=t^{[(p-2)d+1]/(\alpha-2d)}.
\end{equation}
For the same integer $p\ge3$ and $2d+1<\alpha<pd$, Proposition~\ref{prop:SM_A2_constructive_transfer} supplies
\begin{equation}
R_{\mathrm{rel}}(t)\gtrsim
\frac{t^{\,2+(pd-\alpha)/(\alpha-2d)}}
{(\log t)^{\eta_{\mathrm{rel}}}}.
\end{equation}
For $\alpha>2d+1$, $\alpha\ge pd$, and $pd-\alpha+1>0$, the sequential packet circuit gives $R_{\mathrm{seq}}(t)\gtrsim t^{2+pd-\alpha}$.
Each estimate includes preparation, microscopic readout, and all local-control time.
Thus, the leading packet-construction exponent is
\begin{equation}
\beta_{\mathrm{low}}^{(\mathrm{conc})}=
\begin{cases}
\displaystyle\frac{(p-2)d+1}{\alpha-2d},
&2d<\alpha\le2d+1,\\[7pt]
\displaystyle 2+\frac{pd-\alpha}{\alpha-2d},
&2d+1<\alpha<pd,\\[8pt]
2+pd-\alpha,
&\alpha>2d+1,\quad pd\le\alpha<pd+1.
\end{cases}
\end{equation}
This exponent describes the leading power attained by the indicated
packet-based construction, not a universal optimality statement.
For $\alpha\ge pd+1$, the sequential choice does not produce a
growing coarse path; the short-range alternatives below remain available.
The collective relay has a polylogarithmic distance loss, whereas the
other two packet constructions attain their displayed powers without it.

\subsubsection{{Short-range propagation with bounded controls}}
\label{S-II-E-short-range}

The competing short-range construction of Ref.~\cite{kuwahara2022optimal} concentrates bosons transversely onto a one-dimensional information path.
Let $n\ge1$ denote the occupation per mode on that path.
{
For the interlayer hopping Hamiltonian at a path site $x$, $K_x=J_0(b_{x,0}^\dagger b_{x,1}+b_{x,1}^\dagger b_{x,0})$, the enhanced matrix element is
}
\begin{equation}
\bra{n-1,n+1}{K_x}\ket{n,n}
=J_0\sqrt{n(n+1)}.
\label{eq:SM_short_range_hopping_enhancement}
\end{equation}
With this transition isolated exactly, a resonant pulse transfers its population in time $\pi/[2J_0\sqrt{n(n+1)}]$.
The transverse concentration gives the ideal occupation scale $n\asymp t^{d-1}$, so the hopping part of an isolated transition takes ${\cal O}(t^{1-d})$ time.
A sequence of order $t^d$ such operations then has a total ideal hopping time of order $t$.
This explains the exponent $d$ but does not include the time or accuracy of transition selection.
The implementations in Methods Eqs.~(31)--(33) of Ref.~\cite{kuwahara2022optimal} use limits of unbounded control coefficients; they do not establish the required uniform estimates under Sec.~\ref{S-I}.

This is a separate protocol with a different encoding, not an additional stage of the packet construction. The $N^{p-2}$ density-coupling estimate of Sec.~\ref{S-II-C-adjacent} does not constrain the hopping matrix element in Eq.~\eqref{eq:SM_short_range_hopping_enhancement}.

{
A microscopic nearest-neighbor path already gives a linear-distance lower bound for every integer $p\ge2$ and every $d\ge1$.
Place the one-particle layer qubits of Sec.~\ref{S-II-B-encoding} on a straight physical path and use the Mott input pair of Sec.~\ref{S-II-A}.
Three neighboring controlled-NOT gates implement an exact swap.
Let $C_{\mathrm{sw}}>0$ bound its duration, including continuous pulses.
Successive swaps over $M=\lfloor t/(2C_{\mathrm{sw}})\rfloor$ edges transfer the input exactly, restore all other layer qubits to $\ket{1,0}$, and leave the final interval idle.
Consequently,
}
\begin{equation}
|i_{\mathrm{in}}-i_{\mathrm{out}}|
=\left\lfloor\frac{t}{2C_{\mathrm{sw}}}\right\rfloor
\asymp t,
\quad e_{\mathrm{out}}=0,
\label{eq:SM_one_dimensional_short_range}
\end{equation}
where $e_{\mathrm{out}}$ is the trace-norm error relative to the ideal output.
All operations are within the Hamiltonian class for every $p\ge2$.
{
In particular, this proves the exponent $1$ for $p=2$ when $\alpha\ge2d+1$, independently of the spatial dimension, and for $d=1$ at every $p\ge2$.
}

{
For $p\ge3$ and $d\ge2$, the stronger exponent $d$ is attained by a bounded-control short-range construction with a subpolynomial loss. We give the full implementation proof, including transverse preparation, transition isolation, and local readout.
}

\begingroup
\begin{proposition}[Short-range propagation with bounded coefficients]
\label{prop:SM_A3_bounded_short_range}
Fix an integer $p\ge3$, $d\ge2$, $\alpha>2d$,
a finite $\xi\ge1$, and $0<e_*<1$.
There are constants $a_{\mathrm{sr}},c_{\mathrm{sr}}>0$
such that, for every sufficiently large $t$, a finite lattice
supports an admissible protocol with
\begin{equation}
\lambda_{\mathrm{sr}}=3^d,\quad
r_{\mathrm{sr}}=1,\quad
h_{\mathrm{sr}}(t)
=\exp\!\left[a_{\mathrm{sr}}\sqrt{\log t}\right].
\end{equation}
Here $r_{\mathrm{sr}}$ denotes the maximum physical range of nonzero couplings, and $h_{\mathrm{sr}}(t)$ is the subpolynomial distance-loss function.
The inputs are the Mott state pair of Sec.~\ref{S-II-A}, on the chosen lattice.
The same controls act on both inputs and obey Eq.~\eqref{eq:SM_power_law_bounds}.
Preparation, all conditional operations, and local readout are included in $t$.
In particular,
\begin{equation}
|i_{\mathrm{in}}-i_{\mathrm{out}}|
\ge c_{\mathrm{sr}}t^d
\me^{-a_{\mathrm{sr}}\sqrt{\log t}},
\quad
\max_{b=0,1}\left\|
\rho_{\mathrm{out,sr}}^{(b)}
-\ket b_{i_{\mathrm{out}},\mathrm{mic}}
\!\bra b_{i_{\mathrm{out}},\mathrm{mic}}
\right\|_1\le e_*,
\label{eq:SM_A3_short_range_conclusion}
\end{equation}
where $\ket0_{i,\mathrm{mic}}=\ket{1,0}_i$ and $\ket1_{i,\mathrm{mic}}=\ket{0,1}_i$
denote the microscopic code at site $i$.
All constants are independent of $t$ and the ambient lattice size.
\end{proposition}

\begin{proof}
\textit{Local code and density phases.---}
At each site $x$ on the information path, use
\begin{equation}
\ket0_{x,\mathrm{sr}}=\ket{n,n}_x,\quad
\ket1_{x,\mathrm{sr}}=\ket{n-1,n+1}_x,\quad
\mathcal Q_{x,\mathrm{sr}}
=
\operatorname{span}
\{\ket0_{x,\mathrm{sr}},\ket1_{x,\mathrm{sr}}\},
\end{equation}
with an integer $n\ge1$.
Both code states have total occupation $2n$.
All controls below depend on $n$, which is prescribed by the preparation, but not on the encoded input.
Use $J_0$ and $U_*$ from Sec.~\ref{S-II-B-preparation}, with the continuous pulses of
Eq.~\eqref{eq:SM_A1_continuous_pulse}.
Only on-site terms and couplings between neighboring physical sites will be used.

We first record a fast number phase.
On the two-mode sector $\mathcal H_M$ of Eq.~\eqref{eq:SM_A1_number_sectors}, reduce $\theta$ modulo $2\pi$ to $[-\pi,\pi]$ and choose $\omega_M\in[-\pi,\pi]$ with
$\omega_M\equiv\theta M/2\pmod{2\pi}$.
Then
\begin{equation}
\left.
\me^{\mi\omega_M(n_1+n_2)/M}
\me^{\mi\theta(n_1^2-n_2^2)/(2M)}
\right|_{\mathcal H_M}
=\me^{\mi\theta n_1}.
\label{eq:SM_A3_fast_number_phase}
\end{equation}
This follows from $n_1^2-n_2^2=2Mn_1-M^2$.
Every coefficient has area ${\cal O}(1/M)$, so the full phase, including its scalar factor, takes at most $C_Z/M$ time.

For neighboring path sites $x,y$, write $a=n_{x,0}$ and $b=n_{y,0}$ within the following gate calculation.
Let $\mathcal H_{2n,y}$ be the full sector of total occupation $2n$ in the target modes.
On $\mathcal Q_{x,\mathrm{sr}}\otimes\mathcal H_{2n,y}$, define
\begin{equation}
\mathsf E_n
=
\exp\!\left[\frac{\mi\pi a^2b}{2n-1}\right]
\exp\!\left[-\frac{\mi\pi n^2b}{2n-1}\right].
\end{equation}
For control occupation $a=n$, this is the identity on the target. For $a=n-1$, it is $(-1)^b$.
The first factor uses a cubic density monomial with area ${\cal O}(1/n)$.
Its coefficient is assigned to the ordered pair $((x,0),(y,0))$, with the reverse monomial absent.
The monomial itself is Hermitian, so this is permitted by Eq.~\eqref{eq:SM_restricted_Hamiltonian}.
For the second factor, reduce its number-phase angle modulo $2\pi$ and apply Eq.~\eqref{eq:SM_A3_fast_number_phase} with $M=2n$ at $y$.
Thus $\mathsf E_n$ and its inverse each cost at most $C_E/n$.
No coefficient grows with $n$.

\textit{Conditional hopping and gate accuracy.---}
Let
\begin{equation}
K_y
=
J_0\left(
b_{y,0}^\dagger b_{y,1}
+b_{y,1}^\dagger b_{y,0}
\right),
\quad
s_n=(-1)^{n-1}.
\end{equation}
Let $D=D_{2n}$ act on the target modes, with its exact physical realization supplied by
Eq.~\eqref{eq:SM_A1_phase_realization}.
The auxiliary generator
\begin{equation}
K_{\mathrm c}
=
\frac14\left[
K_y+s_nDK_yD^\dagger
-\mathsf E_nK_y\mathsf E_n^\dagger
-s_n\mathsf E_nDK_yD^\dagger\mathsf E_n^\dagger
\right]
\label{eq:SM_A3_conditional_hopping}
\end{equation}
vanishes for control zero.
For control one, target parity reverses the hopping, so $K_{\mathrm c}=(K_y+s_nDK_yD^\dagger)/2$.
This retains the isolated transition $n-1\leftrightarrow n$ that contains the target code.
Evolution for
\begin{equation}
\upsilon_n
=
\frac{\pi}{2J_0\sqrt{n(n+1)}}
\end{equation}
gives an ideal controlled $(-iX)$ on the two codes.
The control correction is
\begin{equation}
\mathsf P_n
=\me^{\mi\chi_n n^2}\me^{-\mi\chi_n a^2},
\quad
\chi_n=\frac{\pi}{2(2n-1)},
\quad
\left.\mathsf P_n\right|_{\mathcal Q_{x,\mathrm{sr}}}
=\operatorname{diag}(1,i).
\label{eq:SM_A3_CNOT_phase_correction}
\end{equation}
Indeed, the phase ratio is $\me^{-\mi\chi_n[(n-1)^2-n^2]}=\mi$.
The $a^2$ pulse has the positive area $\chi_n$.
Reduce $\chi_n n^2$ modulo $2\pi$ and implement its scalar factor by a phase proportional to the control total number $2n$.
The correction takes ${\cal O}(1/n)$ time and converts this ideal evolution into a controlled-NOT gate.
No parity restriction on $n$ is required.

We implement Eq.~\eqref{eq:SM_A3_conditional_hopping} by the product formula in Eq.~\eqref{eq:SM_A1_Suzuki_recursion}, not as an additional instantaneous interaction.
Each summand is a signed conjugate of $K_y/4$; the conjugations use the exact density pulses above and their inverses.
On the invariant space $\mathcal Q_{x,\mathrm{sr}}\otimes\mathcal H_{2n,y}$,
$\|K_y\|=2J_0n$, so the norm sum of the four summands obeys $A_*=2J_0n$ and $A_*\upsilon_n\le\pi$.
For order $2q$, the factor count and weighted norm in Eq.~\eqref{eq:SM_A1_product_formula_error} satisfy
$L_q=\exp[{\cal O}(q)]$ and
$\lambda_q\upsilon_n=\exp[{\cal O}(q)]$.
Consequently, an error at most $0<\varepsilon<1$ is attained with
\begin{equation}
r_{\mathrm{PF}}
\le C \me^{c q}\varepsilon^{-1/(2q)}+1
\end{equation}
for fixed constants $C,c$ independent of $n,q,\varepsilon$.
The product formula contains at most $L_qr_{\mathrm{PF}}$ factors.
Each factor is implemented using a single hopping pulse and a fixed number of density pulses.
The total absolute hopping time is at most $C L_q\upsilon_n$, while the density pulses for each factor cost ${\cal O}(1/n)$ time.
Thus, the total implementation time, including all conjugations, is bounded by
$C L_q\upsilon_n+C L_qr_{\mathrm{PF}}/n$.
Including Eq.~\eqref{eq:SM_A3_CNOT_phase_correction}
and choosing
$q=\lceil\sqrt{\log(b_{\mathrm{gate}}/\varepsilon)}\rceil$,
with a fixed sufficiently large $b_{\mathrm{gate}}>1$, gives
\begin{equation}
t_{\mathrm{CNOT}}(n,\varepsilon)
\le
\frac{C_{\mathrm{gate}}}{n}
\exp\!\left[
a_{\mathrm{gate}}
\sqrt{\log(b_{\mathrm{gate}}/\varepsilon)}
\right].
\label{eq:SM_A3_CNOT_time}
\end{equation}
The error is in the operator norm on the product of the two codes, relative to the ideal controlled-NOT, and remains valid after adjoining any reference system.
All constants in Eq.~\eqref{eq:SM_A3_CNOT_time} are independent of $n$ and $\varepsilon$.
Increasing $q$ changes the pulse count, not the degree of the instantaneous density polynomials.
Negative product-formula coefficients are implemented by reversing the relevant controls, and every physical pulse has zero endpoints.

\textit{Microscopic input and readout.---}
At the source boundary, the control is a one-particle-layer qubit.
Set $a=n_{i_{\mathrm{in}},1}\in\{0,1\}$ and let $b$ be the occupation of the first target mode. Replace $\mathsf E_n$ in Eq.~\eqref{eq:SM_A3_conditional_hopping} by $\mathsf E_{\mathrm{in}}=\me^{\mi\pi ab}$.
It is the identity or target parity according to the input bit, and takes ${\cal O}(1)$ time with a bounded quadratic coefficient.
The target sign operation $D_{2n}$ still takes ${\cal O}(1/n)$.
Since $A_*\upsilon_n$ is uniformly bounded, a fixed product-formula order and repetition count achieve any prescribed fixed boundary error.
The correction $\me^{\mi\pi a/2}$ supplies the control phase.
For the first path site $x_1$, the ideal boundary action is
\begin{equation}
\ket b_{i_{\mathrm{in}},\mathrm{mic}}
\ket0_{x_1,\mathrm{sr}}
\longmapsto
\ket b_{i_{\mathrm{in}},\mathrm{mic}}
\ket b_{x_1,\mathrm{sr}},
\quad b=0,1,
\end{equation}
in time bounded independently of $n$.

At the receiver, let $a=n_{x_K,0}$ be the occupation at the last path site and $b=n_{i_{\mathrm{out}},1}\in\{0,1\}$ that of the microscopic target.
Use
\begin{equation}
\mathsf E_{\mathrm{out}}
=\me^{\mi\pi(a-n)b}
=\me^{\mi\pi ab}\me^{-\mi\pi nb},
\quad
K_{\mathrm{out},\mathrm c}
=
\frac12\left(
K_{\mathrm{mic}}
-\mathsf E_{\mathrm{out}}K_{\mathrm{mic}}
\mathsf E_{\mathrm{out}}^\dagger
\right),
\end{equation}
where $K_{\mathrm{mic}}$ is the hopping between the two receiver modes with coefficient $J_0$.
The angle in the linear factor is reduced modulo $2\pi$, so both factors of $\mathsf E_{\mathrm{out}}$ take ${\cal O}(1)$ time.
The auxiliary generator vanishes for path control zero and equals $K_{\mathrm{mic}}$ for path control one.
Evolve for $\pi/(2J_0)$ using a fixed-accuracy product formula, and apply $\mathsf P_n$ to the path control.
This gives the boundary-controlled NOT in ${\cal O}(1)$ time, uniformly in $n$.
Both boundary estimates are subspace operator-norm estimates and include all phase corrections.

\textit{Transverse preparation.---}
Choose path sites $x_j=(j,0,\ldots,0)$, $1\le j\le K$, and microscopic endpoints $i_{\mathrm{in}}=(0,\ldots,0)$ and $i_{\mathrm{out}}=(K+1,0,\ldots,0)$.
Take a finite lattice containing these sites and the disjoint columns
\begin{equation}
\mathcal V_j
=\{j\}\times\{-\ell,\ldots,\ell\}^{d-1},
\quad
\ell=\lfloor c_\ell t\rfloor.
\end{equation}
Each column has odd volume $(2\ell+1)^{d-1}$.
Leave one outer-corner particle as a spectator and set
\begin{equation}
n=\frac{(2\ell+1)^{d-1}-1}{2}
\asymp\ell^{d-1}.
\label{eq:SM_A3_path_occupation}
\end{equation}
Partition its other $2n$ sites into two prescribed groups of $n$ sites.
Starting with one boson in layer zero at every site, apply local layer flips to the second group.
In each layer, concentrate the resulting zero--one occupation pattern at $x_j$.
The layer controls may use the same physical routes because they act on distinct modes.
No route crosses the spectator, and both microscopic endpoints are left untouched.

The spatial construction of Sec.~\ref{S-II-B-preparation} applies to these fixed patterns in transverse dimension $d-1$, without an initial fanout.
Each merge has a known total number and a prescribed Fock input.
The omitted corner only shortens one row in the first transverse sweep.
To spell out the global accuracy, the total number of selected ladder steps is bounded by
$G_{\mathrm{col}}\le C_dK n\log(2+\ell)$.
With an allowance $0<\varepsilon_{\mathrm{col}}<1$, assign error $\varepsilon_{\mathrm{col}}/G_{\mathrm{col}}$ to each step.
Equations~\eqref{eq:SM_A1_merge_estimate} and \eqref{eq:SM_A1_concentration_time} then give
\begin{equation}
t_{\mathrm{prep,sr}}
\le
C\ell+C_{q_{\mathrm{col}}}\log(2+\ell)
\left(
\frac{G_{\mathrm{col}}}{\varepsilon_{\mathrm{col}}}
\right)^{1/(2q_{\mathrm{col}})}.
\end{equation}
This estimate includes both layers, routing, layer flips, and the known phase corrections at their centers.
For $K\le\ell^{b_{\mathrm{col}}}$ with fixed $b_{\mathrm{col}}$, choose a fixed order
$2q_{\mathrm{col}}>b_{\mathrm{col}}+d-1$.
The second term is $o(\ell)$, so preparation takes at most $C_{\mathrm{col}}\ell$ time and has total norm error at most $\varepsilon_{\mathrm{col}}$ on its prescribed input,
including an arbitrary reference attached to the untouched source.
The ideal path state is $\bigotimes_{j=1}^K\ket0_{x_j,\mathrm{sr}}$; all other active modes in the columns are empty.

\textit{Full protocol and accumulated error.---}
After preparation, apply the input boundary gate, then the $K-1$ path controlled-NOT gates in order, and finally the receiver gate.
For each basis input $b\in\{0,1\}$, the ideal sequence copies the bit along the path and onto the microscopic receiver.
This is the distinguishable local signal needed for the lower bound; no arbitrary-qubit transfer with reset of the path is asserted.
The controls are the same for both inputs and use neither measurements nor postselection.

Choose $\varepsilon_{\mathrm{col}}=e_*/8$, a norm-error allowance $e_*/8$ for each boundary gate, and $\varepsilon=e_*/(8K)$ for each path gate.
Let $\ket{\Psi^{(b)}_{\mathrm{fin}}}$ and $\ket{\Psi^{(b),\mathrm{id}}_{\mathrm{fin}}}$
denote the full actual and ideal output vectors.
The ideal input at every gate lies in its specified code subspace.
The unitary telescope of Sec.~\ref{S-II-D-error} gives
\begin{equation}
\|\ket{\Psi^{(b)}_{\mathrm{fin}}}
-\ket{\Psi^{(b),\mathrm{id}}_{\mathrm{fin}}}\|
\le
\varepsilon_{\mathrm{col}}+\frac{e_*}{4}
+(K-1)\varepsilon
\le\frac{e_*}{2},
\quad b=0,1.
\end{equation}
Earlier leakage is propagated by unitaries; the actual intermediate state need not lie
in the next ideal code subspace.
The trace-norm distance of the full pure states is at most twice the vector distance, and the partial trace cannot increase it.
This proves the output-error bound in Eq.~\eqref{eq:SM_A3_short_range_conclusion}.

By Eq.~\eqref{eq:SM_A3_CNOT_time}, the path gates take at most
\begin{equation}
t_{\mathrm{path}}
\le
\frac{C_{\mathrm{gate}}K}{n}
\exp\!\left[
a_{\mathrm{gate}}
\sqrt{\log(8b_{\mathrm{gate}}K/e_*)}
\right].
\label{eq:SM_A3_path_time}
\end{equation}
Fix $b_{\mathrm{col}}=d+1$ and a preparation order satisfying
$2q_{\mathrm{col}}>2d$.
The column estimate fixes $C_{\mathrm{col}}$ before $c_\ell$ is chosen.
Choose $c_\ell>0$ so that $C_{\mathrm{col}}c_\ell<1/3$, and set
\begin{equation}
K
=
\left\lfloor
c_Kt^d \me^{-a_{\mathrm{sr}}\sqrt{\log t}}
\right\rfloor,
\quad
c_K>0,\quad
a_{\mathrm{sr}}>a_{\mathrm{gate}}\sqrt d.
\label{eq:SM_A3_path_length}
\end{equation}
For every fixed $c_\ell,c_K>0$, this choice satisfies $K\le\ell^{d+1}$ at sufficiently large $t$, verifying the envelope used to obtain $C_{\mathrm{col}}$.
As in Sec.~\ref{S-II-B-preparation}, changing the fixed $c_\ell$ affects this large-time threshold, not the preparation constant.
Equations~\eqref{eq:SM_A3_path_occupation} and \eqref{eq:SM_A3_path_time} imply
\begin{equation}
\frac{t_{\mathrm{path}}}{t}
\le
C\exp\!\left[
-\bigl(a_{\mathrm{sr}}-a_{\mathrm{gate}}\sqrt d\bigr)
\sqrt{\log t}+{\cal O}(1)
\right]
\longrightarrow0.
\end{equation}
The two boundary gates cost ${\cal O}(1)$ for the fixed allowances.
Hence, preparation, propagation, and readout fit within time $t$. Every pulse is continuous and vanishes at its endpoints; setting all coefficients to zero during the remaining interval leaves the output unchanged.
Since $|i_{\mathrm{in}}-i_{\mathrm{out}}|=K+1$, Eq.~\eqref{eq:SM_A3_path_length} proves the distance bound.

Initially, every physical site, including both endpoints and all spectators, contains exactly one particle in its two modes.
The two states differ only by the source layer swap.
Thus, Eq.~\eqref{eq:SM_lower_density_check} applies unchanged with $\lambda_{\mathrm{sr}}=3^d$ for every finite $\xi$.
All couplings have a physical range of at most one.
With $A_{\mathrm{out,sr}}=A_{\mathrm{out}}$ from Eq.~\eqref{eq:SM_destination_readout},
the two ideal expectations are $+1$ and $-1$, so the output error gives a signal at least $2-2e_*$.
Moreover, $K/\ell\to\infty$ for $d\ge2$.
Taking $R=(K+1)/2$ ensures Eq.~\eqref{eq:SM_input_outside_enlargement} at sufficiently large $t$, and Sec.~\ref{S-II-D-witness} gives the same lower bound $1-e_*$ on the uniform local-approximation error.
Finally,
$\log h_{\mathrm{sr}}(t)/\log t
=a_{\mathrm{sr}}/\sqrt{\log t}\to0$.
This establishes the short-range lower bound using bounded controls,
without an additional implementation assumption.
\end{proof}
\endgroup

\subsubsection{{Combined lower bound and matching regime}}
\label{S-II-E-matching}

The constructed protocols provide the following leading exponents.
For $p=2$, the microscopic recursion of Sec.~\ref{S-II-B-preparation} applies when
$2d<\alpha\le2d+1$, and the exact nearest-neighbor path applies when $\alpha\ge2d+1$.
Hence
\begin{equation}
\beta_{\mathrm{low}}
=\beta_{\mathrm{LR}}
=\max\left\{1,\frac{1}{\alpha-2d}\right\},
\quad p=2.
\label{eq:SM_quadratic_branch_crossover}
\end{equation}
For each integer $p\ge3$, select the construction with the largest leading propagation power at each parameter value.
The resulting exponent is
\begin{equation}
\beta_{\mathrm{low}}=
\begin{cases}
\displaystyle\frac{(p-2)d+1}{\alpha-2d},
&2d<\alpha\le2d+1,\\[7pt]
\displaystyle\max\!\left\{
d,\,
2+\frac{pd-\alpha}{\alpha-2d}
\right\},
&2d+1<\alpha<pd,\\[9pt]
\displaystyle\max\!\left\{d,\,2+pd-\alpha\right\},
&\alpha>2d+1,\quad\alpha\ge pd.
\end{cases}
\quad p\ge3.
\label{eq:SM_combined_lower_exponent}
\end{equation}
The last expression includes $\alpha\ge pd+1$, where $2+pd-\alpha\le1\le d$ and the short-range protocol supplies the bound; no growing packet path is asserted in that range.
The maximum compares alternative protocols available within the same total time, rather than adding their durations.
The collective relay contributes a polylogarithmic loss when selected, and the bounded-control $d$-dimensional short-range protocol contributes $\exp[a_{\mathrm{sr}}\sqrt{\log t}]$ to the distance denominator.
The exact microscopic and recursive or sequential packet circuits do not require such losses.

For $p\ge3$ and $2d<\alpha\le2d+1$,
\begin{equation}
\beta_{\mathrm{low}}^{(\mathrm{conc})}-d
=
\frac{(p-2)d+1}{\alpha-2d}-d
\ge(p-3)d+1>0.
\end{equation}
Beyond this interval, the comparison depends on the actual constructive relay or sequential exponent, respectively:
\begin{equation}
\beta_{\mathrm{low}}^{(\mathrm{conc})}-d
=
\begin{cases}
\displaystyle
2-d+\frac{pd-\alpha}{\alpha-2d},
&2d+1<\alpha<pd,\\[7pt]
(p-1)d+2-\alpha,
&\alpha>2d+1,\quad pd\le\alpha<pd+1.
\end{cases}
\end{equation}
These crossovers compare proven lower-bound constructions and are not optimality boundaries.
For $2d+1<\alpha<pd$, the collective-relay construction gives $2+(pd-\alpha)/(\alpha-2d)$ rather than the enhanced-linear packet-transfer exponent $pd-\alpha+2$.

The main-text upper exponent is
\begin{equation}
\beta_{\mathrm{up}}
=
1+\max\{d+1,pd-\alpha+1\}\beta_{\mathrm{LR}}.
\end{equation}
For integer $p\ge4$ and $2d<\alpha\le2d+1$, we have $pd-\alpha+1\ge d+1$.
The recursive lower bound, therefore, matches the main-text upper bound in the leading-power sense:
\begin{equation}
\beta_{\mathrm{opt}}
=\beta_{\mathrm{up}}=\beta_{\mathrm{low}}
=\frac{(p-2)d+1}{\alpha-2d},
\quad p\ge4,\quad 2d<\alpha\le2d+1.
\label{eq:SM_matching_light_cone_exponent}
\end{equation}
The upper-bound estimate is taken as a separate input to this comparison.
For each prescribed positive loss in the upper exponent, the required finite moment order is chosen as in Sec.~\ref{S-IV-E}.
The Mott inputs satisfy the controlled-density condition for every such order.
Since the matching exponent exceeds one, the constructed separation is asymptotically larger than $t$.
The localization witness therefore also holds for the block partition with $\ell=\lceil vt\rceil$ used in the upper bound.
No additional implementation assumption is required for the displayed lower bound.
No optimality for the Bose--Hubbard-type model is inferred outside this interval.
No conclusion for $\alpha\le2d$ is asserted here.

\begin{proposition}[{Constructive lower-bound propagation protocol}]
\label{prop:SM_lower_bound_protocol}
Fix integers $p\ge2$ and $d\ge1$, a decay exponent $\alpha>2d$, a finite moment order $\xi\ge1$, and $0<e_*<1$.
Let $\beta_{\mathrm{low}}$ be given by Eq.~\eqref{eq:SM_quadratic_branch_crossover} for $p=2$ and Eq.~\eqref{eq:SM_combined_lower_exponent} for $p\ge3$.
There exist fixed constants $\lambda,c,c_R>0$ and a function $h(t)\ge1$, with $\log h(t)/\log t\to0$, such that for every sufficiently large $t$, a finite lattice with two bosonic modes per physical site supports an admissible Hamiltonian and two states
$\rho_{\mathrm{in}}^{(b)}\in\mathcal D_\xi^\lambda$,
$b=0,1$, differing by a number-preserving unitary supported on one physical site $i_{\mathrm{in}}$.
The same Hamiltonian acts on both inputs.
Let $\rho_{\mathrm{out}}^{(b)}$ be the reduced state at $i_{\mathrm{out}}$ at time $t$.
There is a Hermitian readout $A_{\mathrm{out}}\in\mathcal A_{\{i_{\mathrm{out}}\}}^0$
with $\|A_{\mathrm{out}}\|=1$ such that
\begin{equation}
\begin{aligned}
|i_{\mathrm{in}}-i_{\mathrm{out}}|
\ge c\,\frac{t^{\beta_{\mathrm{low}}}}{h(t)},\quad
\left|\operatorname{Tr}\!\left[
A_{\mathrm{out}}
(\rho_{\mathrm{out}}^{(0)}-\rho_{\mathrm{out}}^{(1)})
\right]\right|
\ge2-2e_*.
\end{aligned}
\end{equation}
One can choose $R(t)\ge c_Rt^{\beta_{\mathrm{low}}}/h(t)$, with the block partition specified in the proof,
such that for every common
$\widetilde A\in\mathcal A_{X_{R(t)}}$,
with $X=\{i_{\mathrm{out}}\}$,
\begin{equation}
\max_{b=0,1}
\left\|
\rho_{\mathrm{in}}^{(b)}
\left[\tau_{0,t}(A_{\mathrm{out}})-\widetilde A\right]
\right\|_1
\ge1-e_*.
\label{eq:SM_lower_protocol_uniform_witness}
\end{equation}
The function $h$ accounts only for subpolynomial corrections of the selected construction and is independent of the lattice size.
All constants are independent of $t$ and of the ambient lattice size.
The time $t$ includes preparation, propagation, and microscopic readout.
No separate implementation assumptions are imposed.
\end{proposition}

\begin{proof}
For $p=2$, use the microscopic recursion of Sec.~\ref{S-II-B-preparation} for $2d<\alpha\le2d+1$, and the exact nearest-neighbor path of Eq.~\eqref{eq:SM_one_dimensional_short_range} for $\alpha>2d+1$.

Now let $p\ge3$.
The short-range alternative is the exact microscopic path of Eq.~\eqref{eq:SM_one_dimensional_short_range} when $d=1$, and the construction of
Proposition~\ref{prop:SM_A3_bounded_short_range} when $d\ge2$.
For $2d<\alpha\le2d+1$, use the recursive packet protocol of Sec.~\ref{S-II-D-recursive-time}, with preparation and decoding provided by Proposition~\ref{prop:SM_A1_bounded_preparation}.
For $2d+1<\alpha<pd$, choose the collective relay of Proposition~\ref{prop:SM_A2_constructive_transfer} or the short-range alternative, whichever attains the larger leading power.
For $\alpha>2d+1$ and $\alpha\ge pd$, choose the sequential packet construction of Sec.~\ref{S-II-D-linear} when $pd-\alpha+1>0$ and $2+pd-\alpha\ge d$.
Otherwise, use the short-range alternative. At a crossover, either construction gives the stated leading power.

The recursive, sequential, and microscopic constructions can be taken with $h(t)=1$.
The collective relay uses $h(t)=(\log t)^{\eta_{\mathrm{rel}}}$ for sufficiently large $t$.
For $p\ge3$ and $d\ge2$, the bounded-control short-range construction uses $h(t)=\exp[a_{\mathrm{sr}}\sqrt{\log t}]$.
Each of these choices obeys $\log h(t)/\log t\to0$.
Every construction uses the Mott input pair, so the controlled-density estimate holds with $\lambda=3^d$ for every fixed finite moment order.

For packet protocols, the source--destination separation divided by $\ell$ tends to infinity, as shown in Sec.~\ref{S-II-D-witness}.
The same holds for the bounded-control short-range construction when $d\ge2$.
In these cases, choose $R=|i_{\mathrm{in}}-i_{\mathrm{out}}|/2$ and retain the block partition used in the construction.
For the microscopic recursion and the nearest-neighbor path,
choose $\ell=0$ in Eq.~\eqref{eq:SM_block_geometry}, so the blocks are single physical sites, and take the same value of $R$.
These choices give $R+2\ell<|i_{\mathrm{in}}-i_{\mathrm{out}}|$ for sufficiently large $t$, together with the stated lower bound on $R(t)$.
The two inputs then have identical reduced states on $X_R$.
Their output signal and the trace-norm argument of Sec.~\ref{S-II-D-witness} give
Eq.~\eqref{eq:SM_lower_protocol_uniform_witness}.
All protocols use admissible bounded coefficients and complete their operations within the single total time $t$.
\end{proof}

\section{Lieb-Robinson bounds for fixed-$k$-local Hamiltonians}
\label{S-III}

The upper-bound argument requires Lieb-Robinson estimates for bounded interactions whose terms may involve more than two blocks. After particle-number truncation, the interaction picture generated by the commuting short-range density terms can enlarge the support of each interaction term. The resulting Hamiltonian is nevertheless fixed-$k$-local: each term acts on at most $k$ blocks, with $k$ independent of time and system size. This support parameter is distinct from the bosonic interaction order $p$.

For $2d<\alpha<2d+1$, we extend the bound of Tran \emph{et al.}~\cite{PhysRevLett.127.160401} to fixed-$k$ interactions under the pair-summed decay condition stated below. The estimates apply to time-dependent Hamiltonians on finite-dimensional local Hilbert spaces, with constants that may depend on $k$ but not on the system size or the local dimensions. {For $\alpha>2d+1$, we also record the strictly linear bound of Kuwahara and Saito~\cite{PhysRevX.10.031010}.} The marginal point $\alpha=2d+1$ is included with an arbitrarily small loss from linearity by applying the bound at a slightly smaller decay exponent. We also obtain the corresponding estimates for observables of finite support. Their application to the bosonic dynamics and the errors associated with particle-number truncation are treated in Sec.~\ref{S-IV}.

\textit{Overview.---} Under the fixed-$k$ and pair-summed decay assumptions of Sec.~\ref{S-III-A}, let $\Gamma$ be the finite lattice and $g>0$ the interaction scale. For $\alpha>2d$ and every $0<\omega<\min\{1,\alpha-2d\}$ sufficiently close to the upper endpoint, there exist finite exponents $\gamma_1>1$ and $\gamma_2=\omega\gamma_1>d$ such that every bounded observable $O_Y$ supported on a nonempty finite $Y\subseteq\Gamma$ admits a local approximant $\widetilde O_{Y,r}$ satisfying
\[
\begin{gathered}
\operatorname{supp}\widetilde O_{Y,r}\subseteq Y^{[r]},
\quad \|\widetilde O_{Y,r}\|\le\|O_Y\|,\\
\|\tau^\Phi_{s,t}(O_Y)-\widetilde O_{Y,r}\|
\le C|Y|\|O_Y\|\frac{(gT)^{\gamma_1}}{r^{\gamma_2}},
\quad T=1+|t-s|,
\end{gathered}
\]
whenever $gT\ge1$ and $r\ge C_r(gT)^{1/\omega}$.
Here $Y^{[r]}$ is the closed $r$-neighborhood of $Y$ on $\Gamma$, and $\tau^\Phi_{s,t}$ is the dynamics of the bounded interaction.
The constants $C,C_r\ge1$ are independent of $g$, the system size, the local Hilbert-space dimensions, and the temporal variation of the interactions, provided that the interaction parameters and $\omega$ are fixed.
The exponent $1/\omega$ can approach $\beta_{\mathrm{LR}}=\max\{1,1/(\alpha-2d)\}$ from above; for $\alpha>2d+1$, a strictly linear bound is also available.
The branchwise bounds and the unified estimate are stated in Corollaries~\ref{cor:SM_klr_all_alpha} and \ref{cor:SM_klr_unified_input}, respectively.

The proof follows the range-decomposition and bootstrap strategy of Ref.~\cite{PhysRevLett.127.160401}. The main modification required for fixed-$k$ interactions is to localize each interaction term as a whole, rather than expanding it into products of single-site operators. We then bound the resulting many-body interaction norm on an auxiliary coarse-grained lattice and derive the corresponding range-truncated propagation estimate. After adding the omitted long-range interactions, we establish a polynomial starting bound that does not rely on a two-body structure and bootstrap it to the Tran exponents. Finally, we remove the finite-volume dependence of the intermediate estimates and extend the resulting bound from single-site to arbitrary finite-support observables.

\subsection{Assumptions and statement of the fixed-$k$ bound}
\label{S-III-A}

\textit{Lattice and local algebras.---}
Let $\Gamma\subset\mathbb Z^d$ be nonempty and finite, with $d\ge1$, and let
\begin{equation}
\mathcal H_\Gamma\coloneqq\bigotimes_{x\in\Gamma}\mathcal H_x,
\quad 1\le\dim\mathcal H_x<\infty.
\end{equation}
The local dimensions may vary from site to site; all bounds below are uniform in these dimensions. For $Y\subseteq\Gamma$, write $\mathcal H_Y=\bigotimes_{x\in Y}\mathcal H_x$ and let $\mathfrak A_Y$ denote its operator algebra, tensored with the identity on $\mathcal H_{\Gamma\setminus Y}$.
Following Sec.~\ref{S-I}, we use the supremum-norm distance $d_\Gamma(x,y)\coloneqq|x-y|$. For nonempty $Y,Z\subseteq\Gamma$, define
\begin{equation}
d_\Gamma(Y,Z)\coloneqq
\min_{x\in Y,\,y\in Z}d_\Gamma(x,y),
\quad
\operatorname{diam}_\Gamma Z
\coloneqq
\max_{x,y\in Z}d_\Gamma(x,y).
\end{equation}
For $x\in\Gamma$ and $r\ge0$, set
\begin{equation}
B_r^\Gamma(x)\coloneqq
\{y\in\Gamma:d_\Gamma(x,y)\le r\},
\quad
Y^{[r]}\coloneqq
\bigcup_{x\in Y}B_r^\Gamma(x).
\end{equation}
In Sec.~\ref{S-IV}, the sites of $\Gamma$ label the retained blocks $P_x$ in units of their spacing $2\ell+1$, and $d_\Gamma$ corresponds to $d_{\mathrm{cg}}$. The set $Y^{[r]}$ is an enlargement of this abstract lattice, distinct from the physical blockwise enlargement $X_R$ of Sec.~\ref{S-I-C}.

\textit{Interactions and dynamics.---} Fix an integer $k\ge1$, independent of time and system size, and consider
\begin{equation}
H_\Phi(u)\coloneqq
\sum_{\substack{Z\subseteq\Gamma\\1\le|Z|\le k}}\Phi_Z(u),
\quad
\Phi_Z(u)=\Phi_Z(u)^\dagger\in\mathfrak A_Z.
\label{eq:SM_klr_Hamiltonian}
\end{equation}
Each term is measurable and locally integrable in operator norm as a function of $u$; in particular, the continuous interactions used in Sec.~\ref{S-IV} are included. Assume the pair-summed decay condition
\begin{equation}
\operatorname*{ess\,sup}_{u\in\mathbb R}
\sum_{Z\ni x,y}\|\Phi_Z(u)\|
\le
\frac{C_\Phi g}{[1+d_\Gamma(x,y)]^\alpha},
\quad
x\ne y,
\quad
\alpha>2d,
\label{eq:SM_klr_pair_decay}
\end{equation}
for some interaction scale $g>0$ and a dimensionless constant $C_\Phi>0$. The prefactor $C_\Phi$ is fixed independently of $g$, time, system size, and the local Hilbert-space dimensions.
The scale $g$ is uniform in $u$ for the interaction family under consideration. In applications, it may depend on the chosen time scale and occupation cutoff.

The sum in Eq.~\eqref{eq:SM_klr_pair_decay} is taken at each time before the essential supremum; no bound on the sum of the separate time suprema is imposed. For continuous interactions, the essential supremum agrees with the ordinary supremum. This is the pair-summed condition used in Ref.~\cite{PhysRevX.10.031010}, with the distance convention of the main text. The parameter $g$ sets the scale of the interaction norms, rather than the individual bosonic coefficients $J_{ij}$ or $U_{ij}$.

Let $U_\Phi(s,t)$ be the propagator from $s$ to $t$. We retain the convention of Sec.~\ref{S-I-A}:
\begin{equation}
\mi\partial_tU_\Phi(s,t)=H_\Phi(t)U_\Phi(s,t),
\quad
U_\Phi(s,s)=I,
\quad
\tau^\Phi_{s,t}(O)\coloneqq
U_\Phi(s,t)^\dagger O U_\Phi(s,t).
\label{eq:SM_klr_dynamics}
\end{equation}
The differential equation holds almost everywhere for locally integrable interactions, and everywhere for continuous ones.
One-site terms require no uniform norm bound. Their interaction picture is generated by a product of on-site unitaries and preserves the supports and norms of all remaining terms. Since such unitaries preserve each local algebra, the estimates below are unaffected.
We omit the one-site and scalar terms in the proof.

\textit{Localization maps.---} For $Y\subseteq\Gamma$, write $Y^c=\Gamma\setminus Y$ and define the conditional expectation
\begin{equation}
\mathbb E_Y(O)\coloneqq
\frac{\operatorname{Tr}_{\mathcal H_{Y^c}}(O)}
     {\dim\mathcal H_{Y^c}}
\otimes I_{Y^c}.
\label{eq:SM_klr_conditional_expectation}
\end{equation}
It maps onto $\mathfrak A_Y$ and is a contraction in operator norm. Let
\begin{equation}
\mathcal P_{x,r}\coloneqq
\operatorname{id}-\mathbb E_{B_r^\Gamma(x)},
\quad
\|\mathbb E_Y\|_{\mathrm{op}\to\mathrm{op}}=1,
\quad
\|\mathcal P_{x,r}\|_{\mathrm{op}\to\mathrm{op}}\le2,
\end{equation}
where $\operatorname{id}$ is the identity superoperator and $\|\cdot\|_{\mathrm{op}\to\mathrm{op}}$ is the norm induced by the operator norm. For an operator $O_x\in\mathfrak A_{\{x\}}$, $\mathbb E_{B_r^\Gamma(x)}\tau^\Phi_{s,t}(O_x)$ is supported on $B_r^\Gamma(x)$, has norm at most $\|O_x\|$, and its approximation error is $\|\mathcal P_{x,r}\tau^\Phi_{s,t}(O_x)\|$.

We use the dimensionless elapsed time
\begin{equation}
\vartheta\coloneqq g|t-s|.
\end{equation}
This does not change the physical-time notation $T=1+|t-s|$ introduced in Sec.~\ref{S-I-C}. The following theorem extends the bound of Ref.~\cite{PhysRevLett.127.160401} to fixed-$k$ interactions.

\begin{theorem}[Fixed-$k$ extension of the Tran bound]
\label{thm:SM_klr_Tran}
Assume Eqs.~\eqref{eq:SM_klr_Hamiltonian} and \eqref{eq:SM_klr_pair_decay}, with fixed $k$, $C_\Phi$, and $2d<\alpha<2d+1$. For every
\begin{equation}
0<\varepsilon_{\mathrm{LR}}<\varepsilon_0,
\quad
\varepsilon_0\coloneqq
\frac{(\alpha-2d)^2}
     {(\alpha-2d)^2+\alpha-d},
\label{eq:SM_klr_admissible_loss}
\end{equation}
there exist constants $c,C_1,C_2>0$, depending only on $d,\alpha,k,C_\Phi,\varepsilon_{\mathrm{LR}}$, such that for every $x\in\Gamma$, $O_x\in\mathfrak A_{\{x\}}$, $r\ge1$, and $s,t\in\mathbb R$ satisfying
\begin{equation}
\vartheta=g|t-s|
\le
c\,r^{\alpha-2d-\varepsilon_{\mathrm{LR}}},
\end{equation}
one has
\begin{equation}
\left\|
\mathcal P_{x,r}\tau^\Phi_{s,t}(O_x)
\right\|
\le
\|O_x\|
\left[
C_1
\left(
\frac{\vartheta}
{r^{\alpha-2d-\varepsilon_{\mathrm{LR}}}}
\right)^{
\frac{\alpha-d}{\alpha-2d}
-\frac{\varepsilon_{\mathrm{LR}}}{2}}
+
C_2\frac{\vartheta}{r^{\alpha-d}}
\right].
\label{eq:SM_klr_Tran_bound}
\end{equation}
The constants are independent of $g$, the lattice $\Gamma$, all local Hilbert-space dimensions, and the temporal variation of the interactions.
\end{theorem}

For $\vartheta\ge1$, the second term in Eq.~\eqref{eq:SM_klr_Tran_bound} can be absorbed into the first. Keeping both terms also controls the short-time limit $\vartheta\to0$. If
$B\in\mathfrak A_{\Gamma\setminus B_r^\Gamma(x)}$, then
\begin{equation}
\left\|
[\tau^\Phi_{s,t}(O_x),B]
\right\|
\le
2\|B\|
\left\|
\mathcal P_{x,r}\tau^\Phi_{s,t}(O_x)
\right\|,
\end{equation}
because $\mathbb E_{B_r^\Gamma(x)}\tau^\Phi_{s,t}(O_x)$ commutes with $B$. No factor proportional to the size of the support of $B$ is required.

Before proving Theorem~\ref{thm:SM_klr_Tran}, we record the two remaining decay regimes.

\begin{corol}[Fixed-$k$ power-law Lieb--Robinson bounds for all
$\alpha>2d$]
\label{cor:SM_klr_all_alpha}
Assume Eq.~\eqref{eq:SM_klr_Hamiltonian} and the pair-summed decay condition~\eqref{eq:SM_klr_pair_decay}, with fixed $k$, $C_\Phi$, and $\alpha>2d$. Let
$O_x\in\mathfrak A_{\{x\}}$ and $\vartheta=g|t-s|$.

\begin{enumerate}
\item
If $2d<\alpha<2d+1$, the bound \eqref{eq:SM_klr_Tran_bound} holds.

\item
If $\alpha=2d+1$, then for every $0<\omega<1$ there exist constants
$c_\omega,C_\omega,\gamma_\omega>0$ and $B_\omega>d$,
depending only on $d,k,C_\Phi$, and $\omega$, such that
\begin{equation}
\left\|
\mathcal P_{x,r}\tau^\Phi_{s,t}(O_x)
\right\|
\le
C_\omega\|O_x\|
\left[
\left(
\frac{\vartheta}{r^\omega}
\right)^{\gamma_\omega}
+
\frac{\vartheta}{r^{B_\omega}}
\right],
\quad
\vartheta\le c_\omega r^\omega.
\label{eq:SM_klr_marginal_bound}
\end{equation}
Thus, the linear light-cone power is obtained with an arbitrarily small exponent loss.

\item
If $\alpha>2d+1$, the fixed-$k$ theorem of Kuwahara and Saito~\cite{PhysRevX.10.031010} gives constants $v_{\mathrm{LR}},C'_1,C'_2>0$, depending only on $d,\alpha,k,C_\Phi$, such that
\begin{equation}
\left\|
\mathcal P_{x,r}\tau^\Phi_{s,t}(O_x)
\right\|
\le
\|O_x\|
\left[
C'_1
\frac{\vartheta^{d+1}}
     {(r-v_{\mathrm{LR}}\vartheta)^{\alpha-d}}
+
C'_2
\frac{\vartheta}{r^{\alpha-d}}
\right],
\quad
r>v_{\mathrm{LR}}\vartheta.
\label{eq:SM_klr_linear_bound}
\end{equation}
In particular, the light cone is strictly linear in this regime.
\end{enumerate}
\end{corol}

\textit{Derivation of the two additional branches.---} At the marginal point $\alpha=2d+1$, Eq.~\eqref{eq:SM_klr_pair_decay} also holds, with the same $C_\Phi g$, for every smaller exponent $\bar\alpha\in(2d,2d+1)$.
Given $0<\omega<1$, choose $\bar\alpha\in(2d+\omega,2d+1)$ sufficiently close to $2d+\omega$, and set $\bar\varepsilon_{\mathrm{LR}}=\bar\alpha-2d-\omega$. Then
\begin{equation}
\bar\alpha-2d-\bar\varepsilon_{\mathrm{LR}}=\omega,
\quad
0<\bar\varepsilon_{\mathrm{LR}}
<
\frac{(\bar\alpha-2d)^2}
     {(\bar\alpha-2d)^2+\bar\alpha-d}.
\end{equation}
Applying Theorem~\ref{thm:SM_klr_Tran} with $(\bar\alpha,\bar\varepsilon_{\mathrm{LR}})$ gives Eq.~\eqref{eq:SM_klr_marginal_bound}, with
\begin{equation}
\gamma_\omega
=
\frac{\bar\alpha-d}{\bar\alpha-2d}
-\frac{\bar\varepsilon_{\mathrm{LR}}}{2},
\quad
B_\omega=\bar\alpha-d>d.
\end{equation}

For $\alpha>2d+1$, the pair-summed condition \eqref{eq:SM_klr_pair_decay} is the fixed-$k$ decay condition used by Kuwahara and Saito~\cite{PhysRevX.10.031010}, up to the
inessential replacement of $d_\Gamma(x,y)$ by $1+d_\Gamma(x,y)$.
Their fixed-$k$ theorem gives the first term of Eq.~\eqref{eq:SM_klr_linear_bound} for
$\vartheta\ge1$. The standard short-time power-law estimate yields the second term and is valid for $0\le\vartheta\le1$. After increasing the constants if necessary, Eq.~\eqref{eq:SM_klr_linear_bound} therefore holds uniformly for all $\vartheta\ge0$.

Consequently, at the leading-power level used in the main text, the bounded-interaction light-cone exponent is
\begin{equation}
\beta_{\mathrm{LR}}
=
\max\left\{
1,\frac{1}{\alpha-2d}
\right\}.
\end{equation}
For $2d<\alpha<2d+1$, this exponent is obtained with an arbitrarily small loss inherited from $\varepsilon_{\mathrm{LR}}$.
At $\alpha=2d+1$, one obtains any exponent arbitrarily close to the linear value, while for $\alpha>2d+1$ the cone is strictly linear.

The extension from single-site operators to arbitrary finite-support observables, together with the removal of finite-volume prefactors arising in the intermediate estimates, is given in Sec.~\ref{S-III-E}.

\textit{Conventions for the proof.---} For the remainder of the proof, we return to
$2d<\alpha<2d+1$ and prove Theorem~\ref{thm:SM_klr_Tran}. We set
\begin{equation}
\chi\coloneqq\alpha-2d\in(0,1),
\quad
\nu_{\mathrm{tail}}\coloneqq\alpha-d=d+\chi,
\quad
\theta\coloneqq2d+1-\alpha=1-\chi.
\end{equation}
Here $\chi$ has the same meaning as in Sec.~\ref{S-II-D-recursive}; $\nu_{\mathrm{tail}}$ is the exponent of the summed interaction tail.

For the interaction scale $g$ in Eq.~\eqref{eq:SM_klr_pair_decay}, define the monotone envelope
\begin{equation}
F_{\Phi,\Gamma}(\vartheta,r)
\coloneqq
\sup_{\substack{
s\in\mathbb R,\;
0\le u\le\vartheta/g,\;
x\in\Gamma\\
O_x\in\mathfrak A_{\{x\}},\;
\|O_x\|=1,\;
r'\ge r}}
\left\|
\mathcal P_{x,r'}
\tau^\Phi_{s,s+u}(O_x)
\right\|,
\quad
\vartheta\ge0,\quad r>0.
\label{eq:SM_klr_profile}
\end{equation}
Thus $F_{\Phi,\Gamma}\le2$; it is nondecreasing in $\vartheta$ and nonincreasing in $r$. When estimating spatial sums, we set
\begin{equation}
F_{\Phi,\Gamma}(\vartheta,0)=2.
\end{equation}
We usually suppress $\Gamma$. The estimates below are uniform over the interaction class in Eqs.~\eqref{eq:SM_klr_Hamiltonian} and \eqref{eq:SM_klr_pair_decay}, with the same $k$, $C_\Phi$, and $g$.
This class is closed under removing interaction terms, truncating their diameters, restricting the lattice, and shifting or reversing time intervals.

In the proof, we rescale time to set $g=1$, so that $\vartheta$ is the elapsed time in these units. This is the exact change
\begin{equation}
\widehat\Phi_Z(w)
\coloneqq
g^{-1}\Phi_Z(w/g),
\quad
\tau^{\widehat\Phi}_{gs,gt}
=
\tau^\Phi_{s,t},
\label{eq:SM_klr_time_rescaling}
\end{equation}
after which the hats are suppressed. The normalized pair-summed bound retains the prefactor $C_\Phi$, whose dependence is included in the constants.
The parameter $k$ always denotes the maximal interaction support size, while range scales are indexed by $m$.
Constants denoted by $C,c$ may change between estimates and, unless specified otherwise, depend only on $d,\alpha,k,C_\Phi$.
Dependence on $\varepsilon_{\mathrm{LR}}$ is included when the exponent loss is fixed.

\subsection{Localization and coarse-grained interaction estimates}
\label{S-III-B}

We work in the normalized units $g=1$ of Sec.~\ref{S-III-A} and omit one-site terms. Unless stated otherwise, $2d<\alpha<2d+1$, and constants depend only on $d,\alpha,k,C_\Phi$. We first derive localization estimates that do not require a two-body decomposition, and then use them to control the interactions added at each range scale. The intermediate finite-volume factors are kept explicit; their removal is treated in Sec.~\ref{S-III-E}.

\subsubsection{Interaction tails and elementary localization estimates}
\label{S-III-B-localization}

The pair-summed condition~\eqref{eq:SM_klr_pair_decay} implies
\begin{align}
\mathcal J_\Phi
&\coloneqq
\operatorname*{ess\,sup}_{u\in\mathbb R}\max_{x\in\Gamma}
\sum_{Z\ni x}|Z|\|\Phi_Z(u)\|\le C,
\label{eq:SM_klr_local_strength}\\
f_\Phi(a)
&\coloneqq
\operatorname*{ess\,sup}_{u\in\mathbb R}\max_{x\in\Gamma}
\sum_{\substack{Z\ni x\\\operatorname{diam}_\Gamma Z>a}}
\|\Phi_Z(u)\|
\le C a^{-\nu_{\mathrm{tail}}},\quad a\ge1.
\label{eq:SM_klr_diameter_tail}
\end{align}
Since the one-site terms have been removed, $|Z|\le2(|Z|-1)$ for every remaining support. Thus, for almost every $u$,
\begin{equation}
\sum_{Z\ni x}|Z|\|\Phi_Z(u)\|
\le2\sum_{y\ne x}\sum_{Z\ni x,y}\|\Phi_Z(u)\|
\le2C_\Phi\sum_{y\ne x}[1+d_\Gamma(x,y)]^{-\alpha}\le C.
\end{equation}
If $x\in Z$ and $\operatorname{diam}_\Gamma Z>a$, at least one $y\in Z$ satisfies $d_\Gamma(x,y)>a/2$. Consequently,
\begin{equation}
\sum_{\substack{Z\ni x\\\operatorname{diam}_\Gamma Z>a}}
\|\Phi_Z(u)\|
\le\sum_{y:\,d_\Gamma(x,y)>a/2}
\sum_{Z\ni x,y}\|\Phi_Z(u)\|
\le C a^{d-\alpha}.
\end{equation}
The strengths are summed before any further counting of sites in $Z$. These two estimates require only $\alpha>d$.

For a unitary conjugation $\mathcal T$ on $\mathcal B(\mathcal H_\Gamma)$, define
\begin{equation}
f_{\mathcal T}(r)\coloneqq
\sup_{\substack{x\in\Gamma,\;O_x\in\mathfrak A_{\{x\}}\\
\|O_x\|=1,\;r'\ge r}}
\|\mathcal P_{x,r'}\mathcal T(O_x)\|,\quad r>0.
\end{equation}
This profile is nonincreasing in $r$ and at most two.

\begin{lemma}[Localization on an arbitrary support]
\label{lem:SM_klr_localization}
Let $\mathcal T$ be a unitary conjugation, $Y\subseteq\Gamma$ a nonempty finite set, and $\rho>0$. For $O_Y\in\mathfrak A_Y$ and $\Omega\supseteq Y^{[\rho]}$,
\begin{equation}
\|(\operatorname{id}-\mathbb E_\Omega)\mathcal T(O_Y)\|
\le4|Y|\|O_Y\|f_{\mathcal T}(\rho).
\label{eq:SM_klr_localization_bound}
\end{equation}
The set $Y$ need not be connected, and the constant is independent of the local Hilbert-space dimensions.
\end{lemma}

\begin{proof}
Let $V$ be a unitary supported on $\Omega^c$, and set $D=\mathcal T^{-1}(V)$. For $x\in Y$ and a one-site unitary $W_x$, $\mathbb E_{B_\rho^\Gamma(x)}\mathcal T(W_x)$ commutes with $V$. Hence
\begin{equation}
\|[D,W_x]\|=\|[V,\mathcal T(W_x)]\|
\le2f_{\mathcal T}(\rho).
\end{equation}
Average successively over the on-site unitary groups at sites in $Y$, with normalized Haar measure. The averaging maps commute and are contractions, so a telescoping sum gives
\begin{equation}
\|D-\mathbb E_{Y^c}(D)\|
\le\sum_{x\in Y}\sup_{W_x}\|[D,W_x]\|
\le2|Y|f_{\mathcal T}(\rho).
\end{equation}
Since $\mathbb E_{Y^c}(D)$ commutes with $O_Y$,
\begin{equation}
\|[\mathcal T(O_Y),V]\|
=\|[O_Y,D]\|
\le4|Y|\|O_Y\|f_{\mathcal T}(\rho).
\end{equation}
Averaging $V$ over the full unitary group on $\mathcal H_{\Omega^c}$ yields Eq.~\eqref{eq:SM_klr_localization_bound}. This is the conditional-expectation localization argument of Ref.~\cite{PhysRevLett.97.050401}, with the support dependence made explicit.
\end{proof}

We also use an exponential Lieb-Robinson estimate for general many-body interactions~\cite{nachtergaele2010lieb}. For a specified decomposition $K(u)=\sum_S\Psi_S(u)$ on $\Gamma$, define
\begin{equation}
\mathcal M_\mu(K)\coloneqq
\operatorname*{ess\,sup}_u\max_{x\in\Gamma}
\sum_{S\ni x}|S|\me^{\mu\operatorname{diam}_\Gamma S}
\|\Psi_S(u)\|,\qquad\mu>0.
\end{equation}
The essential supremum may be restricted to the time interval under consideration. Write $\tau^K_{s,t}$ for its dynamics with the convention of Eq.~\eqref{eq:SM_klr_dynamics}. If
$\mathcal M_\mu(K)<\infty$, disjoint supports $Y,Z$ satisfy
\begin{equation}
\|[\tau^K_{s,s+\vartheta}(O_Y),B_Z]\|
\le2|Y|\|O_Y\|\|B_Z\|
\bigl(\me^{2\mathcal M_\mu(K)\vartheta}-1\bigr)
 \me^{-\mu d_\Gamma(Y,Z)},\qquad\vartheta\ge0.
\label{eq:SM_klr_weighted_LR}
\end{equation}
There is no restriction on $|S|$ in this estimate beyond the finiteness of the weighted norm. To see the relevant counting, introduce the nonnegative matrix
\begin{equation}
M_{xy}(u)\coloneqq\sum_{S\ni x,y}\|\Psi_S(u)\|,
\quad
\max_x\sum_y \me^{\mu d_\Gamma(x,y)}M_{xy}(u)
\le\mathcal M_\mu(K).
\end{equation}
Selecting an intersection site between consecutive interaction supports overcounts each chain in the commutator expansion. The sum of chains of length $n$ connecting $Y$ to $Z$ is bounded by $|Y|\me^{-\mu d_\Gamma(Y,Z)}\mathcal M_\mu(K)^n$.
The ordered time integrals give $\vartheta^n/n!$; summing the commutator series proves Eq.~\eqref{eq:SM_klr_weighted_LR}.
Only the pointwise row bound is used, not a sum of separate time suprema. Averaging over an arbitrary external unitary gives the corresponding localization estimate without a factor $|Z|$.

For $a\ge1$, let $\Phi_{\le a}$ be the interaction obtained by retaining only supports of diameter at most $a$, and write
\begin{equation}
H_{\le a}(u)\coloneqq
\sum_{\operatorname{diam}_\Gamma Z\le a}\Phi_Z(u),
\quad
\tau^{\le a}_{s,t}\coloneqq\tau^{\Phi_{\le a}}_{s,t},
\quad F_{\le a}\coloneqq F_{\Phi_{\le a},\Gamma}.
\end{equation}
Since $\mathcal M_{1/a}(H_{\le a})\le \me\mathcal J_\Phi$, Eq.~\eqref{eq:SM_klr_weighted_LR} and conditional expectation give
\begin{equation}
F_{\le a}(\vartheta,r)
\le\min\{2,\,2\me^{c_0\vartheta-r/a}\},
\quad\vartheta\ge0,\quad r>0,
\label{eq:SM_klr_finite_range}
\end{equation}
with $c_0>0$ independent of $a$ and $\Gamma$.
The same expansion with the power-law kernel uses
\begin{equation}
\sum_{z\in\Gamma}
\frac{1}{[1+d_\Gamma(x,z)]^\alpha[1+d_\Gamma(z,y)]^\alpha}
\le\frac{C}{[1+d_\Gamma(x,y)]^\alpha},\quad\alpha>d.
\end{equation}
The off-diagonal matrix entries are controlled by Eq.~\eqref{eq:SM_klr_pair_decay}, and the diagonal entries by Eq.~\eqref{eq:SM_klr_local_strength}. Summing the resulting
commutator bound over the complement of $B_r^\Gamma(x)$ yields
\begin{equation}
F_{\Phi,\Gamma}(\vartheta,r)
\le C\frac{\me^{C\vartheta}-1}{r^{\nu_{\mathrm{tail}}}},
\quad\vartheta\ge0,\quad r\ge1.
\label{eq:SM_klr_Hastings_Koma}
\end{equation}
This is the Hastings-Koma estimate~\cite{ref:Hastings2006-ExpDec} in the form needed below. In particular,
\begin{equation}
F_{\Phi,\Gamma}(\vartheta,r)
\le C\vartheta r^{-\nu_{\mathrm{tail}}},
\quad0\le\vartheta\le1,\quad r\ge1.
\label{eq:SM_klr_short_time}
\end{equation}
Equations~\eqref{eq:SM_klr_Hastings_Koma} and \eqref{eq:SM_klr_short_time} also hold for $\alpha\ge2d+1$; their derivation requires only $\alpha>d$.

\subsubsection{Composition and shell decomposition}
\label{S-III-B-composition}

For the remainder of this subsection, keep $\Gamma$ fixed and set
\begin{equation}
N_\Gamma\coloneqq|\Gamma|,
\quad r_*\coloneqq\max\{2,\operatorname{diam}_\Gamma\Gamma\},
\quad \lambda_*\coloneqq\log(\me r_*).
\end{equation}
Here $N_\Gamma\le C r_*^d$. The quantity $\lambda_*$ is an auxiliary logarithmic size factor, not the density parameter $\lambda$ of Sec.~\ref{S-I-B}.
The following estimate keeps finite-volume prefactors separate from short time intervals, as in the composition step of Ref.~\cite{PhysRevLett.127.160401}.

\begin{lemma}[Composition of localized evolutions]
\label{lem:SM_klr_composition}
Let $1\le a\le b$, $P_1,P_2\ge2$, and $w_1,w_2\ge0$.
Suppose two unitary conjugations $\mathcal T_1,\mathcal T_2$ satisfy
\begin{equation}
f_{\mathcal T_1}(r)\le P_1\me^{(w_1-r)/a},
\quad
f_{\mathcal T_2}(r)\le P_2\me^{(w_2-r)/b},\quad r>0.
\end{equation}
There is a geometric constant $C_D\ge1$ such that, with
$D_*\coloneqq C_DN_\Gamma(r_*+3)$,
\begin{equation}
f_{\mathcal T_2\circ\mathcal T_1}(r)
\le D_*P_2P_1^{a/b}\me^{(w_1+w_2-r)/b},\quad r>0.
\label{eq:SM_klr_composition_bound}
\end{equation}
\end{lemma}

\begin{proof}
Fix $x\in\Gamma$, $\|O_x\|=1$, and set $\rho_0=w_1+a\log P_1$.
Decompose $\mathcal T_1(O_x)=\sum_{j\ge0}O_j$, where
\begin{align}
O_0=\mathbb E_{B_{\rho_0}^\Gamma(x)}\mathcal T_1(O_x),
\quad
O_j=\bigl(\mathbb E_{B_{\rho_0+ja}^\Gamma(x)}
-\mathbb E_{B_{\rho_0+(j-1)a}^\Gamma(x)}\bigr)
\mathcal T_1(O_x),\quad j\ge1.
\end{align}
Then $\|O_0\|\le1$ and $\|O_j\|\le2\me^{-(j-1)}$ for $j\ge1$.
At most $r_*+3$ terms are nonzero, and each support has at most $N_\Gamma$ sites. If $r-\rho_0-ja>0$, its closed neighborhood of that radius is contained in $B_r^\Gamma(x)$.
Lemma~\ref{lem:SM_klr_localization} therefore bounds the $j$th contribution, for $j\ge1$, by
\begin{equation}
\|\mathcal P_{x,r}\mathcal T_2(O_j)\|
\le C N_\Gamma P_2 \me^{(w_2+\rho_0-r)/b}
 \me^{-j+ja/b}
\le C N_\Gamma P_2 \me^{(w_2+\rho_0-r)/b}.
\end{equation}
If $r-\rho_0-ja\le0$, the trivial bound $2\|O_j\|$ gives the same estimate. The central term has the same bound without the factor $\me^{-j+ja/b}$. Summing and using $\me^{\rho_0/b}=P_1^{a/b}\me^{w_1/b}$ proves the claim for each $O_x$.
Taking the supremum over $x$, $O_x$, and radii $r'\ge r$ gives Eq.~\eqref{eq:SM_klr_composition_bound}.
If $\rho_0\ge r_*$, the claim follows directly from the trivial bound for $r<\rho_0$ and from zero leakage for $r\ge\rho_0$.
\end{proof}

Assume that, for some $a\ge1$, $v_a>0$, and a prefactor $P_*\ge2$
to be fixed below,
\begin{equation}
F_{\le a}(\vartheta,r)
\le P_*\me^{(v_a\vartheta-r)/a},\quad\vartheta\ge0,\quad r>0.
\label{eq:SM_klr_range_induction}
\end{equation}
The coefficient $v_a$ is an auxiliary propagation velocity for the range cutoff $a$, distinct from the fixed constant \(v\). For $b>a$, let
\begin{equation}
V_{(a,b]}(u)\coloneqq
\sum_{a<\operatorname{diam}_\Gamma Z\le b}\Phi_Z(u),
\quad H_{\le b}(u)=H_{\le a}(u)+V_{(a,b]}(u).
\end{equation}
Choose one representative $x_Z\in Z$ for each support, independently of time. For a fixed initial time $s$, the interaction-picture Hamiltonian on a time interval of length $\Delta$ is
\begin{equation}
V_I^{(s)}(u)\coloneqq
\tau^{\le a}_{s,s+u}\bigl(V_{(a,b]}(s+u)\bigr),
\quad0\le u\le\Delta,
\quad\Delta\coloneqq\frac{b}{4v_a}.
\end{equation}
For fixed $s$, suppress this initial-time label on the individual terms and set $\widetilde\Phi_Z(u)=\tau^{\le a}_{s,s+u}(\Phi_Z(s+u))$.
Decompose each evolved interaction as a whole:
\begin{align}
\widetilde\Phi_Z(u)=\sum_{j\ge0}w_{Z,j}(u),\
w_{Z,0}(u)=\mathbb E_{B_{3b}^\Gamma(x_Z)}\widetilde\Phi_Z(u),
\
w_{Z,j}(u)=\bigl(\mathbb E_{B_{(j+3)b}^\Gamma(x_Z)}
-\mathbb E_{B_{(j+2)b}^\Gamma(x_Z)}\bigr)
\widetilde\Phi_Z(u),\ j\ge1.
\end{align}
The sum is finite because $\Gamma$ is finite.
Since $\operatorname{diam}_\Gamma Z\le b$, $Z^{[jb]}\subseteq B_{(j+1)b}^\Gamma(x_Z)
\subseteq B_{(j+2)b}^\Gamma(x_Z)$.
Lemma~\ref{lem:SM_klr_localization} and Eq.~\eqref{eq:SM_klr_range_induction} give, everywhere on
$0\le u\le\Delta$,
\begin{align}
\|w_{Z,0}(u)\|\le\|\Phi_Z(s+u)\|,
\quad
\|w_{Z,j}(u)\|\le8kP_*\|\Phi_Z(s+u)\|
 \me^{-(j-1/4)b/a},\quad j\ge1.
\end{align}
In particular, if
\begin{equation}
\frac ba\ge4\log(8kP_*)+32,
\label{eq:SM_klr_range_ratio}
\end{equation}
then for all $j\ge0$,
\begin{equation}
\|w_{Z,j}(u)\|\le\|\Phi_Z(s+u)\|\me^{-8j},
\quad
\operatorname{supp}w_{Z,j}(u)
\subseteq B_{(j+3)b}^\Gamma(x_Z).
\label{eq:SM_klr_shell_bound}
\end{equation}
The fixed support bound enters through $|Z|\le k$ in the localization lemma. No expansion into products of single-site operators is required.

\subsubsection{The coarse-grained interaction norm}
\label{S-III-B-coarse}

Take $b$ to be an integer. Partition $\Gamma$ into nonempty cubes
\begin{equation}
Q_z^{(b)}\coloneqq
\bigl(bz+\{0,\ldots,b-1\}^d\bigr)\cap\Gamma,
\quad z\in\mathbb Z^d,
\end{equation}
and denote their collection by $\mathcal G_b$. Equip this auxiliary lattice with
\begin{equation}
d_b(Q_z^{(b)},Q_{z'}^{(b)})\coloneqq|z-z'|,
\quad
\operatorname{diam}_b\mathcal Z\coloneqq
\max_{Q,Q'\in\mathcal Z}d_b(Q,Q').
\end{equation}
These cubes are internal to the present proof and are not the bosonic blocks $P_x$ of Sec.~\ref{S-I-C}. Let $\mathcal Z_{Z,j}\subseteq\mathcal G_b$ contain all cubes meeting $B_{(j+3)b}^\Gamma(x_Z)$. Grouping terms with the same auxiliary support gives the exact decomposition
\begin{equation}
\Psi_{\mathcal Z}^{(b)}(u)\coloneqq
\sum_{(Z,j):\,\mathcal Z_{Z,j}=\mathcal Z}w_{Z,j}(u),
\quad
V_I^{(s)}(u)=\sum_{\mathcal Z\subseteq\mathcal G_b}
\Psi_{\mathcal Z}^{(b)}(u).
\label{eq:SM_klr_coarse_interaction}
\end{equation}
The terms are Hermitian and may involve more than two auxiliary blocks. We control their many-body interaction norm rather than assuming a two-body or fixed-$k$ form on $\mathcal G_b$.

\begin{proposition}[Coarse-grained interaction bound]
\label{prop:SM_klr_coarse_norm}
Under Eq.~\eqref{eq:SM_klr_range_ratio}, the decomposition \eqref{eq:SM_klr_coarse_interaction} satisfies
\begin{align}
\mathcal M_1^{(b)}(V_I^{(s)})
&\coloneqq
\operatorname*{ess\,sup}_{0\le u\le\Delta}
\max_{Q\in\mathcal G_b}
\sum_{\mathcal Z\ni Q}|\mathcal Z|
 \me^{\operatorname{diam}_b\mathcal Z}
\|\Psi_{\mathcal Z}^{(b)}(u)\|
\notag\\
&\le C b^d f_\Phi(a)
\le C b^d a^{-\nu_{\mathrm{tail}}}.
\end{align}
The constant is uniform in $s$, $a$, $b$, $\Gamma$, and all local dimensions.
\end{proposition}

\begin{proof}
Geometry gives
\begin{equation}
|\mathcal Z_{Z,j}|\le C(j+4)^d,
\quad
\operatorname{diam}_b\mathcal Z_{Z,j}\le2j+C.
\end{equation}
If $\mathcal Z_{Z,j}$ contains a fixed cube $Q$, its representative $x_Z$ lies within distance $C(j+4)b$ of $Q$. There are at most $Cb^d(j+4)^d$ such sites. At any one of them,
\begin{equation}
\sum_{\substack{Z:\,x_Z=x\\a<\operatorname{diam}_\Gamma Z\le b}}
\|\Phi_Z(s+u)\|\le f_\Phi(a)
\end{equation}
for almost every $u$. Combining this with Eq.~\eqref{eq:SM_klr_shell_bound} and the triangle inequality gives
\begin{equation}
\mathcal M_1^{(b)}(V_I^{(s)})
\le Cb^df_\Phi(a)
\sum_{j\ge0}(j+4)^{2d}\me^{2j+C}\me^{-8j}
\le Cb^df_\Phi(a).
\end{equation}
The sum converges independently of the volume and local dimensions.
\end{proof}

This counting replaces the independent endpoint sums in the two-body argument of Ref.~\cite{PhysRevLett.127.160401}.
There is no factor $b^{kd}$: the other sites in each original interaction support have already been summed in $f_\Phi(a)$. The exponential diameter weight controls both interactions between distant auxiliary blocks and their local strength.

Let $\tau^{I,s}_{0,\vartheta}$ be the dynamics generated by $V_I^{(s)}(u)$ on $[0,\Delta]$.
Apply Eq.~\eqref{eq:SM_klr_weighted_LR} on $\mathcal G_b$. If $x\in Q_z^{(b)}$ and $y\in Q_{z'}^{(b)}$, then $d_\Gamma(x,y)\le b|z-z'|+b-1$.
Thus, an auxiliary neighborhood of radius $r/b-2$, when positive, is contained in $B_r^\Gamma(x)$. Conditional expectations onto nested regions change the localization bound by at most a factor of two; for $r\le2b$, the trivial bound suffices.
It follows that, for a constant $C_I\ge2$ independent of $a,b,s$ and $\Gamma$,
\begin{equation}
f_{\tau^{I,s}_{0,\vartheta}}(r)
\le C_I \me^{(\Delta v\,\vartheta-r)/b},
\quad
\Delta v\le Cb^{d+1}a^{-\nu_{\mathrm{tail}}},
\quad0\le\vartheta\le\Delta,\quad r>0.
\end{equation}
For the subsequent iteration, we use the weaker estimate
\begin{equation}
\Delta v\le C\frac{b^{2d+1}}{a^\alpha},
\end{equation}
which follows from $a\le b$ and $\nu_{\mathrm{tail}}=\alpha-d$. This retains the range powers of the Tran recursion; no improvement of logarithmic factors is needed.

\subsubsection{Iteration over interaction ranges}
\label{S-III-B-iteration}

Choose the numerical prefactor
\begin{equation}
P_*\coloneqq \me C_I D_*.
\end{equation}
It is polynomial in $r_*$, and $\log P_*\le C\lambda_*$.
The propagator convention in Eq.~\eqref{eq:SM_klr_dynamics} gives
\begin{equation}
\tau^{\le b}_{s,s+\vartheta}
=\tau^{I,s}_{0,\vartheta}\circ\tau^{\le a}_{s,s+\vartheta}.
\end{equation}
Indeed, the corresponding Schr\"odinger propagator factors with the $H_{\le a}$ propagator on the left and the interaction-picture propagator on the right.
If $b/a\ge\log P_*$ as well as Eq.~\eqref{eq:SM_klr_range_ratio}, Lemma~\ref{lem:SM_klr_composition} yields
\begin{equation}
F_{\le b}(\vartheta,r)
\le P_*\me^{((v_a+\Delta v)\vartheta-r)/b},
\quad0\le\vartheta\le\Delta.
\end{equation}
Here $D_*C_IP_*^{a/b}\le \me D_*C_I=P_*$.
All estimates are uniform in the initial time $s$.

For $\vartheta>0$, divide the interval into $n=\lceil\vartheta/\Delta\rceil$ successive intervals of length at most $\Delta$. Apply Lemma~\ref{lem:SM_klr_composition} with equal ranges $a=b$ at each join, in the order specified by $\tau_{s,t}=\tau_{s,u}\circ\tau_{u,t}$.
The prefactor after $n$ intervals is $P_*(D_*P_*)^{n-1}$. Since $n-1\le\vartheta/\Delta$,
\begin{align}
F_{\le b}(\vartheta,r)
\le P_*\me^{(v_b\vartheta-r)/b},
\quad
v_b\coloneqq v_a+\Delta v+
\frac b\Delta\log(D_*P_*)
\le C_{\mathrm{rec}}\lambda_*v_a
+C_{\mathrm{inc}}\frac{b^{2d+1}}{a^\alpha}.
\end{align}
The constants $C_{\mathrm{rec}},C_{\mathrm{inc}}$ are independent of $a,b$ and $\Gamma$. At $\vartheta=0$ the leakage vanishes, so the same estimate holds. Keeping $P_*$ explicit avoids absorbing a time-independent prefactor into $\me^{C\vartheta}$ at arbitrarily short times.

Choose a fixed, sufficiently large constant $C_L$ and define
\begin{equation}
L_{\mathrm{sc}}\coloneqq
\left\lceil(C_L\lambda_*)^{1/\theta}\right\rceil,
\quad a_m\coloneqq L_{\mathrm{sc}}^m,\quad m\ge1.
\end{equation}
These are interaction-range cutoffs on $\Gamma$, not the physical bosonic block radius $\ell$. Since $0<\theta<1$, $C_L$ can be chosen to ensure
\begin{equation}
L_{\mathrm{sc}}\ge4\log(8kP_*)+32,
\quad L_{\mathrm{sc}}\ge\log P_*,
\quad
\frac{C_{\mathrm{rec}}\lambda_*}{L_{\mathrm{sc}}^\theta}
\le\frac12.
\end{equation}
The constant $C_L$ depends only on $d,\alpha,k,C_\Phi$.
Starting from Eq.~\eqref{eq:SM_klr_finite_range} with $v_1=c_0a_1$, iteration gives
\begin{equation}
F_{\le a_m}(\vartheta,r)
\le P_*\me^{(v_m\vartheta-r)/a_m},
\quad
v_{m+1}\le C_{\mathrm{rec}}\lambda_*v_m
+C_{\mathrm{inc}}\frac{a_{m+1}^{2d+1}}{a_m^\alpha}.
\end{equation}
Set $z_m=v_m/a_m^\theta$. Since $2d+1-\theta=\alpha$,
\begin{equation}
z_{m+1}\le\frac12z_m+C_{\mathrm{inc}}L_{\mathrm{sc}}^\alpha,
\quad
z_1=c_0L_{\mathrm{sc}}^\chi\le C L_{\mathrm{sc}}^\alpha.
\end{equation}
Thus $z_m\le C L_{\mathrm{sc}}^\alpha$ uniformly in $m$.
With the nonoptimized logarithmic exponent
\begin{equation}
a_{\log}\coloneqq\frac{2d+1}{\theta},
\end{equation}
we obtain
\begin{equation}
v_m\le C\lambda_*^{a_{\log}}a_m^\theta,
\quad
F_{\le a_m}(\vartheta,r)
\le P_*\exp\!\left[
\frac{C\lambda_*^{a_{\log}}a_m^\theta\vartheta-r}{a_m}
\right].
\label{eq:SM_klr_truncated_bound}
\end{equation}
This is the range-truncated estimate required for the bootstrap.
Its dependence on the cutoff is $a_m^{2d+1-\alpha}$, as in Ref.~\cite{PhysRevLett.127.160401}; the fixed support size affects only constants. The explicit factors $P_*$ and $\lambda_*$ will be removed by the spatial-restriction argument in Sec.~\ref{S-III-E}. The next subsection controls the interactions omitted by the range cutoff and establishes the polynomial starting bound.

\subsection{Omitted interactions and polynomial starting bound}
\label{S-III-C}

We continue in the normalized units $g=1$, with $2d<\alpha<2d+1$, and use the notation of
Secs.~\ref{S-III-A} and \ref{S-III-B}.
We first compare the full evolution with its restriction to interactions of bounded diameter. We then establish a polynomial starting bound that is uniform in the diameter cutoff and the lattice size. This supplies the initial estimate for the bootstrap without assuming a two-body interaction structure.

\subsubsection{The Duhamel estimate}
\label{S-III-C-Duhamel}

For $a\ge1$ and $s\le t$, the propagator convention of Eq.~\eqref{eq:SM_klr_dynamics} gives the exact identity
\begin{align}
\tau^\Phi_{s,t}(O)-\tau^{\le a}_{s,t}(O)
={}&\mi\int_s^t
\tau^\Phi_{s,u}\!\left(
\left[H_\Phi(u)-H_{\le a}(u),
\tau^{\le a}_{u,t}(O)\right]
\right)\,\md u.
\label{eq:SM_klr_Duhamel_identity}
\end{align}
Differentiating $\tau^\Phi_{s,u}\bigl(\tau^{\le a}_{u,t}(O)\bigr)$ with respect to $u$ and integrating from $s$ to $t$ proves Eq.~\eqref{eq:SM_klr_Duhamel_identity}.
The derivative holds almost everywhere. Finite dimensionality and local integrability of the interactions ensure the absolute continuity required for integration.

Fix an initial site $x_0\in\Gamma$ and $O_{x_0}\in\mathfrak A_{\{x_0\}}$ with $\|O_{x_0}\|=1$. For each omitted support $Z$, choose $x(Z)\in Z$ nearest to $x_0$ and set
\begin{equation}
\rho_Z\coloneqq d_\Gamma(x(Z),x_0)
=d_\Gamma(Z,\{x_0\}).
\end{equation}
The choice depends on $x_0$ but not on time; it is distinct from the representative $x_Z$ used in Sec.~\ref{S-III-B}.
If $\rho_Z>0$, then $Z$ is disjoint from the closed ball $B_{\rho_Z/2}^\Gamma(x_0)$. Hence $\Phi_Z(u)$ commutes with $\mathbb E_{B_{\rho_Z/2}^\Gamma(x_0)}
\tau^{\le a}_{u,t}(O_{x_0})$, and
\begin{align}
\|[\Phi_Z(u),\tau^{\le a}_{u,t}(O_{x_0})]\|
&\le 2\|\Phi_Z(u)\|
\|\mathcal P_{x_0,\rho_Z/2}
\tau^{\le a}_{u,t}(O_{x_0})\|
\notag\\
&\le 2\|\Phi_Z(u)\|F_{\le a}(t-u,\rho_Z/2).
\label{eq:SM_klr_omitted_commutator}
\end{align}
For $\rho_Z=0$, the last inequality follows from the trivial commutator bound and the convention $F_{\le a}(\vartheta,0)=2$.
The smaller radius ensures disjointness under the closed-ball convention of Sec.~\ref{S-III-A}.

For almost every $u$, the total strength assigned to a given site $x(Z)=x$ obeys
\begin{equation}
\sum_{\substack{Z:\,x(Z)=x\\
\operatorname{diam}_\Gamma Z>a}}\|\Phi_Z(u)\|
\le f_\Phi(a).
\end{equation}
Apply $\mathcal P_{x_0,r'}$ to Eq.~\eqref{eq:SM_klr_Duhamel_identity}, use
$\|\mathcal P_{x_0,r'}\|_{\mathrm{op}\to\mathrm{op}}\le2$, and sum Eq.~\eqref{eq:SM_klr_omitted_commutator} over the omitted supports. Taking the suprema in
Eq.~\eqref{eq:SM_klr_profile} gives
\begin{align}
F_{\Phi,\Gamma}(\vartheta,r)
\le{}F_{\le a}(\vartheta,r)
+4f_\Phi(a)\int_0^\vartheta
\max_{x_0\in\Gamma}\sum_{x\in\Gamma}
F_{\le a}\!\left(w,\frac{d_\Gamma(x,x_0)}{2}\right)\,\md w,
\quad \vartheta\ge0,\quad r>0.
\label{eq:SM_klr_Duhamel_profile}
\end{align}
The integration variable $w$ is the remaining elapsed time in the truncated evolution. The estimate is uniform in the initial time.
The same argument applies when the full Hamiltonian is itself restricted to diameter at most $a_{\max}>a$, with $F_{\Phi,\Gamma}$ replaced by $F_{\le a_{\max}}$ and the same
upper bound $f_\Phi(a)$ on the omitted strength.
This supplies the omitted-interaction estimate used in the range-decomposition argument of
Ref.~\cite{PhysRevLett.127.160401} without any additional counting factor for the other sites in $Z$.

\subsubsection{A polynomial starting bound}
\label{S-III-C-seed}

The bootstrap requires an initial spatial decay exponent greater than $d$, but its initial time exponent need not be optimal.
We obtain such an estimate by iterating the diameter cutoff, as in the approach of Else \emph{et al.}~\cite{PhysRevA.101.022333}.
The relevant tail exponent is $\nu_{\mathrm{tail}}=\alpha-d$, not $\alpha$.

\begin{lemma}[Polynomial starting bound]
\label{lem:SM_klr_polynomial_seed}
There exist $\beta_0>d$, $\gamma_0\ge\beta_0/\chi$, and $C_{\mathrm{seed}}>0$, depending only on $d,\alpha,k,C_\Phi$, such that every interaction family of Sec.~\ref{S-III-A}, in the normalized units $g=1$, satisfies
\begin{equation}
F_{\Phi,\Gamma}(\vartheta,r)
\le C_{\mathrm{seed}}
\frac{(1+\vartheta)^{\gamma_0}}{r^{\beta_0}},
\quad \vartheta\ge0,\quad r\ge1.
\label{eq:SM_klr_seed_bound}
\end{equation}
The same constants apply to all diameter truncations and spatial restrictions. They are independent of the lattice size, the local Hilbert-space dimensions, and the temporal variation of the interactions.
\end{lemma}

\begin{proof}
Choose $\sigma\in(d/\nu_{\mathrm{tail}},1)$ and set
\begin{equation}
T_{\mathrm{LR}}\coloneqq1+\vartheta.
\end{equation}
This is a shifted time in the normalized units and does not redefine the physical time $T=1+|t-s|$.
We prove estimates for $H_{\le a_{\max}}$ uniformly in the auxiliary diameter cutoff $a_{\max}\ge1$.

\textit{First cutoff estimate.---}
The finite-range bound~\eqref{eq:SM_klr_finite_range} gives
\begin{equation}
\max_{x_0\in\Gamma}\sum_{x\in\Gamma}
F_{\le a}\!\left(\vartheta,\frac{d_\Gamma(x,x_0)}{2}\right)
\le C a^d T_{\mathrm{LR}}^d,
\qquad a\ge1.
\label{eq:SM_klr_seed_initial_sum}
\end{equation}
Split the sum at a sufficiently large fixed multiple of $aT_{\mathrm{LR}}$ and use the exponential bound outside that ball. Lattice volume growth and $F_{\le a}\le2$ control the
interior. The factor $1/2$ in the distance changes only the geometric constants.

For $r\ge1$, set $a=r^\sigma$.
If $a<a_{\max}$, apply Eq.~\eqref{eq:SM_klr_Duhamel_profile} to $H_{\le a_{\max}}$, using
Eqs.~\eqref{eq:SM_klr_diameter_tail}, \eqref{eq:SM_klr_finite_range}, and \eqref{eq:SM_klr_seed_initial_sum}.
The omitted contribution is at most $C a^{d-\nu_{\mathrm{tail}}}T_{\mathrm{LR}}^{d+1}$.
If $a\ge a_{\max}$, the finite-range estimate alone applies, with $r/a_{\max}\ge r^{1-\sigma}$. Combining both cases yields
\begin{equation}
F_{\le a_{\max}}(\vartheta,r)
\le\min\left\{2,\,
C_1\left[
 \me^{c_0\vartheta-r^{1-\sigma}}
 +T_{\mathrm{LR}}^{\gamma^{(1)}}r^{-\beta^{(1)}}
 \mathbf1_{\{r\le a_{\max}^{1/\sigma}\}}
\right]\right\},
\label{eq:SM_klr_seed_first_cutoff}
\end{equation}
where
\begin{equation}
\beta^{(1)}=\sigma(\nu_{\mathrm{tail}}-d),
\quad \gamma^{(1)}=d+1.
\end{equation}
Here $\mathbf1_E$ denotes the indicator of $E$.
The superscripts on $\beta^{(n)},\gamma^{(n)}$ index this cutoff iteration. The indicator confines the polynomial correction to $r\le a_{\max}^{1/\sigma}$.

\textit{Iteration of the spatial exponent.---} Suppose Eq.~\eqref{eq:SM_klr_seed_first_cutoff} holds with $C_1,\beta^{(1)},\gamma^{(1)}$ replaced by $C_n,\beta^{(n)},\gamma^{(n)}$, uniformly in $a_{\max}$.
The exponential part, bounded above by two, has spatial sum at most $C_nT_{\mathrm{LR}}^{d/(1-\sigma)}$.
For the polynomial part, the indicator restricts the sum to $d_\Gamma(x,x_0)\le2a_{\max}^{1/\sigma}$.
If $\beta^{(n)}\ne d$, summing the lattice shells gives
\begin{align}
\max_{x_0\in\Gamma}\sum_{x\in\Gamma}
F_{\le a_{\max}}\!\left(
\vartheta,\frac{d_\Gamma(x,x_0)}{2}\right)
\le C_n'\left[
T_{\mathrm{LR}}^{d/(1-\sigma)}
+T_{\mathrm{LR}}^{\gamma^{(n)}}
a_{\max}^{(d-\beta^{(n)})_+/\sigma}
\right],
\label{eq:SM_klr_seed_spatial_sum}
\end{align}
where $(z)_+\coloneqq\max\{z,0\}$.
The finitely many sites with $d_\Gamma(x,x_0)/2<1$ are controlled by the trivial bound; their number is bounded by a constant depending only on $d$.
The constants may change with the iteration number, but not with $a_{\max}$, $\Gamma$, or the local dimensions.

Apply Eq.~\eqref{eq:SM_klr_Duhamel_profile} once more with $a=r^\sigma<a_{\max}$, using
Eq.~\eqref{eq:SM_klr_seed_spatial_sum} for $F_{\le a}$.
The omitted contribution is now bounded by
\begin{equation}
C_n''\left[
T_{\mathrm{LR}}^{1+d/(1-\sigma)}r^{-\sigma\nu_{\mathrm{tail}}}
+
T_{\mathrm{LR}}^{1+\gamma^{(n)}}
r^{-\sigma\nu_{\mathrm{tail}}+(d-\beta^{(n)})_+}
\right].
\end{equation}
The finite-range term is unchanged, and the case $a\ge a_{\max}$ is treated as before. Thus, the same form of the bound holds at the next iteration with
\begin{equation}
\beta^{(n+1)}
=\sigma\nu_{\mathrm{tail}}-(d-\beta^{(n)})_+,
\quad
\gamma^{(n+1)}
=1+\max\left\{\frac{d}{1-\sigma},\gamma^{(n)}\right\}.
\end{equation}
As long as $\beta^{(n)}<d$, the spatial exponent increases by $\sigma\nu_{\mathrm{tail}}-d>0$. Therefore, $\beta^{(n_*)}>d$ after a finite number $n_*$ of iterations. We choose $\sigma$ outside the at most countable set of values for which an intermediate exponent equals $d$.
Such a choice is possible in $(d/\nu_{\mathrm{tail}},1)$ and avoids logarithmic factors in Eq.~\eqref{eq:SM_klr_seed_spatial_sum}.
The choice of $\sigma$ and the resulting $n_*$ depend only on $d$ and $\alpha$.

\textit{Uniform polynomial bound.---} Set $\beta_0=\beta^{(n_*)}$ and choose
\begin{equation}
\gamma_0>
\max\left\{
\gamma^{(n_*)},\frac{\beta_0}{1-\sigma},
\frac{\beta_0}{\chi}
\right\}.
\end{equation}
If $r^{1-\sigma}\le2c_0T_{\mathrm{LR}}$, then $r^{\beta_0}\le C T_{\mathrm{LR}}^{\beta_0/(1-\sigma)}$, so the trivial bound is at most $C T_{\mathrm{LR}}^{\gamma_0}/r^{\beta_0}$.
Otherwise,
\begin{equation}
\me^{c_0\vartheta-r^{1-\sigma}}
\le \me^{-r^{1-\sigma}/2}
\le C r^{-\beta_0}.
\end{equation}
The polynomial term is bounded by $C T_{\mathrm{LR}}^{\gamma_0}/r^{\beta_0}$ after dropping its indicator. Hence, the same estimate holds uniformly for every
$a_{\max}\ge1$. Taking $a_{\max}\ge\max\{1,\operatorname{diam}_\Gamma\Gamma\}$ proves Eq.~\eqref{eq:SM_klr_seed_bound} for the full interaction.

Every step uses only the uniform interaction-strength and tail bounds of Sec.~\ref{S-III-B}, lattice volume growth, and ordered time integrals. Removing interactions or restricting the lattice preserves these bounds. This proves the asserted uniformity, including for measurable, locally integrable time dependence.
\end{proof}

The polynomial starting bound contains no finite-volume prefactor. Its exponents are not optimized: only $\beta_0>d$ and $\gamma_0\ge\beta_0/\chi$ are needed.
Together with Eq.~\eqref{eq:SM_klr_truncated_bound} and the Duhamel estimate~\eqref{eq:SM_klr_Duhamel_profile}, it provides the inputs to the bootstrap in Sec.~\ref{S-III-D}.

\subsection{Bootstrap to the Tran bound}
\label{S-III-D}

We combine the polynomial starting bound of Lemma~\ref{lem:SM_klr_polynomial_seed} with the range-truncated estimate~\eqref{eq:SM_klr_truncated_bound} and the Duhamel estimate~\eqref{eq:SM_klr_Duhamel_profile}.
Throughout this subsection, $2d<\alpha<2d+1$, $g=1$, and $T_{\mathrm{LR}}=1+\vartheta$. We retain the notation of Secs.~\ref{S-III-A}--\ref{S-III-C}.
Following the bootstrap of Ref.~\cite{PhysRevLett.127.160401}, we improve the spatial and temporal powers while allowing the power of $\lambda_*=\log(\me r_*)$ to increase. This keeps each inductive estimate valid for all $\vartheta\ge0$ and $r\ge1$, without accumulating additional time restrictions.

\subsubsection{One bootstrap step}
\label{S-III-D-step}

\begin{lemma}[Bootstrap step]
\label{lem:SM_klr_bootstrap}
Suppose that, uniformly over the interaction class of Sec.~\ref{S-III-A} in the normalized units $g=1$,
\begin{equation}
F_{\Phi,\Gamma}(\vartheta,r)
\le C_{\mathrm{in}}\lambda_*^{\kappa_{\log}}
\frac{T_{\mathrm{LR}}^\gamma}{r^\beta},
\quad
\vartheta\ge0,\quad r\ge1,
\quad
\beta>d,\quad \gamma\ge\frac{\beta}{\chi},\quad
\kappa_{\log}\ge0.
\end{equation}
For every $0<\eta_{\mathrm{bs}}<1/\nu_{\mathrm{tail}}$, there exist finite constants $C_{\mathrm{out}}>0$ and $\kappa_{\log}'\ge0$ such that
\begin{equation}
F_{\Phi,\Gamma}(\vartheta,r)
\le C_{\mathrm{out}}\lambda_*^{\kappa_{\log}'}
\frac{T_{\mathrm{LR}}^{\gamma'}}{r^{\beta'}},
\quad \vartheta\ge0,\quad r\ge1,
\end{equation}
where
\begin{equation}
\beta'=\nu_{\mathrm{tail}}(1-\eta_{\mathrm{bs}}\chi),
\quad
\gamma'=1+\frac{d\gamma}{\beta}
-\eta_{\mathrm{bs}}\nu_{\mathrm{tail}}.
\label{eq:SM_klr_bootstrap_exponents}
\end{equation}
These exponents satisfy $\beta'>d$ and $\gamma'\ge\beta'/\chi$.
The output constants may depend on $d,\alpha,k,C_\Phi$, the fixed input constants and exponents, and $\eta_{\mathrm{bs}}$, but not on $\Gamma$, the local Hilbert-space dimensions, or the temporal variation of the interactions. The same statement holds
uniformly for diameter truncations and spatial restrictions.
\end{lemma}

The parameter $\eta_{\mathrm{bs}}$ is used only in this bootstrap; it is distinct from the spatial-restriction exponent $\eta$ in the bosonic upper-bound argument.

\begin{proof}
First, note that
\begin{equation}
\beta'-d=\chi(1-\eta_{\mathrm{bs}}\nu_{\mathrm{tail}})>0,
\quad
\gamma'\ge1+\frac d\chi-\eta_{\mathrm{bs}}\nu_{\mathrm{tail}}
=\frac{\beta'}{\chi}.
\end{equation}
Thus, the output exponents remain in the class required for the next iteration.

\textit{Spatial summation and the range cutoff.---} The inductive hypothesis applies to $\Phi_{\le a}$ on the same lattice $\Gamma$, since removing interaction terms preserves the pair-summed bound. Combining it with $F_{\le a}\le2$ gives
\begin{equation}
\max_{x_0\in\Gamma}\sum_{x\in\Gamma}
F_{\le a}\!\left(w,\frac{d_\Gamma(x,x_0)}2\right)
\le C\lambda_*^{d\kappa_{\log}/\beta}
(1+w)^{d\gamma/\beta},
\quad w\ge0,\quad a\ge1.
\label{eq:SM_klr_bootstrap_spatial_sum}
\end{equation}
To obtain this estimate, split the sum at a sufficiently large fixed multiple of $\lambda_*^{\kappa_{\log}/\beta}(1+w)^{\gamma/\beta}$.
Use the trivial bound inside and the inductive bound outside; the exterior sum converges because $\beta>d$.
The sites with $d_\Gamma(x,x_0)/2<1$ contribute only a fixed geometric constant. The factor $1/2$ is the one required by the closed-ball convention in Eq.~\eqref{eq:SM_klr_Duhamel_profile}.

For a grid value $a=a_m$, insert Eq.~\eqref{eq:SM_klr_bootstrap_spatial_sum} into the Duhamel estimate and use Eqs.~\eqref{eq:SM_klr_diameter_tail} and
\eqref{eq:SM_klr_truncated_bound}. This yields
\begin{equation}
F_{\Phi,\Gamma}(\vartheta,r)
\le P_*\me^{(v_a\vartheta-r)/a}
+C a^{-\nu_{\mathrm{tail}}}
\lambda_*^{d\kappa_{\log}/\beta}
T_{\mathrm{LR}}^{1+d\gamma/\beta},
\quad
v_a\le C\lambda_*^{a_{\log}}a^\theta,
\label{eq:SM_klr_bootstrap_start}
\end{equation}
where $v_a=v_m$ when $a=a_m$.

\textit{Choice of the range.---} Set $z=T_{\mathrm{LR}}/r^\chi$ and first consider the region
\begin{equation}
T_{\mathrm{LR}}\le c_{\mathrm{bs}}r^\chi
\lambda_*^{-a_{\log}},
\quad
a_{\mathrm{opt}}\coloneqq
\frac{r}{M_{\mathrm{bs}}\lambda_*}z^{\eta_{\mathrm{bs}}}
\ge L_{\mathrm{sc}},
\label{eq:SM_klr_bootstrap_regular_region}
\end{equation}
with $c_{\mathrm{bs}}>0$ sufficiently small and $M_{\mathrm{bs}}>0$ sufficiently large, both independent of $\vartheta,r$, and $\Gamma$.
Choose the largest grid value $a=a_m\le a_{\mathrm{opt}}$. Then
\begin{equation}
\frac{a_{\mathrm{opt}}}{L_{\mathrm{sc}}}
\le a\le a_{\mathrm{opt}}\le r.
\end{equation}
Since $\chi+\theta=1$, the first condition in Eq.~\eqref{eq:SM_klr_bootstrap_regular_region} implies $v_a\vartheta\le r/4$ after choosing $c_{\mathrm{bs}}$.
Moreover, $\log P_*\le C\lambda_*$ and $r/a\ge M_{\mathrm{bs}}\lambda_*z^{-\eta_{\mathrm{bs}}}$.
Taking $M_{\mathrm{bs}}$ large enough therefore gives
\begin{equation}
P_*\me^{(v_a\vartheta-r)/a}
\le \me^{-r/(2a)}
\le \me^{-c z^{-\eta_{\mathrm{bs}}}}
\le C z^{\gamma'}
\le C\frac{T_{\mathrm{LR}}^{\gamma'}}{r^{\beta'}}.
\end{equation}
The last inequality follows from $\chi\gamma'\ge\beta'$ and $r\ge1$.

For the second term in Eq.~\eqref{eq:SM_klr_bootstrap_start}, use $a\ge a_{\mathrm{opt}}/L_{\mathrm{sc}}$ and $L_{\mathrm{sc}}\le C\lambda_*^{1/\theta}$ to obtain
\begin{align}
a^{-\nu_{\mathrm{tail}}}
\lambda_*^{d\kappa_{\log}/\beta}
T_{\mathrm{LR}}^{1+d\gamma/\beta}
&\le
C\lambda_*^{d\kappa_{\log}/\beta+
\nu_{\mathrm{tail}}(1+1/\theta)}
r^{-\nu_{\mathrm{tail}}}
z^{-\eta_{\mathrm{bs}}\nu_{\mathrm{tail}}}
T_{\mathrm{LR}}^{1+d\gamma/\beta}
\notag\\
&=
C\lambda_*^{d\kappa_{\log}/\beta+
\nu_{\mathrm{tail}}(1+1/\theta)}
\frac{T_{\mathrm{LR}}^{\gamma'}}{r^{\beta'}}.
\end{align}

\textit{Complementary parameter regions.---} If the first condition in Eq.~\eqref{eq:SM_klr_bootstrap_regular_region} fails, then
\begin{equation}
r^{\beta'}
\le C\lambda_*^{a_{\log}\beta'/\chi}
T_{\mathrm{LR}}^{\beta'/\chi}
\le C\lambda_*^{a_{\log}\beta'/\chi}
T_{\mathrm{LR}}^{\gamma'}.
\end{equation}
If $a_{\mathrm{opt}}<L_{\mathrm{sc}}$, then $r^{1-\eta_{\mathrm{bs}}\chi}T_{\mathrm{LR}}^{\eta_{\mathrm{bs}}}
<M_{\mathrm{bs}}\lambda_*L_{\mathrm{sc}}$. Since $T_{\mathrm{LR}}\ge1$, it follows that
\begin{equation}
r^{\beta'}\le
C\lambda_*^{\nu_{\mathrm{tail}}(1+1/\theta)}.
\end{equation}
In either case, the trivial bound $F_{\Phi,\Gamma}\le2$ proves the required estimate after increasing its constant and its logarithmic power. For example, the choice
\begin{equation}
\kappa_{\log}'
=1+\max\left\{
\frac{d\kappa_{\log}}{\beta}
+\nu_{\mathrm{tail}}\left(1+\frac1\theta\right),
\frac{a_{\log}\beta'}{\chi}
\right\}
\end{equation}
works in all parameter regions.
Every estimate uses the same interaction class and the explicit finite-volume parameters from Sec.~\ref{S-III-B}. This proves the stated uniformity.
\end{proof}

\subsubsection{Iteration to the target exponents}
\label{S-III-D-iteration}

The exponent update in Eq.~\eqref{eq:SM_klr_bootstrap_exponents} is the one used in the
Tran bootstrap~\cite{PhysRevLett.127.160401}.
Start from Lemma~\ref{lem:SM_klr_polynomial_seed}, with $\beta_0>d$, $\gamma_0\ge\beta_0/\chi$, and initial logarithmic power zero. Fix $0<\eta_{\mathrm{bs}}<1/\nu_{\mathrm{tail}}$.
After the first application of Lemma~\ref{lem:SM_klr_bootstrap}, the spatial exponent is
\begin{equation}
\beta_{\mathrm{bs}}
\coloneqq\nu_{\mathrm{tail}}(1-\eta_{\mathrm{bs}}\chi)>d
\end{equation}
and remains unchanged at subsequent steps.
Let $\gamma_{\mathrm{bs}}^{(n)}$ denote the time exponent after
$n$ bootstrap steps. Then
\begin{align}
\gamma_{\mathrm{bs}}^{(1)}
=1+\frac{d\gamma_0}{\beta_0}
-\eta_{\mathrm{bs}}\nu_{\mathrm{tail}},
\quad
\gamma_{\mathrm{bs}}^{(n+1)}
=\frac{d}{\beta_{\mathrm{bs}}}\gamma_{\mathrm{bs}}^{(n)}
+1-\eta_{\mathrm{bs}}\nu_{\mathrm{tail}},
\quad n\ge1.
\end{align}
Because $d/\beta_{\mathrm{bs}}<1$, this recurrence converges geometrically to
\begin{equation}
\gamma_{\mathrm{bs}}^{(\infty)}
=\frac{1-\eta_{\mathrm{bs}}\nu_{\mathrm{tail}}}
       {1-d/\beta_{\mathrm{bs}}}
=\frac{\beta_{\mathrm{bs}}}{\chi},
\end{equation}
with
\begin{equation}
\gamma_{\mathrm{bs}}^{(n)}-\frac{\beta_{\mathrm{bs}}}{\chi}
=\left(\frac{d}{\beta_{\mathrm{bs}}}\right)^{n-1}
\left(\gamma_{\mathrm{bs}}^{(1)}
-\frac{\beta_{\mathrm{bs}}}{\chi}\right)\ge0.
\label{eq:SM_klr_bootstrap_convergence}
\end{equation}
At each finite step, the numerical constant and the exponent of $\lambda_*$ remain finite and independent of the lattice size and local dimensions.

Fix the loss parameter $\varepsilon_{\mathrm{LR}}$ from Eq.~\eqref{eq:SM_klr_admissible_loss} and choose
\begin{equation}
\eta_{\mathrm{bs}}=\frac{\varepsilon_{\mathrm{LR}}}
                         {\nu_{\mathrm{tail}}},
\quad
\beta_{\mathrm{bs}}=\nu_{\mathrm{tail}}-\chi\varepsilon_{\mathrm{LR}},
\quad
\gamma_{\mathrm{bs}}^{(\infty)}
=\frac{\nu_{\mathrm{tail}}}{\chi}-\varepsilon_{\mathrm{LR}}.
\end{equation}
The admissible range ensures $\varepsilon_{\mathrm{LR}}<1$, so $0<\eta_{\mathrm{bs}}<1/\nu_{\mathrm{tail}}$ and $\beta_{\mathrm{bs}}>d$.
Equation~\eqref{eq:SM_klr_bootstrap_convergence} permits a finite number of iterations $n_*$ such that
\begin{equation}
\frac{\beta_{\mathrm{bs}}}{\chi}
\le\gamma_{\mathrm{bs}}\coloneqq\gamma_{\mathrm{bs}}^{(n_*)}
\le\frac{\nu_{\mathrm{tail}}}{\chi}
-\frac{\varepsilon_{\mathrm{LR}}}{2}.
\end{equation}
Consequently, there are finite $C_{\mathrm{bs}}>0$ and
$\kappa_{\log}\ge0$ such that
\begin{equation}
F_{\Phi,\Gamma}(\vartheta,r)
\le C_{\mathrm{bs}}\lambda_*^{\kappa_{\log}}
\frac{T_{\mathrm{LR}}^{\gamma_{\mathrm{bs}}}}
     {r^{\beta_{\mathrm{bs}}}},
\quad \vartheta\ge0,\quad r\ge1.
\label{eq:SM_klr_finite_volume_bootstrap_bound}
\end{equation}
The iteration number and all constants depend only on $d,\alpha,k,C_\Phi,\varepsilon_{\mathrm{LR}}$, not on $\vartheta,r$, the lattice size, or the local dimensions.
The same estimate applies to all diameter truncations and spatial restrictions, with $\lambda_*$ evaluated on the lattice on which the dynamics act.

The remaining system-size dependence is the explicit factor $\lambda_*^{\kappa_{\log}}$.
In Sec.~\ref{S-III-E}, a spatial restriction replaces this factor by a logarithm of the localization radius. The strict spatial exponent margin then absorbs that logarithm and yields Eq.~\eqref{eq:SM_klr_Tran_bound}.

\subsection{System-size independence and finite-support observables}
\label{S-III-E}

We complete the proof of Theorem~\ref{thm:SM_klr_Tran} by replacing the finite-volume factor in Eq.~\eqref{eq:SM_klr_finite_volume_bootstrap_bound} with a logarithm of the localization radius. We then extend the bound to arbitrary finite supports and state the uniform form used in the bosonic upper-bound argument. The spatial restriction follows Ref.~\cite{PhysRevLett.127.160401}, with the closed-ball convention of Sec.~\ref{S-III-A}.

\subsubsection{Removing the system-size dependence}
\label{S-III-E-size}

\begin{proof}[Completion of the proof of Theorem~\ref{thm:SM_klr_Tran}]
Work first in the normalized units $g=1$ with $s\le t$ and $\vartheta=t-s$. Fix $\varepsilon_{\mathrm{LR}}$ in the range \eqref{eq:SM_klr_admissible_loss}, and use
$\beta_{\mathrm{bs}},\gamma_{\mathrm{bs}},\kappa_{\log}$ from Sec.~\ref{S-III-D}. In particular,
\begin{equation}
\beta_{\mathrm{bs}}
=\nu_{\mathrm{tail}}-\chi\varepsilon_{\mathrm{LR}}>d,
\quad
\frac{\beta_{\mathrm{bs}}}{\chi}
\le\gamma_{\mathrm{bs}}
\le\frac{\nu_{\mathrm{tail}}}{\chi}
-\frac{\varepsilon_{\mathrm{LR}}}{2}.
\end{equation}
Fix $x_0\in\Gamma$ and a unit-norm $O_{x_0}\in\mathfrak A_{\{x_0\}}$. Consider initially $r\ge2$ and $1\le\vartheta\le r^{\chi-\varepsilon_{\mathrm{LR}}}$. Choose
\begin{equation}
r_{\mathrm{aux}}\coloneqq\lceil r^{\eta_{\mathrm{aux}}}\rceil,
\quad
\eta_{\mathrm{aux}}>
\max\left\{
1,\frac{\beta_{\mathrm{bs}}+\chi-\varepsilon_{\mathrm{LR}}}
        {\beta_{\mathrm{bs}}-d},
\frac{\nu_{\mathrm{tail}}}{\chi}
\right\}.
\label{eq:SM_klr_auxiliary_radius}
\end{equation}
This is an auxiliary spatial restriction on $\Gamma$; neither $r_{\mathrm{aux}}$ nor $\eta_{\mathrm{aux}}$ redefines the physical radius $R$ or the exponent $\eta$ used in Sec.~\ref{S-IV}. We set
\begin{equation}
\Gamma_{\mathrm{in}}\coloneqq B_{r_{\mathrm{aux}}}^\Gamma(x_0),
\quad
H_{\mathrm{in}}(u)\coloneqq
\sum_{Z\subseteq\Gamma_{\mathrm{in}}}\Phi_Z(u),
\quad
H_{\mathrm{out}}(u)\coloneqq H_\Phi(u)-H_{\mathrm{in}}(u),
\end{equation}
and denote the dynamics of $H_{\mathrm{in}}$ by $\tau^{\mathrm{in}}_{s,t}$. Its interaction family belongs to the same class as $\Phi$. On $\Gamma_{\mathrm{in}}$, the size parameter satisfies
\begin{equation}
\log\!\left(\me\max\{2,
\operatorname{diam}_\Gamma\Gamma_{\mathrm{in}}\}\right)
\le C\log(\me r_{\mathrm{aux}}).
\label{eq:SM_klr_restricted_logarithm}
\end{equation}
The evolution under $H_{\mathrm{in}}$ is tensored with the identity outside $\Gamma_{\mathrm{in}}$. For this evolution, normalized partial trace on $\Gamma$ therefore gives the same localization error as partial trace on $\Gamma_{\mathrm{in}}$.
Applying Eq.~\eqref{eq:SM_klr_finite_volume_bootstrap_bound} on that smaller lattice yields
\begin{equation}
\|\mathcal P_{x_0,r}\tau^{\mathrm{in}}_{s,t}(O_{x_0})\|
\le C[\log(\me r_{\mathrm{aux}})]^{\kappa_{\log}}
\frac{(1+\vartheta)^{\gamma_{\mathrm{bs}}}}
     {r^{\beta_{\mathrm{bs}}}}.
\label{eq:SM_klr_restricted_evolution_bound}
\end{equation}

\textit{Interactions crossing the auxiliary boundary.---} Use the Duhamel identity~\eqref{eq:SM_klr_Duhamel_identity}, with $H_{\le a}$ replaced by $H_{\mathrm{in}}$.
Terms supported wholly outside $\Gamma_{\mathrm{in}}$ commute with $\tau^{\mathrm{in}}_{u,t}(O_{x_0})$ and do not contribute.
For each remaining term, choose $x(Z)\in Z$ nearest to $x_0$ and put $\rho_Z=d_\Gamma(x(Z),x_0)$, as in Sec.~\ref{S-III-C}. There are two cases.

If $r_{\mathrm{aux}}/2<\rho_Z\le r_{\mathrm{aux}}$, the commutator is controlled by localization to $B_{\rho_Z/2}^\Gamma(x_0)$, which is disjoint from $Z$.
The total interaction strength assigned to each $x(Z)$ is at most $\mathcal J_\Phi\le C$. There are at most $C r_{\mathrm{aux}}^d$ possible nearest sites, and $\rho_Z/2>r_{\mathrm{aux}}/4$.
The restricted bootstrap bound and the time integral, therefore, bound this contribution to the localization error by
\begin{equation}
C[\log(\me r_{\mathrm{aux}})]^{\kappa_{\log}}
\frac{\vartheta(1+\vartheta)^{\gamma_{\mathrm{bs}}}}
     {r_{\mathrm{aux}}^{\beta_{\mathrm{bs}}-d}}
\le
C[\log(\me r_{\mathrm{aux}})]^{\kappa_{\log}}
\frac{\vartheta^{\gamma_{\mathrm{bs}}+1}}
     {r_{\mathrm{aux}}^{\beta_{\mathrm{bs}}-d}}.
\label{eq:SM_klr_outer_boundary_error}
\end{equation}
The endpoint $\rho_Z=r_{\mathrm{aux}}$ is included because the auxiliary ball is closed.

If $\rho_Z\le r_{\mathrm{aux}}/2$, the crossing term also has a site outside $B_{r_{\mathrm{aux}}}^\Gamma(x_0)$, so its diameter is greater than $r_{\mathrm{aux}}/2$. The strength assigned to a fixed nearest site is consequently at most $f_\Phi(r_{\mathrm{aux}}/3)\le
C r_{\mathrm{aux}}^{-\nu_{\mathrm{tail}}}$.
The trivial commutator bound gives the contribution
\begin{equation}
C\vartheta r_{\mathrm{aux}}^d
f_\Phi(r_{\mathrm{aux}}/3)
\le C\vartheta r_{\mathrm{aux}}^{-\chi}.
\label{eq:SM_klr_inner_boundary_error}
\end{equation}
These estimates include every crossing interaction, regardless of the locations of its other support sites. The factor $\|\mathcal P_{x_0,r}\|_{\mathrm{op}\to\mathrm{op}}\le2$ is included in the constants.

Combining Eqs.~\eqref{eq:SM_klr_restricted_evolution_bound}--
\eqref{eq:SM_klr_inner_boundary_error} and using
$1+\vartheta\le2\vartheta$ gives
\begin{align}
\|\mathcal P_{x_0,r}\tau^\Phi_{s,t}(O_{x_0})\|
\le{}
C[\log(\me r_{\mathrm{aux}})]^{\kappa_{\log}}
\frac{\vartheta^{\gamma_{\mathrm{bs}}}}{r^{\beta_{\mathrm{bs}}}}
+C[\log(\me r_{\mathrm{aux}})]^{\kappa_{\log}}
\frac{\vartheta^{\gamma_{\mathrm{bs}}+1}}
     {r_{\mathrm{aux}}^{\beta_{\mathrm{bs}}-d}}
+C\frac{\vartheta}{r_{\mathrm{aux}}^\chi}.
\end{align}
The radius choice~\eqref{eq:SM_klr_auxiliary_radius} implies
\begin{equation}
\frac{\vartheta r^{\beta_{\mathrm{bs}}}}
     {r_{\mathrm{aux}}^{\beta_{\mathrm{bs}}-d}}\le1,
\quad
r_{\mathrm{aux}}^{-\chi}\le r^{-\nu_{\mathrm{tail}}},
\quad
\log(\me r_{\mathrm{aux}})\le C\log(\me r).
\end{equation}
Thus, for $1\le\vartheta\le r^{\chi-\varepsilon_{\mathrm{LR}}}$,
\begin{equation}
\|\mathcal P_{x_0,r}\tau^\Phi_{s,t}(O_{x_0})\|
\le C[\log(\me r)]^{\kappa_{\log}}
\frac{\vartheta^{\gamma_{\mathrm{bs}}}}{r^{\beta_{\mathrm{bs}}}}
+C\frac{\vartheta}{r^{\nu_{\mathrm{tail}}}}.
\label{eq:SM_klr_size_free_intermediate}
\end{equation}
If $\Gamma_{\mathrm{in}}=\Gamma$, the Duhamel error is zero; Eq.~\eqref{eq:SM_klr_restricted_logarithm} still gives the same bound. No constant depends on the ambient lattice size.

\textit{Absorbing the logarithm and restoring time.---} Set
\begin{equation}
\omega=\chi-\varepsilon_{\mathrm{LR}},
\quad
\gamma_1=\frac{\nu_{\mathrm{tail}}}{\chi}
-\frac{\varepsilon_{\mathrm{LR}}}{2},
\quad
\gamma_2=\omega\gamma_1.
\end{equation}
By construction, $\gamma_{\mathrm{bs}}\le\gamma_1$, and
\begin{equation}
\beta_{\mathrm{bs}}-\gamma_2
=\varepsilon_{\mathrm{LR}}
\left(\frac{\nu_{\mathrm{tail}}}{\chi}
-\frac\chi2-\frac{\varepsilon_{\mathrm{LR}}}{2}\right)>0.
\end{equation}
The strict margin absorbs the logarithm:
$[\log(\me r)]^{\kappa_{\log}}
\le C r^{\beta_{\mathrm{bs}}-\gamma_2}$ for $r\ge1$.
Therefore, Eq.~\eqref{eq:SM_klr_size_free_intermediate} gives
\begin{equation}
\|\mathcal P_{x_0,r}\tau^\Phi_{s,t}(O_{x_0})\|
\le C_1\left(\frac{\vartheta}{r^\omega}\right)^{\gamma_1}
+C_2\frac{\vartheta}{r^{\nu_{\mathrm{tail}}}}.
\end{equation}
For $0\le\vartheta\le1$, use the short-time estimate \eqref{eq:SM_klr_short_time}. The remaining bounded radii are covered by increasing the constants. The same is true when
$\vartheta\ge r^\omega$, since the first term is then at least one before its constant is included. These estimates are uniform in $x_0$, the initial time, and unit-norm $O_{x_0}$.
Their right-hand side increases with $\vartheta$ and decreases with $r$, so taking the suprema in Eq.~\eqref{eq:SM_klr_profile} gives the same bound for $F_{\Phi,\Gamma}$.

Reverse time intervals are treated by reflecting time and reversing the sign of the Hamiltonian, which preserves the pair-summed condition. Finally, undo Eq.~\eqref{eq:SM_klr_time_rescaling} and replace the normalized elapsed time by $\vartheta=g|t-s|$. The prefactor $C_\Phi$ remains included in the constants, not in the definition of $g$.
Restoring the factor $\|O_{x_0}\|$ proves Eq.~\eqref{eq:SM_klr_Tran_bound}, with constants independent of $g$, $\Gamma$, and all local Hilbert-space dimensions.
\end{proof}

\subsubsection{Finite-support observables}
\label{S-III-E-support}

Restore arbitrary $g>0$ and let $\alpha>2d$.
For a chosen branch of Corollary~\ref{cor:SM_klr_all_alpha}, let $\mathcal B_{\mathrm{LR}}(\vartheta,r)$ denote its scalar one-site bound, without the factor $\|O_x\|$ and with any exponent loss fixed. Work in the time-radius range of that branch, with $r\ge1$. Each bound is nonincreasing in $r$ and therefore also
bounds the map profile of $\tau^\Phi_{s,t}$ at $\vartheta=g|t-s|$. Lemma~\ref{lem:SM_klr_localization} gives
\begin{equation}
\|(\operatorname{id}-\mathbb E_{Y^{[r]}})
\tau^\Phi_{s,t}(O_Y)\|
\le4|Y|\|O_Y\|\mathcal B_{\mathrm{LR}}(\vartheta,r),
\quad O_Y\in\mathfrak A_Y,
\label{eq:SM_klr_finite_support_bound}
\end{equation}
for every nonempty finite $Y\subseteq\Gamma$. In particular,
\begin{equation}
\widetilde O_{Y,r}\coloneqq
\mathbb E_{Y^{[r]}}\tau^\Phi_{s,t}(O_Y)
\in\mathfrak A_{Y^{[r]}},
\quad \|\widetilde O_{Y,r}\|\le\|O_Y\|.
\label{eq:SM_klr_local_approximant}
\end{equation}
If $B_Z\in\mathfrak A_Z$ and $d_\Gamma(Y,Z)>r$, this approximant commutes with $B_Z$, so
\begin{equation}
\|[\tau^\Phi_{s,t}(O_Y),B_Z]\|
\le8|Y|\|O_Y\|\|B_Z\|
\mathcal B_{\mathrm{LR}}(\vartheta,r).\label{approximant_commuts}
\end{equation}
The strict separation is appropriate for the closed neighborhood $Y^{[r]}$. There is no factor $|Z|$, no connectedness assumption on $Y$ or $Z$, and no dependence on the local dimensions.

\subsubsection{Unified estimate used in the bosonic upper bound}
\label{S-III-E-unified}

We now put these bounds in the form used in the main text.
The time $T=1+|t-s|$ below is the physical-time variable, not $T_{\mathrm{LR}}=1+\vartheta$.

\begin{corol}[Uniform finite-support localization estimate]
\label{cor:SM_klr_unified_input}
Assume Eqs.~\eqref{eq:SM_klr_Hamiltonian} and
\eqref{eq:SM_klr_pair_decay}, with fixed $k$, $C_\Phi$, and $\alpha>2d$. For every
\begin{equation}
0<\omega<\min\{1,\alpha-2d\}
\end{equation}
sufficiently close to the upper endpoint, there exist finite exponents $\gamma_1>1$ and $\gamma_2=\omega\gamma_1>d$, and constants $C,C_r\ge1$, depending only on $d,\alpha,k,C_\Phi,\omega$, such that
\begin{equation}
\|(\operatorname{id}-\mathbb E_{Y^{[r]}})
\tau^\Phi_{s,t}(O_Y)\|
\le C|Y|\|O_Y\|
\frac{(gT)^{\gamma_1}}{r^{\gamma_2}},
\quad T=1+|t-s|,
\end{equation}
whenever $Y\subseteq\Gamma$ is nonempty and finite, $O_Y\in\mathfrak A_Y$, and
\begin{equation}
gT\ge1,\quad r\ge C_r(gT)^{1/\omega}.
\label{eq:SM_klr_unified_radius_condition}
\end{equation}
All constants are independent of $g$, the lattice size, the local Hilbert-space dimensions, and the temporal variation of the interactions. The approximant in Eq.~\eqref{eq:SM_klr_local_approximant} realizes this bound.
\end{corol}

\begin{proof}
For $2d<\alpha<2d+1$, set $\bar\alpha=\alpha$ and $\bar\varepsilon_{\mathrm{LR}}=\alpha-2d-\omega$.
For $\omega$ sufficiently close to $\alpha-2d$ from below, this is an admissible positive loss.
For $\alpha\ge2d+1$, choose
\begin{equation}
\bar\alpha=2d+\frac{1+\omega}{2},
\quad
\bar\varepsilon_{\mathrm{LR}}=\frac{1-\omega}{2}.
\end{equation}
Then $2d<\bar\alpha<2d+1\le\alpha$, and the same pair-summed bound holds with exponent $\bar\alpha$ and prefactor $C_\Phi g$.
For $\omega$ sufficiently close to one, $\bar\varepsilon_{\mathrm{LR}}$ is admissible.
In both cases, $\bar\alpha-2d-\bar\varepsilon_{\mathrm{LR}}
=\omega$. Applying Theorem~\ref{thm:SM_klr_Tran} at the exponent $\bar\alpha$, and then
Eq.~\eqref{eq:SM_klr_finite_support_bound}, gives
\begin{align}
\|(\operatorname{id}-\mathbb E_{Y^{[r]}})
\tau^\Phi_{s,t}(O_Y)\|
\le C|Y|\|O_Y\|\left[
\left(\frac{\vartheta}{r^\omega}\right)^{\gamma_1}
+\frac{\vartheta}{r^{\bar\alpha-d}}
\right],
\quad
\gamma_1=\frac{\bar\alpha-d}{\bar\alpha-2d}
-\frac{\bar\varepsilon_{\mathrm{LR}}}{2},
\label{eq:SM_klr_unified_two_term_bound}
\end{align}
provided $\vartheta\le c r^\omega$. The exponents satisfy $\gamma_1>1$ and $\gamma_2=\omega\gamma_1<\bar\alpha-d$.
As $\omega$ approaches its upper endpoint, $\gamma_2$ approaches $\alpha-d>d$ in the first case and $d+1>d$ in the second. Hence $\gamma_2>d$ for the choices above.

Choose $C_r$ in Eq.~\eqref{eq:SM_klr_unified_radius_condition} so that $\vartheta\le gT\le c r^\omega$. Since $gT\ge1$, $r\ge1$, $\gamma_1>1$, and $\bar\alpha-d>\gamma_2$,
\begin{equation}
\left(\frac{\vartheta}{r^\omega}\right)^{\gamma_1}
\le\frac{(gT)^{\gamma_1}}{r^{\gamma_2}},
\quad
\frac{\vartheta}{r^{\bar\alpha-d}}
\le\frac{gT}{r^{\bar\alpha-d}}
\le\frac{(gT)^{\gamma_1}}{r^{\gamma_2}}.
\end{equation}
Substitution in Eq.~\eqref{eq:SM_klr_unified_two_term_bound} proves the corollary. No constant depends on the occupation cutoff through the local Hilbert-space dimensions.
\end{proof}

For $2d<\alpha<2d+1$, the fixed-$k$ extension proved in this section preserves the light-cone scaling $\vartheta\lesssim r^{\alpha-2d-\varepsilon_{\mathrm{LR}}}$ and the same near-optimal leakage tail as Ref.~\cite{PhysRevLett.127.160401}.
Corollary~\ref{cor:SM_klr_all_alpha} combines this extension with the strictly linear result of Kuwahara and Saito~\cite{PhysRevX.10.031010} to cover all $\alpha>2d$.
At the marginal point $\alpha=2d+1$, the corollary gives an arbitrarily near-linear cone by lowering the decay exponent before applying Theorem~\ref{thm:SM_klr_Tran}; no new strictly linear endpoint theorem is claimed here.
We also do not establish a zero-loss bound for $2d<\alpha<2d+1$ or an estimate uniform in $k$. Corollary~\ref{cor:SM_klr_unified_input} provides the finite-support localization estimate used in the main text.
These results concern bounded interactions on finite-dimensional local spaces.
Their application to the bosonic blocks $P_x$ {of the Bose--Hubbard-type model}, including the cutoff dependence of the interaction scale $g$ and the errors from removing the occupation cutoff, is treated in {Secs.~\ref{S-IV-A}--\ref{S-IV-E}}.

\section{Upper bound on information propagation}
\label{S-IV}

In this section, we prove Lieb--Robinson bounds for the Bose--Hubbard-type dynamics introduced in Sec.~\ref{S-I}.
For the reader's convenience, we restate Theorem~\ref{thm:SM_main_upper_bound} below.

Recall that $\beta_{\mathrm{LR}}=\max\{1,1/(\alpha-2d)\}$ is the reference leading exponent for bounded interactions discussed in Sec.~\ref{S-III}. Moreover, $d\ge1$  is the spatial dimension and $\alpha>2d$ is the interaction decay exponent.

\SMUpperBound*

The parameter \(\lambda\) controls the initial density, \(\xi\) denotes the highest density moment assumed to be bounded,  \(q_0\) is the maximum change in particle number that the operator \(A\) can induce, and $p$ is the degree of the density interaction.
Recall that $\tau$ denotes the dynamics generated by $H$, and $X[R]$ and $X_R$ denote the enlargement of $X$ with respect to  lattice sites and blocks, respectively.

\textit{Proof strategy.---}
 The proof combines particle-number propagation with occupation truncation and local approximation, using the methods of
Refs.~\cite{Lemm2025-ov,kuwahara2022optimal,LR26}.
Section~\ref{S-IV-A} establishes the moment bounds for  controlled-density initial states.
Section~\ref{S-IV-B} uses them to truncate both the observable and the dynamics.
Section~\ref{S-IV-C} then restricts the truncated Hamiltonian to $X_{2R^\eta}$, where $\eta>1$ controls an auxiliary spatial enlargement and is fixed independently of the system size.
A coarse estimate of interaction strength suffices at this stage.
In the final step, Sec.~\ref{S-IV-D} uses the commuting short-range density terms to define an interaction picture and to establish the improved interaction scale.
Section~\ref{S-IV-E} applies Corollary~\ref{cor:SM_klr_unified_input}, constructs a norm-nonincreasing local approximant, and extends it to the full Fock space while controlling the removal of exterior occupation projectors.

Let $\bar A$ and $\bar H$ denote the particle-truncated observable and Hamiltonian defined in Sec.~\ref{S-IV-B}.
Write $\bar\tau_{s,t}$ for the dynamics of $\bar H$ and $\bar\tau_{s,t}^{X_{2R^\eta}}$ for the dynamics of its spatial restriction $\bar H_{X_{2R^\eta}}$.
The following is a schematic representation of the series of approximations that leads to the desired bound
\[
\tau_{s,t}(A)
\longrightarrow\tau_{s,t}(\bar A)
\longrightarrow\bar\tau_{s,t}(\bar A)
\longrightarrow\bar\tau_{s,t}^{X_{2R^\eta}}(\bar A)
\longrightarrow\widetilde A.
\]

In the final two steps, we use our extension of the Lieb--Robinson bound of Tran \emph{et al.}~\cite{PhysRevLett.127.160401} to $k$-body interactions, established in Sec.~\ref{S-III}, together with the linear light-cone bound of Kuwahara and Saito~\cite{PhysRevX.10.031010}.
To deal with the particle-number projector, we use the particle propagation estimates established in Sec.~\ref{S-IV-A}.
The final choice of $\xi$ makes the occupation errors sufficiently small to be combined with the spatial-restriction and localization errors.

\subsection{Particle-number propagation bounds}
\label{S-IV-A}

Particle propagation bounds provide the occupation-tail estimates needed in Sec.~\ref{S-IV-B}. 
In this section we state and prove such estimates. 
In a recent work~\cite{Lemm2025-ov}, the authors establish particle-propagation bounds for pure states with local particle-number moments controlled from above and below.
Here, we prove the upper-moment bound needed below without the lower-density assumption, and allow mixed states and time-dependent Hamiltonians.

Throughout this section, we only require $\alpha>d+1$, which follows from $\alpha>2d$ for $d\ge1$.

\begin{theorem}[Particle-number propagation from controlled density]
\label{thm:SM_particle_propagation}
Consider $\alpha>d+1$, \(v>0\), $\lambda>0$, and $\xi\ge1$. 
There is a constant \(C_{\mathrm{p}}>0\) such that 
\begin{equation}
\operatorname{Tr}\!\left[
\rho\,\tau_{s,t}\bigl(n_{B_r(i)}^\zeta\bigr)\right]
\le \bigl(C_{\mathrm{p}}(1+\lambda)\zeta R^d\bigr)^\zeta
\end{equation}
for every $\rho\in\mathcal D_\xi^\lambda$, $i\in\Lambda$, $s,t\in\mathbb R$,  $1\le\zeta\le\xi$, and $R>r\ge 1$ with $R-r\ge  v|t-s|$.

\end{theorem}
The constants above are independent of system size or total particle number. 
Moreover, the constants are independent of the density-interaction order \(p\).
Indeed, the density potential commutes with all local number operators and therefore does not contribute to the number-current estimates.
The state \(\rho\) is specified at time \(s\).

The relative moment estimate in~\cite{Lemm2025-ov} requires the velocity parameter to exceed a fixed multiple of the first moment of the hopping matrix (i.e., assuming $v>\kappa$ for some $\kappa$).
However, we can safely remove the assumption by appropriately rescaling the ball radius in Ref.~\cite{Lemm2025-ov}, and our estimate holds for every fixed \(v>0\). 
The constant \(C_{\mathrm p}\) depends on \(v\) and may diverge as \(v\downarrow0\).
This freedom reflects that we control occupation moments in terms of the initial density bound, rather than compare them directly with initial local moments, and does not imply an arbitrarily small propagation speed.

A similar particle-propagation bound was independently obtained using the same method in \cite{Elgart2026}.

\begin{remark}
      The same statement holds if we only require the controlled density assumption to hold for radii larger than a fixed threshold  $r_{\mathrm{cd}}$ provided $R>r\ge  r_{\mathrm{cd}}$.
\end{remark}

We first bound the local particle number by a weighted number operator. 
We then differentiate the weighted particle-number moment along the evolution. 
Its derivative is determined by the commutator of the Hamiltonian with the weighted moment observable. The density potential commutes with this observable, so only the hopping terms contribute.
Notice that the exponential time dependency that arises from the Gronwall argument is harmless, as its rate is proportional to $R^{-1}$.
We first impose a cutoff on the total particle number and prove the estimate with constants independent of the cutoff. We then let the cutoff tend to infinity.

\subsubsection{Weighted-number estimate and proof of propagation}
\label{S-IV-A-proof}

Define $\sigma_{\mathrm w}:=(d+\alpha)/2$. For $i\in\Lambda$ and $L\ge1$, set
\begin{equation}
w_j\coloneqq\left(1+\frac{|i-j|}{L}\right)^{-\sigma_{\mathrm w}},
\quad
N_{\mathrm w}\coloneqq\sum_{j\in\Lambda}w_j n_j.
\end{equation}
The weights satisfy $0<w_j\le1$. They decay away from $i$ on the length scale $L$, defining a weighted particle number rather than an occupation cutoff. Shell counting gives
\begin{equation}
\sum_{j\in\Lambda}w_j
\le C\sum_{m=0}^{\infty}(1+m)^{d-1}
       (1+m/L)^{-\sigma_{\mathrm w}}
\le C L^d.
\label{eq:SM_ppb_weight_sum}
\end{equation}
Indeed, the sum over $m\le L$ is $O(L^d)$, while the remainder is bounded by $CL^{\sigma_{\mathrm w}}\sum_{m>L}m^{d-1-\sigma_{\mathrm w}}
\le CL^d$. Constants below may depend on the fixed exponent $\sigma_{\mathrm w}$.

\begin{lemma}[Weighted hopping estimate]
\label{lem:SM_ppb_weighted_hopping}
There is $C_{\mathrm w}>0$, depending only on $d,\alpha,\sigma_{\mathrm w},C_{J}$, such that
\begin{equation}
\sum_{k\in\Lambda}|J_{jk}(u)|\,|w_j-w_k|
\le\frac{C_{\mathrm w}}{L}w_j
\end{equation}
for every $i,j\in\Lambda$, $u\in\mathbb R$, and $L\ge1$.
\end{lemma}

\begin{proof}
Put $D_j=L+|i-j|$, so $w_j=L^{\sigma_{\mathrm w}}D_j^{-\sigma_{\mathrm w}}$.
The coefficient bound and shell counting imply the tail estimate
\begin{equation}
\sum_{k:\,|j-k|>a}|J_{jk}(u)|\le C a^{d-\alpha},
\quad a\ge\tfrac12.
\label{eq:SM_ppb_hopping_tail}
\end{equation}
For $a\ge1$, this follows by summing $Cm^{d-1}(1+m)^{-\alpha}$ over $m>a$.
The interval $1/2\le a<1$ is covered by increasing $C$.

First consider $|j-k|\le D_j/2$.
The reverse triangle inequality yields $|D_k-D_j|\le|j-k|$ and $D_j/2\le D_k\le3D_j/2$.
Since \(\alpha>d+1\), the decay assumption~\eqref{eq:SM_power_law_bounds} gives \(\sum_{k\in\Lambda}|J_{jk}(u)|\,|j-k|\le C\), uniformly in \(j\),~\(u\), and the lattice size.
Applying the mean-value theorem to $y^{-\sigma_{\mathrm w}}$ gives
\begin{equation}
|w_j-w_k|\le C w_j\frac{|j-k|}{D_j},
\quad
\sum_{|j-k|\le D_j/2}|J_{jk}(u)|\,|w_j-w_k|
\le\frac{C}{L}w_j.
\end{equation}

For $|j-k|>D_j/2$, use $|w_j-w_k|\le w_j+w_k$.
The contribution containing $w_j$ is at most $Cw_jD_j^{d-\alpha}$ by Eq.~\eqref{eq:SM_ppb_hopping_tail}.
To control the contribution containing $w_k$, first suppose $|i-j|\le L$. Then $w_j\ge2^{-\sigma_{\mathrm w}}$ and
\begin{equation}
\sum_{|j-k|>D_j/2}|J_{jk}(u)|w_k
\le C D_j^{d-\alpha}
\le C L^{d-\alpha}
\le C w_j L^{d-\alpha}.
\end{equation}
If $|i-j|>L$, split the remaining sites into $|k-i|\ge|i-j|/2$ and $|k-i|<|i-j|/2$. In the first region $D_k\ge D_j/2$, so $w_k\le2^{\sigma_{\mathrm w}}w_j$.
Its contribution is again at most $Cw_jD_j^{d-\alpha}$.
In the second region $|j-k|>|i-j|/2$. Therefore, Eq.~\eqref{eq:SM_ppb_weight_sum} gives
\begin{equation}
\sum_{|k-i|<|i-j|/2}|J_{jk}(u)|w_k
\le C L^d|i-j|^{-\alpha}
\le C w_jL^{d-\alpha}.
\end{equation}
For the last inequality, $D_j\le2|i-j|$ implies
\begin{equation}
\frac{L^d|i-j|^{-\alpha}}{w_jL^{d-\alpha}}
=\left(\frac{L}{|i-j|}\right)^{\alpha-\sigma_{\mathrm w}}
 \left(\frac{D_j}{|i-j|}\right)^{\sigma_{\mathrm w}}
\le2^{\sigma_{\mathrm w}}.
\end{equation}
Finally, $D_j\ge L\ge1$ and $\alpha>d+1$ imply $D_j^{d-\alpha}\le L^{d-\alpha}\le L^{-1}$.
Combining the near and far contributions proves the lemma.
\end{proof}

\begin{lemma}[Propagation of weighted moments]
\label{lem:SM_ppb_weighted_moments}
For $\zeta\ge1$, set $G\coloneqq N_{\mathrm w}+\zeta I$.
There is $C_{\mathrm{tr}}>0$, depending only on $d,\alpha,\sigma_{\mathrm w},C_{J}$, such that every positive trace-class operator $\rho$ with
$\operatorname{Tr}(n_\Lambda^\zeta\rho)<\infty$ satisfies
\begin{equation}
\operatorname{Tr}\!\left[\rho\,\tau_{s,t}(G^\zeta)\right]
\le \exp\!\left(\frac{C_{\mathrm{tr}}\zeta|t-s|}{L}\right)
\operatorname{Tr}(\rho G^\zeta).
\label{eq:SM_ppb_weighted_moment_evolution}
\end{equation}
The constant is independent of $\zeta,L,i$, the lattice size, and the density potential.
\end{lemma}

\begin{proof}
First, work on the invariant subspace with \(n_\Lambda\le M\), where \(M\) is a temporary total-number bound.
All matrices are finite-dimensional on this subspace.
On the number basis \(\ket{\mathbf n}\), write \(G\ket{\mathbf n}=G_{\mathbf n}\ket{\mathbf n}\).
For a hop $k\to j$ with $j\ne k$ and $n_k\ge1$, let $\mathbf n'=\mathbf n+\mathbf e_j-\mathbf e_k$, where $\mathbf e_j$ is the unit occupation vector at site $j$. Then
$G_{\mathbf n'}-G_{\mathbf n}=w_j-w_k$.
Both eigenvalues are at least $\zeta$, and $|w_j-w_k|\le1$. The mean-value theorem thus yields
\begin{equation}
|G_{\mathbf n'}^\zeta-G_{\mathbf n}^\zeta|
\le \me^{1/2}\zeta|w_j-w_k|
(G_{\mathbf n'}G_{\mathbf n})^{(\zeta-1)/2}.
\end{equation}
In detail, the mean-value bound is $\zeta|w_j-w_k|\max\{G_{\mathbf n'},G_{\mathbf n}\}^{\zeta-1}$. Dividing its last factor by
$(G_{\mathbf n'}G_{\mathbf n})^{(\zeta-1)/2}$ gives at most
\begin{equation}
\left(1+\frac1\zeta\right)^{(\zeta-1)/2}\le \me^{1/2}.
\end{equation}

For a positive matrix $\rho$, the positivity of every $2\times2$ principal minor implies $|\rho_{\mathbf n,\mathbf n'}|
\le(\rho_{\mathbf n,\mathbf n}\rho_{\mathbf n',\mathbf n'})^{1/2}$.
The number-basis matrix element of $b_j^\dagger b_k$ is
$\sqrt{(n_j+1)n_k}$. Therefore
\begin{align}
\left|\operatorname{Tr}\!\left[
\rho[b_j^\dagger b_k,G^\zeta]\right]\right|
&\le\sum_{\mathbf n:\,n_k\ge1}
|\rho_{\mathbf n,\mathbf n'}|
\sqrt{(n_j+1)n_k}\,
|G_{\mathbf n'}^\zeta-G_{\mathbf n}^\zeta|
\notag\\
&\le \me^{1/2}\zeta|w_j-w_k|
\sum_{\mathbf n:\,n_k\ge1}
\left[
\rho_{\mathbf n',\mathbf n'}(n_j+1)G_{\mathbf n'}^{\zeta-1}
\right]^{1/2}
\left[
\rho_{\mathbf n,\mathbf n}n_kG_{\mathbf n}^{\zeta-1}
\right]^{1/2}.
\end{align}
All sums here are restricted to a total number at most $M$.
Applying $2ab\le a^2+b^2$ bounds the last sum by
\begin{equation}
\frac12\sum_{\mathbf n:\,n_k\ge1}
\left[
\rho_{\mathbf n',\mathbf n'}(n_j+1)G_{\mathbf n'}^{\zeta-1}
+\rho_{\mathbf n,\mathbf n}n_kG_{\mathbf n}^{\zeta-1}
\right].
\end{equation}
The map $\mathbf n\mapsto\mathbf n'$ is a bijection between configurations with $n_k\ge1$ and those with $n_j'\ge1$ at the same total number. In the first summand, $n_j+1=n_j'$. After reindexing, the two sums are $\operatorname{Tr}(\rho n_jG^{\zeta-1})$ and $\operatorname{Tr}(\rho n_kG^{\zeta-1})$, respectively. Terms with zero corresponding occupation vanish, so the restrictions $n_j'\ge1$ and $n_k\ge1$ can be removed. It follows that
\begin{equation}
\left|\operatorname{Tr}\!\left[
\rho[b_j^\dagger b_k,G^\zeta]\right]\right|
\le\frac{\me^{1/2}\zeta}{2}|w_j-w_k|
\operatorname{Tr}\!\left[
\rho(n_j+n_k)G^{\zeta-1}\right].
\end{equation}

The density terms and diagonal hopping terms commute with $G$. Hermiticity gives $J_{kj}(u)={\overline{J}_{jk}(u)}$ for $j\ne k$.
Using this symmetry to combine the two occupation terms gives
\begin{align}
\left|\operatorname{Tr}\!\left[\rho \mi[H(u),G^\zeta]\right]\right|
&\le \me^{1/2}\zeta\sum_j
\left(\sum_k|J_{jk}(u)|\,|w_j-w_k|\right)
\operatorname{Tr}(\rho n_jG^{\zeta-1})
\notag\\
&\le\frac{C_{\mathrm{tr}}\zeta}{L}
\operatorname{Tr}(\rho N_{\mathrm w}G^{\zeta-1})
\le\frac{C_{\mathrm{tr}}\zeta}{L}
\operatorname{Tr}(\rho G^\zeta).
\label{eq:SM_ppb_moment_derivative_bound}
\end{align}
The second line uses Lemma~\ref{lem:SM_ppb_weighted_hopping} and $0\le N_{\mathrm w}\le G$, with $N_{\mathrm w}$ commuting with $G$.

Set \(f(u)=\operatorname{Tr}[\tau_{u,s}(\rho)G^\zeta]\).
Differentiating \(f\) and applying~\eqref{eq:SM_ppb_moment_derivative_bound} gives \(|f'(u)|\le(C_{\mathrm{tr}}\zeta/L)f(u)\).
Gronwall's inequality, integrated from $s$ toward $t$ in either time direction, proves Eq.~\eqref{eq:SM_ppb_weighted_moment_evolution} on the finite-total-number subspace.

For a general positive trace-class $\rho$, let $P_M=\mathbf1_{[0,M]}(n_\Lambda)$ and
$\rho_M=P_M\rho P_M$, without renormalization.
Both $G$ and $U(s,t)$ commute with $P_M$.
If $D$ denotes either $G^\zeta$ or $\tau_{s,t}(G^\zeta)$, then
\begin{equation}
\operatorname{Tr}(\rho_M D)
=\sum_{m=0}^M
\operatorname{Tr}(\Pi_{\Lambda,m}\rho\Pi_{\Lambda,m}D)
\mathop{\longrightarrow}_{M\to\infty}
\operatorname{Tr}(\rho D),
\end{equation}
where $\Pi_{\Lambda,m}$ is the spectral projector onto total particle number $m$. Every summand is nonnegative, so convergence is monotone. The initial trace is finite because $G^\zeta\le(n_\Lambda+\zeta I)^\zeta$.
The estimate is uniform in $M$, and its limit bounds the evolved trace as well. This proves the lemma on the full Fock space.
\end{proof}

\begin{proof}[Proof of Theorem~\ref{thm:SM_particle_propagation}]

Choose \(L=R\) and apply the controlled-density assumption directly to \(\rho\).
On $B_L(i)$, the weight is at most one. On $B_{2^{m+1}L}(i)\setminus B_{2^mL}(i)$ it is at most $2^{-m\sigma_{\mathrm w}}$, for $m\ge0$. Consequently,
\begin{equation}
N_{\mathrm w}\le n_{B_L(i)}+
\sum_{m=0}^{\infty}2^{-m\sigma_{\mathrm w}}
 n_{B_{2^{m+1}L}(i)\setminus B_{2^mL}(i)}.
\end{equation}
The number operators commute, so their joint spectral distribution in $\rho$ is a classical probability measure. Minkowski's inequality and monotonicity under inclusion of the annuli in their outer balls give
\begin{align}
\bigl[\operatorname{Tr}(\rho N_{\mathrm w}^\zeta)\bigr]^{1/\zeta}
&\le
\bigl[\operatorname{Tr}(\rho n_{B_L(i)}^\zeta)\bigr]^{1/\zeta}
+\sum_{m=0}^{\infty}2^{-m\sigma_{\mathrm w}}
\bigl[\operatorname{Tr}(\rho
 n_{B_{2^{m+1}L}(i)}^\zeta)\bigr]^{1/\zeta}
\notag\\
&\le\lambda\zeta L^d
\left[1+2^d\sum_{m=0}^{\infty}2^{-m(\sigma_{\mathrm w}-d)}\right]
\le C\lambda\zeta L^d.
\label{eq:SM_ppb_initial_weighted_moment}
\end{align}
The geometric sum converges because $\sigma_{\mathrm w}>d$. A further application of Minkowski's inequality, including the constant shift in $G$, yields
\begin{equation}
\bigl[\operatorname{Tr}(\rho G^\zeta)\bigr]^{1/\zeta}
\le \zeta+C\lambda\zeta L^d
\le C(1+\lambda)\zeta L^d.
\label{eq:SM_ppb_initial_regularized_moment}
\end{equation}
Only upper moments at radii at least $L$ have been used.

Since $r<R=L$, every $j\in B_r(i)$ satisfies $w_j\ge2^{-\sigma_{\mathrm w}}$. Hence $n_{B_r(i)}\le2^{\sigma_{\mathrm w}}N_{\mathrm w}$. The common spectral calculus of these commuting operators implies $n_{B_r(i)}^\zeta\le2^{\sigma_{\mathrm w}\zeta}G^\zeta$.
Combining this with positivity, the weighted-moment estimate, and Eq.~\eqref{eq:SM_ppb_initial_regularized_moment}, we obtain
\begin{align}
\operatorname{Tr}\!\left[
\rho\,\tau_{s,t}(n_{B_r(i)}^\zeta)\right]
&\le2^{\sigma_{\mathrm w}\zeta}
\exp\!\left(\frac{C_{\mathrm{tr}}\zeta|t-s|}{R}\right)
\bigl[C(1+\lambda)\zeta R^d\bigr]^\zeta
\notag\\
&\le\bigl[C_{\mathrm{p}}(1+\lambda)\zeta R^d\bigr]^\zeta.
\end{align}
The time condition gives $|t-s|/R\le1/v$. The exponential is thus at most $[\exp(C_{\mathrm{tr}}/v)]^\zeta$, and the factor $2^{\sigma_{\mathrm w}\zeta}$ is absorbed using $\zeta\ge1$.
All remaining constants depend only on $d,\alpha,C_{J}$ and $v$, with $\sigma_{\mathrm w}=(d+\alpha)/2$ fixed.
This proves the theorem.
The same proof applies when the initial moment bounds hold only for radii at least \(r_{\mathrm{cd}}\ge1\), provided \(R\ge r_{\mathrm{cd}}\), since every initial ball used in~\eqref{eq:SM_ppb_initial_weighted_moment} has radius at least \(R\).
\end{proof}

\textit{Extension to positive trace-class operators.---} Let $\sigma$ be a positive trace-class operator satisfying $\operatorname{Tr}(n_\Lambda^\xi\sigma)<\infty$. Suppose $K_\sigma>0$ obeys
\begin{equation}
\begin{gathered}
\operatorname{Tr}\sigma\le K_\sigma,\quad
\operatorname{Tr}(n_{B_a(i)}^\zeta\sigma)
\le K_\sigma(\lambda\zeta a^d)^\zeta,\quad
a\ge r_{\mathrm{cd}},\quad 1\le\zeta\le\xi,\quad i\in\Lambda.
\end{gathered}
\end{equation}

The operator
\begin{equation}
\rho_\sigma\coloneqq\frac{\sigma}{K_\sigma}
+\left(1-\frac{\operatorname{Tr}\sigma}{K_\sigma}\right)
 |\mathrm{vac}\rangle\langle\mathrm{vac}|
\end{equation}
is positive, has trace one, and satisfies the same upper-moment condition with density parameter $\lambda$.
The vacuum density matrix is invariant under the dynamics, and all its particle-number moments vanish. Therefore
\begin{equation}
\operatorname{Tr}\!\left[
\rho_\sigma\,\tau_{s,t}(n_{B_r(i)}^\zeta)\right]
=\frac1{K_\sigma}\operatorname{Tr}\!\left[
\sigma\,\tau_{s,t}(n_{B_r(i)}^\zeta)\right].
\end{equation}
Applying the preceding weighted-moment argument to $\rho_\sigma$ gives
\begin{equation}
\operatorname{Tr}\!\left[
\sigma\,\tau_{s,t}(n_{B_r(i)}^\zeta)\right]
\le K_\sigma\bigl[C_{\mathrm{p}}(1+\lambda)\zeta R^d\bigr]^\zeta
\label{eq:SM_ppb_positive_data_propagation}
\end{equation}
under the same radius and time conditions, with $R\ge r_{\mathrm{cd}}$.
This form applies to the positive operators obtained after a local observable is inserted in Sec.~\ref{S-IV-B}.

\textit{Moments on the fixed block partition.---} Fix $s,t$ and retain $T=1+|t-s|$ and
$\ell=\lceil v|t-s|\rceil$.
Hold the block partition fixed throughout this time interval.
For each nonempty $P_x$, choose a site $i_x\in P_x$.
The block center $x\in\mathcal C$ need not lie in $\Lambda$, but $|i-i_x|\le|i-x|+|x-i_x|\le2\ell$ for every $i\in P_x$.
Thus $P_x\subseteq B_{2\ell}(i_x)$.
Choose the comparison radii
\begin{equation}
r_{\mathrm B}\coloneqq\max\{2\ell,1\},\quad
R_{\mathrm B}\coloneqq2r_{\mathrm B}+v|t-s|+2.
\end{equation}
For any $u$ between $s$ and $t$,
\begin{equation}
R_{\mathrm B}-r_{\mathrm B}
=r_{\mathrm B}+v|t-s|+2
>v|u-s|.
\end{equation}
These are the conditions of Theorem~\ref{thm:SM_particle_propagation}. Moreover, $\ell\le v|t-s|+1$ implies $R_{\mathrm B}\le CT$.
Monotonicity of commuting number operators and the theorem give
\begin{equation}
\sup_{u\in[\min\{s,t\},\max\{s,t\}]}
\operatorname{Tr}\!\left[
\rho\,\tau_{s,u}(n_{P_x}^\zeta)\right]
\le C_\zeta T^{d\zeta},
\quad x\in\mathcal C,\quad 1\le\zeta\le\xi.
\label{eq:SM_ppb_block_moments}
\end{equation}
The constant $C_\zeta$ may depend on $d,\alpha,C_{J},v,\lambda$ and $\zeta$, but not on the block, time interval, initial state, or lattice size.

For $q>0$, define the high-occupation projector $\Pi_{P_x,>q}\coloneqq\mathbf1_{(q,\infty)}(n_{P_x})$.
The scalar inequality $\mathbf1_{(q,\infty)}(n)\le(n/q)^\zeta$ for $n\ge0$ gives $\Pi_{P_x,>q}\le q^{-\zeta}n_{P_x}^\zeta$ by spectral calculus. Since unitary evolution preserves positivity, Eq.~\eqref{eq:SM_ppb_block_moments} implies
\begin{equation}
\sup_{u\in[\min\{s,t\},\max\{s,t\}]}
\operatorname{Tr}\!\left[
\rho\,\tau_{s,u}(\Pi_{P_x,>q})\right]
\le\min\left\{1,\ C_\zeta
\left(\frac{T^d}{q}\right)^\zeta\right\}.
\label{eq:SM_ppb_block_tail}
\end{equation}
This is the occupation-tail estimate used in Sec.~\ref{S-IV-B}, where the thresholds are chosen blockwise, and their errors are summed. Its proof uses only particle-number propagation and is independent of the fixed-$k$ Lieb-Robinson estimates of Sec.~\ref{S-III}.

\subsection{Particle-number truncation and dynamical comparison}
\label{S-IV-B}

We prove the first two approximations in the upper-bound argument,
\begin{equation}
\tau_{s,t}(A)\longrightarrow\tau_{s,t}(\bar A)
\longrightarrow\bar\tau_{s,t}(\bar A).
\end{equation}
The proof follows the occupation-truncation method of Refs.~\cite{kuwahara2022optimal,LR26}, using the particle-number estimates established in Sec.~\ref{S-IV-A}. Throughout this subsection, fix $\rho\in\mathcal D_\xi^\lambda$, a nonempty finite $X\subseteq\Lambda$, and $A\in\mathcal A_X^{q_0}$. The case $A=0$ is immediate, so we assume $\|A\|>0$ when normalizing a moment
bound. Retain
\begin{equation}
T=1+|t-s|,\quad \ell=\lceil v|t-s|\rceil,\quad v>0,
\end{equation}
and hold the resulting block partition fixed throughout every comparison on the interval between $s$ and $t$. Constants may depend on $d,\alpha,C_{J},v,\lambda,q_0,\xi$ and the fixed cutoff-growth exponent $\delta$, but not on the lattice size or shape, $X$, $A$, the times, the base cutoff, or the particular state. No estimate below depends on the strength or order of the density potential.

\subsubsection{Block cutoffs and elementary estimates}
\label{S-IV-B-cutoffs}

\begin{figure}[t]
\centering
\includegraphics[width=0.88\linewidth]{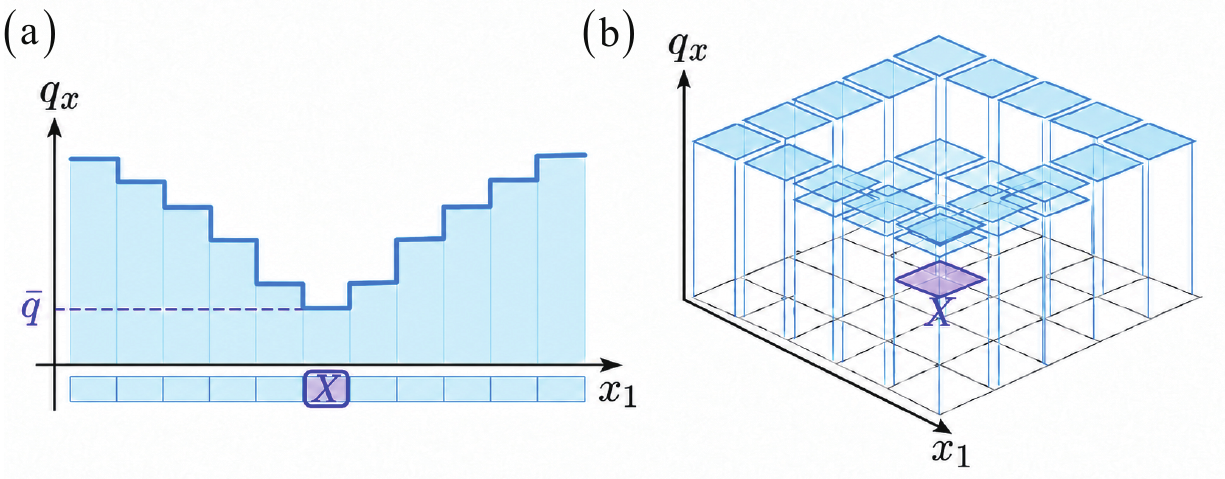}
\caption{The position-dependent particle-number cutoff.
The cutoff is constant on each block and takes its minimum value on the blocks that intersect $X$.
{It increases with the distance from $X$, according to $q_x=\lfloor\bar q[1+d(P_x,X)]^\delta\rfloor$.}
Panel (a) shows a one-dimensional spatial section, while panel (b) represents the corresponding stepped profile over a two-dimensional lattice.}
\label{fig:SM_block_cutoff_profile}
\end{figure}

Throughout the paper we will consider a bounded operator $A$ supported on some compact region $X\subset\Lam$. 
Our main goal is to approximate its time evolution with an operator supported on some enlargement of $X$. 
To this end, we define the following truncation parameter for every $y\in\cC$
 \begin{align}\label{qi def}
	q_y=\lfloor\bar{q}\, (d(\partset{y},X)+1)^{\delta}\rfloor,
\end{align}
where $\bar q>1$ and $0<\delta\le 1$ to be determined later on. 
Recall that \(\cC\) is the set of block centers defined in~\eqref{eq:SM_block_geometry}.
For $x\in\cC$, write 
\begin{equation}
      \Pi_x \coloneqq \Pi_{P_x,\le q_x} \quad \text{and}\quad \Pi_x^\perp=I-\Pi_x.
    \notag
\end{equation}
Here \(\Pi_x\) projects onto the subspace with at most \(q_x\) particles in \(P_x\), and \(\Pi_x^\perp\) onto its orthogonal complement.
Furthermore, we define 
\begin{equation}
 \Pi \coloneqq \prod_{x\in\mathcal C}\Pi_x,
 \quad \text{and}\quad \Pi^\perp \coloneqq I-\Pi.
\label{eq:SM_global_cutoff}
\end{equation}
 The position-dependent particle-number cutoff is illustrated in Fig.~\ref{fig:SM_block_cutoff_profile}.

As in the main text, for an operator $O$ on Fock space, write 
\begin{align}
      \bar O:=\Pi O \Pi.
\end{align}
Essentially, we are sandwiching the operator $O$ between projectors that truncate particles across the lattice. 
We consider the dynamics generated by the truncated Hamiltonian, $\bar H$, and we denote it by $\bar \tau_{s,t}$.

\begin{figure}[t]
\centering
\includegraphics[width=0.68\linewidth]{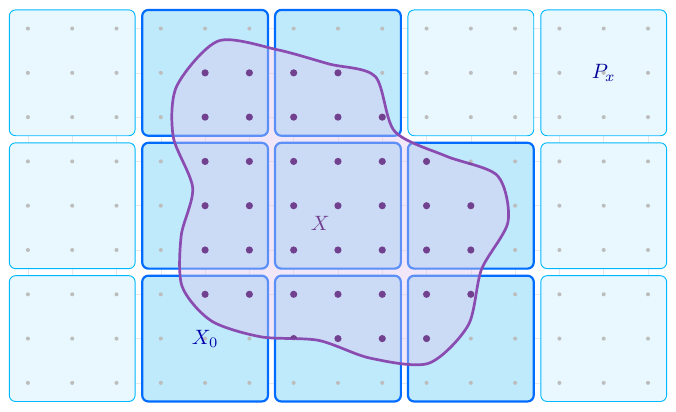}
\caption{An illustration of the coarse-grained lattice formed by the blocks $P_x$, $x\in\mathcal C$. Given a set $X\subset\Lambda$, shown in lilac, its blockwise enlargement $X_0$ is the union of the darker blue blocks, namely all $P_x$ that intersect $X$.}
\label{fig:SM_block_fattening}
\end{figure}

For clarity, \(\bar A\) is not generally supported on \(X\) as an operator on the full Fock space.
Consider $r\ge 0$ and $Y\subseteq\Lambda$, and recall that  \(Y_r\), defined by~\eqref{eq:SM_block_enlargement}, is the $r$-enlargement of \(Y\) in terms of blocks.
In particular, $Y_0$ is the union of the blocks intersecting $Y$.
We define
\begin{equation}
    \Pi_{Y_r}
    \coloneqq
    \prod_{\substack{x\in\cC\\d(P_x,Y)\le r}}\Pi_x,
    \quad
    \Pi_{Y_r^c}
    \coloneqq
    \prod_{\substack{x\in\cC\\d(P_x,Y)>r}}\Pi_x
    .
\end{equation}
Since $A$ is supported on $X\subseteq X_0$, it commutes with $\Pi_{X_0^c}$. Thus, using $\Pi=\Pi_{X_0}\Pi_{X_0^c}$, we have
\begin{equation}
\bar A=(\Pi_{X_0}A\Pi_{X_0})\Pi_{X_0^c}.
\label{eq:SM_global_cutoff_support}
\end{equation}
The retained space factorizes as
\begin{equation}
 \operatorname{Ran}\Pi
 =\bigotimes_{x\in\mathcal C}\operatorname{Ran}\Pi_x,
\end{equation}
where each factor is taken in the Fock space of its block.
On this finite-dimensional tensor product, the exterior projectors are identities. The restriction of $\bar A$ is supported on the blocks intersecting $X$.
The removal of exterior projectors on full Fock space is a separate step in Sec.~\ref{S-IV-E}. Figure~\ref{fig:SM_block_fattening} illustrates the block
decomposition and the blockwise fattening of $X$.

The blocks are disjoint, so their number projectors commute.
Therefore for every finite collection of centers $(x_k)_{k=1}^m\in \cC$ we have 
\begin{equation}\label{eq:SM_cutoff_union_bound}
    \one-\prod_{k=1}^m\Pi_{x_k}
    \le
    \sum_{k=1}^m\Pi_{x_k}^{\perp}
    .
\end{equation}
Indeed, for commuting projections one can show we have 
\begin{align}
    \del{\prod_{k=1}^{n}\Pi_{x_k}}^\perp
   = \sum_{j=1}^{n}\Pi_{x_j}^{\perp}
        \prod_{k=1}^{j-1}\Pi_{x_k}.
\end{align}

We will repeatedly use the following geometric sums. For every
$\mu>d$,
\begin{align}
 \sum_{x\in\mathcal C}(1+d(P_x,X))^{-\mu}\le C_\mu|X|,
    \quad
 \sup_{x\in\mathcal C}
 \sum_{y\in\mathcal C}(1+d(P_x,P_y))^{-\mu}\le C_\mu.
\label{eq:SM_cutoff_geometric_sums}
\end{align}
Importantly $C_\mu$ is independent of $\ell$.
 To prove the inequalities above, notice that for every $x\in\cC$ 
 \begin{align}
      \abs{y\in\mathcal C:d(P_x,P_y)\le R}
    \le 3^d(1+R)^d.
 \end{align}
In fact, if $d(P_x,P_y)\le R$, then by the triangle inequality, choosing $i\in P_x$ and $j\in P_y$ to realize the distance
\begin{align}
      |x-y|
    \le |x-i|+|i-j|+|j-y|
    \le R+2\ell.
\end{align}
Thus the possible centers \(y\) lie in a cube of radius \(R+2\ell\). Since their spacing is \(2\ell+1\), their number is at most
\begin{equation}
    \left(1+\frac{2(R+2\ell)}{2\ell+1}\right)^d
    \le (3+2R)^d
    \le 3^d(1+R)^d.
\end{equation}
By a change of summation index, it then follows
\begin{align}
      \sum_{x\in\mathcal C}(1+d(P_x,X))^{-\mu}\le C3^d\abs{X}\sum_{k=1}^\infty k^{d-1-\mu}\le C_\mu \abs{X}.
\end{align}
Similarly, we obtain for every $x\in\cC$
\begin{align}
      \sum_{y\in\mathcal C}(1+d(P_x,P_y))^{-\mu}\le 3^d \sum_{k=1}^\infty k^{d-1-\mu}\le C_\mu.
\end{align}

The triangle inequality also gives
\begin{equation}
    \label{eq:SM_cutoff_weight_comparison}
    \begin{aligned}
        1+d(P_y,X)
        &\le
        1+d(P_x,X)+d(P_x,P_y)+2\ell
        \\&\le
        (2\ell+1)\,(1+d(P_x,X))(1+d(P_x,P_y))
        \\&\le
        CT\,(1+d(P_x,X))(1+d(P_x,P_y))
        .
    \end{aligned}
\end{equation}
For the first inequality, connect a point of $P_x$ nearest $X$ to a pair realizing $d(P_x,P_y)$ inside $P_x\times P_y$.
The additional displacement inside $P_x$ is at most $2\ell$.

Combining Eqs.~\eqref{eq:SM_cutoff_union_bound}, \eqref{eq:SM_cutoff_geometric_sums}, and \eqref{eq:SM_ppb_block_tail} yields, whenever $\delta\xi>d$,
\begin{align}
    \label{eq:SM_original_state_cutoff_tail}
 \operatorname{Tr}[\rho\,\tau_{s,t}(\Pi^\perp)]
 \le\sum_x q_x^{-\xi}
       \operatorname{Tr}[\rho\,\tau_{s,t}(n_{P_x}^{\xi})]
 \le C\frac{T^{d\xi}}{\bar q^{\xi}}
          \sum_x (1+d(P_x,X))^{-\delta\xi}
 \le C|X|\frac{T^{d\xi}}{\bar q^{\xi}}.
\end{align}

\subsubsection{Local insertions and truncation of the observable}
\label{S-IV-B-observable}

We first control how an insertion of $A$ changes the number moments.
The following estimate is a convenient version of the local insertion bound used in Ref.~\cite{kuwahara2022optimal}.

\begin{lemma}
\label{lem:SM_cutoff_insertion}
For every integer $q_0\ge 0$ and $X,Y\subset\Lambda$, every operator \(A\in\mathcal A_X^{q_0}\), and every real \(\zeta\ge1\),
\begin{equation}
    \label{eq:SM_cutoff_insertion_operator}
    A^\dagger n_Y^\zeta A
    \preceq
    \norm{A}^2 \, (n_{X\cup Y}+q_0 I)^\zeta
    .
\end{equation}
The inequality is understood in the sense of quadratic forms.
\end{lemma}

\begin{proof}
Since \(A\) is supported on \(X\), the number-change restriction~\eqref{eq:SM_q0_observable} also holds for \(n_{X\cup Y}\).
For every integer \(m\ge0\), this implies \(\Pi_{Y,>m}A=\Pi_{Y,>m}A\Pi_{X\cup Y,>m-q_0}\), and hence
\begin{equation}
    A^\dagger\Pi_{Y,>m}A
    \preceq
    \norm{A}^2 \, \Pi_{X\cup Y,>m-q_0}.
\end{equation}
Multiply this inequality by \((m+1)^\zeta-m^\zeta\) and sum over \(m\ge0\).
The spectral identity
\begin{equation}
    n_Y^\zeta
    =
    \sum_{m=0}^{\infty}\bigl[(m+1)^\zeta-m^\zeta\bigr] \, \Pi_{Y,>m},
\end{equation}
and the corresponding identity for \(n_{X\cup Y}+q_0 I\) give~\eqref{eq:SM_cutoff_insertion_operator}.
The sums converge monotonically in the sense of quadratic forms.
\end{proof}

Write
\begin{equation}
    \label{eq:SM_cutoff_inserted_data}
    \rho_A
    \coloneqq
    A^\dagger\Pi\tau_{t,s}(\rho)\Pi A,
    \quad
    \bar\rho_A
    \coloneqq
    \Pi\rho_A\Pi
    =
    \bar A^\dagger\tau_{t,s}(\rho)\bar A
    .
\end{equation}
\begin{lemma}
    \label{lem:SM_cutoff_inserted_tails}
    Consider $\alpha>d+1$, integer \(q_0\ge0\),~\(\xi\ge1\),~\(\lambda>0\), and \(0<\delta\le1\) with \(\delta\xi>d\).
    There is \(C>0\) such that, for every \(\bar q>1\),~\(s\),~\(t\in\R\), nonempty \(X\subseteq\Lambda\),~\(A\in\cA_X^{q_0}\), and \(\rho\in\cD_\xi^\lambda\), the operators~\eqref{eq:SM_cutoff_inserted_data} satisfy
    \begin{align}
        \label{eq:SM_cutoff_inserted_tail}
        \Tr\!\sbr{\rho_A\Pi^\perp}
        &\le
        C\norm{A}^2\abs{X}^{1+\xi}\frac{T^{d\xi}}{\bar q^\xi}
        ,
        \\
        \label{eq:SM_cutoff_inserted_evolved_tail}
        \Tr\!\sbr{\tau_{s,t}(\rho_*)\Pi^\perp}
        &\le
        C\norm{A}^2\abs{X}^{1+\xi}\frac{T^{d\xi}}{\bar q^\xi}
        ,
    \end{align}
    for every \(\rho_*\in\{\rho_A,\bar\rho_A\}\).
\end{lemma}

\begin{proof}
    The operators in \eqref{eq:SM_cutoff_inserted_data} are positive and need not be normalized.
    Their traces are at most $\|A\|^2$.
    They have finite $\xi$-th total-number moments.
    Indeed, apply Eq.~\eqref{eq:SM_cutoff_insertion_operator} with $Y=\Lambda$, use number conservation, and recall that $\Pi$ commutes with $n_\Lambda$.

    Set
    \begin{equation}
     L\coloneqq2\ell+1.
    \end{equation}
    This scale satisfies \(L\ge v|t-s|+1\), \(L\ge2\ell\), and \(L\le CT\).
    For any \(r\ge L\), Theorem~\ref{thm:SM_particle_propagation}, with inner radius \(r\) and outer radius \(2r\), gives
    \begin{equation}
     \bigl[\operatorname{Tr}(\tau_{t,s}(\rho) n_{B_r(i)}^\zeta)\bigr]^{1/\zeta}
     \le C(1+\lambda)\zeta r^d,
     \quad i\in\Lambda,\quad 1\le\zeta\le\xi.
    \end{equation}
    All radius conditions hold because \(r\ge L\ge1\) and \(v|t-s|<L\le r\).

    If $B_r(i)\cap X=\varnothing$, then $A$ commutes with its number operator, and
    \begin{align}
     \operatorname{Tr}(\rho_A n_{B_r(i)}^\zeta)
     =\operatorname{Tr}
           (\tau_{t,s}(\rho)\Pi A n_{B_r(i)}^\zeta A^\dagger\Pi)
     \le\|A\|^2
           \operatorname{Tr}(\tau_{t,s}(\rho) n_{B_r(i)}^\zeta).
    \end{align}
    In general, Lemma~\ref{lem:SM_cutoff_insertion}, applied to $A^\dagger$, gives
    \begin{align}
     \operatorname{Tr}(\rho_A n_{B_r(i)}^\zeta)
     &\le \|A\|^2
     \operatorname{Tr}\!\left[
     \tau_{t,s}(\rho)(n_{B_r(i)}+n_X+q_0 I)^\zeta\right].
    \end{align}
    The projectors have been removed only after being bounded by a positive function of the commuting number operators.
    Since $n_X\le\sum_{j\in X}n_{B_r(j)}$, Minkowski's inequality for their joint spectral distribution yields
    \begin{align}
     \left\{\operatorname{Tr}\!\left[
     \tau_{t,s}(\rho)(n_{B_r(i)}+n_X+q_0 I)^\zeta\right]\right\}^{1/\zeta}
     &\le q_0+
     \bigl[\operatorname{Tr}(\tau_{t,s}(\rho) n_{B_r(i)}^\zeta)\bigr]^{1/\zeta}
     +\sum_{j\in X}
     \bigl[\operatorname{Tr}(\tau_{t,s}(\rho) n_{B_r(j)}^\zeta)\bigr]^{1/\zeta}
     \notag\\
     &\le q_0+C(1+\lambda)\zeta(1+|X|)r^d.
    \end{align}
    Consequently, a constant $C_{q_0}$ can be chosen so that
    \begin{equation}
     \operatorname{Tr}(\rho_A n_{B_r(i)}^\zeta)
     \le\|A\|^2(\lambda_A\zeta r^d)^\zeta,
     \quad
     \lambda_A \coloneqq C_{q_0}(1+\lambda)|X|,
     \quad r\ge L.
    \end{equation}
    The same inequality holds for $\bar\rho_A$, because $\Pi n_{B_r(i)}^\zeta\Pi\le n_{B_r(i)}^\zeta$.

    Apply Eq.~\eqref{eq:SM_ppb_positive_data_propagation} with $K_\sigma=\|A\|^2$, density parameter $\lambda_A$, and initial control scale $r_{\mathrm{cd}}=L$.
    For each block choose $i_x\in P_x$, so that $P_x\subseteq B_{2\ell}(i_x)$.
    The inner radius \(r_B=\max\{2\ell,1\}\) and outer radius $R_B'=2L+v|t-s|+2$ satisfy the required conditions for propagation from $t$ to any $u$ between $s$ and $t$, and $R_B'\le CT$.
    Thus, for $\rho_*\in\{\rho_A,\bar\rho_A\}$,
    \begin{equation}
     \operatorname{Tr}\!\left[
     \tau_{u,t}(\rho_*) n_{P_x}^\zeta\right]
     \le C_\zeta\|A\|^2|X|^\zeta T^{d\zeta}.
    \label{eq:SM_cutoff_inserted_block_moments}
    \end{equation}
    Finally, the union bound, Markov's inequality, and~\eqref{eq:SM_cutoff_geometric_sums}, applied at \(u=s\), give~\eqref{eq:SM_cutoff_inserted_evolved_tail}.
    Taking \(u=t\) and \(\rho_*=\rho_A\) gives~\eqref{eq:SM_cutoff_inserted_tail}.
\end{proof}

\begin{proposition}[Truncation of the observable]
\label{prop:SM_observable_truncation}
Consider $\alpha>d+1$.
For every  $q_0\ge 0$, $X\subset \Lambda$, $A\in\mathcal{A}_X^{q_0}$, $\xi\ge 1$ and $\delta\in(0,1)$ such that  $\delta\xi>d$, then, for every $\lambda>0$, initial  state $\rho\in\mathcal{D}_\xi^\lambda$, and $s,t\in\R$,
\begin{equation}
 \bigl\|\rho\,\tau_{s,t}(A-\bar A)\bigr\|_1
 \le C\|A\|\,|X|^{(1+\xi)/2}
             \frac{T^{d\xi/2}}{\bar q^{\xi/2}}.
\label{eq:SM_observable_truncation_error}
\end{equation}
\end{proposition}

\begin{proof}
Fix a bounded probe \(B\).
Then \(\norm{\tau_{t,s}(B)}=\norm{B}\) and
\begin{equation}
 \operatorname{Tr}[\rho\,\tau_{s,t}(A-\bar A)B]
 =\operatorname{Tr}[\tau_{t,s}(\rho)(A-\bar A)\tau_{t,s}(B)].
\end{equation}
The decomposition
\begin{equation}
 A-\bar A=\Pi^\perp A\Pi+\Pi A\Pi^\perp
                    +\Pi^\perp A\Pi^\perp
\end{equation}
separates the two placements of the high-occupancy projector.
 Cauchy--Schwarz implies
\begin{align}
 |\operatorname{Tr}(\tau_{t,s}(\rho)\Pi^\perp A\Pi \tau_{t,s}(B))|
 +|\operatorname{Tr}(\tau_{t,s}(\rho)\Pi^\perp A\Pi^\perp \tau_{t,s}(B))|
 \le2\|A\|\|B\|
       [\operatorname{Tr}(\tau_{t,s}(\rho)\Pi^\perp)]^{1/2}.
\end{align}
For the remaining term, apply Cauchy--Schwarz to
$\tau_{t,s}(\rho)^{1/2}\Pi A\Pi^\perp$ and $\tau_{t,s}(B)\tau_{t,s}(\rho)^{1/2}$:
\begin{align}
 |\operatorname{Tr}(\tau_{t,s}(\rho)\Pi A\Pi^\perp \tau_{t,s}(B))|
 \le\|B\|
 [\operatorname{Tr}
       (\tau_{t,s}(\rho)\Pi A\Pi^\perp A^\dagger\Pi)]^{1/2}
 =\|B\|[\operatorname{Tr}(\rho_A\Pi^\perp)]^{1/2}.
\end{align}
Substitute Eqs.~\eqref{eq:SM_original_state_cutoff_tail} and \eqref{eq:SM_cutoff_inserted_tail}, and use $|X|\ge1$.
Taking the supremum over $\|B\|\le1$ proves the result by trace-norm duality.
\end{proof}

\subsubsection{Comparison of full and truncated dynamics}
\label{S-IV-B-comparison}

\begin{proposition}[Particle-number truncation of the dynamics]
    \label{prop:SM_dynamical_truncation}
    Fix \(\lambda>0\), an integer \(q_0\ge0\),~\(v>0\),~\(0<\delta\le1\), and \(\xi>3\) such that \(\delta(\xi-3)>2d\).
    Assume \(\alpha>2d+\delta\).
    There exists a constant \(C>0\), uniform in \(\Lambda\), such that, for every nonempty \(X\subseteq\Lambda\),~\(\bar q>1\),~\(A\in\cA_X^{q_0}\),~\(\rho\in\cD_\xi^\lambda\), and \(s\),~\(t\in\R\),
    \begin{equation}
        \label{eq:SM_dynamical_truncation_error}
        \begin{aligned}
            \norm{\rho\del{\tau_{s,t}(\bar A)-\bar\tau_{s,t}(\bar A)}}_1
            &\le
            C\norm{A}\abs{X}^{(1+\xi)/2}\bigg(
            \frac{T^{d(\xi-1)/2}}{\bar q^{(\xi-1)/2}}
            +
            \frac{\abs{t-s}\,T^{d(\xi-1)/2+\delta}}{\bar q^{(\xi-3)/2}}
            \bigg)
            .
        \end{aligned}
    \end{equation}
    Here \(T=1+\abs{t-s}\),~\(\ell=\lceil v\abs{t-s}\rceil\), and the cutoffs are given by~\eqref{qi def} and~\eqref{eq:SM_global_cutoff}.
\end{proposition}

\begin{proof}
    We use the particle-propagation and insertion bounds at moment order \(\xi-1\), since \(\cD_\xi^\lambda\subseteq\cD_{\xi-1}^\lambda\).
    Fix a bounded probe \(B\) with \(\norm{B}=1\), and let \(\bar U(s,t)\) be the propagator of \(\bar H\).
    By reversing time if necessary, assume \(s\le t\).
    Recall~\eqref{eq:SM_dynamics}.
    By the triangle inequality, we can upper-bound the left-hand side of~\eqref{eq:SM_dynamical_truncation_error} by trace-norm duality, using
    \begin{equation}
        \label{eq:SM_cutoff_two_sided_difference}
        \begin{aligned}
            &\abs{\Tr\sbr{\rho\del{\tau_{s,t}(\bar A)-\bar\tau_{s,t}(\bar A)}B}}
            \\&\quad\le
            \abs{\Tr\sbr{\rho\del{U(s,t)^\dagger-\bar U(s,t)^\dagger}\bar A\bar U(s,t)B}}
            +
            \abs{\Tr\sbr{\rho U(s,t)^\dagger\bar A\del{U(s,t)-\bar U(s,t)}B}}
            .
        \end{aligned}
    \end{equation}

    Step 1.
    We focus first on the first term in~\eqref{eq:SM_cutoff_two_sided_difference}.
    Since \(\Pi^\perp\bar A=0\), we apply Duhamel's formula to obtain
    \begin{equation}
        \label{eq:SM_cutoff_Duhamel_propagator}
        \begin{aligned}
            &\abs{\Tr\sbr{\rho\del{\bar U(s,t)^\dagger-U(s,t)^\dagger}\bar A\bar U(s,t)B}}
            \\&\quad\le
            \int_s^t\md u\,
            \abs{\Tr\sbr{\rho U(s,u)^\dagger\del{H(u)\Pi-\bar H(u)}\bar U(u,t)^\dagger\bar A\bar U(s,t)B}}
            .
        \end{aligned}
    \end{equation}
    Notice that \(\Pi^\perp V_{ij}(u)\Pi=0\) for all \(i\),~\(j\in\Lambda\) and \(u\in[s,t]\).
    If \(i\in P_x\) and \(j\in P_y\) with \(x\ne y\), the hopping operator \(T_{ij}\) can only increase the particle number in \(P_x\).
    Furthermore, all projectors associated with blocks other than \(P_x\) and \(P_y\) commute with \(T_{ij}\), and \(T_{ij}\) preserves the particle number in \(P_x\) whenever \(i\),~\(j\in P_x\).
    Thus,
    \begin{equation}
        \label{eq:SM_cutoff_leakage_operators}
        \begin{aligned}
            H(u)\Pi-\bar H(u)
            & =
            \Pi^\perp H(u)\Pi
             =
            \sum_{\substack{x,y\in\cC\\x\ne y}}
            \Pi_x^\perp\sum_{\substack{i\in P_x\\j\in P_y}}T_{ij}(u)\Pi
            .
        \end{aligned}
    \end{equation}
    Together with~\eqref{eq:SM_cutoff_Duhamel_propagator}, the triangle inequality and Cauchy--Schwarz imply
    \begin{equation}
        \label{eq:SM_cutoff_first_dynamical_term}
        \begin{aligned}
            &\abs{\Tr\sbr{\rho\del{\bar U(s,t)^\dagger-U(s,t)^\dagger}\bar A\bar U(s,t)B}}
            \\&\le
            \norm{A}\norm{B}\int_s^t\md u
            \sum_{\substack{x,y\in\cC\\x\ne y}}
            \del{\Tr\sbr{\tau_{u,s}(\rho)\Pi_x^\perp}}^{1/2}
           \bigg \Vert\sum_{\substack{i\in P_x\\j\in P_y}}T_{ij}(u)\Pi\bigg\Vert
            .
        \end{aligned}
    \end{equation}
    Since \(q_x\ge\bar q\del{1+d(P_x,X)}^\delta/2\), the particle-propagation bound~\eqref{eq:SM_ppb_block_tail} and~\eqref{qi def} give
    \begin{equation}
        \label{eq:SM_cutoff_original_block_tail}
        \Tr\sbr{\tau_{u,s}(\rho)\Pi_x^\perp}
        \le
        C\frac{T^{d(\xi-1)}}{\bar q^{\xi-1}\del{1+d(P_x,X)}^{\delta(\xi-1)}}
        .
    \end{equation}
    Notice that, for \(i\in P_x\),
    \begin{equation}
        b_i^\dagger\Pi_{P_x,\le q_x}
        =
        \Pi_{P_x,\le q_x+1}b_i^\dagger
        .
    \end{equation}
    Therefore, for any normalized \(\psi\),~\(\phi\in\cF\) with \(\phi=\Pi_{P_x,\le q_x+1}\phi\), Cauchy--Schwarz gives
    \begin{align}
        \big \vert\bra{\phi}\sum_{\substack{i\in P_x\\j\in P_y}}T_{ij}(u)\Pi\ket{\psi}\big \vert
        &\le
        \sum_{\substack{i\in P_x\\j\in P_y}}\abs{J_{ij}(u)}\,\norm{b_i\phi}\,\norm{b_j\Pi\psi}
        \nonumber\\&\le
        \bigg(\sum_{\substack{i\in P_x\\j\in P_y}}\abs{J_{ij}(u)}\,\norm{b_i\phi}^2\bigg)^{1/2}
        \bigg(\sum_{\substack{i\in P_x\\j\in P_y}}\abs{J_{ij}(u)}\,\norm{b_j\Pi\psi}^2\bigg)^{1/2}
        \\&\le
        (q_x+1)\sup_{i\in P_x}\sum_{j\in P_y}\abs{J_{ij}(u)}
        +q_y\sup_{j\in P_y}\sum_{i\in P_x}\abs{J_{ij}(u)}\nonumber
        .
    \end{align}
    The assumption of polynomial decay~\eqref{eq:SM_power_law_bounds} implies
    \begin{align}
        \sup_{i\in P_x}\sum_{j\in P_y}\abs{J_{ij}(u)}
        &\le
        C_J\sup_{i\in P_x}\sum_{j\in P_y}(1+\abs{i-j})^{-\alpha}
       \nonumber \\&\le
        C\sum_{k\ge d(P_x,P_y)+1}k^{d-1-\alpha}
        \\&\le
        C\del{1+d(P_x,P_y)}^{d-\alpha}.\nonumber
    \end{align}
    The same bound holds for the column sum.
    Taking the supremum over the unit vectors above and using \(q_x\ge1\), we obtain
    \begin{equation}
        \bigg \Vert\sum_{\substack{i\in P_x\\j\in P_y}}T_{ij}(u)\Pi\bigg \Vert
        \le
        C\frac{q_x+q_y}{\del{1+d(P_x,P_y)}^{\alpha-d}}.
    \end{equation}
    Combining this with~\eqref{eq:SM_cutoff_first_dynamical_term},~\eqref{eq:SM_cutoff_original_block_tail}, and~\eqref{qi def} gives
    \begin{equation}
        \label{eq:SM_cutoff_hopping_pair_sum}
        \begin{aligned}
            &\abs{\Tr\sbr{\rho\del{\bar U(s,t)^\dagger-U(s,t)^\dagger}\bar A\bar U(s,t)B}}
            \\ &\quad \le
            C\norm{A}\norm{B}\frac{(t-s)T^{d(\xi-1)/2}}{\bar q^{(\xi-3)/2}}
            \sum_{\substack{x,y\in\cC\\x\ne y}}
            \frac{\del{1+d(P_x,X)}^\delta+\del{1+d(P_y,X)}^\delta}
            {\del{1+d(P_x,P_y)}^{\alpha-d}\del{1+d(P_x,X)}^{\delta(\xi-1)/2}}
            .
        \end{aligned}
    \end{equation}
    Since \(\alpha>2d\) and \(\delta(\xi-3)>2d\), the geometric sums~\eqref{eq:SM_cutoff_geometric_sums} give
    \begin{equation}
        \label{eq:SM_cutoff_hopping_first_sum}
        \sum_{\substack{x,y\in\cC\\x\ne y}}
        \frac{1}{\del{1+d(P_x,P_y)}^{\alpha-d}\del{1+d(P_x,X)}^{\delta(\xi-3)/2}}
        \le
        C\abs{X}
        .
    \end{equation}
    For the second sum,~\eqref{eq:SM_cutoff_weight_comparison} implies
    \begin{equation}
        \label{eq:SM_cutoff_hopping_second_sum}
        \begin{aligned}
            \sum_{\substack{x,y\in\cC\\x\ne y}}
            \frac{\del{1+d(P_y,X)}^\delta}
            {\del{1+d(P_x,P_y)}^{\alpha-d}\del{1+d(P_x,X)}^{\delta(\xi-1)/2}}
            &\le
            CT^\delta\sum_{x,y\in\cC}
            \frac{1}{\del{1+d(P_x,P_y)}^{\alpha-d-\delta}\del{1+d(P_x,X)}^{\delta(\xi-3)/2}}
            \\&\le
            C\abs{X}T^\delta
            .
        \end{aligned}
    \end{equation}
    Both sums are bounded uniformly in the block size and system size because \(\alpha-d-\delta>d\) and \(\delta(\xi-3)/2>d\).
    Combining~\eqref{eq:SM_cutoff_hopping_pair_sum},~\eqref{eq:SM_cutoff_hopping_first_sum}, and~\eqref{eq:SM_cutoff_hopping_second_sum} gives
    \begin{equation}
        \label{eq:SM_cutoff_original_propagator_error}
        \abs{\Tr\sbr{\rho\del{U(s,t)^\dagger-\bar U(s,t)^\dagger}\bar A\bar U(s,t)B}}
        \le
        C\norm{A}\norm{B}\abs{X}\frac{(t-s)T^{d(\xi-1)/2+\delta}}{\bar q^{(\xi-3)/2}}
        .
    \end{equation}

    Step 2.
    To control the remaining term in~\eqref{eq:SM_cutoff_two_sided_difference}, write \(B=\Pi B+\Pi^\perp B\).
    By the triangle inequality and \(\bar A\bar U(s,t)\Pi^\perp=0\), we have
    \begin{equation}
        \label{eq:SM_cutoff_second_dynamical_term}
        \begin{aligned}
            &\abs{\Tr\sbr{\rho U(s,t)^\dagger\bar A\del{U(s,t)-\bar U(s,t)}B}}
            \\&\quad\le
            \abs{\Tr\sbr{\rho U(s,t)^\dagger\bar A\del{U(s,t)-\bar U(s,t)}\Pi B}}
            +
            \abs{\Tr\sbr{\rho U(s,t)^\dagger\bar A U(s,t)\Pi^\perp B}}
            .
        \end{aligned}
    \end{equation}
    For the first summand, Duhamel's formula and~\eqref{eq:SM_cutoff_leakage_operators} give
    \begin{equation}
        \begin{aligned}
            \del{U(s,t)-\bar U(s,t)}\Pi
            & =
            -\mi\int_s^t\md u\,U(u,t)\del{H(u)\Pi-\bar H(u)}\bar U(s,u)
            \\& =
            -\mi\int_s^t\md u\sum_{\substack{x,y\in\cC\\x\ne y}}
            U(u,t)\Pi_x^\perp\sum_{\substack{i\in P_x\\j\in P_y}}T_{ij}(u)\Pi\bar U(s,u).
        \end{aligned}
    \end{equation}
    Using the triangle inequality and Cauchy--Schwarz, we obtain
    \begin{equation}
        \label{eq:SM_cutoff_inserted_Cauchy_Schwarz}
        \begin{aligned}
            &\abs{\Tr\sbr{\rho U(s,t)^\dagger\bar A\del{U(s,t)-\bar U(s,t)}\Pi B}}
            \le
            \norm{B}\int_s^t\md u
            \sum_{\substack{x,y\in\cC\\x\ne y}}
            \bigg \Vert\sum_{\substack{i\in P_x\\j\in P_y}}T_{ij}(u)\Pi\bigg \Vert
            \del{\Tr\sbr{\tau_{u,t}(\bar\rho_A)\Pi_x^\perp}}^{1/2}
            .
        \end{aligned}
    \end{equation}
    The norm of the hopping term is bounded as in Step 1.
    For the trace, Markov's inequality,~\eqref{qi def}, and~\eqref{eq:SM_cutoff_inserted_block_moments} imply
    \begin{equation}
        \label{eq:SM_cutoff_inserted_block_tail}
        \begin{aligned}
            \Tr\sbr{\tau_{u,t}(\bar\rho_A)\Pi_x^\perp}
            &\le
            q_x^{-(\xi-1)}\Tr\sbr{\tau_{u,t}(\bar\rho_A)n_{P_x}^{\xi-1}}
            \\&\le
            C\norm{A}^2\abs{X}^{\xi-1}\frac{T^{d(\xi-1)}}{\bar q^{\xi-1}\del{1+d(P_x,X)}^{\delta(\xi-1)}}
            .
        \end{aligned}
    \end{equation}
    Combining~\eqref{eq:SM_cutoff_inserted_block_tail} with~\eqref{eq:SM_cutoff_inserted_Cauchy_Schwarz} and the sums~\eqref{eq:SM_cutoff_hopping_first_sum}--\eqref{eq:SM_cutoff_hopping_second_sum}, we obtain
    \begin{equation}
        \label{eq:SM_cutoff_inserted_propagator_error}
        \abs{\Tr\sbr{\rho U(s,t)^\dagger\bar A\del{U(s,t)-\bar U(s,t)}\Pi B}}
        \le
        C\norm{A}\norm{B}\abs{X}^{(1+\xi)/2}\frac{(t-s)T^{d(\xi-1)/2+\delta}}{\bar q^{(\xi-3)/2}}
        .
    \end{equation}
    For the second summand in~\eqref{eq:SM_cutoff_second_dynamical_term}, write \(\bar A=\Pi A-\Pi A\Pi^\perp\).
    The triangle inequality and Cauchy--Schwarz, together with~\eqref{eq:SM_cutoff_inserted_data}, give
    \begin{equation}
        \begin{aligned}
            \abs{\Tr\sbr{\rho U(s,t)^\dagger\bar A U(s,t)\Pi^\perp B}}
            &\le
            \abs{\Tr\sbr{\rho U(s,t)^\dagger\Pi A U(s,t)\Pi^\perp B}}
            +
            \abs{\Tr\sbr{\rho U(s,t)^\dagger\Pi A\Pi^\perp U(s,t)\Pi^\perp B}}
            \\&\le
            \norm{B}\bigg(
            \del{\Tr\sbr{\tau_{s,t}(\rho_A)\Pi^\perp}}^{1/2}
            +\del{\Tr\sbr{\rho_A\Pi^\perp}}^{1/2}
            \bigg).
        \end{aligned}
    \end{equation}
    Thus,~\eqref{eq:SM_cutoff_inserted_evolved_tail} and~\eqref{eq:SM_cutoff_inserted_tail} imply
    \begin{equation}
        \label{eq:SM_cutoff_inserted_exterior_error}
        \abs{\Tr\sbr{\rho U(s,t)^\dagger\bar A U(s,t)\Pi^\perp B}}
        \le
        C\norm{A}\norm{B}\abs{X}^{\xi/2}\frac{T^{d(\xi-1)/2}}{\bar q^{(\xi-1)/2}}
        .
    \end{equation}
    Combining~\eqref{eq:SM_cutoff_original_propagator_error},~\eqref{eq:SM_cutoff_inserted_propagator_error}, and~\eqref{eq:SM_cutoff_inserted_exterior_error}, and taking the supremum over \(\norm{B}=1\), proves~\eqref{eq:SM_dynamical_truncation_error}.
    Here the Step 1 bound is absorbed into the second term, and \(\abs{X}^{\xi/2}\le\abs{X}^{(1+\xi)/2}\) because \(\abs{X}\ge1\).
    The Duhamel identities may first be checked on fixed-total-number sectors, where the matrices are finite-dimensional.
    Since \(\Pi\) vanishes when \(n_\Lambda>\sum_xq_x\), only finitely many such sectors enter the projected identities above.
    For \(t<s\), reverse the time variable and replace \(t-s\) by \(\abs{t-s}\).
\end{proof}

\textit{Choice of the base cutoff.---} Both error estimates hold uniformly for every bounded probe $B$, since $|\operatorname{Tr}(\rho DB)|\le\|\rho D\|_1\|B\|$. With the main-text choice
\begin{equation}
 \bar q=R^\delta T^d,\quad R>1,
\end{equation}
Eqs.~\eqref{eq:SM_observable_truncation_error} and~\eqref{eq:SM_dynamical_truncation_error} give
\begin{align}
 \|\rho\,\tau_{s,t}(A-\bar A)\|_1
 &\le C\|A\|\,|X|^{(1+\xi)/2}R^{-\delta\xi/2},
\label{eq:SM_cutoff_final_observable_error}\\
 \|\rho\del{\tau_{s,t}(\bar A)-\bar\tau_{s,t}(\bar A)}\|_1
 &\le C\|A\|\,|X|^{(1+\xi)/2}
          T^{d+1+\delta}R^{-\delta(\xi-3)/2}.
\label{eq:SM_cutoff_final_dynamics_error}
\end{align}
The second inequality uses \(\abs{t-s}\le T\) and \(R>1\).
The sufficient moment condition in the End Matter,
\begin{equation}
 \xi>3+\frac{2(d+2)}{\delta},
\end{equation}
implies $\delta(\xi-3)>2d$ and hence every summability condition used here. With $\delta$ chosen sufficiently small and then $\xi$ chosen above this threshold, these errors decay at the final radii specified in Sec.~\ref{S-IV-E}.
The required condition \(\alpha>2d+\delta\) follows from the later choice \(\delta p<\alpha-2d\), since \(p\ge1\).
Their support dependence is bounded by  $|X|^{(1+\xi)/2}+|X|^2$.
We next approximate $\bar\tau_{s,t}(\bar A)$ by its spatially restricted counterpart $\bar\tau_{s,t}^{X_{2R^\eta}}(\bar A)$ in Sec.~\ref{S-IV-C}.

\subsection{Spatial restriction of the truncated dynamics}
\label{S-IV-C}

We next approximate $\bar\tau_{s,t}(\bar A)$ by $\bar\tau_{s,t}^{X_{2R^\eta}}(\bar A)$, as in the End Matter. The comparison is in operator norm and is therefore independent of the initial state. We retain the fixed block partition, $T=1+|t-s|$, $\ell=\lceil v|t-s|\rceil$, and the cutoffs $q_x=\lfloor\bar q[1+d(P_x,X)]^\delta\rfloor$ from Sec.~\ref{S-IV-B}.
In particular, the global projector $\Pi$ is unchanged when the Hamiltonian is spatially restricted. The proof separates interactions with an endpoint close to $X$ from those with an interior endpoint in the outer shell. Polynomial interaction tails control the former, whereas the bounded-interaction estimates of Sec.~\ref{S-III} control the latter.

\subsubsection{Restricted Hamiltonian and statement}
\label{S-IV-C-statement}

For a union $Y$ of complete blocks of the fixed partition, define
\begin{equation}
 \bar H_Y(u) \coloneqq \Pi H_Y(u)\Pi,
 \quad
 \bar\tau^Y_{s,t}(O)
 \coloneqq \bar U^Y(s,t)^\dagger O\bar U^Y(s,t),
 \quad
 \mi\partial_t\bar U^Y(s,t)=\bar H_Y(t)\bar U^Y(s,t),
\end{equation}
where $\bar U^Y(s,s)=I$, and $H_Y$ is the restriction defined in Eq.~\eqref{eq:SM_restricted_Hamiltonian}.
Boundary blocks of $\Lambda$ need not be full cubes; a complete block here means that the entire set $P_x$ is retained.
We use $Y=X_{2R^\eta}$ and $Y^c=\Lambda\setminus Y$ below.
Recall that \(L=2\ell+1\).
Then $1\le L\le C_vT$.

To fix the ordered-pair convention in $\bar H=\sum_{i,j}\bar h_{ij}$, take
\begin{equation}
 h_{ij}(u) \coloneqq \frac14\bigl[
 T_{ij}(u)+T_{ji}(u)+V_{ij}(u)+V_{ji}(u)
 +\mathrm{H.c.}\bigr],
 \quad
 \bar h_{ij}(u) \coloneqq \Pi h_{ij}(u)\Pi.
\label{eq:SM_spatial_hermitian_pair_terms}
\end{equation}
The adjoint in this expression applies to all four preceding terms. Thus, $h_{ij}=h_{ji}=h_{ij}^\dagger$ and $H_Y=\sum_{i,j\in Y}h_{ij}$ for every restricted Hamiltonian.
This groups the two orientations and their adjoints without changing $H_Y$ or its dynamics. In particular, the interaction crossing the spatial boundary is exactly
\begin{equation}
 \bar H-\bar H_Y-\bar H_{Y^c}
 =2\sum_{i\in Y,\,j\in Y^c}\bar h_{ij}.
\end{equation}
This convention accounts for the constant multiplying the interior--exterior sum in the main-text Duhamel estimate.
The estimates below apply separately to the original hopping and density terms in Eq.~\eqref{eq:SM_spatial_hermitian_pair_terms}.

Choose $\omega$ as in Corollary~\ref{cor:SM_klr_unified_input}, with the exponents constructed in its proof. They satisfy
\begin{equation}
 0<\omega<\min\{1,\alpha-2d\},\quad
 \gamma_1>1,\quad
 d<\gamma_2=\omega\gamma_1<d+1\le2d.
\label{eq:SM_spatial_LR_exponents}
\end{equation}
Indeed, Eq.~\eqref{eq:SM_klr_unified_two_term_bound} gives $\gamma_2<\bar\alpha-d$, where $2d<\bar\alpha<2d+1$ in both cases of that proof. The upper bound on $\gamma_2$ will allow us to retain the time powers stated in the End Matter when converting coarse-grained distances to physical distances.

\begin{proposition}[Spatial restriction after occupation truncation]
\label{prop:SM_spatial_restriction}
Let $p\ge1$ and $\alpha>2d$. Fix $\eta>1$ and $0<\delta\le1$ such that
\begin{equation}
 \delta p<\alpha-2d,
 \quad
 \delta p(\gamma_1+1)<\gamma_2-d.
\label{eq:SM_spatial_summability_conditions}
\end{equation}
There are constants $C,C_{\mathrm{geom}}>0$ such that, for all $X\subset \Lambda$, $A\in\mathcal{A}_X$, $R>1$, $\bar q>1$, and $R^\eta\ge C_{\mathrm{geom}}L$,
\begin{align}
 \bigl\|\bar\tau_{s,t}(\bar A)
       -\bar\tau_{s,t}^{X_{2R^\eta}}(\bar A)\bigr\|
 \le C\|A\|\,|X|\,\bar q^pT^{d+1}
 \left[
 \frac{1}{R^{\eta(\alpha-2d-\delta p)}}
 +\frac{\bar q^{p\gamma_1}T^{\gamma_1+2d}}
 {R^{\eta[\gamma_2-d-\delta p(\gamma_1+1)]}}
 \right].
\label{eq:SM_spatial_restriction_before_cutoff_choice}
\end{align}
In particular, for $\bar q=R^\delta T^d$ and
\begin{equation}
 h=(\eta+1)\delta,\quad
 a_0=pd+d+1,\quad
 a_1=pd+3d+1+(pd+1)\gamma_1,
    \notag
\end{equation}
this becomes
\begin{align}
 \bigl\|\bar\tau_{s,t}(\bar A)
       -\bar\tau_{s,t}^{X_{2R^\eta}}(\bar A)\bigr\|
 \le C\|A\|E_{\mathrm{far}},
 \quad
 E_{\mathrm{far}}
 \coloneqq |X|\left[
 \frac{T^{a_0}}{R^{\eta(\alpha-2d)-hp}}
 +\frac{T^{a_1}}{R^{\eta(\gamma_2-d)-hp(\gamma_1+1)}}
 \right].
\label{eq:SM_spatial_restriction_main_error}
\end{align}
\end{proposition}

\subsubsection{Duhamel formula and decomposition of boundary-crossing interactions}
\label{S-IV-C-Duhamel}

We first make the support statements on the retained space explicit. For a block union $Y$, let
\begin{equation}
 \mathcal C_Y \coloneqq \{x\in\mathcal C:P_x\subseteq Y\},\quad
 \mathcal F_{\mathbf q,Y} \coloneqq
 \bigotimes_{x\in\mathcal C_Y}
 \operatorname{Ran}\bigl(\Pi_x|_{\mathcal F(P_x)}\bigr),
\end{equation}
where $\mathcal F(P_x)$ is the Fock space of $P_x$.
As explained in Sec.~\ref{S-IV-B-cutoffs},
\begin{equation}
 \mathcal F_{\mathbf q} \coloneqq \operatorname{Ran}\Pi
 =\mathcal F_{\mathbf q,Y}\otimes\mathcal F_{\mathbf q,Y^c}.
\label{eq:SM_spatial_retained_factorization}
\end{equation}
On this tensor product, $\bar H_Y$ and $\bar H_{Y^c}$ act on different factors. Also, the restriction of $\bar A$ is supported on the blocks in $X_0\subseteq Y$.
Consequently, for every pair of times $u,w$,
\begin{equation}
 [\bar H_Y(u),\bar H_{Y^c}(w)]=0,
 \quad
 [\bar A,\bar H_{Y^c}(u)]=0.
\label{eq:SM_spatial_exterior_commutation}
\end{equation}
These identities also hold on the full Fock space: all compressed operators vanish on $\mathcal F_{\mathbf q}^{\perp}$.
Both evolutions of $\bar A$ preserve this zero extension, so
\begin{equation}
 \bigl\|\bar\tau_{s,t}(\bar A)-\bar\tau^Y_{s,t}(\bar A)\bigr\|_{
 \mathcal B(\mathcal F)}
 =\bigl\|[\bar\tau_{s,t}(\bar A)-\bar\tau^Y_{s,t}(\bar A)]
           |_{\mathcal F_{\mathbf q}}\bigr\|_{
 \mathcal B(\mathcal F_{\mathbf q})}.
\end{equation}
In particular, no locality assertion on full Fock space is being made for $\bar A$ itself.

Suppose first that $s\le t$. Define $F(u) \coloneqq \bar\tau_{s,u}(\bar\tau^Y_{u,t}(\bar A))$ for $s\le u\le t$. Differentiation with respect to the initial time of the inner evolution gives
\begin{equation}
 \partial_u\bar\tau^Y_{u,t}(\bar A)
 =-\mi[\bar H_Y(u),\bar\tau^Y_{u,t}(\bar A)].
\end{equation}
Combining this with the final-time derivative of $\bar\tau_{s,u}$ yields
\begin{align}
 F'(u)
 =\mi\bar\tau_{s,u}\!\bigl(
 [\bar H(u)-\bar H_Y(u),\bar\tau^Y_{u,t}(\bar A)]\bigr)
 =2\mi\bar\tau_{s,u}\!\Big(
 \Big[\sum_{i\in Y,\,j\in Y^c}\bar h_{ij}(u),
       \bar\tau^Y_{u,t}(\bar A)\Big]\Big).
\end{align}
The exterior Hamiltonian has dropped out by Eq.~\eqref{eq:SM_spatial_exterior_commutation}.
Since $F(t)=\bar\tau_{s,t}(\bar A)$ and $F(s)=\bar\tau^Y_{s,t}(\bar A)$, integration and unitary invariance of the norm give
\begin{align}
 \bigl\|\bar\tau_{s,t}(\bar A)-\bar\tau^Y_{s,t}(\bar A)\bigr\|
 \le2\int_s^t \md u\,
 \Big\lVert\Big[\sum_{i\in Y,\,j\in Y^c}\bar h_{ij}(u),
                 \bar\tau^Y_{u,t}(\bar A)\Big]\Big\rVert
 =2\int_s^t \md u\,
 \Big\lVert\Big[
       \sum_{i\in Y,\,j\in Y^c}
               \bar\tau^Y_{t,u}(\bar h_{ij}(u)),\bar A
             \Big]\Big\lVert.
\label{eq:SM_spatial_Duhamel_bound}
\end{align}
The last equality conjugates the commutator by $\bar U^Y(u,t)$ and uses $\bar U^Y(t,u)=\bar U^Y(u,t)^\dagger$.
The same argument for $t<s$ gives the integral over the interval between the two times, with the same norm estimates below.
All these differentiations take place on finite-dimensional retained spaces and require only the continuous coefficients assumed in Sec.~\ref{S-I}.

\begin{figure}[t]
\centering
\includegraphics[width=0.58\linewidth]{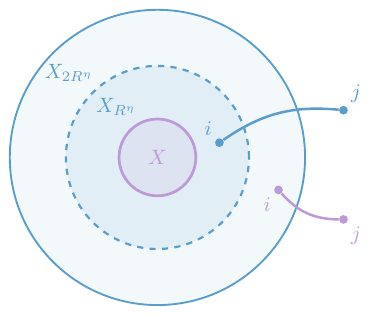}
\caption{The two contributions to the commutator. For the blue interaction, $i\in X_{R^\eta}$ and $j\in X_{2R^\eta}^{c}$, whereas the lilac interaction has $i\in X_{2R^\eta}\setminus X_{R^\eta}$ and $j\in X_{2R^\eta}^{c}$.}
\label{fig:SM_boundary_crossing}
\end{figure}

Partition the interior endpoints into
\begin{equation}
 X_{R^\eta}
 \quad\hbox{and}\quad
 \widetilde X_{2R^\eta} \coloneqq X_{2R^\eta}\setminus X_{R^\eta}.
\end{equation}
Define the two contributions
\begin{align}
 \mathcal E_{\mathrm{near}}(u)
  \coloneqq \sum_{\substack{i\in X_{R^\eta}\\j\in X_{2R^\eta}^c}}
 \bigl\|[\bar\tau^Y_{t,u}(\bar h_{ij}(u)),\bar A]\bigr\|,
 \quad
 \mathcal E_{\mathrm{shell}}(u)
 \coloneqq \sum_{\substack{i\in\widetilde X_{2R^\eta}\\j\in X_{2R^\eta}^c}}
 \bigl\|[\bar\tau^Y_{t,u}(\bar h_{ij}(u)),\bar A]\bigr\|.
\end{align}
The integrand in Eq.~\eqref{eq:SM_spatial_Duhamel_bound} is bounded by $\mathcal E_{\mathrm{near}}(u) +\mathcal E_{\mathrm{shell}}(u)$. If $X_{2R^\eta}=\Lambda$, both contributions vanish and the comparison is exact.
Otherwise, we distinguish boundary-crossing interactions according to whether their interior endpoint lies in $X_{R^\eta}$ or in $X_{2R^\eta}\setminus X_{R^\eta}$,
as illustrated in Fig.~\ref{fig:SM_boundary_crossing}.

We will use the block-volume estimate, valid for $r\ge1$,
\begin{equation}
 |X_r|\le C_d|X|L^d r^d.
\label{eq:SM_spatial_enlargement_volume}
\end{equation}
Indeed, every block included in $X_r$ contains a site at distance at most $r$ from $X$. Choose one such site per block.
The chosen sites are distinct and lie in $\bigcup_{i\in X}B_r(i)$, which has at most
$|X|(2r+1)^d$ sites. Multiplication by the maximal block volume $L^d$ proves the estimate. This deliberately {coarse-bound} requires neither connectedness of $X$ nor full boundary blocks.

\subsubsection{Interactions between the inner region and the exterior}
\label{S-IV-C-near}

For $i\in P_{x_i}$ and $j\in P_{x_j}$ in distinct blocks, compression and the occupation cutoffs give
\begin{equation}
 \|\bar h_{ij}(u)\|
 \le\frac{C(q_{x_i}^p+q_{x_j}^p)}{(1+|i-j|)^\alpha}
 \le C\bar q^p
 \frac{[1+d(P_{x_i},X)]^{\delta p}+[1+d(P_{x_j},X)]^{\delta p}}
      {(1+|i-j|)^\alpha}.
\label{eq:SM_spatial_microscopic_term_bound}
\end{equation}
For hopping, the bound follows from $\|\Pi b_i^\dagger b_j\Pi\|\le
\sqrt{(q_{x_i}+1)q_{x_j}}\le C(q_{x_i}+q_{x_j})$
and $q_x\ge1$, $p\ge1$.
For the representative density term, use $q_{x_i}^{p/2}q_{x_j}^{p/2}\le(q_{x_i}^p+q_{x_j}^p)/2$.
More generally, a polynomial monomial $n_i^a n_j^c$ of total degree $a+c\le p$ has norm at most $q_{x_i}^a q_{x_j}^c\le C(q_{x_i}^p+q_{x_j}^p)$.
The fixed finite sum of monomials and the orientations in Eq.~\eqref{eq:SM_spatial_hermitian_pair_terms} only change $C$.

If $i\in X_{R^\eta}$ and $j\notin X_{2R^\eta}$, then
\begin{equation}
 d(i,X)\le R^\eta+2\ell,
 \quad d(j,X)>2R^\eta,
 \quad |i-j|>R^\eta-2\ell\ge\frac{R^\eta}{2},
\end{equation}
where $C_{\mathrm{geom}}$ is chosen sufficiently large.
The $2\ell$ term is necessary because the enlargement is blockwise. Moreover,
\begin{equation}
 1+d(P_{x_i},X)\le1+R^\eta\le2R^\eta,
 \quad
 1+d(P_{x_j},X)\le1+d(j,X)
 \le1+R^\eta+2\ell+|i-j|
 \le C(1+|i-j|).
\end{equation}
The trivial commutator bound, therefore, yields
\begin{equation}
 \mathcal E_{\mathrm{near}}(u)
 \le C\|A\|\bar q^p
 \sum_{\substack{i\in X_{R^\eta}\\j\in X_{2R^\eta}^c}}
 \frac{[1+d(P_{x_i},X)]^{\delta p}+[1+d(P_{x_j},X)]^{\delta p}}
      {(1+|i-j|)^\alpha}.
\label{eq:SM_spatial_near_initial_bound}
\end{equation}
We estimate the two weights separately, retaining the microscopic sums. Shell counting around a fixed $i$ and Eq.~\eqref{eq:SM_spatial_enlargement_volume} give
\begin{align}
 \sum_{\substack{i\in X_{R^\eta}\\j\in X_{2R^\eta}^c}}
 \frac{[1+d(P_{x_i},X)]^{\delta p}}{(1+|i-j|)^\alpha}
 &\le C|X|L^d R^{\eta(d+\delta p)}
       \sum_{m\ge R^\eta/2} (1+m)^{d-1-\alpha}
 \le C|X|L^dR^{\eta(2d+\delta p-\alpha)},
\\
 \sum_{\substack{i\in X_{R^\eta}\\j\in X_{2R^\eta}^c}}
 \frac{[1+d(P_{x_j},X)]^{\delta p}}{(1+|i-j|)^\alpha}
 &\le C|X|L^dR^{\eta d}
       \sum_{m\ge R^\eta/2}(1+m)^{d-1-\alpha+\delta p}
 \le C|X|L^dR^{\eta(2d+\delta p-\alpha)}.
\end{align}
The second series converges because $\alpha-\delta p>2d\ge d$; the first requires only $\alpha>d$.
Inserting these estimates into Eq.~\eqref{eq:SM_spatial_near_initial_bound} gives
\begin{equation}
 \mathcal E_{\mathrm{near}}(u)
 \le C\|A\|\,|X|\bar q^pT^d
       R^{-\eta(\alpha-2d-\delta p)}.
\label{eq:SM_spatial_near_final_bound}
\end{equation}

\subsubsection{Bounded interactions on the retained block spaces}
\label{S-IV-C-bounded-input}

To control the shell contribution, we verify the interaction assumptions before applying a Lieb--Robinson bound.
The relevant finite lattice is $\mathcal C_Y$, identified with a subset of $\mathbb Z^d$ by $x\mapsto x/L$, with metric $d_{\mathrm{cg}}(P_x,P_y)=|x-y|/L$.
Its local Hilbert spaces are the truncated block spaces in Eq.~\eqref{eq:SM_spatial_retained_factorization}, not the individual microscopic Fock spaces.

For every block contained in $Y=X_{2R^\eta}$,
\begin{equation}
 q_x\le\bar q(1+2R^\eta)^\delta
 \le Q,\quad
 Q \coloneqq 3^\delta\bar q R^{\eta\delta}>1.
\end{equation}
On $\mathcal F_{\mathbf q,Y}$, remove only the scalar part of each pair term when verifying the interaction bounds. Namely, let $c_{ij}(u):=\langle\mathrm{vac}|h_{ij}(u)|\mathrm{vac}\rangle$ and write
\begin{equation}
 \widehat h_{ij}(u) \coloneqq
 \bar h_{ij}(u)|_{\mathcal F_{\mathbf q,Y}}-c_{ij}(u)I,
 \quad i,j\in Y.
\end{equation}
The hat here denotes removal of this scalar only; the particle cutoff is unchanged. Group the remaining interactions as
\begin{equation}
 \Phi_{\{x\}}(u) \coloneqq \sum_{i,j\in P_x}\widehat h_{ij}(u),
 \quad
 \Phi_{\{x,y\}}(u) \coloneqq
       2\sum_{i\in P_x,\,j\in P_y}\widehat h_{ij}(u),\quad x\ne y.
\end{equation}
Each unordered pair $\{x,y\}$ occurs once. Thus, the sum of these interactions equals $\bar H_Y|_{\mathcal F_{\mathbf q,Y}}$ minus $[\sum_{i,j\in Y}c_{ij}(u)]I$ and generates exactly the same Heisenberg dynamics. This removes the degree-zero polynomial components, if present, but retains every component of positive total degree.

For an on-block or neighboring-block hopping Hamiltonian, the one-particle matrix has norm bounded by a constant, because its row and column sums are bounded for $\alpha>d$.
On its $N$-particle sector, its second quantization is the restriction of $\sum_{j=1}^N J^{(j)}$, with norm at most $N\|J\|$. Since one or two retained blocks contain at most $2Q$ particles, the hopping norm is at most $CQ$.
For a density monomial of positive total degree $a+c\le p$, Young's inequality and the integer number spectrum imply
\begin{equation}
 n_i^a n_j^c\le C(n_i^p+n_j^p).
\end{equation}
This includes $a=c=p/2$ and terms with one exponent zero.
For two blocks, the bounded row and column sums then give
\begin{align}
 \sum_{i\in P_x,\,j\in P_y}
 \frac{n_i^a n_j^c}{(1+|i-j|)^\alpha}
 \le C\left(\sum_{i\in P_x}n_i^p
              +\sum_{j\in P_y}n_j^p\right)
 \le C(n_{P_x}^p+n_{P_y}^p).
\end{align}
The last inequality uses $p\ge1$ and the joint spectral calculus of the number operators. The same argument applies within one block. Thus,
\begin{equation}
 \|\Phi_{\{x\}}(u)\|\le CQ^p,
 \quad
 \|\Phi_{\{x,y\}}(u)\|\le CQ^p
 \quad\text{if }d_{\mathrm{cg}}(P_x,P_y)=1.
\label{eq:SM_spatial_near_block_strength}
\end{equation}

For coarse distance $m=d_{\mathrm{cg}}(P_x,P_y)\ge2$, Eq.~\eqref{eq:SM_block_geometry} gives $|i-j|\ge L(m-1)+1\ge Lm/2$.
There are at most $L^{2d}$ microscopic pairs. Therefore, Eq.~\eqref{eq:SM_spatial_microscopic_term_bound} yields
\begin{equation}
 \|\Phi_{\{x,y\}}(u)\|
 \le C Q^p L^{2d-\alpha}m^{-\alpha}
 \le \frac{CQ^p}{(1+m)^\alpha},
\end{equation}
where $\alpha>2d$ and $L\ge1$ were used in the last step.
Together with Eq.~\eqref{eq:SM_spatial_near_block_strength}, this verifies Eq.~\eqref{eq:SM_klr_pair_decay} with $k=2$ and
\begin{equation}
 g_{\mathrm{big}} \coloneqq C_0 Q^p
 \le C\bar q^pR^{\eta\delta p},\quad C_0\ge1.
\end{equation}
The constant $C_0$ is fixed independently of the cutoffs, $R$, the block radius, and the system size.
For example, summing the pair estimate over coarse shells also gives $\sup_x\sum_{y\ne x}\|\Phi_{\{x,y\}}(u)\|
\le CQ^p$.
The scale $g_{\mathrm{big}}$ is a preliminary bound for this spatial restriction. It is not the improved interaction scale obtained in Sec.~\ref{S-IV-D}.

Corollary~\ref{cor:SM_klr_unified_input} combines the results of Refs.~\cite{PhysRevLett.127.160401,PhysRevX.10.031010} into the polynomial localization bound used here.
Although the corollary requires $r\ge C_r(g_{\mathrm{big}}T)^{1/\omega}$, the same bound extends to all coarse radii $r\ge1$ with an adjusted constant.
For $1\le r<C_r(g_{\mathrm{big}}T)^{1/\omega}$, the relation $\gamma_2=\omega\gamma_1$ implies
\begin{equation}
 \frac{(g_{\mathrm{big}}T)^{\gamma_1}}{r^{\gamma_2}}
 > C_r^{-\gamma_2}.
\label{eq:SM_spatial_small_radius_absorption}
\end{equation}
In this regime, the polynomial bound absorbs the trivial localization estimate $2\|O\|$.
For larger radii, the corollary applies directly.
The condition $g_{\mathrm{big}}T\ge1$ holds by construction.
This extension provides a uniform upper bound without implying a small localization error inside the light cone.

Let $O_x$ be any bounded operator on the retained space of one block, and let $B$ be supported on a set of retained blocks at coarse distance strictly greater than $r$ from $x$.
The conditional-expectation approximant to the evolution of $O_x$ is supported in the closed coarse ball of radius $r$ and therefore commutes with $B$. Consequently,
\begin{equation}
 \|[\bar\tau^Y_{t,u}(O_x),B]\|
 \le C\|O_x\|\|B\|
       \frac{(g_{\mathrm{big}}T)^{\gamma_1}}{r^{\gamma_2}},
 \quad r\ge1,
\label{eq:SM_spatial_block_commutator}
\end{equation}
for every $u$ between $s$ and $t$.
We used $1+|t-u|\le T$. The evolved support in the corollary is a single coarse site, so there is no factor counting the blocks in the support of $B$. The estimate also applies to non-Hermitian operators and is independent of all local Hilbert-space dimensions.

\subsubsection{Interactions with an endpoint in the outer shell}
\label{S-IV-C-shell}

Let $i\in\widetilde X_{2R^\eta}$ and $j\in X_{2R^\eta}^c$.
On $\mathcal F_{\mathbf q}$, write $\bar b_i=\Pi b_i\Pi$ and $\overline{n_i^a}=\Pi n_i^a\Pi$ for their restrictions.
For different blocks, the compressed hopping and density terms factor into the corresponding block operators. Since $j$ is outside $Y$, its factor is fixed by $\bar\tau^Y_{t,u}$ and commutes with $\bar A$. Thus,
\begin{align}
 [\bar\tau^Y_{t,u}(\overline{T_{ij}(u)}),\bar A]
 &=J_{ij}(u)\bar b_j
       [\bar\tau^Y_{t,u}(\bar b_i^\dagger),\bar A], \quad
 [\bar\tau^Y_{t,u}(\overline{T_{ji}(u)}),\bar A]
 =J_{ji}(u)\bar b_j^\dagger
       [\bar\tau^Y_{t,u}(\bar b_i),\bar A],
 \notag\\
 [\bar\tau^Y_{t,u}(\overline{V_{ij}(u)}),\bar A]
 &=U_{ij}(u)\overline{n_j^{p/2}}
       [\bar\tau^Y_{t,u}(\overline{n_i^{p/2}}),\bar A]
\label{eq:SM_spatial_factor_exterior_operator}
\end{align}
for the representative density term.
The same factorization holds for the adjoints and for each additional density monomial. This factorization is valid on the retained tensor product.
The exterior global projectors are not replaced by identities on the full Fock space.

In particular, each commutator on the right acts only on $\mathcal F_{\mathbf q,Y}$. Its full-space norm equals its norm on that retained interior factor. Explicitly, for any such commutator $C_Y$,
\begin{equation}
 \|C_Y\otimes I_{\mathbf q,Y^c}\|
 =\|C_Y\|
 =\sup_{\substack{\psi\in\mathcal F_{\mathbf q,Y}\\\|\psi\|=1}}
       \|C_Y\psi\|.
\end{equation}
This formula does not require the commutator to be Hermitian.

The block $P_{x_i}$ lies in the outer shell, so $d(P_{x_i},X)>R^\eta$.
For any block $P_z$ intersecting $X$, choose $k\in P_z\cap X$. The inequalities
$|i-k|>R^\eta$ and $|i-k|\le|x_i-z|+2\ell$ give
\begin{equation}
 d_{\mathrm{cg}}(P_{x_i},P_z)
 >\frac{R^\eta-2\ell}{L}
 \ge\frac{R^\eta}{2L}.
\end{equation}
Choose $C_{\mathrm{geom}}\ge8$ and apply Eq.~\eqref{eq:SM_spatial_block_commutator} at radius $r=R^\eta/(4L)\ge1$, with $B=\bar A|_{\mathcal F_{\mathbf q}}$. Since
$\|\bar b_i\|,\|\bar b_i^\dagger\|\le q_{x_i}^{1/2}$ and $\|\overline{n_i^a}\|\le q_{x_i}^{a}$, we obtain
\begin{align}
 \|[\bar\tau^Y_{t,u}(\bar b_i^\dagger),\bar A]\|
 +\|[\bar\tau^Y_{t,u}(\bar b_i),\bar A]\|
 &\le C\|A\|q_{x_i}^{1/2}
       \frac{L^{\gamma_2}(g_{\mathrm{big}}T)^{\gamma_1}}
            {R^{\eta\gamma_2}},
 \notag\\
 \|[\bar\tau^Y_{t,u}(\overline{n_i^{p/2}}),\bar A]\|
 &\le C\|A\|q_{x_i}^{p/2}
       \frac{L^{\gamma_2}(g_{\mathrm{big}}T)^{\gamma_1}}
            {R^{\eta\gamma_2}}.
\label{eq:SM_spatial_shell_local_commutators}
\end{align}
The factor $L^{\gamma_2}$ records the conversion from coarse distance to physical distance. It cannot be discarded before the time powers are counted.
For hopping, $p\ge1$ and $q_x\ge1$ imply $(q_{x_i}q_{x_j})^{1/2}\le(q_{x_i}q_{x_j})^{p/2}$.
Combining Eqs.~\eqref{eq:SM_spatial_factor_exterior_operator} and \eqref{eq:SM_spatial_shell_local_commutators}, including the orientations in Eq.~\eqref{eq:SM_spatial_hermitian_pair_terms}, therefore, yields
\begin{align}
 \|[\bar\tau^Y_{t,u}(\bar h_{ij}(u)),\bar A]\|
 \le
 C\|A\|\bar q^{p(\gamma_1+1)}T^{\gamma_1}L^{\gamma_2}
 \frac{[1+d(P_{x_i},X)]^{\delta p/2}[1+d(P_{x_j},X)]^{\delta p/2}}
      {(1+|i-j|)^\alpha}
 \frac{1}{R^{\eta(\gamma_2-\delta p\gamma_1)}}
\end{align}
for the representative density interaction and hopping.

To sum over the exterior endpoint, use $1+d(P_{x_i},X)\le1+2R^\eta\le3R^\eta$ and
\begin{align}
 [1+d(P_{x_j},X)]
 \le1+d(j,X)
 \le1+d(i,X)+|i-j|
 \le1+2R^\eta+2\ell+|i-j|
 \le CR^\eta(1+|i-j|).
\label{eq:SM_spatial_shell_exterior_weight}
\end{align}
Consequently,
\begin{align}
 \sum_{\substack{i\in\widetilde X_{2R^\eta}\\j\in X_{2R^\eta}^c}}
 \frac{[1+d(P_{x_i},X)]^{\delta p/2}[1+d(P_{x_j},X)]^{\delta p/2}}
      {(1+|i-j|)^\alpha}
 \le CR^{\eta\delta p}
       \sum_{i\in\widetilde X_{2R^\eta}}
       \sum_{j\in\Lambda}(1+|i-j|)^{-\alpha+\delta p/2}
 \le C|X|L^dR^{\eta(d+\delta p)}.
\end{align}
The microscopic row sum is finite for $\alpha-\delta p/2>d$, as follows from Eq.~\eqref{eq:SM_spatial_summability_conditions}.

For completeness, the same estimate covers every additional polynomial component without assuming balanced powers.
For a monomial $n_i^a n_j^c$, $a,c\ge0$, $a+c\le p$, the factorization in Eq.~\eqref{eq:SM_spatial_factor_exterior_operator} gives the factor $q_{x_i}^{a}q_{x_j}^{c}$ instead.
If $a=0$, the commutator of its interior factor is zero.
Otherwise, Eqs.~\eqref{eq:SM_spatial_shell_exterior_weight} and \eqref{eq:SM_spatial_enlargement_volume} imply
\begin{align}
 \sum_{\substack{i\in\widetilde X_{2R^\eta}\\j\in X_{2R^\eta}^c}}
 \frac{q_{x_i}^{a}q_{x_j}^{c}}{(1+|i-j|)^\alpha}
 \le C\bar q^{a+c}R^{\eta\delta(a+c)}
       \sum_{i\in\widetilde X_{2R^\eta}}
       \sum_{j\in\Lambda}(1+|i-j|)^{-\alpha+\delta c}
 \le C\bar q^p|X|L^dR^{\eta(d+\delta p)},
\end{align}
where $\alpha-\delta c\ge\alpha-\delta p>d$, $\bar q>1$, and $R>1$.
Reversing the endpoints interchanges $a$ and $c$ and obeys the same bound. Thus, the finite sum of polynomial components changes only the overall constant.

Combining the pair estimates and the spatial sums gives
\begin{align}
 \mathcal E_{\mathrm{shell}}(u)
 &\le C\|A\|\,|X|\,
       \bar q^{p(\gamma_1+1)}T^{\gamma_1}L^{d+\gamma_2}
       R^{-\eta[\gamma_2-d-\delta p(\gamma_1+1)]}
 \notag\\
 &\le C\|A\|\,|X|\,
       \bar q^{p(\gamma_1+1)}T^{\gamma_1+3d}
       R^{-\eta[\gamma_2-d-\delta p(\gamma_1+1)]}.
\label{eq:SM_spatial_shell_final_bound}
\end{align}
For the second line, we used $L\le C_vT$ and $\gamma_2<d+1\le2d$ from Eq.~\eqref{eq:SM_spatial_LR_exponents}. Thus, the coarse-to-physical conversion is absorbed within the quadratic block-size factor $T^{2d}$ retained in the main-text estimate; no cutoff-dependent Hilbert-space dimension enters the constant.

\subsubsection{Total error and parameter choice}
\label{S-IV-C-conclusion}

\begin{proof}[Completion of the proof of Proposition~\ref{prop:SM_spatial_restriction}]
Insert Eqs.~\eqref{eq:SM_spatial_near_final_bound} and \eqref{eq:SM_spatial_shell_final_bound} into Eq.~\eqref{eq:SM_spatial_Duhamel_bound}.
The estimates are uniform for every intermediate time.
The integration interval has length $|t-s|\le T$, so
\begin{align}
 \bigl\|\bar\tau_{s,t}(\bar A)
       -\bar\tau_{s,t}^{X_{2R^\eta}}(\bar A)\bigr\|
 \le C\|A\|\,|X|
 \left[
 \bar q^pT^{d+1}R^{-\eta(\alpha-2d-\delta p)}
 +\bar q^{p(\gamma_1+1)}T^{\gamma_1+3d+1}
       R^{-\eta[\gamma_2-d-\delta p(\gamma_1+1)]}
 \right].
\end{align}
This proves Eq.~\eqref{eq:SM_spatial_restriction_before_cutoff_choice}.
Substitute $\bar q=R^\delta T^d$. The first term becomes
\begin{equation}
 \bar q^pT^{d+1}R^{-\eta(\alpha-2d-\delta p)}
 =\frac{T^{pd+d+1}}
 {R^{\eta(\alpha-2d)-(\eta+1)\delta p}}.
\end{equation}
For the second term,
\begin{align}
 \bar q^{p(\gamma_1+1)}T^{\gamma_1+3d+1}
       R^{-\eta[\gamma_2-d-\delta p(\gamma_1+1)]}
 =
 \frac{T^{pd(\gamma_1+1)+\gamma_1+3d+1}}
 {R^{\eta(\gamma_2-d)-(\eta+1)\delta p(\gamma_1+1)}}
 =\frac{T^{a_1}}
 {R^{\eta(\gamma_2-d)-hp(\gamma_1+1)}}.
\end{align}
These are precisely the two contributions in Eq.~\eqref{eq:SM_spatial_restriction_main_error}.
All spatial sums were bounded by sums over $\mathbb Z^d$ or the rescaled block lattice. The fixed-$k$ constants are independent of the retained dimensions. This proves the claimed uniformity in system size and occupation cutoffs.
\end{proof}

\textit{Compatibility with the final radius.---} Choose $\omega$ first, and then choose $\eta>1$ as in the End Matter, so that
\begin{equation}
 (d+2)\eta(\alpha-2d)>a_0,\quad
 (d+2)\eta(\gamma_2-d)>a_1,\quad
 (d+2)\eta\omega>pd+1.
\end{equation}
All three choices are possible because $\alpha>2d$, $\gamma_2>d$, and $\omega>0$.
Next choose $\delta>0$ sufficiently small to satisfy Eq.~\eqref{eq:SM_spatial_summability_conditions} and
\begin{align}
 e_0& \coloneqq (d+2)[\eta(\alpha-2d)-hp]-a_0>0,
 \\
 e_1& \coloneqq (d+2)[\eta(\gamma_2-d)-hp(\gamma_1+1)]-a_1>0, \quad
 (d+2)(\eta\omega-hp)>pd+1.
\end{align}
The value of $\delta$ may be further reduced in Sec.~\ref{S-IV-E} without violating these strict inequalities. At every radius $R\ge C_RT^{d+2}$, with $C_R$ sufficiently large,
\begin{equation}
 E_{\mathrm{far}}\le C|X|(T^{-e_0}+T^{-e_1}),
 \quad
 R^{\eta\omega-hp}\ge CT^{pd+1}.
\end{equation}
Such radii also satisfy $R^\eta\ge C_{\mathrm{geom}}L$.
The final radii in Sec.~\ref{S-IV-E} are at least this large, since $\beta_{\mathrm{up}}\ge d+2$.

The actual coarse radius used above is $R^\eta/(4L)$.
Although the all-radius extension \eqref{eq:SM_spatial_small_radius_absorption} made a separate cone condition unnecessary in the proposition, the final parameter choices place this radius outside the bounded-system cone as well. Indeed,
\begin{align}
 (d+2)(\eta\omega-hp)-(pd+1+\omega)
 =
 \frac{e_1+(d+2)\eta d+(d+2)hp+pd+3d+1-\gamma_2}
      {\gamma_1}>0.
\end{align}
Together with $g_{\mathrm{big}}T\le CT^{pd+1}R^{hp}$ and $L\le C_vT$, this implies
$R^\eta/(4L)\ge C_r(g_{\mathrm{big}}T)^{1/\omega}$ at the final radii after increasing $C_R$. No additional condition on the final light-cone exponent is required.

We have therefore restricted the truncated evolution to $X_{2R^\eta}$ with the error stated in the End Matter. The uniform occupation bound on this region is $Q\le CT^dR^h$. In Sec.~\ref{S-IV-D}, we use the commuting short-range density terms to replace the preliminary scale $g_{\mathrm{big}}$ by the sharper effective interaction scale needed for the final local approximation.

\subsection{Interaction-picture reduction and effective interaction scale}
\label{S-IV-D}

We now analyze the dynamics restricted to $Y=X_{2R^\eta}$.
The short-range density terms can have norms of order $Q^p$, but their mutual commutativity allows us to treat their evolution exactly. After this change of picture, the interaction strength is controlled by hopping and the density couplings between well-separated blocks. We follow the block decomposition of the
main text and the interaction-picture method of Refs.~\cite{kuwahara2022optimal,LR26,LemmWessel2025Commuting}.
The support and norm estimates are provided below for the time-dependent Hamiltonian used here.

Retain $T=1+|t-s|$, $\ell=\lceil v|t-s|\rceil$, $L=2\ell+1$, and the global projector $\Pi$ of Sec.~\ref{S-IV-B}. As in Sec.~\ref{S-IV-C}, we work on
\begin{equation}
 \mathcal F_{\mathbf q,Y}
 =\bigotimes_{x\in\mathcal C_Y}
   \operatorname{Ran}\bigl(\Pi_x|_{\mathcal F(P_x)}\bigr),
 \quad
 Q=3^\delta\bar q R^{\eta\delta},
 \quad q_x\le Q\quad(x\in\mathcal C_Y).
\label{eq:SM_ip_retained_space}
\end{equation}
Restrictions of barred operators to this tensor product are understood throughout the support and interaction-norm calculations. The corresponding full-Fock-space operators retain the global projectors. In particular, no exterior projector is removed in this subsection. The fixed value $v>0$ gives
\begin{equation}
 c_vT\le L\le C_vT,
 \quad
 c_v=\min\{1,2v\},\quad C_v=3+2v.
\label{eq:SM_ip_block_spacing_time}
\end{equation}
Indeed, $1+2v|t-s|\le L\le3+2v|t-s|$.
All spatially restricted Hamiltonians are taken to be Hermitian, as required for the unitary dynamics in Sec.~\ref{S-IV-C}.
Constants in the interaction estimates depend only on the fixed model parameters and on $v,\eta,\delta$. They are independent of $\bar q,R,s,t,X$, the lattice size or shape, and the dimensions of the retained block spaces.

\subsubsection{Block decomposition and the interaction picture}
\label{S-IV-D-decomposition}

To keep the ordered-pair convention of Eq.~\eqref{eq:SM_spatial_hermitian_pair_terms}, separate its Hermitian hopping and density parts as
\begin{equation}
 h^0_{ij}(u)\coloneqq\frac14
 [T_{ij}(u)+T_{ji}(u)+\mathrm{H.c.}],
 \quad
 v_{ij}(u)\coloneqq h_{ij}(u)-h^0_{ij}(u).
\end{equation}
Here, the adjoint applies to both preceding hopping terms.
Both components are Hermitian and symmetric under $i\leftrightarrow j$.
For one block and for an unordered pair of distinct blocks, define
\begin{align}
 H_x^0(u)&\coloneqq\sum_{i,j\in P_x}h^0_{ij}(u), \quad
 V_x(u)\coloneqq\sum_{i,j\in P_x}v_{ij}(u),\notag\\
 H_{xy}^0(u)&\coloneqq2\sum_{i\in P_x,\,j\in P_y}h^0_{ij}(u),\quad
 V_{xy}(u)\coloneqq2\sum_{i\in P_x,\,j\in P_y}v_{ij}(u).
\label{eq:SM_ip_block_terms}
\end{align}
The factor of two includes both orientations of a distinct-block pair. Write
\begin{equation}
 \begin{split}
 \mathcal P_{\mathrm{near}}
 &\coloneqq\{\{x,y\}\subseteq\mathcal C_Y:
                     x\ne y,\ d_{\mathrm{cg}}(P_x,P_y)=1\},\\
 \mathcal P_{\mathrm{long}}
 &\coloneqq\{\{x,y\}\subseteq\mathcal C_Y:
                     x\ne y,\ d_{\mathrm{cg}}(P_x,P_y)\ge2\}.
 \end{split}
\end{equation}
Neighbors are defined by the coarse-grained centers, not by the minimum distance between the occupied parts of boundary blocks.
This convention also applies to incomplete blocks.

The three contributions in the main text are
\begin{align}
 H^0_{\mathrm{on\text{-}site}}(u)&\coloneqq
                   \sum_{x\in\mathcal C_Y}H_x^0(u),\quad
 V_{\mathrm{on\text{-}site}}(u)\coloneqq
                   \sum_{x\in\mathcal C_Y}V_x(u),\quad
 H^0_{\mathrm{near}}(u)\coloneqq
                  \sum_{\{x,y\}\in\mathcal P_{\mathrm{near}}}H_{xy}^0(u),\\
 V_{\mathrm{near}}(u)&\coloneqq
                  \sum_{\{x,y\}\in\mathcal P_{\mathrm{near}}}V_{xy}(u), \quad
 H_{\mathrm{long}}(u)\coloneqq
       \sum_{\{x,y\}\in\mathcal P_{\mathrm{long}}}
                    [H_{xy}^0(u)+V_{xy}(u)].
\end{align}
Set
\begin{equation}
 V_{\mathrm{short}}
 =V_{\mathrm{on\text{-}site}}+V_{\mathrm{near}},
 \quad
 H^0_{\mathrm{short}}
 =H^0_{\mathrm{on\text{-}site}}+H^0_{\mathrm{near}}.
\end{equation}
The definitions give the exact identity
\begin{equation}
 \bar H_Y(u)=\bar V_{\mathrm{short}}(u)
             +\bar H^0_{\mathrm{short}}(u)
             +\bar H_{\mathrm{long}}(u).
\end{equation}
In this decomposition, ``on-site'' refers to a single coarse block. It includes interactions between different microscopic sites within that block.

Write $\mathcal V_{\{x\}}=\bar V_x$ and $\mathcal V_{\{x,y\}}=\bar V_{xy}$ for the density terms in $\bar V_{\mathrm{short}}$. Every such term and every occupation projector is a function of the commuting number operators. Hence,
\begin{equation}
 [\mathcal V_Z(u),\mathcal V_{Z'}(w)]=0
\label{eq:SM_ip_density_commutation}
\end{equation}
for all one-block or neighboring-block density contributions, at arbitrary times $u,w$. For fixed initial time $s$, define
\begin{equation}
 U^V(s,u)\coloneqq
 \exp\left[-\mi\int_s^u\bar V_{\mathrm{short}}(w)\,\md w\right].
\end{equation}
The integral commutes with its derivative, so $\mi\partial_uU^V(s,u)=\bar V_{\mathrm{short}}(u)U^V(s,u)$ and $U^V(s,s)=I$. This formula is valid in either time direction. Introduce
\begin{equation}
 U^{\mathrm{int}}(s,u)\coloneqq
        U^V(s,u)^\dagger\bar U^Y(s,u),
 \quad
 H^{\mathrm{int}}(u)\coloneqq
 U^V(s,u)^\dagger
 [\bar H^0_{\mathrm{short}}(u)+\bar H_{\mathrm{long}}(u)]
 U^V(s,u).
\label{eq:SM_ip_generator}
\end{equation}
The dependence of $H^{\mathrm{int}}$ on the fixed anchor $s$ is suppressed. Differentiating the product gives
\begin{align}
 \mi\partial_uU^{\mathrm{int}}(s,u)
 &=-U^V(s,u)^\dagger\bar V_{\mathrm{short}}(u)\bar U^Y(s,u)
   +U^V(s,u)^\dagger\bar H_Y(u)\bar U^Y(s,u)\notag\\
 &=H^{\mathrm{int}}(u)U^{\mathrm{int}}(s,u),
 \quad U^{\mathrm{int}}(s,s)=I.
\end{align}
Thus, $\bar U^Y(s,t)=U^V(s,t)U^{\mathrm{int}}(s,t)$. With
\begin{equation}
 \tau^V_{s,t}(O)\coloneqq U^V(s,t)^\dagger O U^V(s,t),
 \quad
 \tau^{\mathrm{int}}_{s,t}(O)\coloneqq
 U^{\mathrm{int}}(s,t)^\dagger O U^{\mathrm{int}}(s,t),
\end{equation}
we obtain
\begin{equation}
 \bar\tau^Y_{s,t}(\bar A)
 =\tau^{\mathrm{int}}_{s,t}(\bar A_V),
 \quad
 \bar A_V\coloneqq\tau^V_{s,t}(\bar A).
\label{eq:SM_ip_observable_identity}
\end{equation}
The order of the two Heisenberg maps is $\bar\tau^Y_{s,t}=\tau^{\mathrm{int}}_{s,t}\circ\tau^V_{s,t}$.
For $s=0$, Eq.~\eqref{eq:SM_ip_generator} is the interaction-picture Hamiltonian in the main text.

\subsubsection{Support under the commuting density evolution}
\label{S-IV-D-support}

For $S\subseteq\mathcal C_Y$, use the coarse neighborhood
\begin{equation}
 S^{[1]}\coloneqq
 \{y\in\mathcal C_Y:\min_{x\in S}d_{\mathrm{cg}}(P_x,P_y)\le1\},
 \quad \mathcal N_x\coloneqq\{x\}^{[1]}.
\end{equation}
Then $|\mathcal N_x|\le3^d$ and $|S^{[1]}|\le3^d|S|$.
Using the density terms $\mathcal V_Z$ defined above, Eq.~\eqref{eq:SM_ip_density_commutation} gives
\begin{equation}
 U^V(s,u)=\prod_Z W_Z(s,u),
 \quad
 W_Z(s,u)\coloneqq\exp\left[-\mi\int_s^u\mathcal V_Z(w)\,\md w\right],
\end{equation}
where the product order is immaterial. If $O_S$ is supported on the retained blocks in $S$, all factors with $Z\cap S=\varnothing$ commute with $O_S$. They also commute with the remaining factors.
Reordering the product and canceling these factors gives
\begin{equation}
 U^V(s,u)^\dagger O_SU^V(s,u)
 =\left(\prod_{Z\cap S\ne\varnothing}W_Z(s,u)\right)^\dagger
   O_S
  \left(\prod_{Z\cap S\ne\varnothing}W_Z(s,u)\right).
\end{equation}
Every remaining $Z$ lies in $S^{[1]}$. In particular,
\begin{equation}
 \operatorname{supp}_{\mathrm{cg}}
       [U^V(s,u)^\dagger O_SU^V(s,u)]\subseteq S^{[1]},
 \quad
 \|U^V(s,u)^\dagger O_SU^V(s,u)\|=\|O_S\|,
\label{eq:SM_ip_exact_support_enlargement}
\end{equation}
where $\operatorname{supp}_{\mathrm{cg}}$ denotes support in the retained block tensor product. The same statements hold for conjugation by $U^V(s,u)$ instead of its adjoint. No bound on the accumulated phases is used. Commutativity prevents the support enlargement from iterating through successive neighboring blocks.

For the observable, define
\begin{equation}
 \mathcal S_0\coloneqq\{x\in\mathcal C_Y:P_x\cap X\ne\varnothing\},
 \quad \mathcal S_V\coloneqq\mathcal S_0^{[1]}.
\label{eq:SM_ip_observable_support_sets}
\end{equation}
Equation~\eqref{eq:SM_global_cutoff_support} identifies the restriction of $\bar A$ as an operator on $\mathcal S_0$.
Each such block contains a site of $X$, and the blocks are disjoint. It follows that
\begin{equation}
 |\mathcal S_V|\le3^d|X|,
 \quad
 \operatorname{supp}_{\mathrm{cg}}\bar A_V\subseteq\mathcal S_V,
 \quad \|\bar A_V\|=\|\bar A\|\le\|A\|.
\label{eq:SM_ip_observable_support_bound}
\end{equation}
To compare with physical distance, let $z\in\mathcal S_V$ and choose $x\in\mathcal S_0$ with $d_{\mathrm{cg}}(P_z,P_x)\le1$. For $i\in P_x\cap X$ and $j\in P_z$,
\begin{equation}
 |i-j|\le|j-z|+|z-x|+|x-i|
 \le2\ell+L<2L.
\end{equation}
Hence $\bigcup_{z\in\mathcal S_V}P_z\subseteq X_{2L}$.
This is an $O(T)$ enlargement in physical distance, with a constant number of coarse layers. These are retained-space support statements. The full-Fock-space operator still has the exterior projector described in Sec.~\ref{S-IV-B-cutoffs}.

\subsubsection{Hopping bounds within and between blocks}
\label{S-IV-D-hopping}

We estimate the block hopping before conjugation, since Eq.~\eqref{eq:SM_ip_exact_support_enlargement} preserves each term norm. Set
\begin{equation}
 J^{\mathrm h}_{ij}(u)\coloneqq
          \frac{J_{ij}(u)+{\overline{J}_{ji}(u)}}2.
\end{equation}
This Hermitian matrix generates the hopping terms in Eq.~\eqref{eq:SM_ip_block_terms} and obeys the same power-law coefficient bound. Its absolute row and column sums satisfy
\begin{equation}
 \sup_{u,i}\sum_j|J^{\mathrm h}_{ij}(u)|
 \le C_{J}\sum_{k\in\mathbb Z^d}(1+|k|)^{-\alpha}
 \le C,
 \quad \alpha>d.
\label{eq:SM_ip_hopping_row_sum}
\end{equation}
For a finite-particle vector $\psi$ and a Hermitian hopping matrix $K$ on one or two blocks, Cauchy--Schwarz gives
\begin{align}
 \left|\left\langle\psi,
           \sum_{i,j}K_{ij}b_i^\dagger b_j\psi\right\rangle\right|
 \le\sum_{i,j}|K_{ij}|\|b_i\psi\|\|b_j\psi\|
 \le\frac12\sum_{i,j}|K_{ij}|
        \langle\psi,(n_i+n_j)\psi\rangle.
\end{align}
Applying Eq.~\eqref{eq:SM_ip_hopping_row_sum} to a single block or a neighboring-block pair yields
\begin{equation}
 \|\bar H_x^0(u)\|\le Cq_x\le CQ,
 \quad
 \|\bar H_{xy}^0(u)\|\le C(q_x+q_y)\le CQ
 \quad(\{x,y\}\in\mathcal P_{\mathrm{near}}).
\end{equation}
The quadratic-form bound applies to both signs of the Hermitian hopping operator. Compressing it between $\Pi$ bounds the operator norm by the retained total occupation. This does not require $H_{xy}^0$ to preserve the individual block cutoffs.

\begin{figure}[t]
\centering
\includegraphics[width=0.98\linewidth]{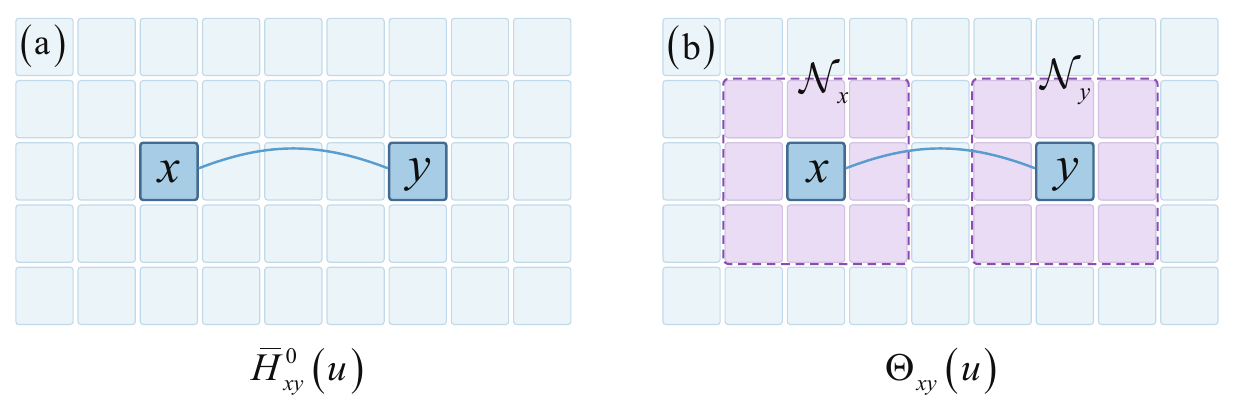}
\caption{Support enlargement under the commuting density evolution on the retained block spaces. (a) The two darker blue blocks support the hopping term $\bar H_{xy}^{0}(u)$. (b) Its interaction-picture form $\Theta_{xy}(u)$ is supported within $\mathcal N_x\cup\mathcal N_y$. The lilac regions indicate the one-layer neighborhoods. The connecting line denotes the hopping coupling, not the support on intervening blocks.}
\label{fig:SM_ip_support}
\end{figure}

For a far pair let $m=d_{\mathrm{cg}}(P_x,P_y)\ge2$. The fixed block geometry gives
\begin{equation}
 |i-j|\ge d(P_x,P_y)
 \ge Lm-2\ell=L(m-1)+1\ge\frac{Lm}{2},
 \quad i\in P_x,\quad j\in P_y.
\end{equation}
Let $J^{\mathrm h}_{xy}$ be the rectangular matrix with indices in $P_x\times P_y$. The Schur test implies
\begin{align}
 \|J^{\mathrm h}_{xy}(u)\|
 \le\left[
    \left(\sup_{i\in P_x}\sum_{j\in P_y}|J^{\mathrm h}_{ij}(u)|\right)
    \left(\sup_{j\in P_y}\sum_{i\in P_x}|J^{\mathrm h}_{ij}(u)|\right)
        \right]^{1/2}
 \le C\sqrt{|P_x||P_y|}(Lm)^{-\alpha}
 \le CL^{d-\alpha}m^{-\alpha}.
\end{align}
The one-particle matrix for $H_{xy}^0$ is
\begin{equation}
 K_{xy}(u)=
 \begin{pmatrix}0&J^{\mathrm h}_{xy}(u)\\
                 J^{\mathrm h}_{xy}(u)^\dagger&0\end{pmatrix},
 \quad \|K_{xy}(u)\|=\|J^{\mathrm h}_{xy}(u)\|.
\end{equation}
On its $N$-particle sector, the second quantization is the restriction of $\sum_{j=1}^NK_{xy}^{(j)}$ and has norm at most $N\|K_{xy}\|$. The retained pair has at most $q_x+q_y\le2Q$ particles. Another compression cannot increase this norm, so
\begin{equation}
 \|\bar H_{xy}^0(u)\|
 \le C Q L^{d-\alpha}m^{-\alpha},
 \quad m\ge2.
\label{eq:SM_ip_long_hopping_bound}
\end{equation}
Combining the near and far estimates with coarse shell counting gives the row bound
\begin{align}
 \|\bar H_x^0(u)\|+\sum_{y\ne x}\|\bar H_{xy}^0(u)\|
 &\le CQ\left[1+L^{d-\alpha}
                    \sum_{m\ge2}m^{d-1-\alpha}\right]
 \le CQ.
\label{eq:SM_ip_hopping_block_row_bound}
\end{align}
We used $L\ge1$ and $\alpha>d$. Define the transformed hopping terms by
\begin{equation}
 \Theta_x(u)\coloneqq U^V(s,u)^\dagger\bar H_x^0(u)U^V(s,u),
 \quad
 \Theta_{xy}(u)\coloneqq U^V(s,u)^\dagger\bar H_{xy}^0(u)U^V(s,u).
\end{equation}
They obey exactly the norm estimates above. Their supports are contained in $\mathcal N_x$ and $\mathcal N_x\cup\mathcal N_y$, respectively. The on-block contribution has a coarse diameter of at most two, and a neighboring-block contribution has a diameter of at most three. A far-pair contribution has diameter at most $m+2$. In every case, its support contains at most $2\cdot3^d$ blocks. Figure~\ref{fig:SM_ip_support} illustrates the support enlargement under the commuting density evolution.

\subsubsection{Long-range density interactions}
\label{S-IV-D-density}

We first consider $p\ge2$. The case $1\le p<2$ is treated separately below. To retain the correct support of general polynomial components, split each far-block density operator as
\begin{equation}
 V_{xy}=V_{xy}^{\mathrm{mix}}
         +V_{x|y}^{\mathrm{loc}}+V_{y|x}^{\mathrm{loc}}+c_{xy}I,
\label{eq:SM_ip_density_support_decomposition}
\end{equation}
where $V_{xy}^{\mathrm{mix}}$ contains the monomials with positive powers at both microscopic endpoints. Here, $V_{x|y}^{\mathrm{loc}}$ and $V_{y|x}^{\mathrm{loc}}$ contain terms depending only on occupations in $P_x$ and $P_y$, respectively. The scalar is $c_{xy}$.
For example, this splitting can be performed on each real two-site density function $f$ by writing
\begin{align}
 f(n_i,n_j)
 =\bigl[f(n_i,n_j)-f(n_i,0)-f(0,n_j)+f(0,0)\bigr]
 +[f(n_i,0)-f(0,0)]+[f(0,n_j)-f(0,0)]+f(0,0).
\end{align}
The four terms are the mixed, first-site, second-site, and scalar contributions. This is an exact regrouping and does not change the Hamiltonian. For the representative interaction $n_i^{p/2}n_j^{p/2}$, only the mixed part is present.

A termwise Young inequality gives the preliminary estimate
\begin{align}
 \|\overline{V_{xy}^{\mathrm{mix}}}(u)\|
 \le\frac{C}{[1+d(P_x,P_y)]^\alpha}
       \left\|\Pi\sum_{i\in P_x,\,j\in P_y}
                          (n_i^p+n_j^p)\Pi\right\|
 \le\frac{C\bigl(|P_y|q_x^p+|P_x|q_y^p\bigr)}
               {[1+d(P_x,P_y)]^\alpha}.
\end{align}
This calculation retains a block-volume factor. The sharper bound follows by keeping the powers at the two endpoints together. For any $a\ge1$, joint spectral calculus gives
\begin{equation}
 \sum_{i\in P_x}n_i^a\le n_{P_x}^a.
\end{equation}
In particular, for $a=p/2$,
\begin{align}
 \left\|\Pi\sum_{i\in P_x,\,j\in P_y}
        u_{ij}(u)n_i^{p/2}n_j^{p/2}\Pi\right\|
 &\le\frac{C}{(Lm)^\alpha}
       \left\|\Pi\left(\sum_{i\in P_x}n_i^{p/2}\right)
                    \left(\sum_{j\in P_y}n_j^{p/2}\right)\Pi\right\|\notag\\
 &\le\frac{Cq_x^{p/2}q_y^{p/2}}{(Lm)^\alpha}
 \le C Q^p L^{-\alpha}m^{-\alpha},
 \quad m\ge2,
\label{eq:SM_ip_density_volume_free}
\end{align}
where $|u_{ij}(u)|\le C(1+|i-j|)^{-\alpha}$ includes the fixed coefficients and the Hermitian orientations. The first inequality uses positivity of the number functions after replacing each coefficient by its absolute value.

Every mixed polynomial monomial has integer powers $a,c\ge1$ with $a+c\le p$. The same argument gives
\begin{align}
 \left\|\Pi\sum_{i\in P_x,\,j\in P_y}
             u_{ij}^{a,c}(u)n_i^a n_j^c\Pi\right\|
 \le C(Lm)^{-\alpha}
       \|\Pi n_{P_x}^a n_{P_y}^c\Pi\|
 \le C Q^{a+c}L^{-\alpha}m^{-\alpha}
 \le C Q^p L^{-\alpha}m^{-\alpha}.
\end{align}
Summing the fixed finite collection of components yields
\begin{equation}
 \|\overline{V_{xy}^{\mathrm{mix}}}(u)\|
 \le\frac{C Q^p L^{-\alpha}}{(1+m)^\alpha},
 \quad \{x,y\}\in\mathcal P_{\mathrm{long}}.
\label{eq:SM_ip_long_density_bound}
\end{equation}
No lower bound on $|P_x|$ or $|P_y|$ has been used.

The one-block terms in Eq.~\eqref{eq:SM_ip_density_support_decomposition} must not be estimated as two-block interactions. Collect these contributions as
\begin{equation}
 D_x(u)\coloneqq\sum_{y:\{x,y\}\in\mathcal P_{\mathrm{long}}}
                          V_{x|y}^{\mathrm{loc}}(u),
 \quad
 c(u)\coloneqq\sum_{\{x,y\}\in\mathcal P_{\mathrm{long}}}c_{xy}(u).
\end{equation}
Each $D_x$ is supported on $P_x$ and remains a function of its number operators. Hence
\begin{equation}
 U^V(s,u)^\dagger\overline{V_{xy}^{\mathrm{mix}}}(u)U^V(s,u)
       =\overline{V_{xy}^{\mathrm{mix}}}(u),
 \quad
 U^V(s,u)^\dagger\bar D_x(u)U^V(s,u)=\bar D_x(u).
\end{equation}
Their support does not enlarge. Section~\ref{S-III-A} allows arbitrary one-site terms on the coarse lattice without a uniform norm bound. They can be removed by a product of one-block unitaries that preserve the supports and norms of the remaining interactions. Thus, the possibly large norms of $\bar D_x$ do not enter the pair-summed interaction scale. Scalar terms have no effect on the retained-space Heisenberg evolution.

At the cutoff eventually used in the main text, $\bar q=R^\delta T^d$, Eq.~\eqref{eq:SM_ip_block_spacing_time} implies $Q\ge c L^d$. For $p\ge2$ this gives
$QL^{d-\alpha}\le C Q^p L^{-\alpha}$.
Combining Eqs.~\eqref{eq:SM_ip_long_hopping_bound} and \eqref{eq:SM_ip_long_density_bound}, the genuine far-block couplings then satisfy
\begin{equation}
 \|\bar H_{xy}^0(u)\|
 +\|\overline{V_{xy}^{\mathrm{mix}}}(u)\|
 \le\frac{C Q^p L^{-\alpha}}{(1+m)^\alpha}.
\end{equation}
For arbitrary $\bar q>1$, the hopping estimate remains bounded by $CQ(1+m)^{-\alpha}$ because $L^{d-\alpha}\le1$.
Both forms will lead to the same maximum of the hopping and density scales below.

\subsubsection{Fixed-$k$ structure and pair-summed decay}
\label{S-IV-D-pair-decay}

On the retained space, the exact decomposition is
\begin{equation}
 H^{\mathrm{int}}(u)
 =\sum_x\Theta_x(u)+\sum_{\{x,y\}}\Theta_{xy}(u)
   +\sum_{\{x,y\}\in\mathcal P_{\mathrm{long}}}
                         \overline{V_{xy}^{\mathrm{mix}}}(u)
   +\sum_x\bar D_x(u)+c(u)I.
\label{eq:SM_ip_transformed_decomposition}
\end{equation}
The second sum includes all unordered pairs in $\mathcal C_Y$.
Assign $\Theta_x$ to the fixed support $\mathcal N_x$, $\Theta_{xy}$ to $\mathcal N_x\cup\mathcal N_y$, $\overline{V_{xy}^{\mathrm{mix}}}$ to $\{x,y\}$, and $\bar D_x$ to $\{x\}$. Terms assigned to the same set are added and denoted by $\Phi_Z^{\mathrm{int}}(u)$. Each term is Hermitian, continuous in time, and supported on at most
\begin{equation}
 k=2\cdot3^d
\end{equation}
blocks. The scalar in Eq.~\eqref{eq:SM_ip_transformed_decomposition} may be omitted from the generator when applying Sec.~\ref{S-III}.
The value of $k$ does not depend on $p$, the block radius, the time interval, or the occupation cutoff.

We next verify the pair-summed condition for $p\ge2$.
Fix distinct retained block centers $z,z'$, and put $r=d_{\mathrm{cg}}(P_z,P_{z'})$.
For $r\ge4$, the two centers cannot lie in a single $\mathcal N_x$, whose diameter is at most two.
If both lie in $\mathcal N_x\cup\mathcal N_y$, they must be assigned to different neighborhoods. Up to exchanging $x,y$,
\begin{equation}
 x\in\mathcal N_z,\quad y\in\mathcal N_{z'},\quad
 m=d_{\mathrm{cg}}(P_x,P_y)\ge r-2\ge r/2\ge2.
\end{equation}
There are at most $2(3^d)^2$ such ordered choices.
Equation~\eqref{eq:SM_ip_long_hopping_bound} now gives
\begin{equation}
 \sum_{\{x,y\}:\ z,z'\in\mathcal N_x\cup\mathcal N_y}
       \|\Theta_{xy}(u)\|
 \le\frac{C Q L^{d-\alpha}}{(1+r)^\alpha}
 \le\frac{C Q}{(1+r)^\alpha}.
\label{eq:SM_ip_distant_hopping_sum}
\end{equation}
An unchanged two-block density term can contain both centers only when its endpoints are $z,z'$. For $r\ge4$, neither the one-block terms nor the dressed on-block hopping terms can have both $z$ and $z'$ in their support.
The remaining contributions are bounded by Eqs.~\eqref{eq:SM_ip_distant_hopping_sum} and \eqref{eq:SM_ip_long_density_bound}. Applying the triangle inequality when collecting terms into $\Phi_Z^{\mathrm{int}}(u)$ gives
\begin{equation}
 \sum_{Z\ni z,z'}\|\Phi_Z^{\mathrm{int}}(u)\|
 \le\frac{C[Q+Q^p L^{-\alpha}]}{(1+r)^\alpha},
 \quad r\ge4.
\label{eq:SM_ip_distant_pair_sum}
\end{equation}

For $1\le r\le3$, bound the hopping contributions by all such terms whose enlarged support contains $z$:
\begin{align}
 \sum_{Z\ni z,z'}\|\Phi_Z^{\mathrm{int}}(u)\|
 &\le\sum_{x\in\mathcal N_z}\|\Theta_x(u)\|
    +\sum_{x\in\mathcal N_z}\sum_{y\ne x}\|\Theta_{xy}(u)\|
    +\sum_{y:\{z,y\}\in\mathcal P_{\mathrm{long}}}
                    \|\overline{V_{zy}^{\mathrm{mix}}}(u)\|\notag\\
 &\le C Q+C Q^p L^{-\alpha}
                  \sum_{m\ge2}m^{d-1-\alpha}\\
 &\le C[Q+Q^p L^{-\alpha}].
    \notag
\end{align}
The first two sums use $|\mathcal N_z|\le3^d$ and Eq.~\eqref{eq:SM_ip_hopping_block_row_bound}. The last series converges for $\alpha>d$. The one-block density terms do not contribute because $z\ne z'$. Since $1+r\le4$, this estimate has the same form as Eq.~\eqref{eq:SM_ip_distant_pair_sum}, after increasing the constant by a fixed factor.

We have proved
\begin{equation}
 \sup_u\sum_{Z\ni z,z'}\|\Phi_Z^{\mathrm{int}}(u)\|
 \le\frac{C\max\{Q,Q^p L^{-\alpha}\}}
              {[1+d_{\mathrm{cg}}(P_z,P_{z'})]^\alpha},
 \quad z\ne z',\quad p\ge2.
\label{eq:SM_ip_pair_decay_bound}
\end{equation}
All sums are bounded by coarse-lattice shell sums, independently of $Y$ and the ambient volume. The conjugations preserve norms at every time. In particular, rapid short-range density phases introduce neither a larger interaction scale nor a time-derivative assumption. The local block dimensions do not occur in any constant. This verifies precisely the fixed-$k$ and pair-summed hypotheses used in Sec.~\ref{S-III}.

\subsubsection{Effective scale and the remaining parameter range}
\label{S-IV-D-effective-scale}

For $p\ge2$, insert $\bar q=R^\delta T^d$ and $h=(\eta+1)\delta$ into Eq.~\eqref{eq:SM_ip_pair_decay_bound}. The occupation scale in Eq.~\eqref{eq:SM_ip_retained_space} is $Q=3^\delta T^dR^h$, and Eq.~\eqref{eq:SM_ip_block_spacing_time} gives
\begin{equation}
 \max\{Q,Q^p L^{-\alpha}\}
 \le C\max\{T^dR^h,T^{pd-\alpha}R^{hp}\}.
\end{equation}
Thus, the interaction strength can be chosen as
\begin{equation}
 g_*\coloneqq C_*\max\{T^dR^h,T^{pd-\alpha}R^{hp}\},
\label{eq:SM_ip_effective_strength}
\end{equation}
with a fixed $C_*$, as in the End Matter.
Up to the small cutoff powers, the hopping and long-range density scales are $T^d$ and $T^{pd-\alpha}$.
Equation~\eqref{eq:SM_ip_density_volume_free} is the step that removes the additional block-volume factor from the density contribution. At the leading power, this gives the main-text scale $g=\max\{t^d,t^{pd-\alpha}\}$ for $s=0$ and large $t$.

\textit{Density powers below two.---} The representative function $n_i^{p/2}n_j^{p/2}$ requires a different block estimate when $1\le p<2$. Put $\nu=p/2<1$. Concavity of $x\mapsto x^\nu$ gives
\begin{equation}
 \sum_{i\in P_x}n_i^\nu
 \le |P_x|^{1-\nu}n_{P_x}^\nu.
\end{equation}
This is a scalar inequality for each joint-number eigenstate, and hence an operator inequality. It follows that
\begin{align}
 \|\overline{V_{xy}^{\mathrm{mix}}}(u)\|
 \le C(Lm)^{-\alpha}
       |P_x|^{1-p/2}|P_y|^{1-p/2}q_x^{p/2}q_y^{p/2}
 \le C Q^p L^{(2-p)d-\alpha}m^{-\alpha}.
\label{eq:SM_ip_concave_density_bound}
\end{align}
Polynomial components of degree at most $p<2$ have no mixed monomials with positive integer powers at both endpoints.
Their remaining one-block terms are handled as above.
Repeating the pair sums with Eq.~\eqref{eq:SM_ip_concave_density_bound} proves, at arbitrary cutoffs,
\begin{equation}
 \sup_u\sum_{Z\ni z,z'}\|\Phi_Z^{\mathrm{int}}(u)\|
 \le\frac{C\max\{Q,Q^p L^{(2-p)d-\alpha}\}}
              {[1+d_{\mathrm{cg}}(P_z,P_{z'})]^\alpha},
 \quad 1\le p<2.
\label{eq:SM_ip_concave_pair_bound}
\end{equation}
For $p=1$, the second scale is no larger than $Q$ because $\alpha>d$. Therefore, Eq.~\eqref{eq:SM_ip_effective_strength} remains valid without an additional restriction.

For real $1<p<2$, the extra factor $L^{(2-p)d}$ in the density bound must be retained at general cutoffs. It is not removed by the inequality used for $p\ge2$.
At the final cutoff, its ratio to the hopping scale satisfies
\begin{equation}
 \frac{Q^p L^{(2-p)d-\alpha}}{Q}
 \le C T^{d-\alpha}R^{h(p-1)}.
\label{eq:SM_ip_concave_scale_ratio}
\end{equation}
In particular, the scale $g_*$ in Eq.~\eqref{eq:SM_ip_effective_strength} applies whenever $R^{h(p-1)}\le T^{\alpha-d}$, up to a fixed multiplicative constant in this condition.
For any prescribed exponent loss $\epsilon>0$ and the final radius $R=C_R T^{\beta_{\mathrm{up}}+\epsilon}$, it is enough to choose $\delta$ so that
\begin{equation}
 h(p-1)(\beta_{\mathrm{up}}+\epsilon)<\alpha-d.
\label{eq:SM_ip_concave_delta_condition}
\end{equation}
With $C_R$ fixed, Eq.~\eqref{eq:SM_ip_concave_scale_ratio} then decays as a power of $T$. The condition is compatible with the parameter order in the End Matter: choose $\omega$ and $\eta$ first, decrease $\delta$ if needed, and choose $\xi$ last. It preserves all the strict inequalities established in Sec.~\ref{S-IV-C}. In this range $pd-\alpha<0$, so the leading scale is the hopping scale and the upper-bound exponent remains unchanged. Bounds at larger requested localization radii can use the approximant constructed at the stated final radius, without increasing the cutoff again. For integer interaction order $p\ge1$, this additional noninteger case is absent.

The identity \eqref{eq:SM_ip_observable_identity}, together with the support and norm bounds \eqref{eq:SM_ip_observable_support_bound} and the fixed-$k$ interaction estimates, allows us to apply Corollary~\ref{cor:SM_klr_unified_input} to $\tau^{\mathrm{int}}_{s,t}(\bar A_V)$.
For $1<p<2$, the additional condition \eqref{eq:SM_ip_concave_delta_condition} is imposed when choosing the final radius.
In Sec.~\ref{S-IV-E}, we use these results to obtain the localization error and construct a local approximant on the full Fock space.

\subsection{Local approximation and upper-bound exponent}
\label{S-IV-E}

We complete the approximation of $\tau_{s,t}(A)$ by an operator supported on $X_R$. The construction first localizes the interaction-picture evolution on the retained block spaces. We then extend the resulting operator to the full Fock space and bound the error from removing its exterior occupation projectors. Combining this error with the estimates of Secs.~\ref{S-IV-B} and \ref{S-IV-C} gives the upper bound {for the Bose--Hubbard-type model} stated at the beginning of Sec.~\ref{S-IV}. The localization uses the conditional expectations of Sec.~\ref{S-III} and Refs.~\cite{doi:10.1063/1.5095769,LR26}.

Throughout, the density interactions are the representative term in Eq.~\eqref{eq:SM_density_potential} and the finite polynomial components of total degree at most $p$ treated in Sec.~\ref{S-IV-D}. We retain all the coefficient and Hermiticity assumptions of Sec.~\ref{S-I}. Fix a nonempty finite $X\subseteq\Lambda$, $A\in\mathcal A_X^{q_0}$, and $s,t\in\mathbb R$.
Write $T=1+|t-s|$, $L=2\ell+1$, $Y=X_{2R^\eta}$, and $h=(\eta+1)\delta$ as before. The cutoff remains
\begin{equation}
 q_x=\lfloor\bar q[1+d(P_x,X)]^\delta\rfloor,
 \quad \bar q=R^\delta T^d.
\end{equation}
The block partition and these projectors are fixed throughout each comparison. The parameter choices needed below are made explicit in Sec.~\ref{S-IV-E-parameters}.

\subsubsection{Localization on the retained block spaces}
\label{S-IV-E-retained}

Use $\mathcal S_V$ and $\bar A_V$ from Eqs.~\eqref{eq:SM_ip_observable_support_sets} and \eqref{eq:SM_ip_observable_identity}. On $\mathcal F_{\mathbf q,Y}$, set
\begin{equation}
 \begin{gathered}
 O_{\mathbf q}(s,t)\coloneqq
 \tau^{\mathrm{int}}_{s,t}(\bar A_V),\quad
 \bar\tau^Y_{s,t}(\bar A)|_{\mathcal F_{\mathbf q}}
 =O_{\mathbf q}(s,t)\otimes I_{\mathbf q,Y^c}.
 \end{gathered}
\label{eq:SM_final_retained_evolution}
\end{equation}
The second identity follows from Sec.~\ref{S-IV-D} and the retained-space tensor factorization. The bounds already proved are
\begin{equation}
 \operatorname{supp}_{\mathrm{cg}}\bar A_V\subseteq\mathcal S_V,
 \quad |\mathcal S_V|\le3^d|X|,
 \quad \|\bar A_V\|\le\|A\|.
\end{equation}
Although $\bar A_V$ depends on the fixed endpoints $s,t$, Corollary~\ref{cor:SM_klr_unified_input} is uniform over all operators with the stated support and norm. It applies to this operator for each fixed pair of times.

Assume initially that $R\ge16L$ and define
\begin{equation}
 r\coloneqq\left\lfloor\frac{R}{8L}\right\rfloor,
 \quad
 \mathcal S_r\coloneqq(\mathcal S_V)^{[r]}
                   \subseteq\mathcal C_Y,
 \quad K_r\coloneqq\bigcup_{x\in\mathcal S_r}P_x.
\end{equation}
Then $r\ge1$ and
\begin{equation}
 \frac{R}{16L}\le r\le\frac{R}{8L}.
\label{eq:SM_final_coarse_radius}
\end{equation}
We check that the resulting physical support lies inside $X_R$.
For $y\in\mathcal S_r$, choose $z\in\mathcal S_V$ with $d_{\mathrm{cg}}(P_y,P_z)\le r$ and $x\in\mathcal S_0$ with $d_{\mathrm{cg}}(P_z,P_x)\le1$.
For $i\in P_x\cap X$ and $j\in P_y$,
\begin{align}
 |i-j|
 &\le |j-y|+|y-z|+|z-x|+|x-i|\notag\\
 &\le 2\ell+Lr+L
 < Lr+2L
 \le\frac{R}{4}.
\end{align}
Since $K_r$ is a union of whole blocks, this proves $K_r\subseteq X_R\subseteq Y$.
Only upper bounds on block diameters are used, so incomplete boundary blocks cause no change.

Let $\mathbb E_{\mathcal S_r}^{Y}$ be the conditional expectation on $\mathcal B(\mathcal F_{\mathbf q,Y})$ onto the algebra supported on $\mathcal S_r$. Explicitly,
\begin{equation}
 \mathbb E_{\mathcal S_r}^{Y}(O)
 \coloneqq
 \frac{\operatorname{Tr}_{\mathcal F_{\mathbf q,Y\setminus K_r}}O}
      {\dim\mathcal F_{\mathbf q,Y\setminus K_r}}
 \otimes I_{\mathbf q,Y\setminus K_r}.
\label{eq:SM_final_conditional_expectation}
\end{equation}
The empty tensor factor has dimension one. This is the map of Eq.~\eqref{eq:SM_klr_conditional_expectation}, now on the retained block spaces. It is a contraction in operator norm. Define
\begin{equation}
 \widetilde A_{\mathbf q}
 \coloneqq\mathbb E_{\mathcal S_r}^{Y}(O_{\mathbf q}(s,t)),
 \quad
 \|\widetilde A_{\mathbf q}\|\le\|A\|.
    \notag
\end{equation}
The dimension in Eq.~\eqref{eq:SM_final_conditional_expectation} normalizes the partial trace. It does not produce a prefactor in the contractivity or localization estimates.

First, consider $p\ge2$ or $p=1$. Section~\ref{S-IV-D} supplies the interaction scale
\begin{equation}
 g_*=C_*\max\{T^dR^h,T^{pd-\alpha}R^{hp}\},
 \quad C_*\ge1.
\end{equation}
For real $1<p<2$, we use this scale only after imposing the additional condition specified below. Fix the exponents of Corollary~\ref{cor:SM_klr_unified_input}, with $\gamma_2=\omega\gamma_1>d$. If
\begin{equation}
 r\ge C_r(g_*T)^{1/\omega},
\label{eq:SM_final_LR_radius_condition}
\end{equation}
then $g_*T\ge1$ and the corollary gives
\begin{align}
 \|O_{\mathbf q}(s,t)-\widetilde A_{\mathbf q}\|
 \le C|\mathcal S_V|\|\bar A_V\|
               \frac{(g_*T)^{\gamma_1}}{r^{\gamma_2}}
 \le C\|A\||X|
               \frac{(g_*T)^{\gamma_1}}{(R/T)^{\gamma_2}}
 \le C\|A\|E_{\mathrm{loc}},
 \quad
 E_{\mathrm{loc}}\coloneqq
 |X|^2\frac{(g_*T)^{\gamma_1}}{(R/T)^{\gamma_2}}.
\end{align}
The second line uses $L\le C_vT$ and Eq.~\eqref{eq:SM_final_coarse_radius}.
The last line uses $|X|\ge1$ and keeps the support prefactor stated in the End Matter. The estimates here actually give a linear support prefactor at this step.
All constants are independent of the retained local dimensions.

For $\alpha>2d+1$, the stronger estimate in Eq.~\eqref{eq:SM_klr_linear_bound} is available. Increasing the fixed radius prefactor ensures $r\ge2v_{\mathrm{LR}}g_*T$, so that the same approximant satisfies
\begin{equation}
 \|O_{\mathbf q}(s,t)-\widetilde A_{\mathbf q}\|
 \le C\|A\|E_{\mathrm{loc}}^{\mathrm{lin}},
 \quad
 E_{\mathrm{loc}}^{\mathrm{lin}}
 \coloneqq |X|^2\frac{(g_*T)^{d+1}}{(R/T)^{\alpha-d}}.
\end{equation}
Here the second term of Eq.~\eqref{eq:SM_klr_linear_bound} is absorbed because $g_*T\ge1$. The radius condition follows from Eq.~\eqref{eq:SM_final_LR_radius_condition}, with an enlarged fixed constant, since $\omega<1$. For $1<p<2$, we first use $g_{<2}$ and then the absorption established in Eq.~\eqref{eq:SM_final_low_power_absorption} below.

The commuting density evolution is already included in $\bar A_V$. Its one-layer enlargement was included in $\mathcal S_V$ before choosing $r$.
No further conjugation by $U^V$ is applied to $\widetilde A_{\mathbf q}$, because the exact order of the Heisenberg maps is the one in Eq.~\eqref{eq:SM_ip_observable_identity}.

\subsubsection{Extension to full Fock space and exterior projectors}
\label{S-IV-E-extension}

Write
\begin{equation}
 \widetilde A_{\mathbf q}
 =a_{\mathbf q,r}\otimes I_{\mathbf q,Y\setminus K_r},
 \quad a_{\mathbf q,r}\in\mathcal B(\mathcal F_{\mathbf q,K_r}).
\end{equation}
On $\mathcal F_{\mathbf q,X_R}$ define
\begin{equation}
 a_{\mathbf q,R}
 \coloneqq a_{\mathbf q,r}\otimes I_{\mathbf q,X_R\setminus K_r}.
\end{equation}
Let $\iota_R:\mathcal F_{\mathbf q,X_R}\to\mathcal F(X_R)$ be the natural isometric inclusion. Define the full-space operator
\begin{equation}
 \widetilde A\coloneqq
 \bigl(\iota_R a_{\mathbf q,R}\iota_R^\dagger\bigr)
       \otimes I_{\mathcal F(X_R^c)}.
\end{equation}
The interior factor is extended by zero on the orthogonal complement of $\mathcal F_{\mathbf q,X_R}$ in $\mathcal F(X_R)$.
The exterior factor is the identity on the full exterior Fock space, not an occupation projector. Hence
\begin{equation}
 \widetilde A\in\mathcal A_{X_R},
 \quad \|\widetilde A\|\le\|A\|,
 \quad \Pi_{X_R}\widetilde A\Pi_{X_R}=\widetilde A,
\label{eq:SM_final_full_space_properties}
\end{equation}
where, for any block union $Z$, $\Pi_Z\coloneqq\prod_{x\in\mathcal C_Z}\Pi_x$.
If $A$ is Hermitian, so is $\widetilde A$.
The local approximant $\widetilde A$ is independent of the particular initial state $\rho\in\mathcal D_\xi^\lambda$.

On the globally retained space, its compression satisfies
\begin{equation}
 (\Pi\widetilde A\Pi)|_{\mathcal F_{\mathbf q}}
 =\widetilde A_{\mathbf q}\otimes I_{\mathbf q,Y^c}.
\label{eq:SM_final_compressed_extension}
\end{equation}
Both $\bar\tau^Y_{s,t}(\bar A)$ and $\Pi\widetilde A\Pi$ vanish on $\mathcal F_{\mathbf q}^{\perp}$. Their difference has full operator norm equal to the norm on $\mathcal F_{\mathbf q}$.
Equations~\eqref{eq:SM_final_retained_evolution} and \eqref{eq:SM_final_compressed_extension} give
\begin{align}
 \|\bar\tau^Y_{s,t}(\bar A)-\Pi\widetilde A\Pi\|
 =\|O_{\mathbf q}(s,t)-\widetilde A_{\mathbf q}\|
 \le C\|A\|E_{\mathrm{loc}}.
\label{eq:SM_final_compressed_local_error}
\end{align}
This is an operator-norm comparison with the compressed local operator. Removing its exterior projectors requires a separate state-dependent estimate.

The factorization $\Pi=\Pi_{X_R}\Pi_{X_R^c}$ and locality of $\widetilde A$ imply
\begin{equation}
 \Pi\widetilde A\Pi=\widetilde A\Pi_{X_R^c},
 \quad
 \Pi\widetilde A\Pi-\widetilde A
 =-\widetilde A(I-\Pi_{X_R^c}).
\end{equation}
Put $P=I-\Pi_{X_R^c}$. Since $[P,\widetilde A]=0$, Schatten H\"older and Cauchy--Schwarz inequalities give
\begin{align}
 \|\rho(\Pi\widetilde A\Pi-\widetilde A)\|_1
 &=\|\rho P\widetilde A\|_1
 \le\|A\|\|\rho P\|_1\notag\\
 &\le\|A\|\|\rho^{1/2}\|_2\|\rho^{1/2}P\|_2
 =\|A\|[\operatorname{Tr}(\rho P)]^{1/2}.
\end{align}
This argument does not require $\rho$ to commute with $P$.

We bound the last trace at the initial time.
For a nonempty block $P_x$, choose $i_x\in P_x$ and set $a_B=\max\{1,2\ell\}$. Then $P_x\subseteq B_{a_B}(i_x)$, $a_B\le C_vT$, and the controlled-density assumption gives
\begin{equation}
 \operatorname{Tr}(\rho n_{P_x}^{\xi})
 \le (\lambda\xi a_B^d)^\xi\le C_\xi T^{d\xi}.
\end{equation}
The commuting-projector union bound, Markov's inequality, and Eq.~\eqref{eq:SM_cutoff_geometric_sums} yield
\begin{align}
 \operatorname{Tr}[\rho(I-\Pi_{X_R^c})]
 &\le\sum_{x\in\mathcal C_{X_R^c}}
                \operatorname{Tr}(\rho\Pi_{P_x,>q_x})
 \le\sum_{x\in\mathcal C_{X_R^c}}
                 q_x^{-\xi}\operatorname{Tr}(\rho n_{P_x}^\xi)\notag\\
 &\le C_\xi\frac{T^{d\xi}}{\bar q^\xi}
          \sum_{x\in\mathcal C}[1+d(P_x,X)]^{-\delta\xi}
 \le C|X|\frac{T^{d\xi}}{\bar q^\xi},
 \quad \delta\xi>d.
\end{align}
Consequently,
\begin{equation}
 \|\rho(\Pi\widetilde A\Pi-\widetilde A)\|_1
 \le C\|A\||X|^{1/2}
       \frac{T^{d\xi/2}}{\bar q^{\xi/2}}
 =C\|A\||X|^{1/2}R^{-\delta\xi/2}.
\label{eq:SM_final_exterior_removal_error}
\end{equation}
The square-root support factor is appropriate for this exterior-projector error. It is distinct from the larger support factor in the dynamical-truncation estimate of Sec.~\ref{S-IV-B}.

\subsubsection{Combination of the approximation errors}
\label{S-IV-E-errors}

For any bounded $D$, $\|\rho D\|_1\le\|D\|$ because $\|\rho\|_1=1$. Insert the intermediate operators in the main-text approximation chain and retain the exterior-projector step explicitly:
\begin{align}
 \|\rho[\tau_{s,t}(A)-\widetilde A]\|_1
 &\le\|\rho\,\tau_{s,t}(A-\bar A)\|_1
 +\|\rho[\tau_{s,t}(\bar A)-\bar\tau_{s,t}(\bar A)]\|_1\notag\\
 &+\|\bar\tau_{s,t}(\bar A)-\bar\tau^Y_{s,t}(\bar A)\|
 +\|\bar\tau^Y_{s,t}(\bar A)-\Pi\widetilde A\Pi\|
 +\|\rho(\Pi\widetilde A\Pi-\widetilde A)\|_1.
\end{align}
Use Eqs.~\eqref{eq:SM_cutoff_final_observable_error},
\eqref{eq:SM_cutoff_final_dynamics_error},
\eqref{eq:SM_spatial_restriction_main_error},
\eqref{eq:SM_final_compressed_local_error}, and
\eqref{eq:SM_final_exterior_removal_error}.
Since $|X|^{1/2}\le|X|^{(1+\xi)/2}$, the exterior error is absorbed into the first occupation error. We obtain
\begin{align}
 \|\rho[\tau_{s,t}(A)-\widetilde A]\|_1
 \le C\|A\|\bigl[|X|^{(1+\xi)/2}R^{-\delta\xi/2}
 +|X|^{(1+\xi)/2}T^{d+1+\delta}
                    R^{-\delta(\xi-3)/2}
 +E_{\mathrm{far}}+E_{\mathrm{loc}}\bigr].
\label{eq:SM_final_combined_bound}
\end{align}
Here, exactly as in the End Matter,
\begin{equation}
 E_{\mathrm{far}}
 =|X|\left[
 \frac{T^{a_0}}{R^{\eta(\alpha-2d)-hp}}
 +\frac{T^{a_1}}{R^{\eta(\gamma_2-d)-hp(\gamma_1+1)}}
 \right],
\label{eq:SM_final_far_error}
\end{equation}
with $a_0=pd+d+1$ and $a_1=pd+3d+1+(pd+1)\gamma_1$.

For the localization term, the definition of $g_*$ and $\gamma_2=\omega\gamma_1$ gives
\begin{align}
 E_{\mathrm{loc}}
 &\le C|X|^2\left[
 T^{(d+1)\gamma_1+\gamma_2}R^{-\gamma_2+h\gamma_1}
 +T^{(pd-\alpha+1)\gamma_1+\gamma_2}
                           R^{-\gamma_2+hp\gamma_1}
 \right]\notag\\
 &=C|X|^2\left[
 \frac{T^{\gamma_1(\omega+d+1)}}
      {R^{\gamma_1(\omega-h)}}
 +\frac{T^{\gamma_1(\omega+pd-\alpha+1)}}
      {R^{\gamma_1(\omega-hp)}}
 \right].
\label{eq:SM_final_local_error_expanded}
\end{align}
The constant in the last line may be increased from that in the preceding inequality. Equations~\eqref{eq:SM_final_combined_bound}--
\eqref{eq:SM_final_local_error_expanded} separate occupation, spatial-restriction, and bounded-interaction localization errors.
No constant grows with the ambient volume or the truncated local dimensions.
For $\alpha>2d+1$, we replace $E_{\mathrm{loc}}$ by $E_{\mathrm{loc}}^{\mathrm{lin}}$. Writing $M\coloneqq\max\{d+1,pd-\alpha+1\}$, we have $g_*T\le CT^M R^{hp}$ and hence
\begin{equation}
 E_{\mathrm{loc}}^{\mathrm{lin}}
 \le C|X|^2
 \frac{T^{(d+1)M+\alpha-d}}
      {R^{\alpha-d-hp(d+1)}}.
 \label{eq:SM_final_linear_error_expanded}
\end{equation}

\subsubsection{Localization radius and parameter choices}
\label{S-IV-E-parameters}

We first recover the finite-cutoff exponent of the End Matter in the range where $g_*$ applies. Assume $\omega-hp>0$. Since $p\ge1$, this also gives $\omega-h>0$.
Equations~\eqref{eq:SM_final_coarse_radius} and \eqref{eq:SM_ip_block_spacing_time} show that Eq.~\eqref{eq:SM_final_LR_radius_condition} follows from
\begin{equation}
 R^\omega\ge C T^\omega g_*T.
\end{equation}
Using the two terms in $g_*$, it is enough to require
\begin{equation}
 R^{\omega-h}\ge CT^{\omega+d+1},
 \quad
 R^{\omega-hp}\ge CT^{\omega+pd-\alpha+1}.
\end{equation}
Both conditions hold for a sufficiently large fixed prefactor in $R\ge CT^{\beta(\omega,\delta)}$, where
\begin{equation}
 \beta(\omega,\delta)
 \coloneqq\max\left\{
 \frac{\omega+d+1}{\omega-h},
 \frac{\omega+pd-\alpha+1}{\omega-hp}
 \right\},
 \quad h=(\eta+1)\delta.
\label{eq:SM_final_finite_moment_exponent}
\end{equation}
This is the exponent in the main text.
A radius at the threshold need not cause the localization error to decay over time. Decay follows from a strictly larger power, as we now verify together with the other errors.

Fix $\epsilon>0$, write $M\coloneqq\max\{d+1,pd-\alpha+1\}$, and prescribe any decay exponent
\begin{equation}
 0<\nu<\nu_\star(\epsilon),
 \qquad
 \nu_\star(\epsilon)
 \coloneqq (\alpha-d)\epsilon+\max\{\alpha-2d-1,0\}M.
\end{equation}
Put
\begin{equation}
 \omega_*\coloneqq\min\{1,\alpha-2d\},
 \quad
 \beta_\epsilon\coloneqq\beta_{\mathrm{up}}+\epsilon,
 \quad
 \beta_{\mathrm{up}}
 =1+\frac{\max\{d+1,pd-\alpha+1\}}{\omega_*}.
\end{equation}
In particular, $\beta_\epsilon>d+2$.
Choose the parameters in the following order.

\textit{Bounded-interaction exponent.---} Choose $\omega<\omega_*$ sufficiently close to $\omega_*$ that Corollary~\ref{cor:SM_klr_unified_input} applies and
\begin{equation}
 \beta(\omega,0)
 =1+\frac{\max\{d+1,pd-\alpha+1\}}{\omega}
 <\beta_{\mathrm{up}}+\frac{\epsilon}{2}.
\end{equation}
If $2d<\alpha\le2d+1$, choose $\omega$ close enough to its endpoint that, in addition,
\begin{equation}
 \gamma_1[(\beta_\epsilon-1)\omega-M]>\nu.
\end{equation}
This is possible by Eq.~\eqref{eq:SM_klr_unified_two_term_bound}, since
\begin{equation}
 \lim_{\omega\uparrow\omega_*}
 \gamma_1[(\beta_\epsilon-1)\omega-M]
 =(\alpha-d)\epsilon=\nu_\star(\epsilon)
 \qquad (2d<\alpha\le2d+1).
\end{equation}
Fix the corresponding $\gamma_1,\gamma_2$.
They are finite for fixed $\omega$ and satisfy $\gamma_2=\omega\gamma_1>d$.

\textit{Auxiliary spatial region.---} Choose $\eta>1$ as in Sec.~\ref{S-IV-C-conclusion}, with
\begin{equation}
 (d+2)\eta(\alpha-2d)>a_0+\nu,
 \quad
 (d+2)\eta(\gamma_2-d)>a_1+\nu,
 \quad
 (d+2)\eta\omega>pd+1.
\end{equation}
The positive quantities $\alpha-2d$, $\gamma_2-d$, and $\omega$ make a finite choice possible.

\textit{Cutoff-growth exponent.---}
Choose $0<\delta<1$ sufficiently small that, with
$h=(\eta+1)\delta$,
\begin{gather}
 hp<\omega,
 \quad \delta p<\alpha-2d,
 \quad \delta p(\gamma_1+1)<\gamma_2-d,
 \quad \beta(\omega,\delta)<\beta_\epsilon,
\\
 (d+2)[\eta(\alpha-2d)-hp]>a_0+\nu,
\label{eq:SM_final_delta_far_zero}\\
 (d+2)[\eta(\gamma_2-d)-hp(\gamma_1+1)]>a_1+\nu,
\label{eq:SM_final_delta_far_one}\\
 (d+2)(\eta\omega-hp)>pd+1.
\label{eq:SM_final_delta_large_ball}
\end{gather}
Require also the corresponding localization margin:
\begin{gather}
 \gamma_1[(\beta_\epsilon-1)\omega-M-\beta_\epsilon hp]>\nu
 \qquad (2d<\alpha\le2d+1),\\
 \nu_\star(\epsilon)-\beta_\epsilon hp(d+1)>\nu
 \qquad (\alpha>2d+1).
 \label{eq:SM_final_delta_target_decay}
\end{gather}
All these inequalities are strict at $\delta=0$ after the choices above. Continuity gives a common positive choice of $\delta$. If $1<p<2$, impose in addition
\begin{equation}
 h(p-1)\beta_\epsilon<\alpha-d.
\end{equation}
This is Eq.~\eqref{eq:SM_ip_concave_delta_condition} and is also satisfied by decreasing $\delta$.

We make the latter case explicit before choosing the radius prefactor. Equation~\eqref{eq:SM_ip_concave_pair_bound} gives, for all cutoffs in this range, the admissible strength
\begin{equation}
 g_{<2}=C_0\max\{T^dR^h,T^{2d-\alpha}R^{hp}\},
 \quad 1<p<2,
\end{equation}
with $C_0$ independent of $R,T$. Define the positive margins
\begin{equation}
 m_{\mathrm{hop}}=\beta_\epsilon(\omega-h)-(\omega+d+1)>0,
 \quad
 m_{\mathrm{den}}=\beta_\epsilon(\omega-hp)
                 -(\omega+pd-\alpha+1)>0.
\label{eq:SM_final_localization_margins}
\end{equation}
Their positivity follows from $\beta(\omega,\delta)<\beta_\epsilon$. For $2d<\alpha\le2d+1$, Eq.~\eqref{eq:SM_final_delta_target_decay} gives the stronger bound
\begin{equation}
 \gamma_1\min\{m_{\mathrm{hop}},m_{\mathrm{den}}\}
 \ge\gamma_1[(\beta_\epsilon-1)\omega-M-\beta_\epsilon hp]
 >\nu.
\end{equation}
For the second term of $g_{<2}$, the corresponding margin is
\begin{align}
 m_{<2}
 \coloneqq\beta_\epsilon(\omega-hp)
              -(\omega+2d-\alpha+1)
 =m_{\mathrm{hop}}+\alpha-d-\beta_\epsilon h(p-1)
 >m_{\mathrm{hop}}>0.
\label{eq:SM_final_low_power_margin}
\end{align}
At $R=C_RT^{\beta_\epsilon}$, the two sufficient coarse-radius conditions for $g_{<2}$ reduce to
\begin{equation}
 C_R^{\omega-h}T^{m_{\mathrm{hop}}}\ge C,
 \quad
 C_R^{\omega-hp}T^{m_{<2}}\ge C.
    \notag
\end{equation}
The constants on the right are independent of $C_R$.
Since the powers of $C_R$ and $T$ on the left are positive, a fixed sufficiently large $C_R$ enforces them for all $T\ge1$.
The resulting localization error is bounded by $C|X|^2(T^{-\gamma_1m_{\mathrm{hop}}}
       +T^{-\gamma_1m_{<2}})$. Also,
\begin{equation}
 T^{d-\alpha}R^{h(p-1)}
 =C_R^{h(p-1)}
  T^{-[\alpha-d-\beta_\epsilon h(p-1)]}
 \le C_R^{h(p-1)}.
\label{eq:SM_final_low_power_absorption}
\end{equation}
Thus, once $C_R$ is fixed, $g_{<2}$ is bounded by a fixed multiple of the hopping term in $g_*$ along this final-radius curve. The notation $E_{\mathrm{loc}}$ in the End Matter remains valid with an enlarged fixed constant. The direct use of $g_{<2}$ above ensures that selecting $C_R$ does not presuppose a constant already depending on $C_R$.
No claim is made that the $p\ge2$ volume-free density estimate holds for $1<p<2$ at arbitrary cutoffs.

\textit{Moment order and radius prefactor.---} Choose a finite threshold
\begin{equation}
 \xi_0(\epsilon,\nu)>3+\frac{2(d+2+\nu)}{\delta}
\label{eq:SM_final_moment_threshold}
\end{equation}
and fix any finite $\xi\ge\xi_0(\epsilon,\nu)$. Then $\delta\xi>d$ and $\delta(\xi-3)>2d$, as required in Sec.~\ref{S-IV-B}. Dependence of $\xi_0$ on the fixed model parameters is suppressed. Finally choose $C_R>1$ sufficiently large and set
\begin{equation}
 R=C_RT^{\beta_\epsilon}.
\label{eq:SM_final_radius_choice}
\end{equation}
For $p\ge2$ or $p=1$, the radius conditions are enforced by $C_R^{\omega-h}\ge C$ and $C_R^{\omega-hp}\ge C$, using Eq.~\eqref{eq:SM_final_localization_margins}.
For $1<p<2$, use Eq.~\eqref{eq:SM_final_low_power_margin} as above. Increase $C_R$ further to ensure $R\ge16L$ and $R^\eta\ge C_{\mathrm{geom}}L$.
The spatial-restriction conditions hold because $\beta_\epsilon>d+2$ and Eqs.~\eqref{eq:SM_final_delta_far_zero}--\eqref{eq:SM_final_delta_large_ball} hold.
There are only finitely many constant inequalities to satisfy, and all relevant powers of $C_R$ are positive. None of these choices depends on the volume or on the
particular state.

\subsubsection{Final estimate and leading light-cone exponent}
\label{S-IV-E-final}

At the radius in Eq.~\eqref{eq:SM_final_radius_choice}, define
\begin{align}
 \nu_A&=\frac{\beta_\epsilon\delta\xi}{2},
 \nu_{\mathrm{dyn}}=
       \frac{\beta_\epsilon\delta(\xi-3)}2-(d+1+\delta),
 \nu_{\mathrm{far},0}=
       \beta_\epsilon[\eta(\alpha-2d)-hp]-a_0,\\
 \nu_{\mathrm{far},1}&=
       \beta_\epsilon[\eta(\gamma_2-d)-hp(\gamma_1+1)]-a_1,
 \nu_{\mathrm{hop}}=\gamma_1m_{\mathrm{hop}},
 \nu_{\mathrm{den}}=\gamma_1m_{\mathrm{den}}.
\end{align}
The relevant localization exponent is
\begin{equation}
 \nu_{\mathrm{loc}}\coloneqq
 \begin{cases}
 \min\{\nu_{\mathrm{hop}},\nu_{\mathrm{den}}\},
      &2d<\alpha\le2d+1,\\
 \nu_\star(\epsilon)-\beta_\epsilon hp(d+1),
      &\alpha>2d+1.
 \end{cases}
\end{equation}
The second branch follows from Eq.~\eqref{eq:SM_final_linear_error_expanded}, since $\beta_{\mathrm{up}}=1+M$ in that regime.
Each exponent entering the final error bound exceeds the prescribed $\nu$.
For the occupation errors, Eq.~\eqref{eq:SM_final_moment_threshold} gives $\delta(\xi-3)/2>d+2+\nu$; using $\beta_\epsilon>1$ and $\delta<1$ proves $\nu_A,\nu_{\mathrm{dyn}}>\nu$.
The spatial errors exceed $\nu$ by Eqs.~\eqref{eq:SM_final_delta_far_zero} and \eqref{eq:SM_final_delta_far_one}, since $\beta_\epsilon>d+2$.
The localization exponent exceeds $\nu$ by Eq.~\eqref{eq:SM_final_delta_target_decay}.
For $1<p<2$, Eq.~\eqref{eq:SM_final_low_power_margin} controls the direct radius conditions, and Eq.~\eqref{eq:SM_final_low_power_absorption} justifies the same localization estimates after $C_R$ is fixed.
Consequently,
\begin{equation}
 \min\{\nu_A,\nu_{\mathrm{dyn}},\nu_{\mathrm{far},0},
        \nu_{\mathrm{far},1},\nu_{\mathrm{loc}}\}>\nu>0.
\end{equation}
The powers of the fixed prefactor $C_R$ are absorbed into $C$.
Since $|X|\le|X|^{(1+\xi)/2}$, the combined estimate becomes
\begin{equation}
 \sup_{\rho\in\mathcal D_\xi^\lambda}
 \|\rho[\tau_{s,t}(A)-\widetilde A]\|_1
 \le C\|A\|F_\xi(X)T^{-\nu},
 \quad F_\xi(X)=|X|^{(1+\xi)/2}+|X|^2.
\label{eq:SM_final_uniform_upper_bound}
\end{equation}
The same $\widetilde A$ is used for every state in this supremum.
Equations~\eqref{eq:SM_final_full_space_properties} and \eqref{eq:SM_final_uniform_upper_bound} prove the upper-bound statement {for the Bose--Hubbard-type model} in Sec.~\ref{S-IV}, including all $s,t\in\mathbb R$.
For any bounded probe $B$, trace-norm duality gives
\begin{equation}
 \left|\operatorname{Tr}\!\left[
 \rho[\tau_{s,t}(A)-\widetilde A]B\right]\right|
 \le C\|A\|\|B\|F_\xi(X)T^{-\nu}.
\end{equation}
Neither $B$ nor $\rho$ requires a support restriction.
The locality and the norm bound concern $\widetilde A$ on the full Fock space.

The finite-cutoff expression \eqref{eq:SM_final_finite_moment_exponent} has the limit
\begin{equation}
 \lim_{\omega\uparrow\omega_*}
 \lim_{\delta\downarrow0}\beta(\omega,\delta)
 =1+\frac{\max\{d+1,pd-\alpha+1\}}{\omega_*}
 =\beta_{\mathrm{up}}.
\end{equation}
For each fixed $\omega$, $\eta$ is selected before sending $\delta$ toward zero. This limit identifies the leading power, not a zero-loss bound at a fixed finite moment order. The regimes are
\begin{equation}
 \beta_{\mathrm{up}}=
 \begin{cases}
 \displaystyle 1+\frac{\max\{d+1,pd-\alpha+1\}}{\alpha-2d},
       &2d<\alpha<2d+1,\\[6pt]
 \max\{d+2,pd-\alpha+2\},&\alpha\ge2d+1.
 \end{cases}
\end{equation}
At $\alpha=2d+1$, the result retains the arbitrarily small positive loss prescribed above. No strictly zero-loss endpoint estimate is asserted. For $\alpha>2d+1$, the displayed leading power is $\max\{d+2,pd-\alpha+2\}$, including the absence of an additional block-volume factor in the density branch.

For any requested radius $R'\ge C_RT^{\beta_\epsilon}$, the same operator constructed at $R=C_RT^{\beta_\epsilon}$ belongs to $\mathcal A_{X_{R'}}$ and satisfies the same error bound. One need not increase the cutoffs with $R'$.

To pass from blockwise to sitewise enlargement, apply the preceding construction at radius \(C_0T^{\beta_\epsilon}\).
Since every block has diameter at most \(2\ell\), we have
\begin{equation}
    X_{C_0T^{\beta_\epsilon}}
    \subseteq
    X[C_0T^{\beta_\epsilon}+2\ell].
\end{equation}
Using \(\ell=\lceil v|t-s|\rceil\) and \(\beta_\epsilon>1\), we obtain
\begin{equation}
    C_0T^{\beta_\epsilon}+2\ell
    \le
    (C_0+2v+2)T^{\beta_\epsilon}.
\end{equation}
Thus, choosing the theorem's radius prefactor $C_R\ge C_0+2v+2$, the same approximant belongs to $\mathcal A_{X[R]}$ for every $R\ge C_RT^{\beta_\epsilon}$, with the same norm and error bounds.

\textit{Two modes per physical site.---}
{The upper bound also applies to the two-mode Bose--Hubbard-type model of Sec.~\ref{S-II-A}.}
Keep the same physical distances and blocks, and replace the number operator of a physical region by
\begin{equation}
 n_Z=\sum_{i\in Z}\sum_{a=0}^{1}n_{i,a}.
\end{equation}
The coefficient bounds are imposed on every pair of modes using the physical distance between their sites, as in that construction. Each microscopic row sum acquires at most a factor of two. Assigning equal spatial weights to the two modes at each site leaves the weighted-number estimates unchanged apart from constants. The number-sector, cutoff, and local-insertion arguments use the total occupation of a region and remain valid with this definition. The block-power estimates sum over both modes. For $p\ge2$,
\begin{equation}
 \sum_{i\in P_x}\sum_{a=0}^{1}n_{i,a}^{p/2}
 \le n_{P_x}^{p/2},
\end{equation}
and for $1\le p<2$ the concavity estimate replaces $|P_x|$ by $2|P_x|$, producing only a fixed additional factor.
The retained Hilbert-space dimension changes, but the fixed-$k$ constants do not depend on it. The same spatial exponents and the same norm-nonincreasing local construction follow, with adjusted constants independent of the volume.

{
For integer $p\ge4$, the upper and lower bounds for the Bose--Hubbard-type model match in leading power throughout $2d<\alpha\le2d+1$, as stated in Eq.~\eqref{eq:SM_matching_light_cone_exponent}.
}

\bibliography{Reference}